\documentclass{elsart}
\usepackage{amsmath}
\usepackage[pdftex]{graphicx}

\usepackage{multirow}

\newcommand{\graph}[2]{{\text{#1}\overline{\text{#2}}}}
\newcommand{\graphr}[2]{{\widehat{\text{#1}\overline{\text{#2}}}}}
\def\be{\begin{equation}}
\def\ee{\end{equation}}

\begin{document}
\begin{frontmatter}
\title{\bf 
$p_\perp$-broadening and production processes versus 
dipole/quadrupole
amplitudes at next-to-leading order
}

\author{A. H. Mueller}

\address{Department of Physics, Columbia University, New York, USA}

\author{S. Munier}

\address{Centre de physique th\'eorique, \'Ecole Polytechnique, 
CNRS, Palaiseau, France}

\begin{abstract}
Through the systematic inspection of graphs
in the framework of lightcone perturbation theory,
we demonstrate that an identity between 
the evolution of
$p_\perp$-broadening amplitudes with the energy
and the evolution of forward scattering amplitudes
of color dipoles off nuclei
holds at next-to-leading order accuracy.
In the general case, 
the relation is not a graph-by-graph correspondence, neither does it
hold strictly speaking for definite values of the 
momenta:
Instead, it relates classes of graphs 
of similar topologies,
and in some cases, the matching
requires an analytical continuation
in the appropriate longitudinal momentum variable.
We check that the same kind of
relation is also true at next-to-leading order between amplitudes
for the production of dijets
and quadrupole forward amplitudes.
\end{abstract}

\end{frontmatter}


\section{Introduction}

Observables such as dijet correlations 
\cite{Kharzeev:2004bw,Marquet:2007vb}
or other production processes
in proton-nucleus scattering
are outstanding probes of the high-density regime of QCD,
and are likely to become one of the major focusses of the
QCD community in the next few years.

Preliminary phenomenological works \cite{Albacete:2010pg} 
have shown that
saturation models have the ability to describe
successfully the present data.
In order
to confirm these results and
provide accurate predictions for the Large Hadron Collider,
firm theoretical foundations for the calculation
of these production processes are now needed.

One outstanding problem in this program is the energy 
(or equivalently, the Bjorken-$x$)
dependence
of the considered observables.
The formalism to compute the
energy evolution of forward amplitudes has reached
a certain degree of sophistication:
The evolution is known to be
given by the Balitsky-Fadin-Kuraev-Lipatov (BFKL)
equation \cite{Lipatov:1976zz,Kuraev:1977fs,Balitsky:1978ic}, whose
kernel is established at next-to-leading
order in $\ln 1/x$ \cite{Ciafaloni:1998gs,Fadin:1998py}
see e.g. \cite{Ciafaloni:2007gf,Altarelli:2008aj} for 
references and for the state of the art of the phenomenology of
deep-inelastic scattering. But
production processes have not received the same attention so
far.

Interesting relations between production 
processes and forward amplitudes
were conjectured some time ago (see e.g. Ref.~\cite{Zakharov:1996fv})
and were proved at leading order \cite{Kovchegov:2001sc} in particular
cases.
In this paper, we establish the relation 
between $p_\perp$-broadening cross sections
and dipole forward amplitudes \cite{Mueller:1993rr} 
at next-to-leading order.

To this aim, we examine essentially all possible
lightcone perturbation theory diagrams \cite{Lepage:1980fj} 
which contribute
to these two processes,
in a similar manner as was done to better understand 
the leading-order evolution of
dipoles \cite{Chen:1995pa}, and compare them.
More precisely, our work consists in the discussion of 
the corrections to the wave functions up to order
$\alpha_s^2 \ln 1/x$ and in the large number of colors ($N_c$) limit.
It is then clear that the arguments may be promoted, by
iteration, to a fully resummed next-to-leading order result,
namely to all diagrams up to order $\alpha_s (\alpha_s \ln 1/x)^k$ for
arbitrary $k$ (with the further approximation that 
quarks are left out of the evolution):
Indeed, the graphs we review actually form the kernel
of the energy evolution equation, which is known to be
the BFKL equation
on the dipole side \cite{Mueller:1993rr}.

Many graphs contribute at next-to-leading order
in lightcone perturbation theory
(although the large-$N_c$ limit reduces significantly
their number by eliminating the nonplanar ones).
This is a drawback of time-ordered perturbation theory
with respect to a covariant formalism.
However, this formalism is the simplest and the most natural one
to formulate the BFKL evolution, and thus, we believe
that it is the most adequate for our purpose.

The outline goes as follows.
Section~\ref{sec2} presents a general discussion of the relations between
production cross sections and scattering amplitudes.
We then specialize to $p_\perp$-broadening
cross sections versus dipole amplitudes: Sec.~\ref{sec:leadingorder}
is devoted to a comprehensive review of their relation when a 
leading-order quantum correction is added,
while the main new results of the paper on the
relation at next-to-leading order are presented in Sec.~\ref{sec4}.
We briefly explain how these results may go over to other
processes in Sec.~\ref{sec5}, and
conclude in Sec.~\ref{sec6}.


\section{\label{sec2}
Production cross sections and scattering amplitudes}

In this section, we shall discuss in a general way
how certain production cross sections may be related to
particular scattering amplitudes. There is of course
one well-known case where this happens. Namely, the total
cross section for the collision of two particles, $A$ and $B$,
is given by the imaginary part of the forward elastic
scattering amplitude for particles $A$ and $B$. This
is the optical theorem and is valid in any quantum field
theory. The relations that we are investigating in this paper,
between cross sections for jet $p_\perp$-broadening
and jet production at high energy with dipole
and quadrupole amplitudes, are specific to gauge theories as we shall
see a little later on. There are also well-known relationships
between inclusive particle production and
particular discontinuities of higher-particle
amplitudes. For example the cross section for 
$A+B\rightarrow C+\text{anything}$
is, in general, given by a particular discontinuity of the elastic
forward scattering amplitude $A+B+\bar C\rightarrow A+B+\bar C$.
This type of relation is true for the processes we are interested in
here, but it is not this type of relationship that we are after.
We are looking for relations between cross sections and the
values of scattering amplitudes and in general there is no
simple relationship between particular
discontinuities of scattering amplitudes and the values of
the amplitudes. To see in more detail the relationships we
are investigating, it is perhaps useful to work through a simple
example and then turn to the more general
issues.

\subsection{$p_\perp$ broadening in the McLerran-Venugopalan model}

The simplest example of the type of relationship we are investigating
is well illustrated in a formula which gives the $p_\perp$-broadening
of a quark (jet) as it passes through nuclear matter in terms of
the scattering amplitude for a quark-antiquark dipole with the same nuclear
matter. In the McLerran-Venugopalan model \cite{McLerran:1993ni} 
the nucleons in a nucleus
are uncorrelated and the gluon distribution of a nucleon
$xG(x,Q^2)$ is evaluated at lowest order in DGLAP
evolution with no $x$-evolution at all. 

Suppose a high-energy quark passes through a length $L$ of nuclear matter.
If the quark enters the nuclear matter with no transverse
momentum ($p_\perp=0$), and exits the nuclear matter with transverse
momentum $p_\perp$, then the distribution of transverse
momenta is given by
\be
\frac{dN}{d^2p_\perp}=\int \frac{d^2x_\perp}{(2\pi)^2}
e^{-i p_\perp x_\perp}S(x_\perp)
\label{eq:correspondence}
\ee
where 
\be
S(x_\perp)=\frac{1}{N_c}\text{tr}
\left\{
\sum_n\langle n|V_{0_\perp}|A\rangle^*\langle n|V_{x_\perp}|A\rangle
\right\}
\label{eq:S}
\ee
with
\be
V_{x_\perp}=T e^{ig\int_{-\infty}^{+\infty}dx_+ A_-(x_\perp,x_+)}
\label{eq:V}
\ee
The state $|A\rangle$ is the ground state of the nucleus
while $|n\rangle$ represents any final state after the quark
has passed through the nucleus. The quark is represented by the
Wilson line $V$ which we take to move along the $x_-=0$ lightcone.
$A_-$ and $V$ are color matrices with the trace in Eq.~(\ref{eq:S})
referring to the matrix indices in $V_{x_\perp}$ and $V_{0_\perp}$.
The time-ordering (path-ordering) in Eq.~(\ref{eq:V})
is such that $A$'s at later times are placed to the left of $A$'s
at earlier times.
At the moment $S(x_\perp)$ is simply defined by~(\ref{eq:S}),
however, we shall soon see that $S$ is in fact the $S$-matrix
for dipole scattering on the nucleus.

\begin{figure}
\begin{center}
\begin{tabular}{c|c|c}
\includegraphics[width=4cm]{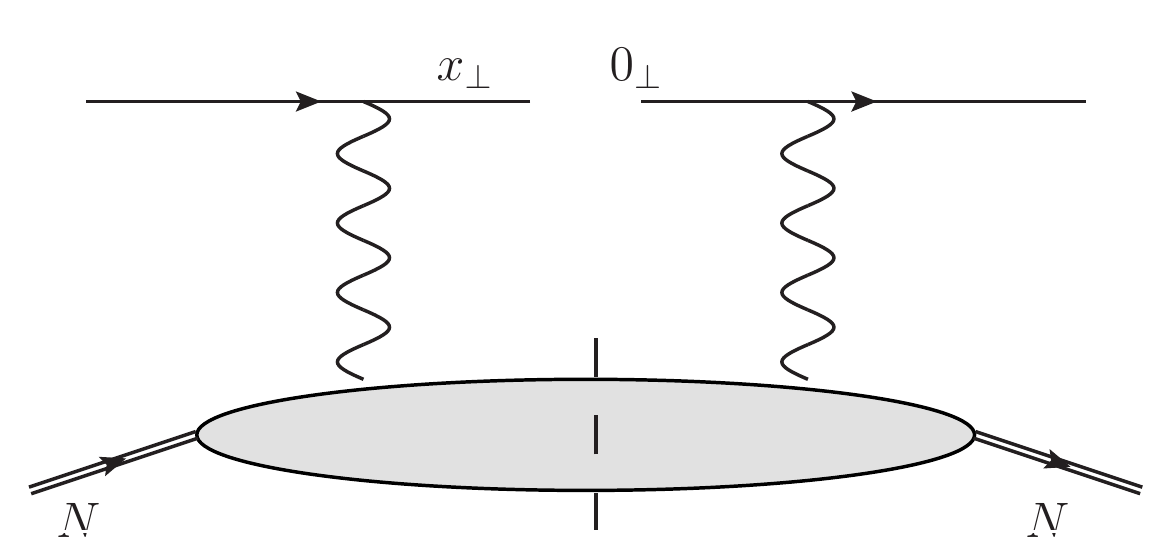}&
\includegraphics[width=4cm]{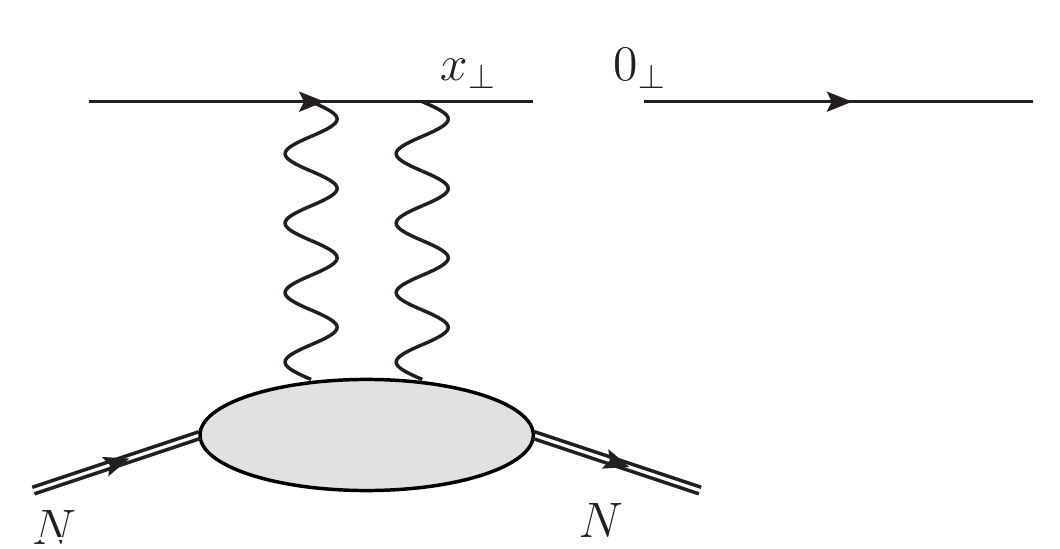}&
\includegraphics[width=4cm]{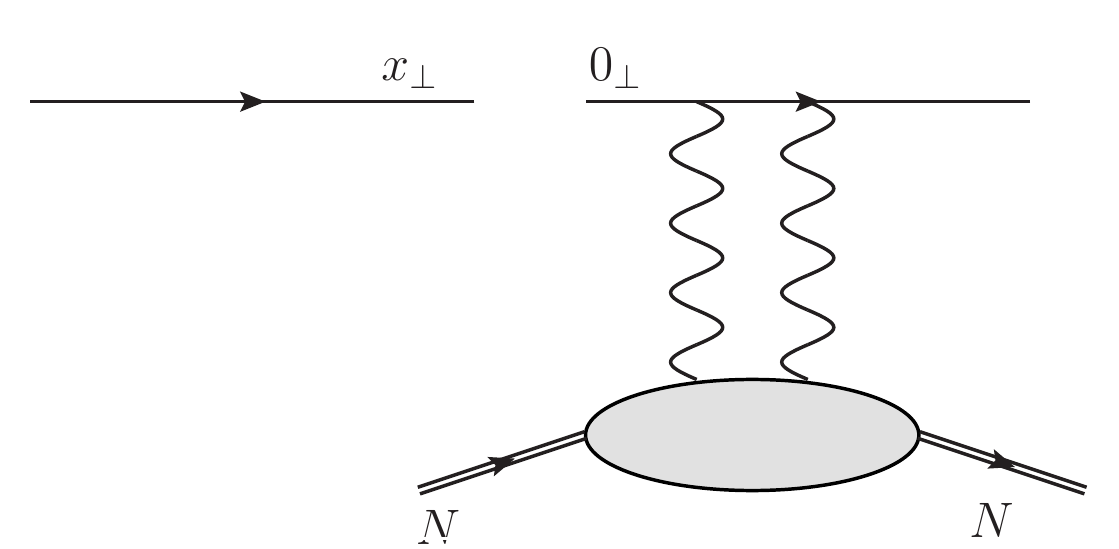}\\
$(a)$&$(b)$&$(c)$
\end{tabular}
\end{center}
\caption{\label{fig1}
The three different types of interactions of a quark and a nucleus:
inelastic (a), elastic in the amplitude (b) and 
elastic in the complex-conjugate
amplitude (c). This illustrates the 3 terms in the r.h.s. of
Eq.~(\ref{eq:S1}).
}
\end{figure}

Because the nucleons in the nucleus are uncorrelated in the
McLerran-Venugopalan model one has, for a sufficiently large
nucleus,
\be
S(x_\perp)=e^{S_1(x_\perp)}
\label{eq:SfromS1}
\ee
where $S_1(x_\perp)$ is the term in Eq.~(\ref{eq:S})
involving scattering with only one nucleon in the nucleus.
(The scattering in~(\ref{eq:S}) occurs at a definite impact
parameter although, for simplicity, we suppress the dependence.)
Expanding Eq.~(\ref{eq:S}) to order $g^2$ and using completeness of the states
$n$ one gets
\begin{multline}
S_1(x_\perp)=g^2\int dx_+ dx^\prime_+ 
\langle A|
\frac{1}{N_c}\text{tr}\bigg\{
A_-(0_\perp,x_+^\prime)A_-(x_\perp,x_+)\\
-\frac12
T(A_-(x_\perp,x_+)A_-(x_\perp,x_+^\prime))\\
-\frac12 \bar T(A_-(0_\perp,x_+)A_-(0_\perp,x_+^\prime))\bigg\}
|A\rangle
\label{eq:S1}
\end{multline}
The three terms in the brackets $\{\}$ correspond to an inelastic
collision with a nucleon in the nucleus, an elastic scattering in the
amplitude and an elastic scattering in the complex conjugate amplitude 
respectively as illustrated in Fig.~\ref{fig1}.
The state $N$ in the figure is that of a single nucleon in the nucleus
while the $x_+$ and $x_+^\prime$ integrals in Eq.~(\ref{eq:S1})
go over the extent of the nucleus.

In the McLerran-Venugopalan model the quark-nucleon scattering
amplitude is purely absorptive so that
\begin{multline}
\langle A|
\frac{1}{N_c}\text{tr}A_-(x_{1\perp},x_{1+})A_-(x_{2\perp},x_{2+})
|A\rangle\\
=
\langle A|
\frac{1}{N_c}\text{tr}T\left[A_-(x_{1\perp},x_{1+})A_-(x_{2\perp},x_{2+})\right]
|A\rangle
\end{multline}
matching onto conventional notations when $x_\perp^2$ is small
\begin{multline}
\langle A|
\frac{1}{N_c}\text{tr}\bigg\{
A_-(0_\perp,x_+^\prime)A_-(x_\perp,x_+)\\
-\frac12
A_-(x_\perp,x_+^\prime)A_-(x_\perp,x_+)
-\frac12 A_-(0_\perp,x_+^\prime)A_-(0_\perp,x_+)\bigg\}
|A\rangle\\
=-\frac{\pi}{8N_c}x_\perp^2
\rho\, xG(x,1/x_\perp^2)\frac{1}{\sqrt{2}}\delta(x_+-x_+^\prime)
\end{multline}
so that
\be
S_1=-\frac{x_\perp^2 Q_s^2}{4}
\label{eq:S1ex}
\ee
with
\be
Q_s^2=\frac{4\pi^2\alpha C_F}{N_c^2-1}L\rho\, xG(x,1/x_\perp^2)
\ee
and where $Q_s$ is the quark saturation momentum and $\rho$
the density of nucleons in the nucleus. $L$ is the path
length of nuclear matter at the impact parameter at which
the quark passes through the nucleus.

Using Eq.~(\ref{eq:S1ex}) in Eq.~(\ref{eq:SfromS1})
gives the relation
\be
S(x_\perp)=e^{-x_\perp^2 Q_s^2/4}
\label{eq:Sqqnucleus}
\ee
which is recognized as the elastic $S$-matrix for a quark-antiquark
dipole passing through the nucleus.
Using Eq.~(\ref{eq:Sqqnucleus}) in Eq.~(\ref{eq:correspondence})
tells us that the distribution of transverse momenta that
a quark picks up in passing through a nucleus is simply related,
by Fourier transform, to the elastic dipole-nucleus $S$-matrix
element. If one integrates $dN/d^2p$ over all transverse momenta
the result must be 1,
\be
\int d^2p_\perp \frac{dN}{d^2p_\perp}=1
\ee
since the quark must pick up some value of the transverse
momentum. In Eq.~(\ref{eq:correspondence}) this relationship\
corresponds to $S(0_\perp)=1$.
From Eq.~(\ref{eq:S}) $S(0_\perp)=1$ corresponds to the unitarity of the
operator $V_{0_\perp}$ while from Eq.~(\ref{eq:Sqqnucleus})
it corresponds to color transparency.
But color transparency is a property of gauge theories.
Thus the relationships that we are investigating will not
be true in a general field theory but are specific to gauge theories.
Finally, we have obtained Eq.~(\ref{eq:Sqqnucleus})
assuming that the dipole-nucleus scattering amplitude is purely
absorptive. While this is natural in the McLerran-Venugopalan
model, where the cross-section has no energy dependence, the assumption
is not necessary as the second and third terms on the right-hand
side of Eq.~(\ref{eq:S1}) combine to give an absorptive part
in contrast to the individual terms which are
time-ordered.

\subsection{$p_\perp$-broadening more generally}

We now turn to a more general discussion of the possible relationship
between $p_\perp$-broadening and dipole scattering.
We shall restrict our discussion to the case where gluons giving
quantum contributions to $p_\perp$-broadening have
a longitudinal momentum much less than 
that of the quark, thus allowing the quark motion to be given
by a Wilson line with a straight line trajectory.
This restriction should not limit our ability to investigate and
compare the energy evolution of $p_\perp$-broadening with that of
dipole scattering. A complete study of $p_\perp$-broadening vs dipole
scattering would require an understanding of how to include the 
high-momentum gluons into (evolution-independent) impact factors and
is beyond our present aim.

Equation~(\ref{eq:correspondence}) is general but, when quantum
gluon corrections are included, Eq.~(\ref{eq:S}) should be
changed to
\be
\tilde S(x_\perp)=\frac{1}{N_c}
\sum_n
\text{tr}
\bigg\{
\langle n|T\left(V_{0_\perp}e^{i\int d^4y {\cal L}_I(y)}\right)
|A\rangle^*
\langle n|T\left(V_{x_\perp}e^{i\int d^4y {\cal L}_I(y)}\right)
|A\rangle
\bigg\},
\label{eq:Stilde}
\ee
where ${\cal L}_I$ is the QCD interaction Lagrangian,
say in the lightcone gauge.
$S(x_\perp)$ now has an energy dependence which,
for simplicity, we suppress.
Using completeness of the states $|n\rangle$, which
now include arbitrary numbers of quarks and gluons,
in addition to nuclear break-up, one can write Eq.~(\ref{eq:Stilde})
as
\be
\tilde S(x_\perp)=\frac{1}{N_c}
\text{tr}
\langle A|
\bar T
\left(
V_{0_\perp}^\dagger e^{-i\int d^4y {\cal L}_I(y)}
\right)
T
\left(
V_{x_\perp} e^{i\int d^4y {\cal L}_I(y)}
\right)
|A\rangle.
\label{eq:Stilde2}
\ee
On the other hand, the $S$-matrix for dipole-nucleus
scattering is
\be
S(x_\perp)=\frac{1}{N_c}\text{tr}
\langle A|
T\left(V_{0_\perp}^\dagger V_{x_\perp} e^{i\int d^4y {\cal L}_I(y)}\right)
|A\rangle.
\label{eq:Sdipolenucleus}
\ee
Equation~(\ref{eq:Sdipolenucleus})
is given as a conventional time-ordered product while 
Eq.~(\ref{eq:Stilde2}) is a time-ordered product in a
Keldysh-Schwinger \cite{Keldysh:1964ud,Schwinger:1960qe} 
sense where the vertices in the $T()$,
the second factor in the matrix element in~(\ref{eq:Stilde2}),
occur in the upper $(x_{1+})$ part of the time contour
\begin{figure}
\begin{center}
\includegraphics[width=12cm]{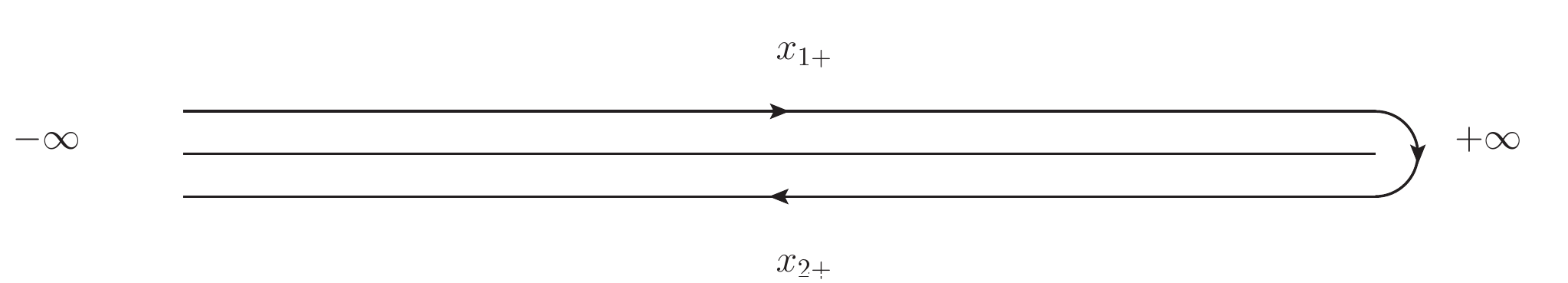}
\end{center}
\caption{\label{fig2}Keldysh-Schwinger
contour.}
\end{figure}
illustrated in Fig.~\ref{fig2}, while the vertices in $\bar T()$,
the first factor in the matrix element in Eq.~(\ref{eq:Stilde2}),
occur on the lower $(x_{2+})$ part of the time contour.\footnote{%
The Keldysh-Schwinger formalism has also been used in e.g.
Ref.~\cite{Gelis:2006yv} in a non-thermal context.
}%
Feynman lines connecting points on the $x_{1+}$-contour have ordinary
$i\varepsilon$ factors, lines connecting points on the $x_{2+}$ part
of the contour have the sign of it changed, corresponding
to the anti-time-ordering indicated in Eq.~(\ref{eq:Stilde2}).
Feynman lines connecting points on the $x_{1+}$ part of the
contour to points on the $x_{2+}$ part of the contour are put
on shell.

$\tilde S$ and $S$ given by Eq.~(\ref{eq:Stilde2}) 
and~(\ref{eq:Sdipolenucleus}) respectively will not in general
be the same.
However, Kovchegov  and Tuchin have shown 
\cite{Kovchegov:2001sc} (see Ref.~\cite{JalilianMarian:2005jf}
for a broader review, see also Ref.~\cite{Kharzeev:2003wz})
that the $x$-evolution
(energy evolution) of production cross sections is the
same as dipole scattering at the leading logarithmic level.
Our goal in this paper is to show that this equality
persists through next-to-leading order in $x$-evolution.

\begin{figure}
\begin{center}\includegraphics[width=8cm]{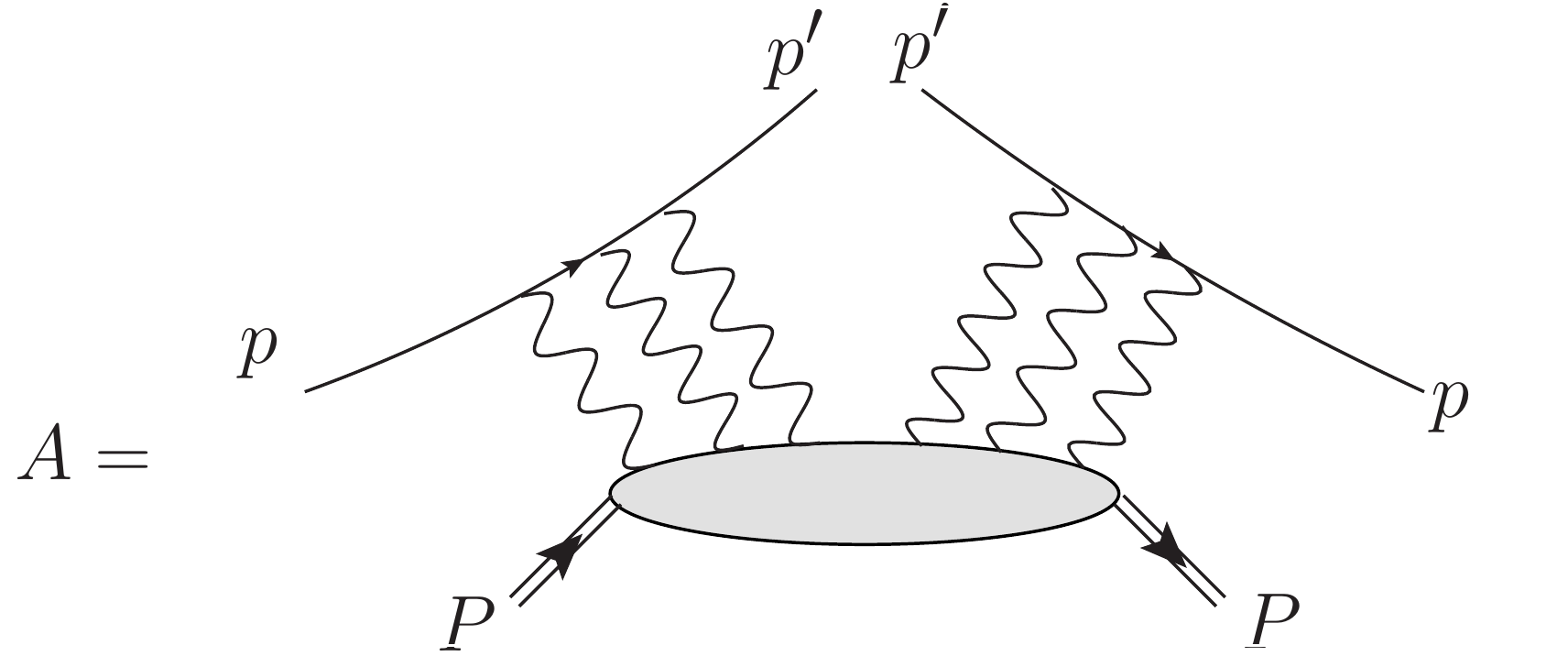}
\end{center}
\caption{\label{fig3}
The Feynman amplitude described by Eq.~(\ref{eq:feynmanamp}).
}
\end{figure}

Finally, we note that~(\ref{eq:Stilde2}) can be obtained as a particular
discontinuity of an analytically continued Feynman amplitude.
To see this we must keep track also of longitudinal
momenta of the initial and final quarks as illustrated for the
amplitude $A$ in Fig.~\ref{fig3}.

\begin{figure}
\begin{center}\includegraphics[width=9cm]{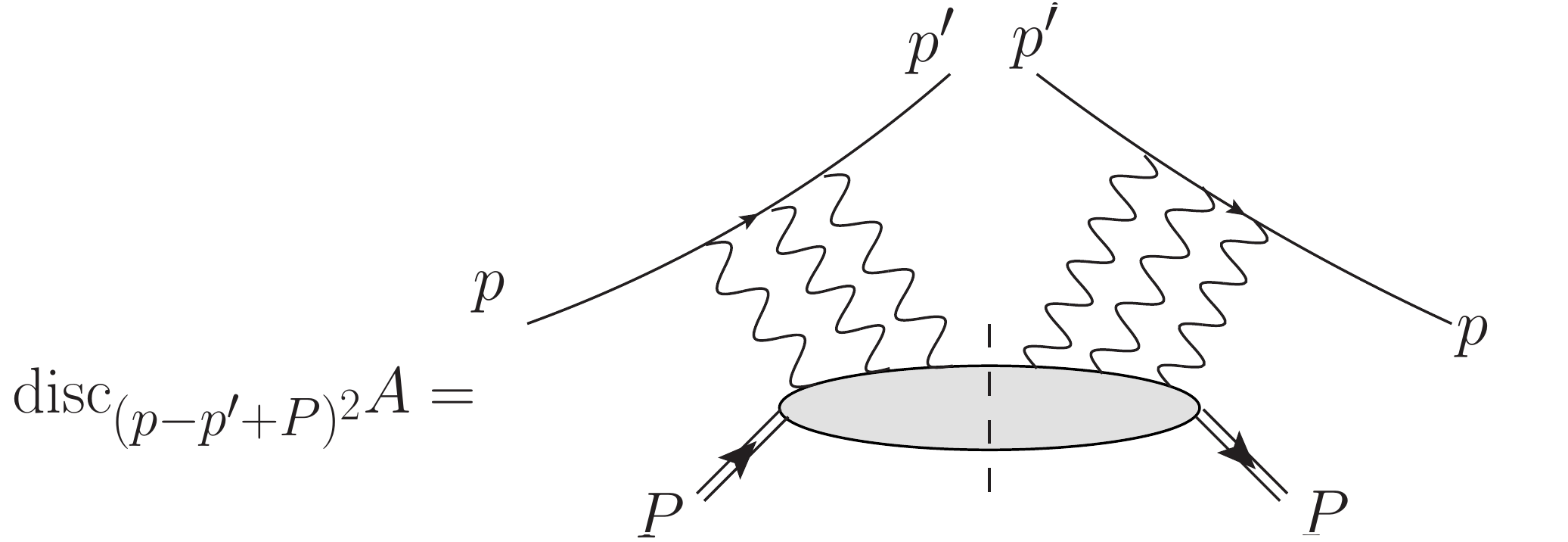}
\end{center}
\caption{\label{fig4}
Graphical illustration of the r.h.s. of Eq.~(\ref{eq:disc}).
}
\end{figure}

$P$ is the momentum of the target nucleus with $p$ and $p^\prime$
the momenta of the quark. One can view $A$ as having the following
dependence:
\be
A=A\left((p+P)^2,(p-p^\prime+P)^2,(p_\perp-p^\prime_\perp)^2\right).
\label{eq:feynmanamp}
\ee
Then
\be
1-\tilde S=\frac{1}{2i}\text{disc}_{(p-p^\prime+P)^2} A,
\label{eq:disc}
\ee
as illustrated in Fig.~\ref{fig4}.

\begin{figure}
\begin{center}\includegraphics[width=6cm]{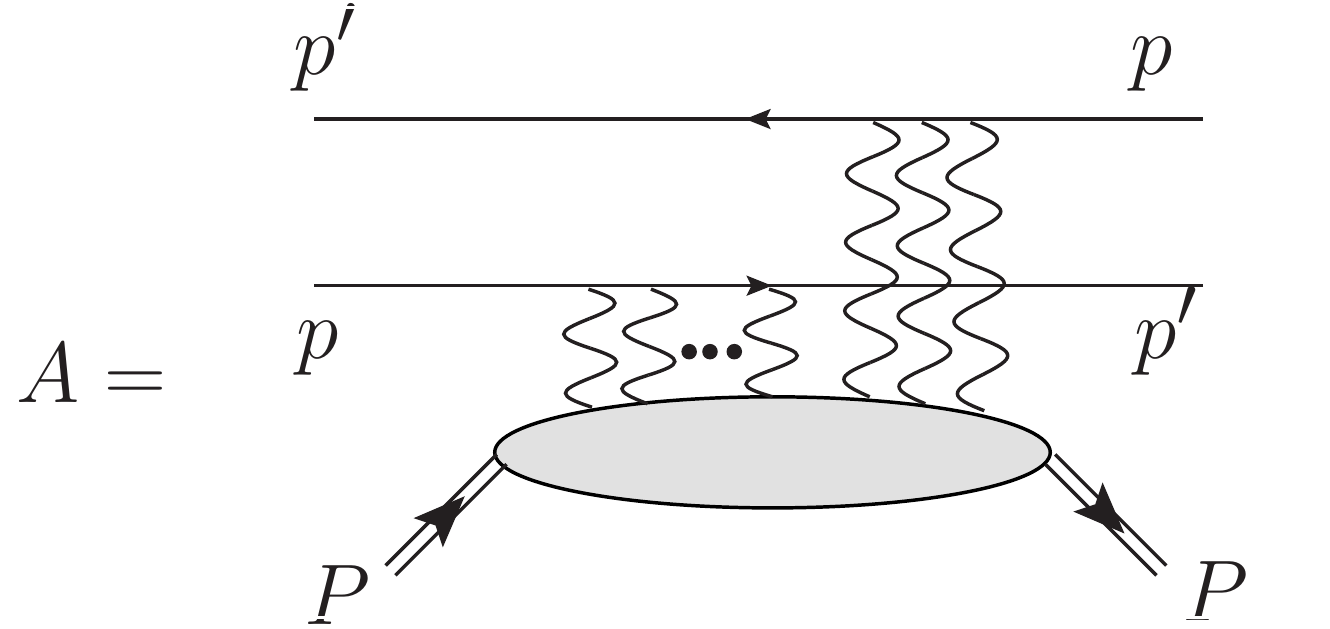}
\end{center}
\caption{\label{fig5}
Amplitude $A$ of Fig.~\ref{fig3}
viewed as a dipole scattering amplitude.
}
\end{figure}

$A$ can be viewed as a dipole scattering amplitude, as shown in
Fig.~\ref{fig5},
 with the quark and the antiquark momenta,
$(p,p^\prime)$ and $(p^\prime,p)$ respectively,
given by
\be
\begin{split}
p&=(p_0,p_x,p_y,p_z)=(p,0,0,p)\\
p^\prime&=(\sqrt{(p^\prime)^2+(p^\prime_\perp)^2},p_\perp,p^\prime).
\end{split}
\ee
The arrows on the quark lines in Fig.~\ref{fig5}
refer to the flow of baryon number while the momenta are all
directed to the right. The discontinuity in Eq.~(\ref{eq:disc})
is not just the imaginary part of the dipole
scattering so that, in general,
one cannot expect a simple relationship between the
dipole scattering in Fig.~\ref{fig5} and the particular
discontinuity illustrated in Fig.~\ref{fig4}.


\section{\label{sec:leadingorder}Leading-order calculation}

In this section, we review the correspondence between
the evolution of $p_\perp$-broadening amplitudes
and dipole cross sections
at leading order. Our aim is also to introduce the basic formalism
and conventions that will be used 
for all calculations
throughout this paper.

\begin{figure}
\begin{center}\includegraphics[width=12cm]{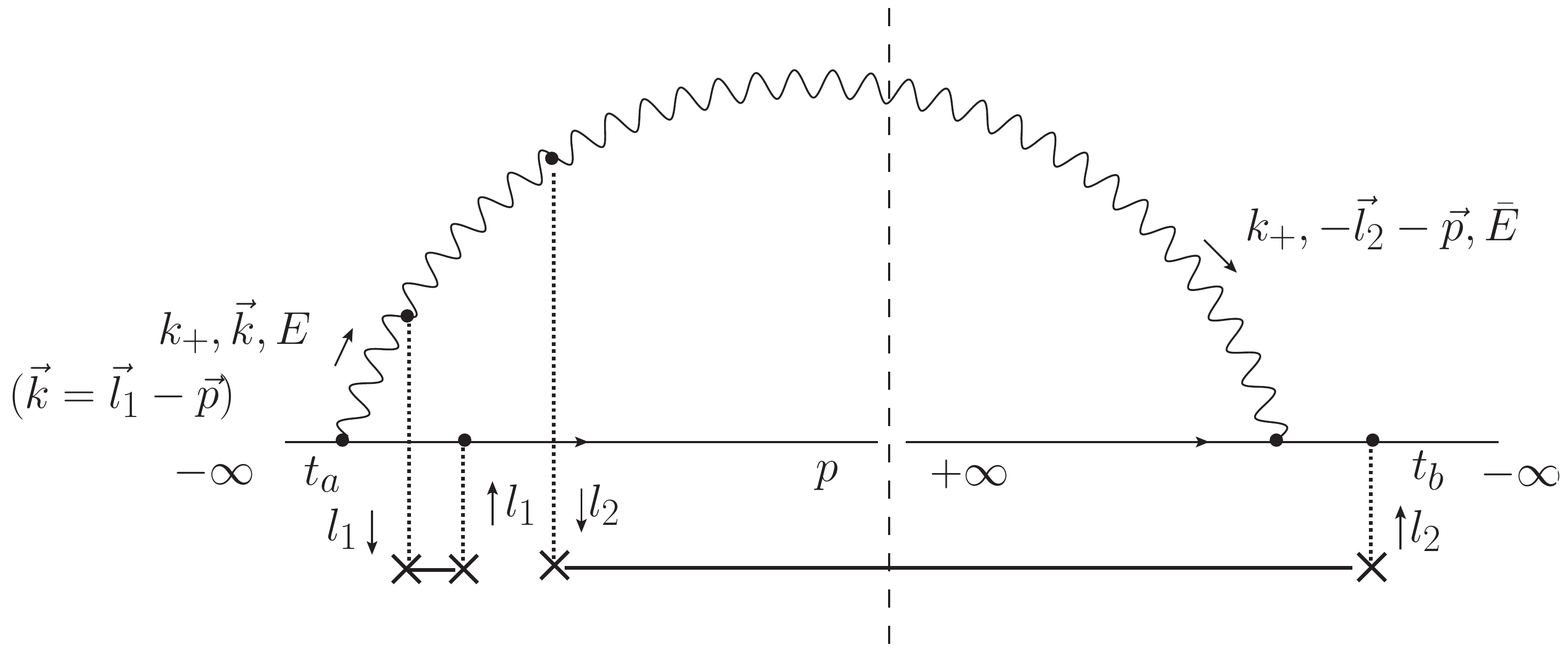}
\end{center}
\caption{\label{lo_labeled}
Graph contributing to the leading-order $p_\perp$-broadening
amplitude.
The freed gluon is emitted in the initial state
in the amplitude and in the final state in the complex-conjugate
amplitude. At the level of the nucleus, we chose one
elastic interaction (momentum transfer $l_1$) and
one inelastic one (momentum transfer $l_2$).
The exchanged gluons are drawn as dotted lines in order
to simplify the graph.
}
\end{figure}

Let us explain the method to evaluate 
a given lightcone perturbation theory graph such as the one
shown in Fig.~\ref{lo_labeled}.
There are essentially four elements: 
the product of the energy denominators $D$,
the vertex factors, the gluon
polarization tensors, and the overall integration.

In order to compute the energy denominators, we first
assign a lightcone time $t$ to each vertex. We
then evaluate $D$ using
the following formula:
\be
D=\int \Pi 
\left[dt\, e^{i(E_\text{outgoing}-E_\text{incoming})t\pm\varepsilon t}\right],
\label{eq:D}
\ee
where the product goes over all vertices and the integration
takes into account the ordering of the different times.
$\varepsilon$ is a small positive parameter (with
dimensions of an energy) that acts as an 
adiabatic regulator:
It is set to zero at the end of the calculation.
The sign in front of it is chosen in order to insure the convergence
of the corresponding time integral at 
(plus or minus) infinity,
depending whether the considered vertex is in the initial
or in the final state.
The energy of a state is the sum of the energies of all the particles
in the state. The energy $E$ 
of a particle of 4-momentum $p=(p_+,p_-,\vec p)$
reads
\be
E=\frac{\vec p^2}{2p_+}.
\ee
The quark-gluon vertices are
evaluated in the eikonal approximation
throughout this paper. 
For incoming/outgoing fermions
of helicities $\alpha$, $\alpha^\prime$, colors $a$, $b$, 
momentum $p$, and a gluon 
which carries the Lorentz index $\mu$ and the color $C$,
the vertex is represented by the factor
\be
\pm 2i g t_{ba}^C p_\mu\delta^{\alpha\alpha^\prime},
\label{eq:eikq}
\ee
where the sign is $+$ for a quark and $-$ for an antiquark.
The triple-gluon vertices
which couple the gluons in the wave function with the nucleus
will also have to be evaluated in the eikonal approximation. 
When the fast gluon line carries
the Lorentz indices $\alpha$, $\beta$,
the colors $A$, $B$,
and the momentum $p$, the vertex factor reads
\be
2g f_{ABC} p_\gamma g_{\alpha\beta}.
\label{eq:eikg}
\ee
The triple-gluon vertex must however
be computed exactly when it appears in the quantum 
evolution of the wave functions.
For incoming momenta $p_1$, $p_2$ and $p_3$ ordered counterclockwise with
respective Lorentz indices $\alpha$, $\beta$ and $\gamma$
and color indices $A$, $B$ and $C$, the corresponding
vertex factor reads
\be
g f_{ABC}
\left[g_{\alpha\gamma}(p_3-p_1)_\beta+g_{\beta\alpha}(p_1-p_2)_\gamma
+g_{\gamma\beta}(p_2-p_3)_\alpha\right]
(2\pi)^3\delta^3(p_1+p_2+p_3).
\ee
At next-to-leading order accuracy, we will never need to consider
four-gluon vertices.

In addition, each leg of vertex through which the
momentum $p$ flows
has a factor 
\be
\frac{1}{\sqrt{2p_+}}.
\label{eq:vertexfactor}
\ee
The components of the
polarization tensor for the gluon are of 3 different types.
For a gluon of momentum $p=(p_+,p_-,p_1,p_2)$, the latter read
\be
d_{--}(p)=\frac{\vec p^2}{p_+^2}\ ,\ 
d_{-i}(p)=\frac{p_i}{p_+}\ ,\
d_{ij}(p)=-g_{ij}=\delta_{ij}\ \ (i=1,2).
\ee
An integration over all $+$ and $\perp$ components of the
loop momenta may then be performed, although, as we will discover,
a momentum-by-momentum identification holds for some classes of graphs.
The integration over the on-shell momentum $k$ is performed as
\be
\int \frac{{dk_+} d^2\vec k}{(2\pi)^3}.
\ee

We are now ready to address the calculation of the diagram
depicted in Fig.~\ref{lo_labeled}.
We interpret it as 
the representation of
a $T$-product, like all other
diagrams in this work.
Using Eq.~(\ref{eq:D}), we find that the energy denominators 
read
\be
D=\frac{1}{E\bar E},
\ee
where $E=(\vec l_{1}-\vec p)^2/(2k_+)$, 
$\bar E=(\vec l_{2}+\vec p)^2/(2k_+)$,
while the polarizations read
\be
\sum_{ij}
d_{-i}(l_1-p)\cdot d_{ij}\cdot d_{j-}(-l_2-p)
=\frac{(\vec p-\vec l_{1})\cdot(\vec p+\vec l_{2})}{k_+^2}.
\ee
Including the vertex factors, we get, averaging over the quark color
and helicity:
\be
\frac12 g^6 N_c^2 C_F
\frac{(\vec p-\vec l_{1})\cdot(\vec p+\vec l_{2})}
{(\vec p-\vec l_{1})^2(\vec p+\vec l_{2})^2}
\frac{1}{k_+}\ ,
\label{eq:lo_c}
\ee
where $C_F=(N_c^2-1)/(2N_c)$.
The transverse momentum conservation
is enforced with the help of the additional factor 
$\delta^2(\vec k+\vec p-\vec l_{1})$.

From the $T$-product, we may compute the 
contribution of the particular graph represented in
Fig.~\ref{lo_labeled} to the
distribution of
transverse momenta of the quark.
To this aim, we need to integrate over the 3-momentum of the gluon $k$
and to convolute with appropriate unintegrated gluon distributions
which appear in the description of the scattering with the nucleus.
The factor which comes with each two-gluon exchange with a particular nucleon
in the nucleus involving the transfer of
the momentum $l$
reads
\be
\pm\int\frac{d^2\vec l}{\vec l^2}\frac{x g(x,\vec l^2)}{N_c^2-1} \rho L
\label{eq:unint}
\ee
where the sign is $+1$ if the nucleon scatters elastically
and $-1$ if it scatters inelastically.
When the exchanged
gluons attach to the same (quark or gluon) line, a combinatorial
factor $\frac12$ must be taken into account.
Another overall combinatorial factor takes care of the time ordering of the
scatterings off the different nucleons.
(For the particular graph in Fig.~\ref{lo_labeled}, none of these
factors is needed).
All in all, we get:
\begin{multline}
\left.\frac{dN}{d^2p}\right|_{
\text{\scriptsize graph
in Fig.~\ref{lo_labeled}}}
=
-\frac{\alpha_s N_c}{N_c^2-1}\int_0^{+\infty}\frac{dk_+}{k_+}\\
\times
\int \frac{d^2\vec l_{1}}{\vec l_{1}^2} \frac{d^2\vec l_{2}}{\vec l_{2}^2}
\frac{(\vec p-\vec l_{1})\cdot(\vec p+\vec l_{2})}
{(\vec p-\vec l_{1})^2(\vec p+\vec l_{2})^2}\\
\times\left[
\alpha_s x g(x,\vec l_{1}^2)\rho L
\right]
\left[
\alpha_s x g(x,\vec l_{2}^2)\rho L
\right].
\end{multline}
A priori, we need to sum over any combination
of elastic and inelastic exchanges.
Actually, they would all look like the ones
shown in Fig.~\ref{lo_labeled} and bring factors
which are identical in the $p_\perp$-broadening
case and in the dipole case. Hence it is
enough to consider a small number of scatterings:
We need only make sure that the case we single out
be sufficiently general.

\begin{figure}
\begin{center}\includegraphics[width=8cm]{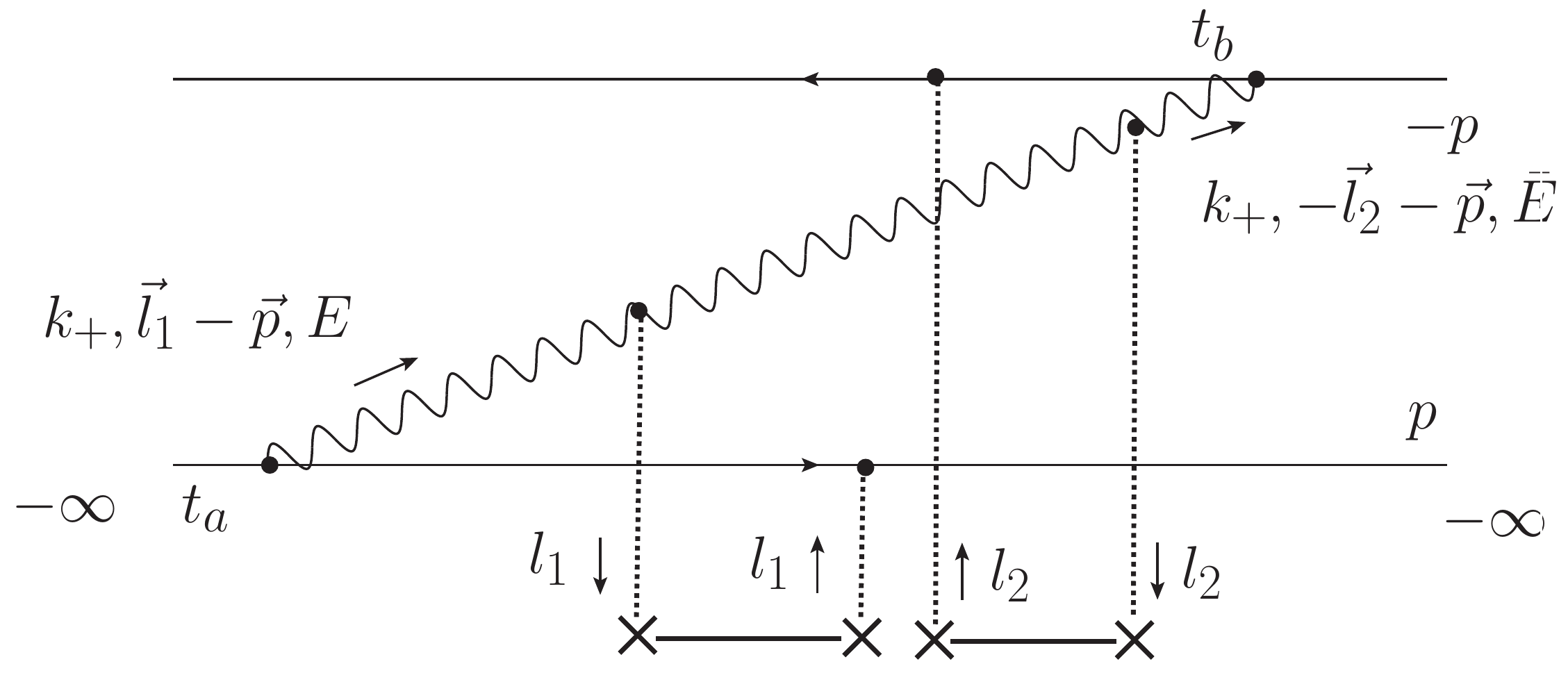}
\end{center}
\caption{\label{lod_labeled}
Graph contributing to the lowest-order quantum
correction to the dipole forward scattering amplitude.
Here there are 2 dipoles interacting with the target nucleus.
This graph corresponds to the $p_\perp$-broadening graph in 
Fig.~\ref{lo_labeled}.
}
\end{figure}

We can compute in the same way the dipole graph
in Fig.~\ref{lod_labeled}.
The only differences with respect to the previous
case are: {\em (i)} the energy denominators come with a
minus sign, {\em (ii)} the 
antiquark-gluon vertices
have opposite sign
with respect to the quark-gluon vertices and 
{\em (iii)} the scatterings with the nucleus
all come with an overall ``+'' sign in Eq.~(\ref{eq:unint}).
All the extra minus signs compensate, and we find an exact 
momentum-by-momentum identity
between the expressions for the
graphs of Fig.~\ref{lo_labeled} and~\ref{lod_labeled}.

We see that we may adopt the following rule as for the relative
signs between $p_\perp$-broadening amplitudes and dipole amplitudes:
Since the extra minus sign which comes from the 
gluons attaching to the antiquarks
is compensated by the change of sign of the
discontinuity of the scattering with the nucleus
when one goes from $p_\perp$ broadening to dipoles,
it is enough to take into account only the extra minus signs 
stemming from the couplings of the gluons in the wave functions
to the antiquark.

We have seen in detail how the particular graphs in Fig.~\ref{lo_labeled}
and~\ref{lod_labeled} are related.
We need to extend the discussion to all possible graphs.
First, we note that the $p_\perp$-broadening graphs
in which the gluon is not produced in the final state
trivially correspond to dipole graphs for which the quark
or the antiquark is a spectator.
So we shall not discuss this case.
As for the form of the interaction with the target, we choose
to consider one elastic and one inelastic scattering.
The gives the most general momentum flow, and since all vertices
are eikonal, it would be straightforward to extend the
discussion to any form of the scattering.

\begin{figure}
\begin{center}
\begin{tabular}{cc}
\includegraphics[width=5.5cm]{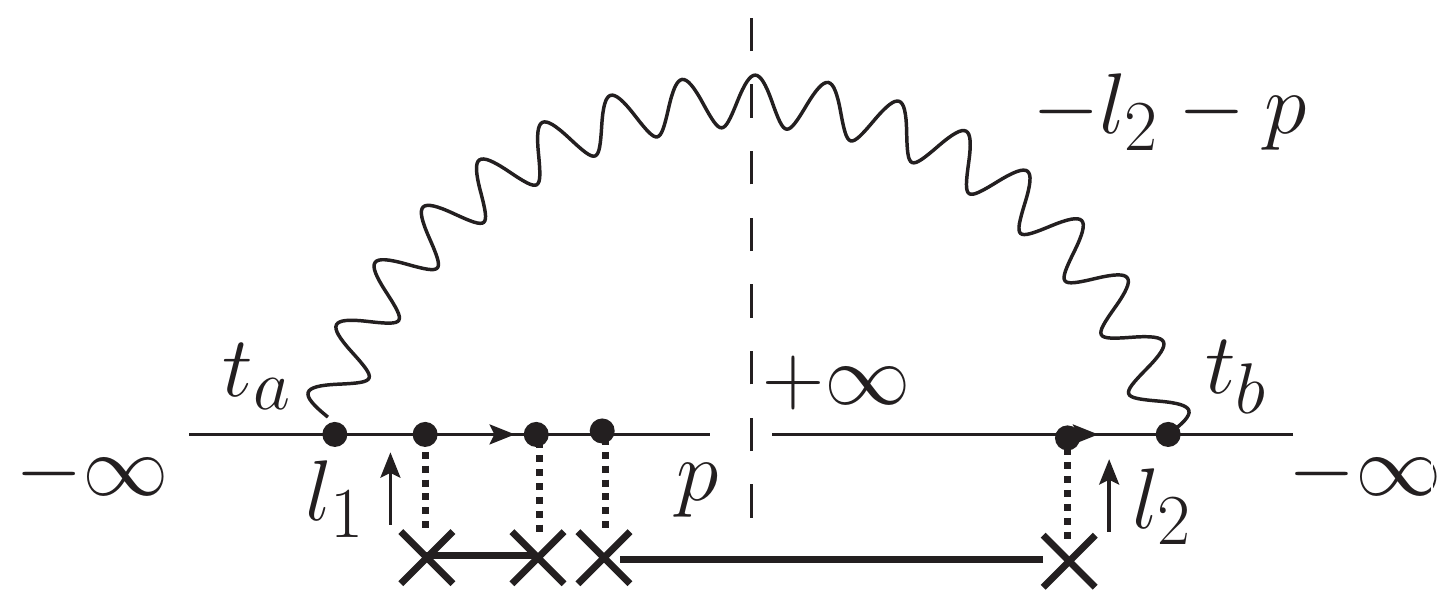}&
\includegraphics[width=6cm]{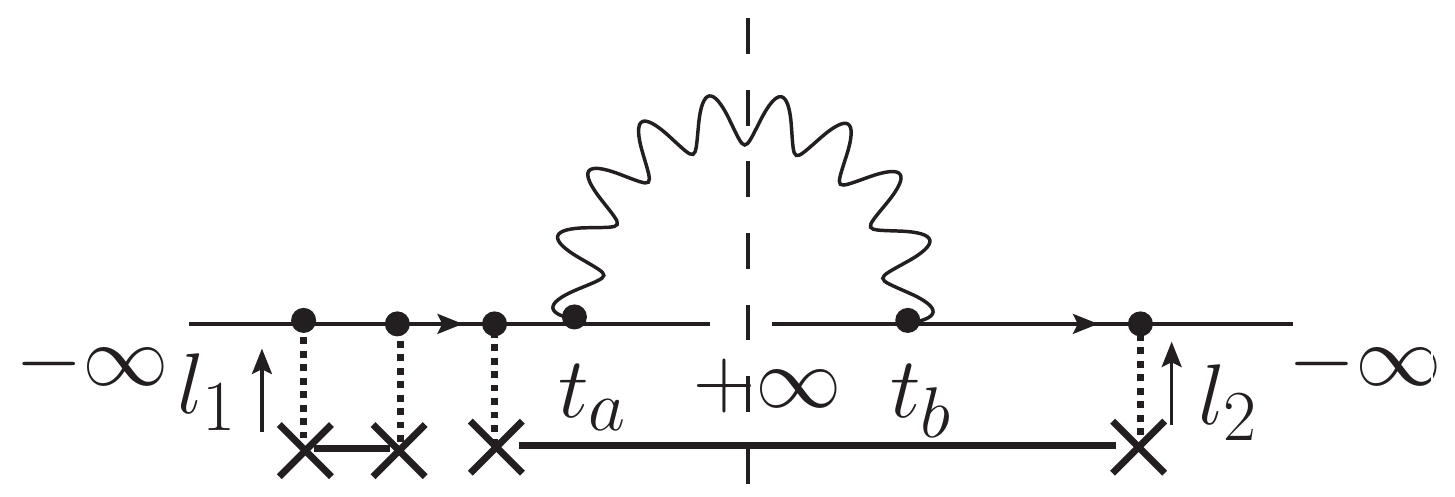}\\
$a$&$b$\\
\\
\includegraphics[width=6cm]{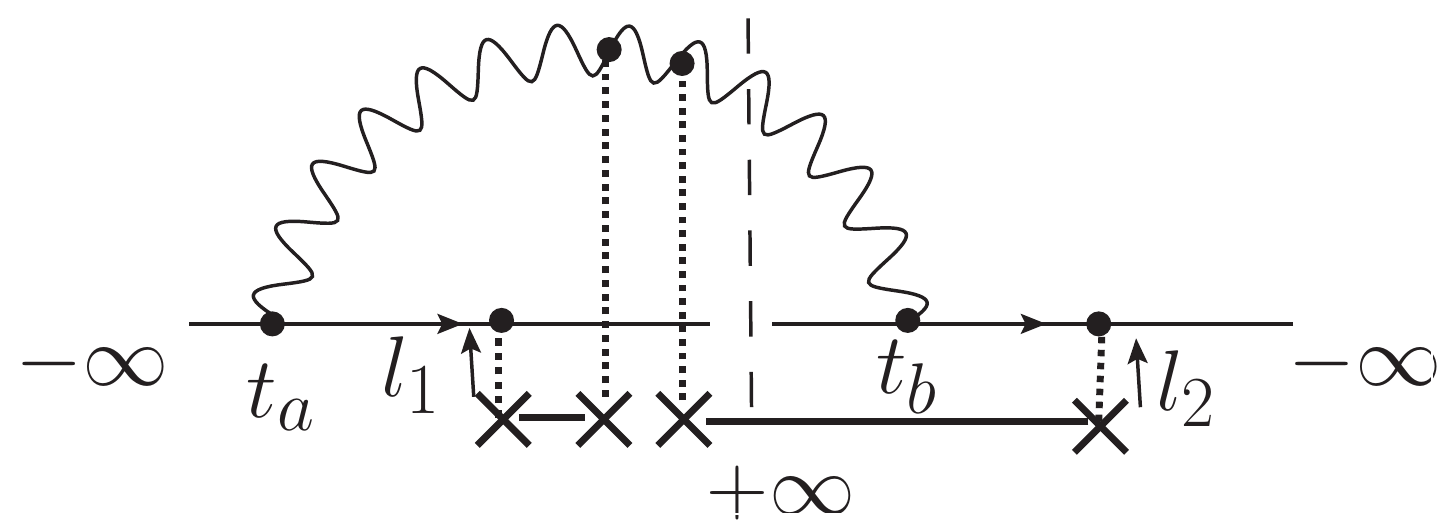}&
\includegraphics[width=5cm]{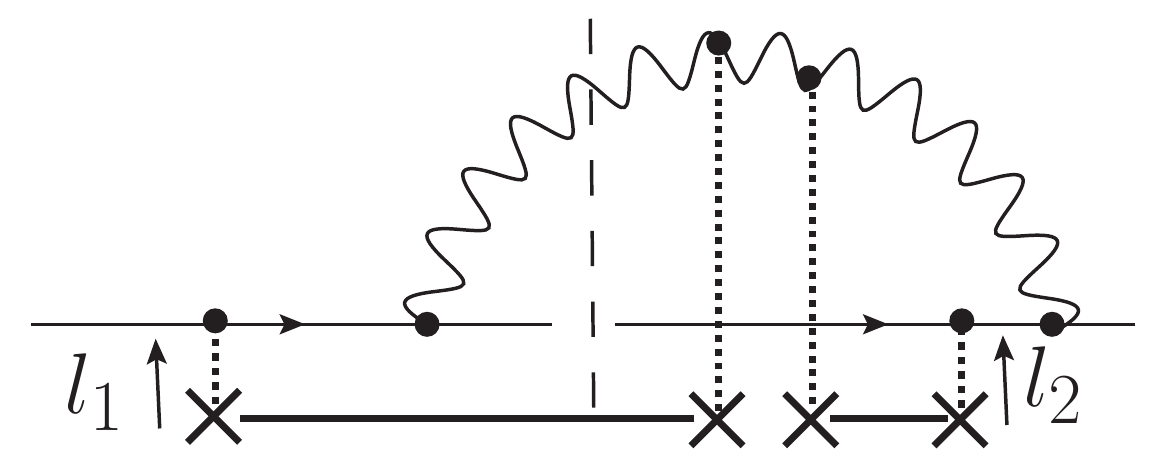}\\
$c$&$d$
\end{tabular}
\end{center}
\caption{\label{lo}
Full set of the lowest-order corrections to the $p_\perp$-broadening
amplitude.
Graph $c$ is the one already represented in Fig.~\ref{lo_labeled},
and graph $d$ has the same expression since it is related by
a simple symmetry.
}
\end{figure}

\begin{figure}
\begin{center}
\begin{tabular}{cc|c}
\includegraphics[width=4cm]{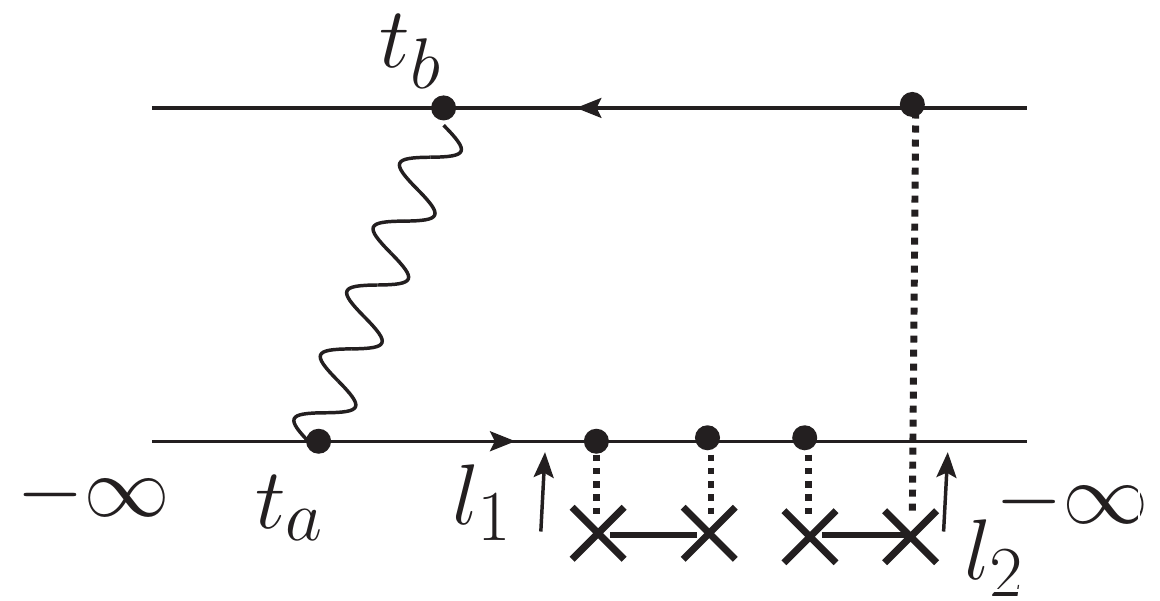}&
\includegraphics[width=4cm]{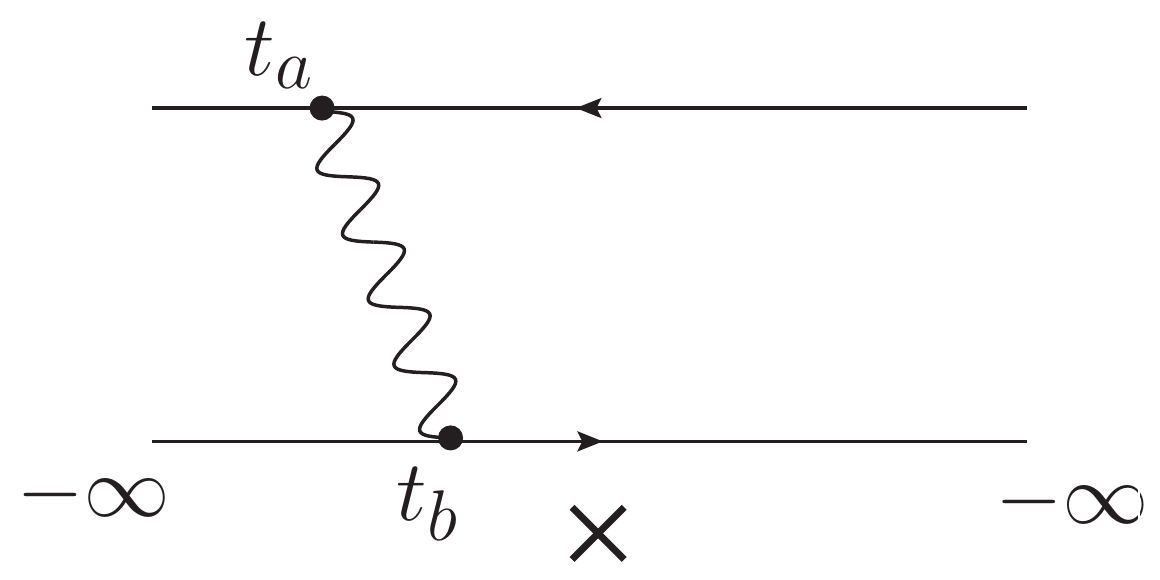}&
\includegraphics[width=3.2cm]{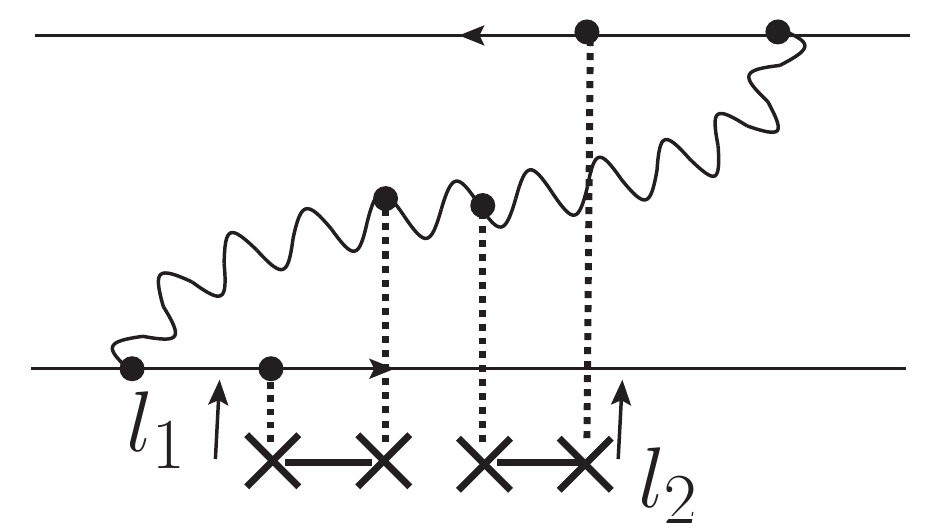}\\
$\alpha$&$\beta$&$\gamma$\\
& &\\
\includegraphics[width=4cm]{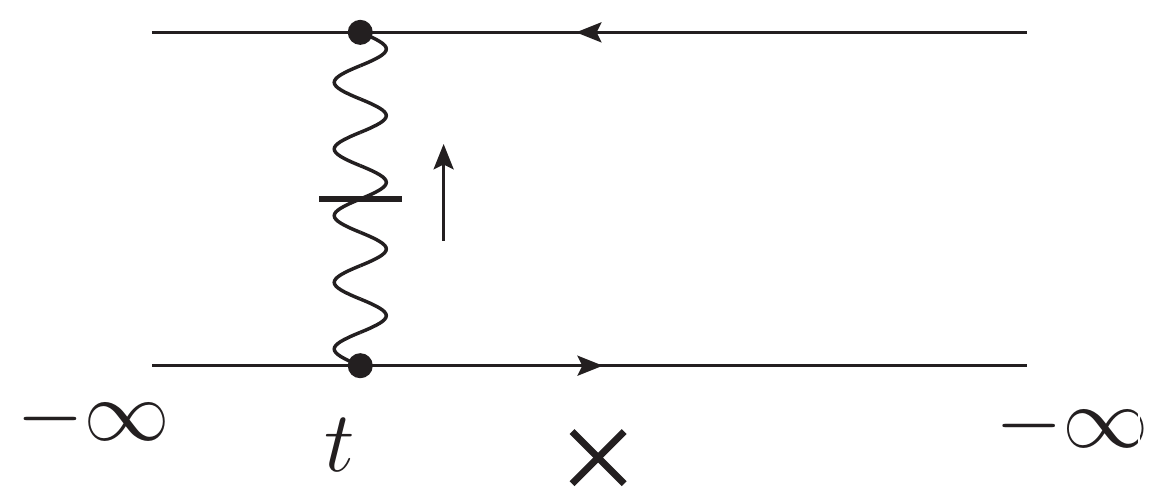}&
\includegraphics[width=4cm]{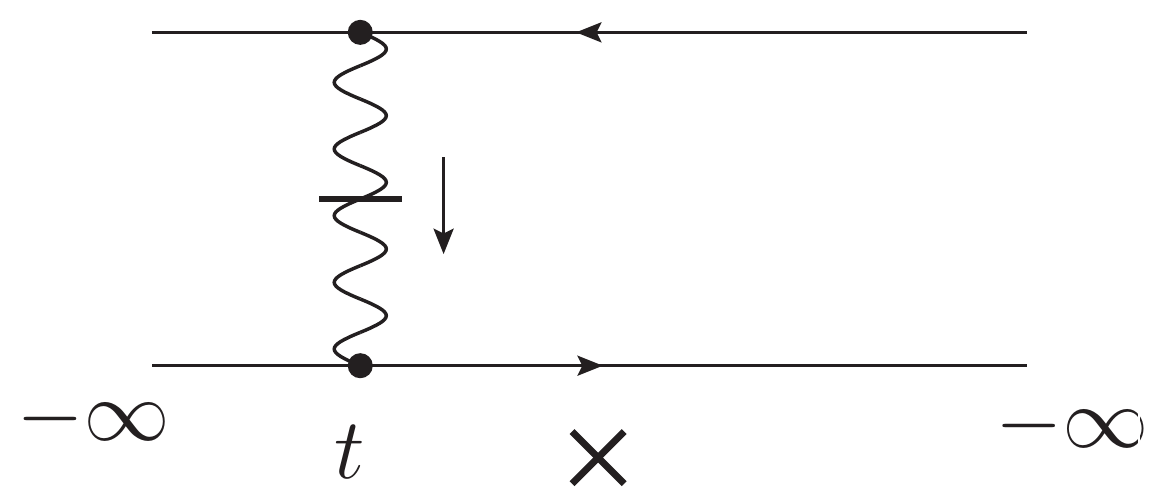}&
\includegraphics[width=3.2cm]{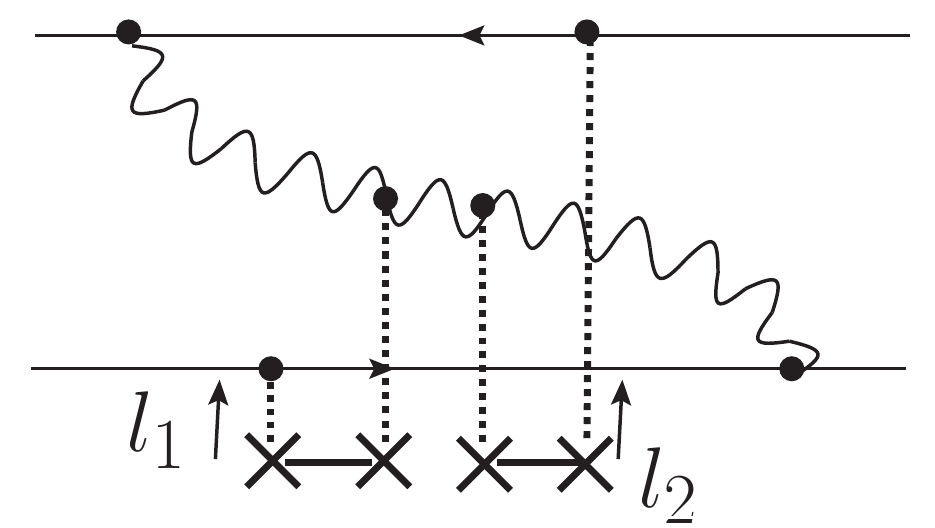}\\
$\alpha^\prime$&$\beta^\prime$&$\delta$
\end{tabular}
\end{center}
\caption{\label{lod}
Set of leading-order virtual (left) and real (right)
corrections to the
dipole amplitude.
In the graphs $\alpha$ and $\beta$, the exchanged gluon is
causal, whereas in $\alpha^\prime$ and $\beta^\prime$,
it is exchanged instantaneously.
Graph $\gamma$ is equivalent to the graph in Fig.~\ref{lod_labeled},
and $\delta$ has the same expression.
}
\end{figure}

In the case of $p_\perp$-broadening, there are four different
nontrivial graphs, 
depending on whether the gluon is emitted after or before the
time of the interaction, in the amplitude and in the
complex conjugate amplitude.
They  are depicted in Fig.~\ref{lo}.
The full set of dipole graphs at leading order
is represented in Fig.~\ref{lod}.

We have computed graph $c$ (Fig.~\ref{lo_labeled} and~\ref{lo})
and shown that its expression
is the same as the expression for the dipole graph $\gamma$
(Fig.~\ref{lod_labeled} and~\ref{lod}).
Graph $d$ is similar to $c$ and 
it is straightforward to check that
its expression is the same
as the expression for the dipole graph in which the gluon
attaches to the antiquark in the amplitude and to the quark
in the complex conjugate amplitude. In other words,
\be
c=\gamma\ \ \text{and}\ \ d=\delta.
\ee

We turn to graph $a$ in Fig.~\ref{lo}.
The scattering occurs after the emission of the
gluon both in the amplitude and in the complex conjugate amplitude.
The gluons exchanged with the nucleus may be chosen to
scatter with the quark line only:
It is easy to check that the graphs in which
the fast gluon scatters cancel between each other.
The polarization factor for the gluon is then 
$d_{--}(-l_2-p)$.
We keep two scatterings with the quark, 
one elastic and one inelastic.
Leaving out the factors describing
the interaction with the target and
the integration over $k_{+}$, the expression for the graph reads
(compare to Eq.~(\ref{eq:lo_c}))
\be
a=-2 g^6 C_F^3
\frac{1}{(\vec p+\vec l_{2})^2}\frac{1}{k_+}.
\label{eq:loi(i)}
\ee

The corresponding dipole graphs would be
$\alpha$, $\alpha^\prime$, $\beta$ and $\beta^\prime$ 
in Fig.~\ref{lod}.
We see that these graphs are purely virtual:
Therefore, for the correspondence to work, it
is crucial that all interactions 
with the nucleus
happen with the quarks
and not with the gluon in the $p_\perp$-broadening case,
which is effectively verified
since scatterings with
the gluons cancel among themselves.

For this process, 
there are also graphs with instantaneous gluon exchanges
(graphs $\alpha^\prime$ and $\beta^\prime$).
We take them into account by modifying the $d_{--}$
component of the polarization tensor 
in the corresponding causal graphs
($\alpha$ and $\beta$ resp.)
as follows \cite{Lepage:1980fj}:
\be
d_{--}(k)\rightarrow d_{--}(k)
\left(
1-\frac{\Sigma}{E_{\text{gluon}}}
\right)
\equiv d_{--}(k)\times F\ ,
\label{eq:BLtrick}
\ee
where $E_{\text{gluon}}$ is the energy carried by the gluon,
and $\Sigma$ is the factor in the energy denominators
associated to the emission of this gluon.
The second equality defines the $F$-factor.
Since the graphs containing gluons which are exchanged
instantaneously may be effectively taken into account
through the above modification in the expressions of
the causal graphs, we will not draw them systematically
in the following sections.
We shall just perform the substitution~(\ref{eq:BLtrick})
whenever the kinematics allow for instantaneous exchanges.

Let us evaluate the dipole graph $\alpha$ in Fig.~\ref{lod}.
The energy denominators read
\be
D=\frac{1}{2i\varepsilon(\bar E-i\varepsilon)}\ ,
\ee
where $\bar E=(\vec p+\vec l_{2})^2/(2k_+)$.
The divergence 
encoded in the $1/\varepsilon$ singularity
may be traced to the absence of
time scale for the exchange of the gluon between
the quark and the antiquark, making it equally likely
for all times between $t=-\infty$ and $t=0$.
The expression for this graph reads, before integration over
the momenta,
\be
\alpha=-\frac12 g^6C_F^3\frac{(\vec p+\vec l_{2})^2}{k_+^3}
\frac{1}{2i\varepsilon}\frac{1}{\bar E-i\varepsilon}\ ,
\ee
and hence this graph gives a divergent contribution.
Adding graph $\alpha^\prime$ turns out to cancel the divergence
(a fact which has already been noticed by Kovchegov \cite{kov}).
To take to latter into account,
we multiply $\alpha$ by the factor 
$F$ defined in Eq.~(\ref{eq:BLtrick}),
the energy of the gluon being $\bar E$, and 
$\Sigma\equiv\bar E-i\varepsilon$:
\be
F_\alpha=1-\frac{\bar E-i\varepsilon}{\bar E}
=\frac{i\varepsilon}{\bar E}.
\ee
The sum of the graphs $\alpha$ and $\alpha^\prime$ reads
\be
\alpha+\alpha^\prime=-g^6 C_F^3\frac{1}{(\vec p+\vec l_{2})^2}\frac{1}{k_+},
\ee
which is now finite.
The sum of the graphs 
$\beta+\beta^\prime$ has exactly the same expression.
Comparing to Eq.~(\ref{eq:loi(i)}), we see that the
following identity holds:
\be
a=\alpha+\alpha^\prime+\beta+\beta^\prime.
\ee
The purely final state graph $b$ corresponds, as
we may check by explicit calculation, to the complex
conjugate graphs:
\be
b=\bar\alpha+\bar\alpha^\prime+\bar\beta+\bar\beta^\prime.
\ee
This completes the proof that at leading logarithmic order, the evolution
kernel of the $p_\perp$-broadening amplitude is identical
to the evolution kernel of the dipole $S$-matrix.

\section{\label{sec4}Proof of the equivalence at next-to-leading order}

We consider systematically all relevant
graphs in the view of proving
the equivalence of the evolution kernels at next-to-leading order,
namely, that the identity~(\ref{eq:correspondence})
is preserved by the energy evolution
at that accuracy.
Therefore, we keep treating the quark-gluon vertices
in the eikonal approximation, but the triple gluon vertices
which contribute to the evolution kernel
are computed exactly.

We classify the graphs according to the number
of quark-gluon vertices. There are two, three of four such vertices.
In the first case, the diagrams contain a gluon loop, but it turns out
that we do not need to evaluate it explicitly to prove the equivalence.
In the second case instead, 
the three-gluon vertex has to be
computed and
the different gluon polarizations fully discussed.
Furthermore,
we find that a momentum-by-momentum identification
no longer holds, and the identification results from
an analytical continuation of the integration over the longitudinal
component of the momenta.
Finally, in the third case, the classes of diagrams
that have to be considered together for the identification
to happen are larger. Moreover, while it is enough to analyze the
energy denominators in order to prove the equivalence,
which makes each calculation quite easy, 
the graphs are numerous, and
analytical continuation is also needed for some diagrams.


\subsection{Two quark-gluon vertices}

We will first analyze the $p_\perp$-broadening
graphs, then the dipole graphs, and the comparison
will be drawn in the last subsection.

\subsubsection{$p_\perp$-broadening graphs}

\begin{figure}
\begin{center}
\includegraphics[width=8cm]{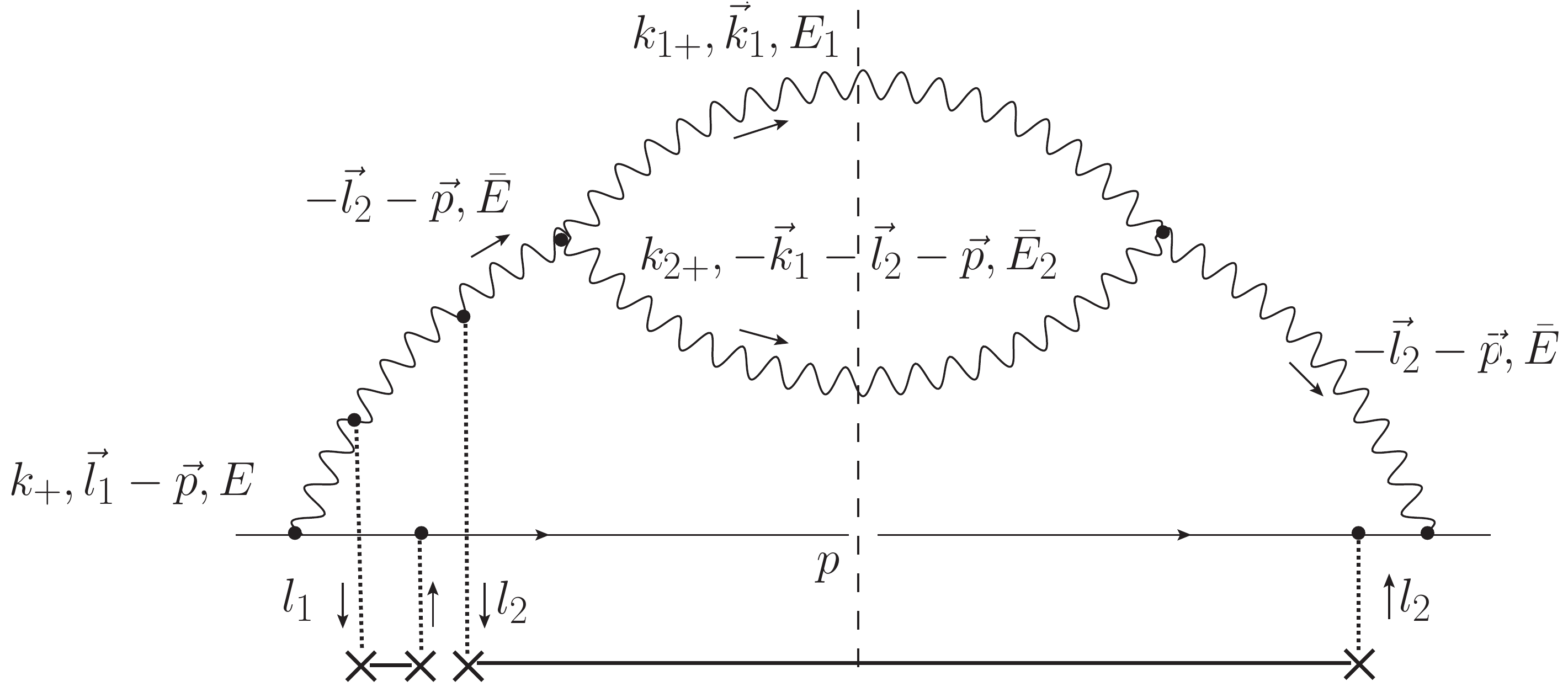}
\end{center}
\caption{\label{q2ifif_labeled}
One particular graph 
contributing to $p_\perp$-broadening at next-to-leading order
and exhibiting two quark-gluon vertices exactly,
as far as quantum evolution is concerned,
with one elastic and one inelastic interaction with
the target.
The energies and momenta of the particles are
labeled explicitly.
}
\end{figure}

With two quark-gluon vertices exactly,
the next-to-leading order graphs contain
a gluon loop. 
One such graph 
contributing to $p_\perp$ broadening
is represented in Fig.~\ref{q2ifif_labeled}.
The other graphs only differ from that one
by the chronology of the 
interactions.

Hence we group the graphs according to the time at which
the quark-gluon and gluon-gluon branchings occur, 
in the amplitude and in the complex conjugate amplitude,
relatively to the time at which
the interaction with the target nucleus occurs.
Each such class of (nonvanishing) $p_\perp$-broadening 
graphs corresponds
to a particular set of dipole graphs, which we will discuss
in Sec.~\ref{sec:dipoles}
after having reviewed all  $p_\perp$-broadening graphs.

We anticipate the fact that the only 
nontrivial factors to compare
between
the contribution of a given set of
$p_\perp$-broadening graphs and of dipole graphs
are the energy denominators: One may therefore
avoid the
discussion
of the gluon loop.

\begin{figure}
\begin{center}
\begin{tabular}{ccc}
\includegraphics[width=4cm]{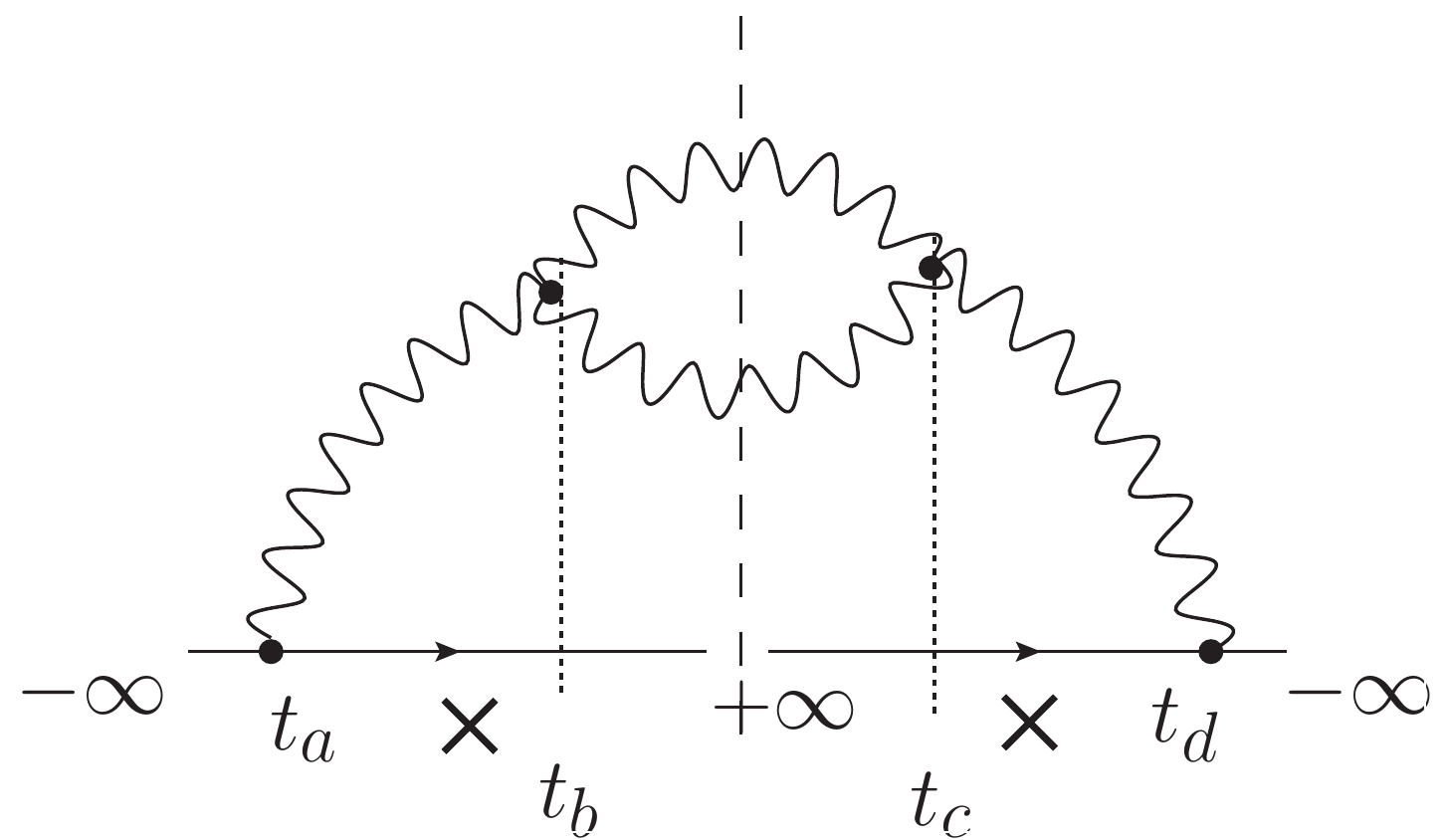}&
\includegraphics[width=4cm]{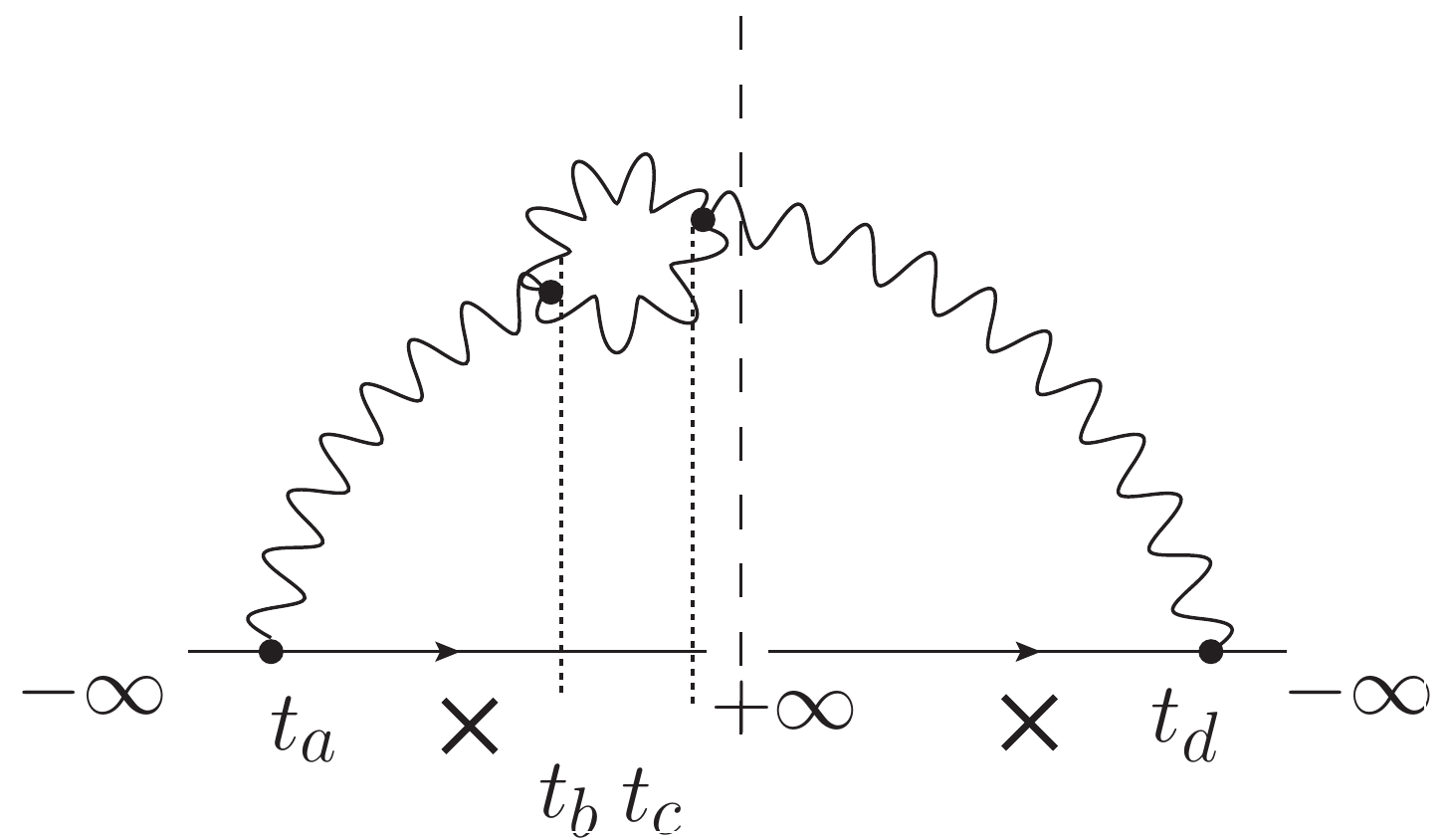}&
\includegraphics[width=4cm]{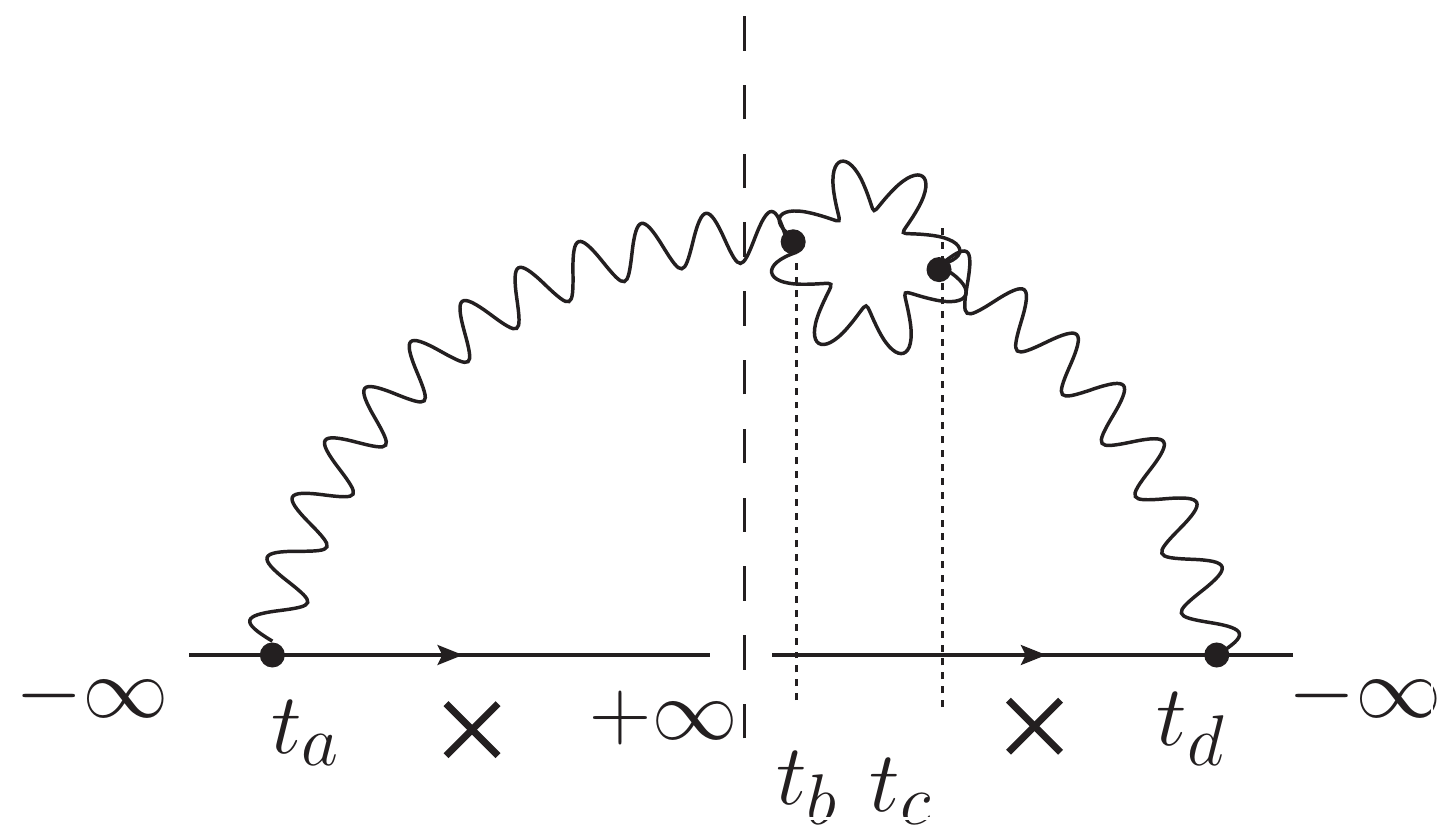}\\
$\graph{IF}{IF}$ & $\graph{IFF}{I}$ & $\graph{I}{IFF}$
\end{tabular}
\end{center}
\caption{\label{q2ii}
Complete set of graphs for which the $qg$ vertices are in the
initial state and the 3-gluon vertices in the final state.
The interactions with the nucleus are understood since
they are the same as in
Fig.~\ref{q2ifif_labeled} for all three graphs.
}
\end{figure}

We shall name the $p_\perp$-broadening graphs
with the help of the ordering of their vertices
with respect to the scattering. For example
$\graph{IF}{IF}$ (Fig.~\ref{q2ii})
means one $qg$ vertex at early times
and one 3-gluon vertex at late times in the amplitude (left of the cut),
and the same in the complex-conjugate amplitude (right of the cut).

\paragraph*{Both quark-gluon vertices in the initial state.}

Let us first consider the case in which the $qg$ branchings
are in the initial state both in the amplitude and in the complex
conjugate amplitude (namely the $qg$ vertices occur at
negative times).
The $gg$ vertices may then occur either
at positive or negative times.

The first case we examine is when
the $gg$ branchings
both occur in the final state.
The full set of graphs is shown
in Fig.~\ref{q2ii}.
We see that we can label 
all momenta exactly as in Fig.~\ref{q2ifif_labeled}
since the topology of all the graphs is the same.
Therefore the expressions for these graphs will differ only
in the energy denominators.
Let us evaluate
the latter using Eq.~(\ref{eq:D}).

We label the time at the branchings as $t_a<t_b$ for
the $qg$ and $gg$ branchings in the amplitude respectively,
and $t_d<t_c$ for the same branchings in the complex conjugate amplitude.
The energy denominators corresponding to the graph
in Fig.~\ref{q2ifif_labeled} 
(or equivalently the leftmost graph in Fig.~\ref{q2ii})
are obtained from the evaluation of the
following integral (see Eq.~\ref{eq:D}):
\begin{multline}
D_{\graph{IF}{IF}}=\int_{-\infty}^0 dt_a \int_0^{+\infty} dt_b \int_{+\infty}^0 dt_c
\int_{0}^{-\infty} dt_d\, \\
\exp\left\{i\left[E t_a+(E_1+\bar E_2-\bar E )(t_b-t_c)
-\bar E t_d\right]+\varepsilon(t_a-t_b-t_c+t_d)\right\}.
\end{multline}
We have defined (see Fig.~\ref{q2ifif_labeled})
\begin{equation}
E =\frac{(\vec p-\vec l_{1})^2}{2k_+},\
\bar E =\frac{(\vec p+\vec l_{2})^2}{2k_+},\
E_1=\frac{\vec k_{1}^2}{2k_{1+}},\ 
\bar E_2=\frac{(\vec k_{1}+\vec l_2+\vec p)^2}{2k_{2+}},
\end{equation}
and from the conservation of the longitudinal momentum,
$k_{2+}=k_+-k_{1+}$.
(Of course, the ``bar'' does not mean complex conjugate in these
equations).
The result of the integrations over the different
times reads
\begin{equation}
D_{{\graph{IF}{IF}}}=\frac{1}{E \bar E 
\left(E_1+\bar E_2-\bar E \right)^2}.
\end{equation}
We perform a similar calculation for the two other graphs in
this group. The one in which the loop is in the amplitude
(Fig.~\ref{q2ii}, middle)
has the following energy denominators, after expansion
for $\varepsilon\rightarrow 0$:
\begin{multline}
D_{{\graph{IFF}{I}}}=-\frac{1}{2i\varepsilon}\frac{1}{E \bar E 
\left(E_1+\bar E_2-\bar E \right)}
-\frac12\bigg[
\frac{1}{E \bar E 
\left(E_1
+\bar E_2-\bar E \right)^2}\\
+\frac{1}{E ^2\bar E 
\left(E_1+\bar E_2-\bar E \right)}
-\frac{1}{E \bar E ^2
\left(E_1+\bar E_2-\bar E \right)}
\bigg].
\label{eq:q2iffi}
\end{multline}
This time, there is a divergent term
when $\varepsilon\rightarrow 0$, which is however imaginary.
The graph in which the loop is in the complex conjugate amplitude
has a similar expression up to sign changes, in such a way
that in the sum $D_{{\graph{IFF}{I}}}+D_{{\graph{I}{IFF}}}$, 
the divergent imaginary part cancels
as well as the two last terms in Eq.~(\ref{eq:q2iffi}) while the second
term gains a factor 2.
The sum of the 3 denominators then cancels,
\be
D_{{\graph{IF}{IF}}}+D_{{\graph{IFF}{I}}}+D_{{\graph{I}{IFF}}}=0,
\ee
and therefore, the total contribution of these graphs is zero.\footnote{%
Such cancellations, which are related to probability
conservation and occur when one sums over a sufficiently
inclusive subset of graphs,
appear in many different contexts, see e.g.
Ref.~\cite{Kovchegov:2007vf}.
}

\begin{figure}
\begin{center}
\begin{tabular}{c|c}
\includegraphics[width=5cm]{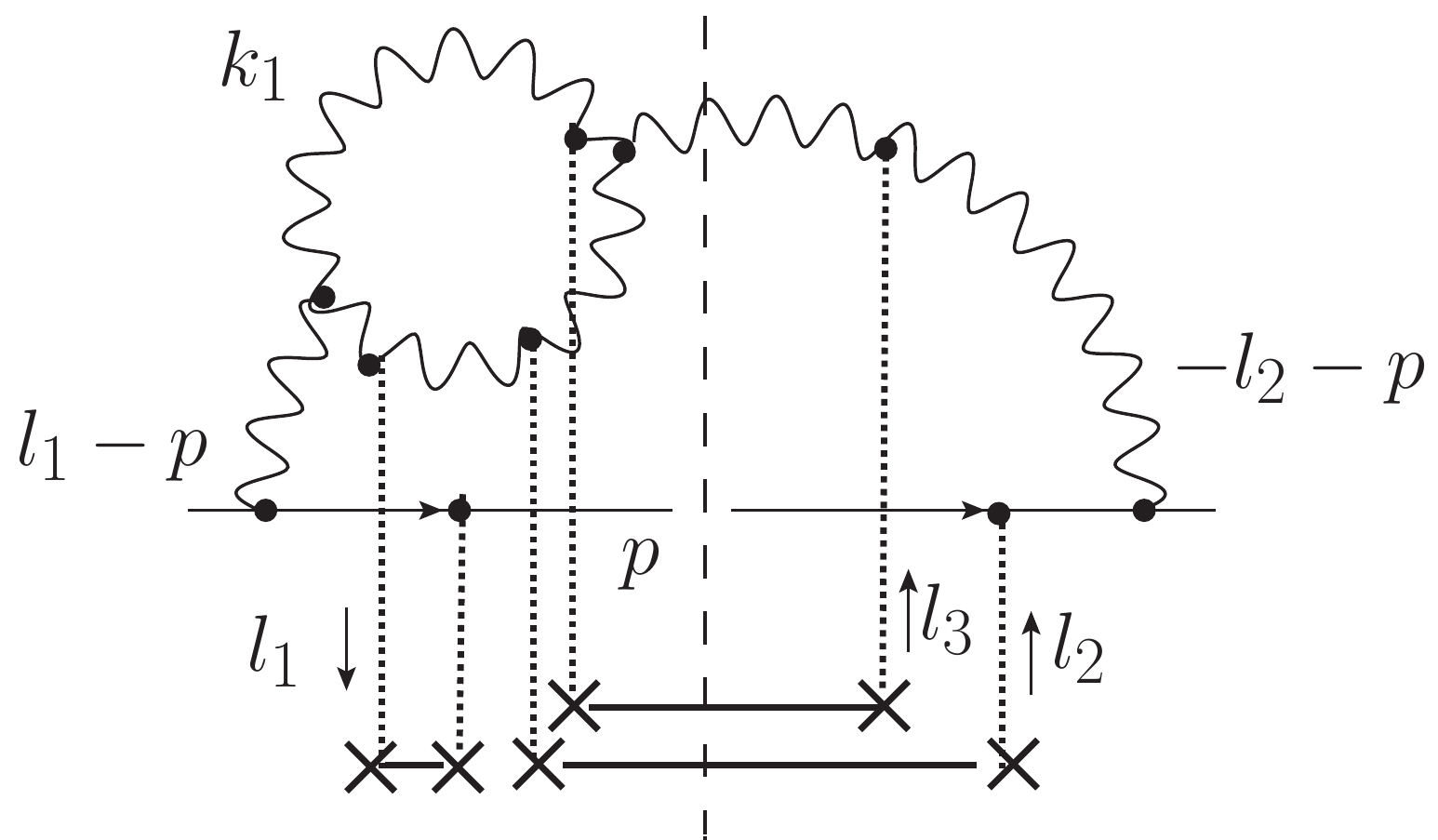}&
\includegraphics[width=4cm]{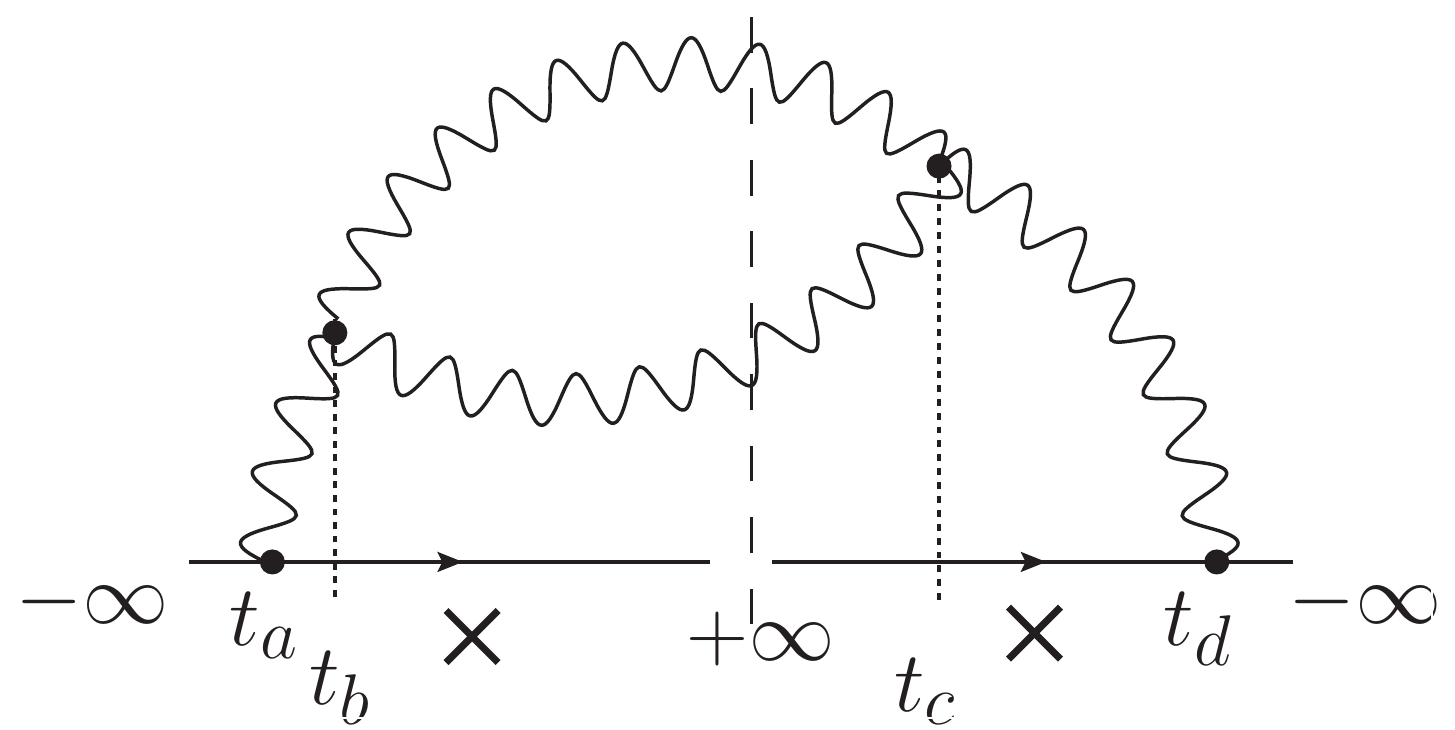}\\
\\
\includegraphics[width=4cm]{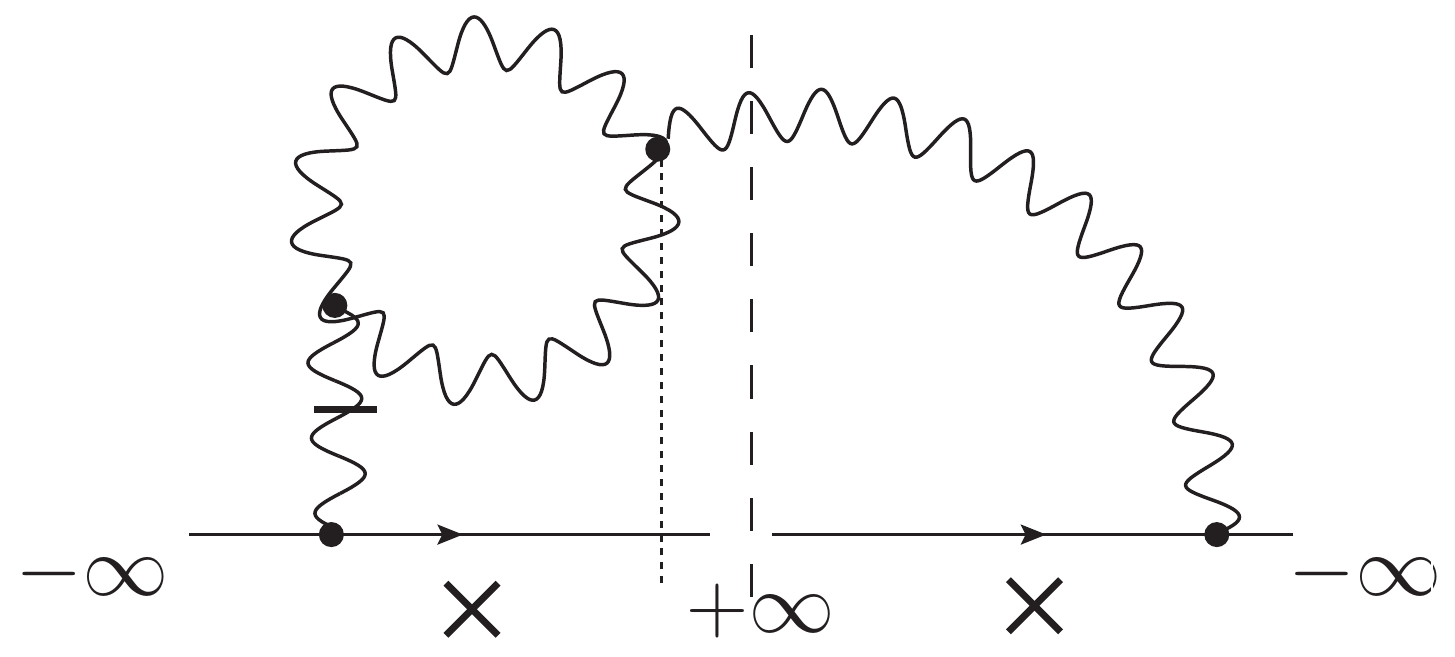}&
\includegraphics[width=4cm]{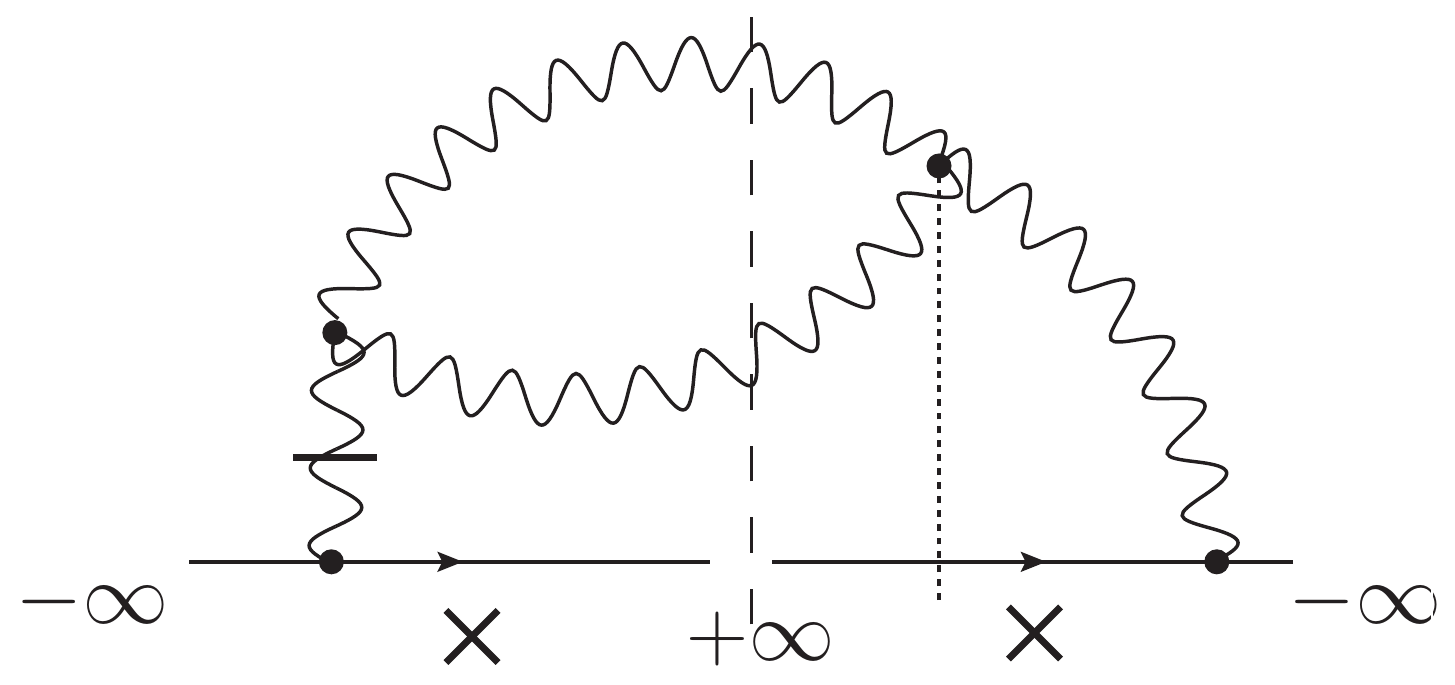}\\
$\graph{IIF}{I}$ & $\graph{II}{IF}$
\end{tabular}
\end{center}
\caption{\label{q2iifi}
Graphs in which the two quark-gluon vertices are
in the initial state, one 3-gluon vertex is in the initial state
in the amplitude, and the other one is in the final state,
either in the amplitude or in the complex-conjugate amplitude.
This set should be supplemented with the symmetric ones, 
obtained by taking
the complex conjugate.
}
\end{figure}
Let us turn to the case in which the leftmost $gg$ branching occurs
before the time of the scattering in the amplitude
while the rightmost one occurs in the final state, either
in the amplitude or in the complex conjugate amplitude.
Then there are the four diagrams depicted 
in Fig.~\ref{q2iifi}.
The Lorentz structure is the same for the 
graphs~$\graph{IIF}{I}$ and~$\graph{II}{IF}$.
We evaluate the energy denominators 
of the causal graphs (first row in Fig.~\ref{q2iifi})
as
\be
D_{\graph{IIF}{I}}=\frac{1}{E
\bar E(E_1+E_2)
(E_1^\prime+\bar E^\prime_2-\bar E ^\prime)},
\ee
where
\be
E_2=\frac{(\vec p-\vec l_1+\vec k_1)^2}{2k_{2+}},\
E_1^\prime=\frac{(\vec k_1-\vec l_3)^2}{2k_{1+}},\
\bar E_2^\prime=\frac{(\vec p+\vec k_1+\vec l_2)^2}{2k_{2+}},\
\bar E ^\prime=\frac{(\vec p+\vec l_2+\vec l_3)^2}{2k_+},
\ee
see Fig.~\ref{q2iifi}.
Computing in the same way $D_{\graph{II}{IF}}$,
we find that $D_{\graph{II}{IF}}=-D_{\graph{IIF}{I}}$.
Since all other factors are the same for these graphs,
their sum cancels, independently 
for each polarization of
the gluons.

When the polarization $(--)$ is chosen for
the leftmost gluon in $\graph{IIF}{I}$ and $\graph{II}{IF}$, then
each of these graphs also cancels with
the same graphs where the latter gluon is exchanged instantaneously
(graphs in the second row in Fig.~\ref{q2iifi}).
Indeed, according to the rule~(\ref{eq:BLtrick}), 
the polarization tensor $d_{--}$
for this gluon
has to be multiplied by the factor 
\be
F_{\graph{IIF}{I},\graph{II}{IF}}\equiv
1-\frac{E -i\varepsilon}{E }
=\frac{i\varepsilon}{E }.
\ee
The product of this factor with the corresponding 
energy denominators vanishes
since $D_{\graph{IIF}{I},\graph{II}{IF}}$ 
are finite for $\varepsilon\rightarrow 0$.
So there is no global 
contribution of the instantaneous-exchange graphs either.

\begin{figure}
\begin{center}
\begin{tabular}{ccc}
\includegraphics[height=2.5cm]{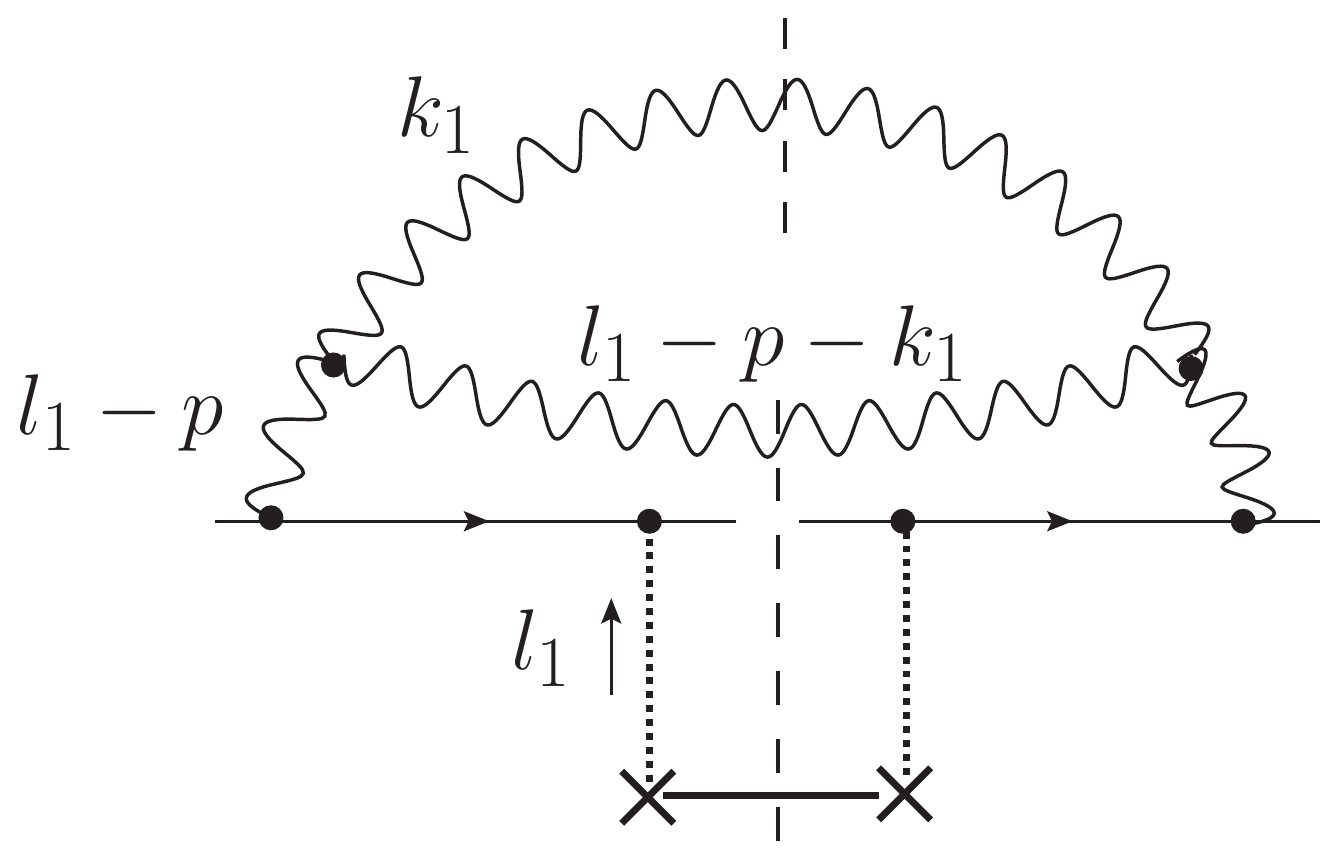}&
\includegraphics[height=2.5cm]{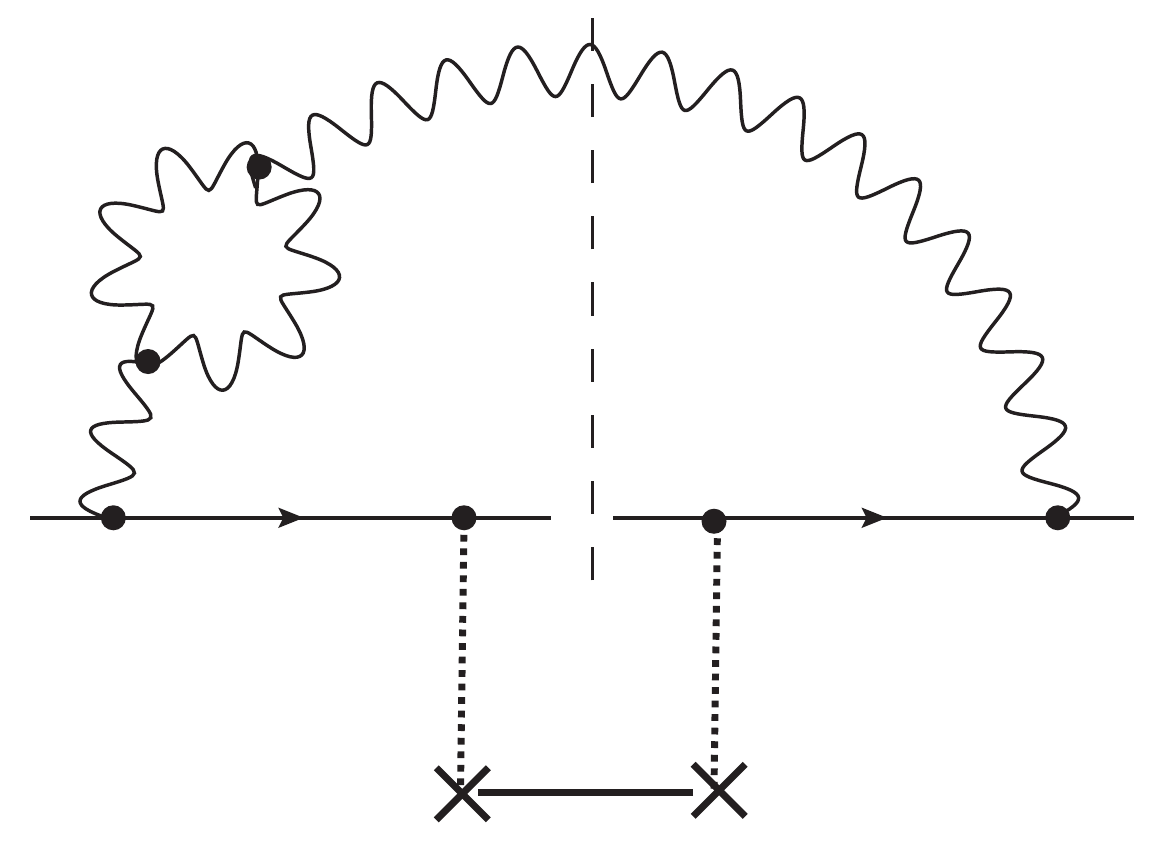}&
\includegraphics[height=2.5cm]{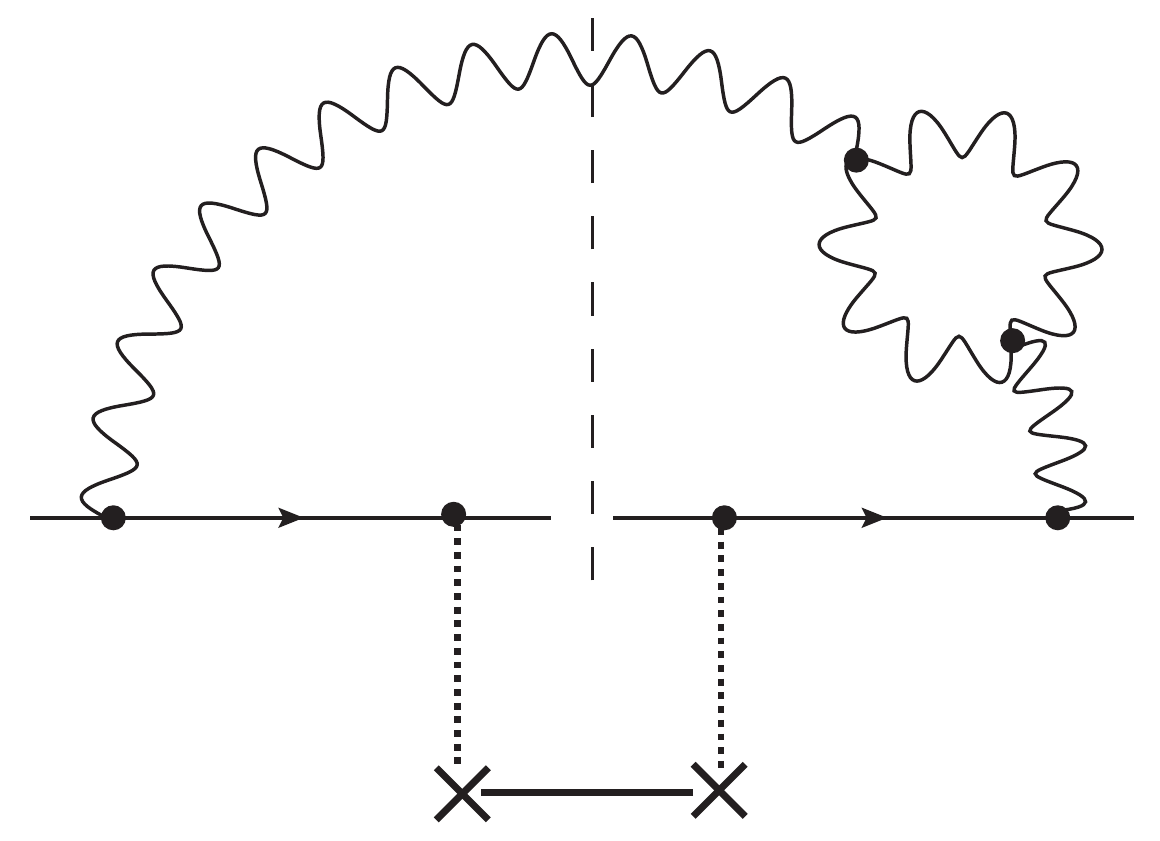}\\
$\graph{II}{II}$ & $\graph{III}{I}$ & $\graph{I}{III}$
\end{tabular}
\end{center}
\caption{\label{q2iiii}
Causal graphs in which all vertices are in the initial state.
The scattering with the target occurs with the quark: Graphs
in which one of the gluons scatters
cancel among each other.
}
\end{figure}

\begin{figure}
\begin{center}
\begin{tabular}{ccc}
\includegraphics[height=2.5cm]{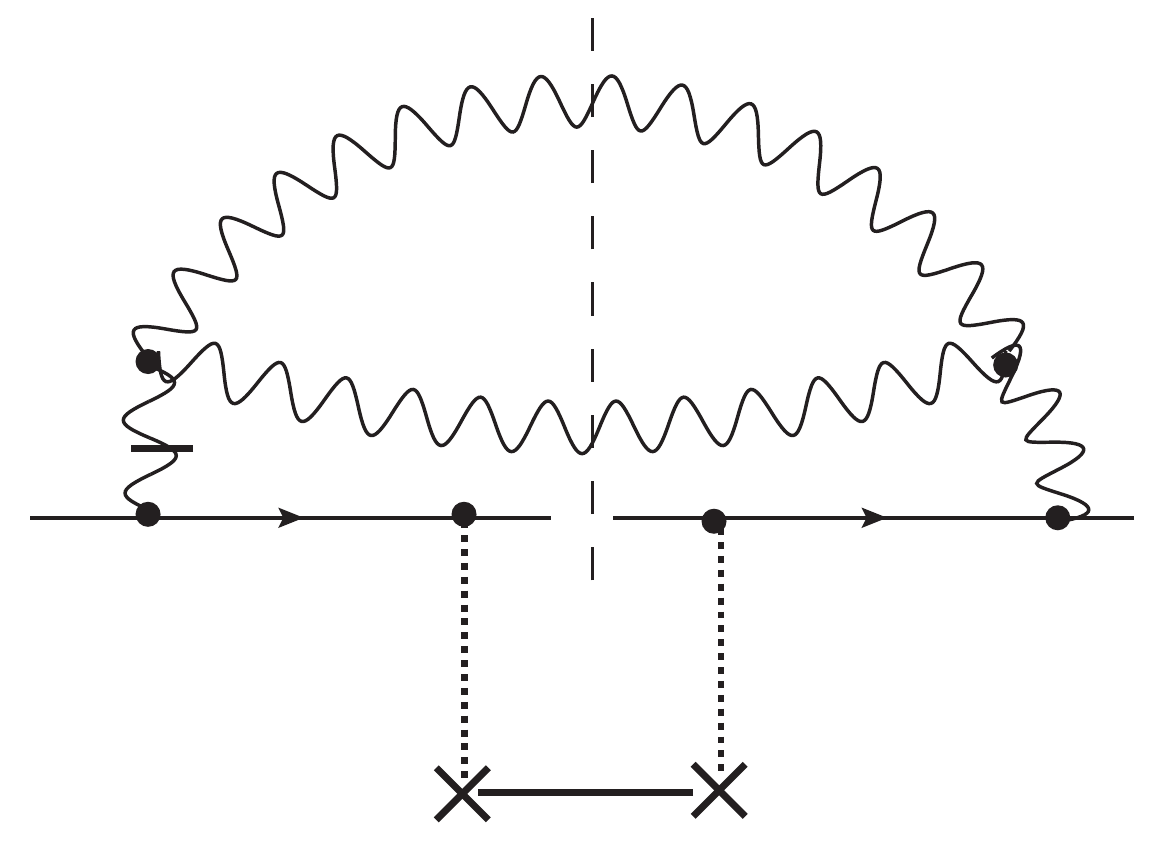}&
\includegraphics[height=2.5cm]{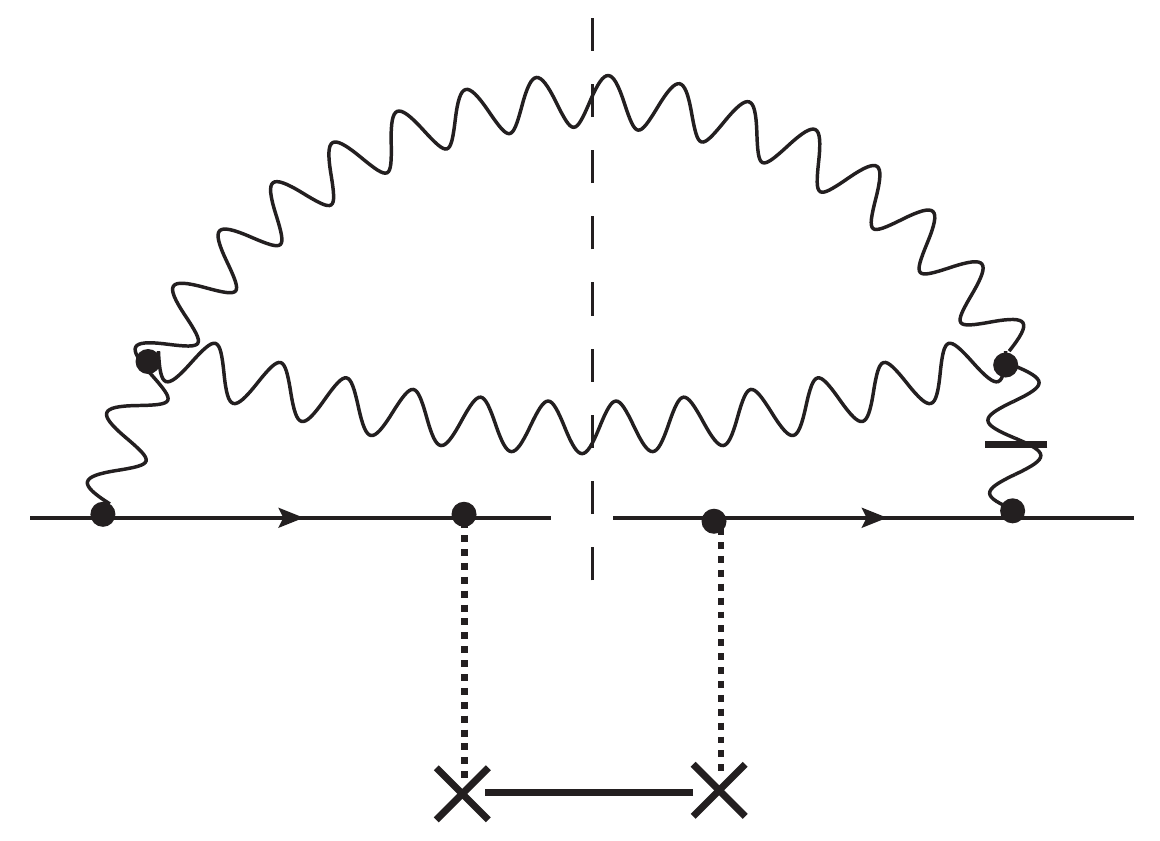}&
\includegraphics[height=2.5cm]{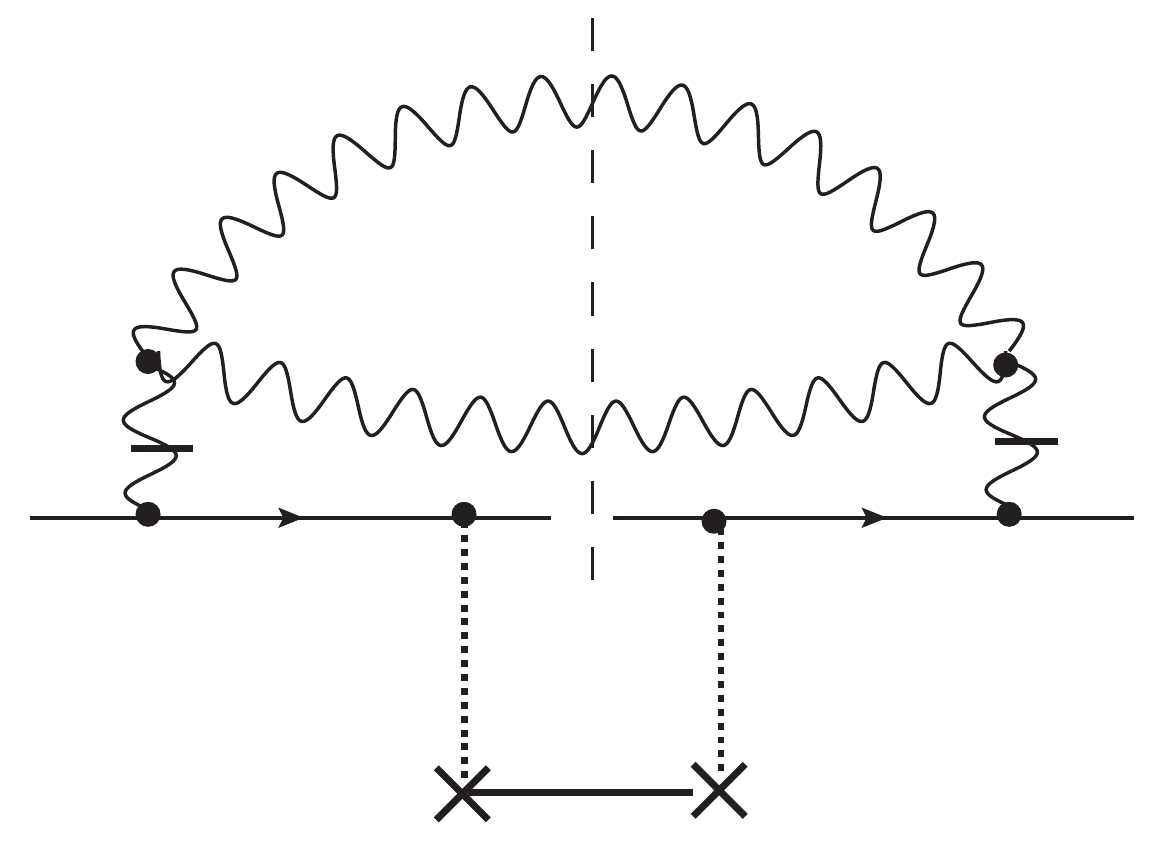}
\end{tabular}
\end{center}
\caption{\label{q2iiiis}
Instantaneous-exchange graphs corresponding to graph 
$\graph{II}{II}$ in Fig.~\ref{q2iiii}.
They are effectively taken into account in our calculation
by multiplying $\graph{II}{II}$ for the longitudinal
polarization of the gluons which attach to the quarks
by two $F$-factors (see Eq.~(\ref{eq:BLtrick})).
}
\end{figure}

\begin{figure}
\begin{center}
\begin{tabular}{ccc}
\includegraphics[height=2.5cm]{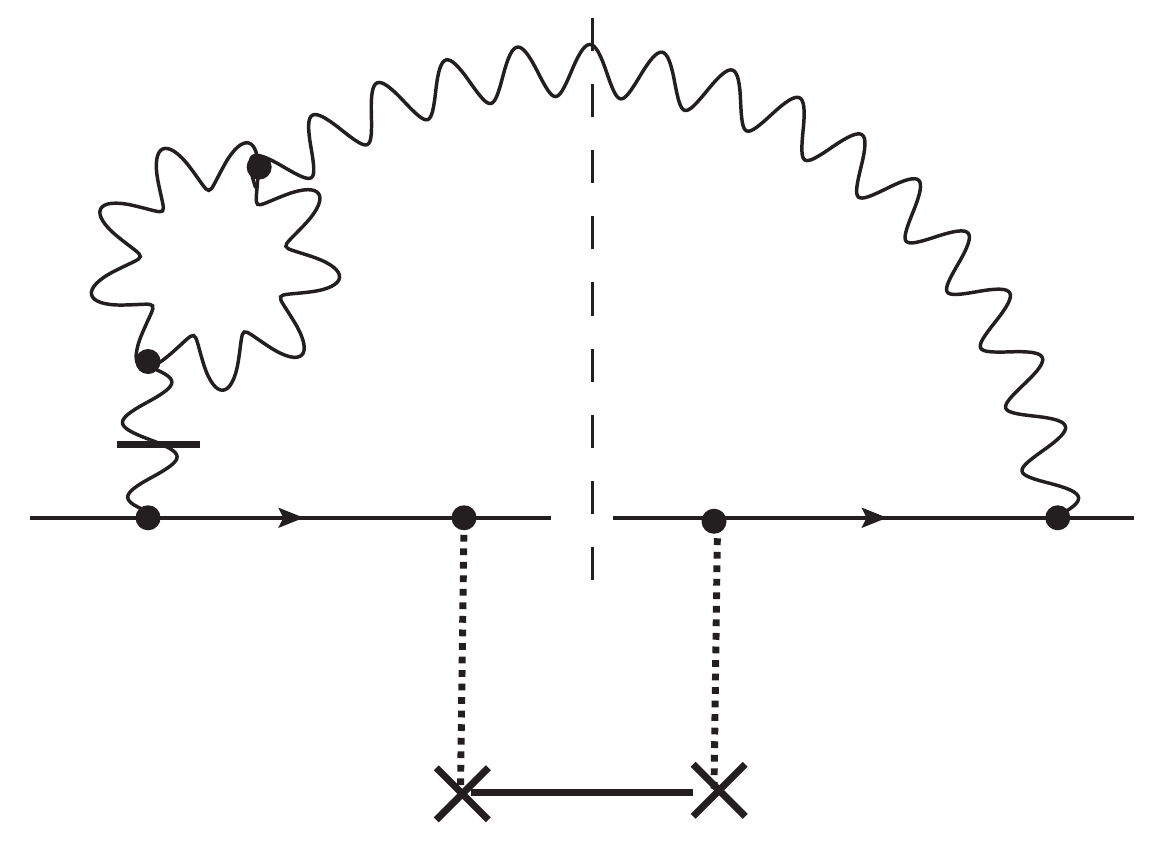}&
\includegraphics[height=2.5cm]{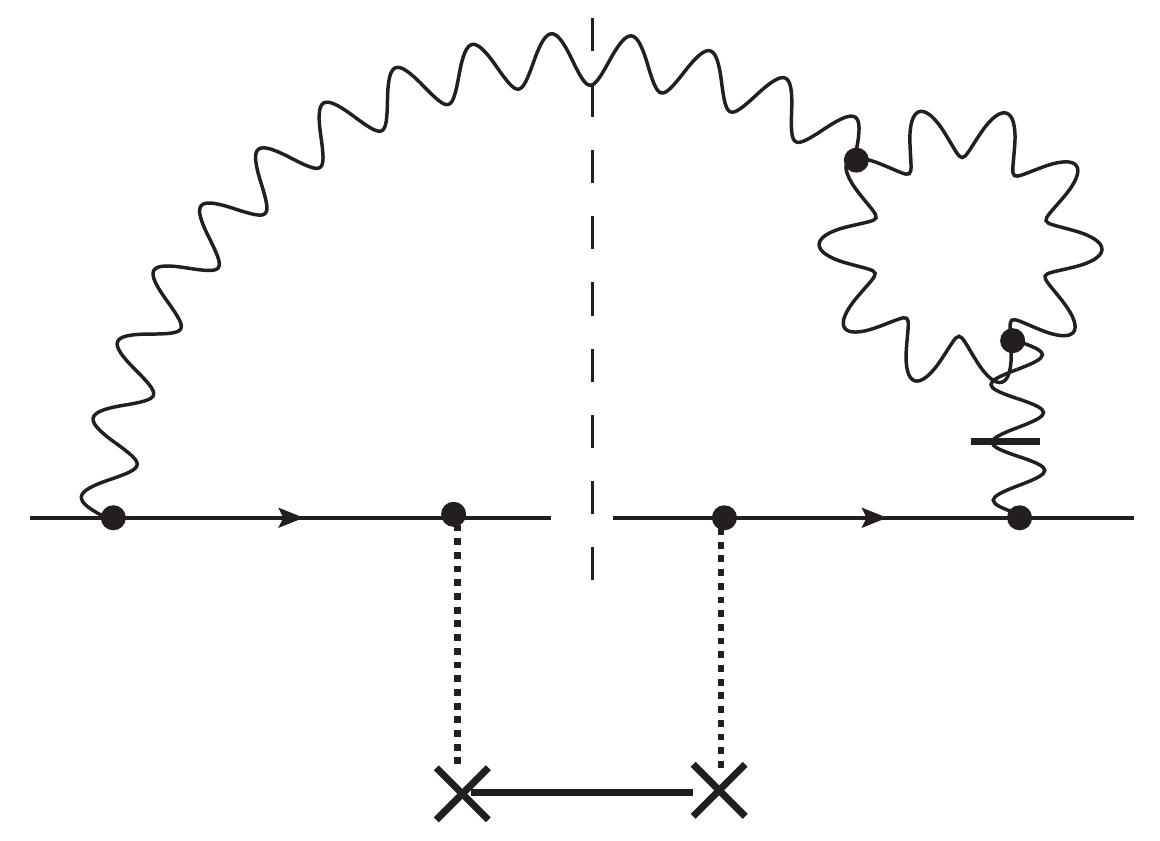}
\end{tabular}
\end{center}
\caption{\label{q2iiiiss}
Instantaneous-exchange graphs corresponding to graphs 
$\graph{III}{I}$ and $\graph{I}{III}$
in Fig.~\ref{q2iiii}.
}
\end{figure}

So far, we have not found any contribution to $p_\perp$-broadening
of the graphs that we have examined.
We turn to the last class of graphs in which both three-gluon vertices
are in the initial state, either in the amplitude or in the
complex-conjugate amplitude.
Again in this case, 
as for the leading-order graph $a$ in Fig.~\ref{lo},
all graphs in which one of the gluons in the
wave functions scatters with the nucleus
cancel among themselves.
The complete set of graphs in which the gluons are causal is 
shown in Fig.~\ref{q2iiii}.
There are in addition graphs in which at least one gluon is replaced
by a contact interaction, see Fig.~\ref{q2iiiis} and~\ref{q2iiiiss}.

Before we proceed to the computation of the energy denominators,
we need to discuss the helicity structure of the gluons.
The leg of the gluons which attached to the quark always
carries
the longitudinal polarization ``$-$'' since the 
$qg$ vertices are eikonal.
The other leg of these gluons is either longitudinally, or transversely
polarized. Actually, in 
these graphs in which no interaction occurs between the gluons in
the wave function and the target, 
the leg which attaches to the loop of both of these
gluons must
have the same polarization.
Hence both gluons come either with a $d_{--}$ or with a $d_{-\perp}$ 
polarization factor.

In order to easily incorporate the instantaneous-exchange 
graphs using the modification~(\ref{eq:BLtrick}) of $d_{--}$,
we will distinguish the two possible polarizations:
In the $(-\perp)$ case, it will be enough to consider the
sum of the
energy denominators of the relevant graphs, 
while in the $(--)$ case,
we will have to weight them by 
appropriate $F$-factors 
computed with the help of
Eq.~(\ref{eq:BLtrick}).

We go back to the graphs of Fig.~\ref{q2iiii}.
The energy denominators read
\be
D_{\graph{II}{II}}=\frac{1}{E ^2(E_1+E_2)^2},\
D_{\graph{III}{I}}=D_{\graph{I}{III}}=
\frac{1}{E ^3(E_1+E_2)}
\ee
The 
sum of the denominators
reads
\be
D_{\graph{II}{II}}+D_{\graph{III}{I}}+D_{\graph{I}{III}}=
\frac{1}{E ^2(E_1+E_2)^2}+\frac{2}{E ^3(E_1+E_2)}.
\label{eq:resq2iiii}
\ee

The incorporation of the graphs of Fig.~\ref{q2iiiis}
and~\ref{q2iiiiss} require to multiply $D_{\graph{III}{I}}$
and $D_{\graph{I}{III}}$ by
\be
F_{\graph{III}{I}}=1-\frac{E -i\varepsilon}{E }
=\frac{i\varepsilon}{E }\ ,\ \
F_{\graph{I}{III}}
=-\frac{i\varepsilon}{E }
\ee
and $D_{\graph{II}{II}}$ by $F_{\graph{III}{I}}\times F_{\graph{I}{III}}$.
So because of the finiteness
of the energy denominators,
once we add the instantaneous-exchange graphs, the contribution
of the $(--)$ polarization vanishes.


\paragraph*{Both quark-gluon vertices in the final state.}

\begin{figure}
\begin{center}
\begin{tabular}{ccc}
\includegraphics[height=2.5cm]{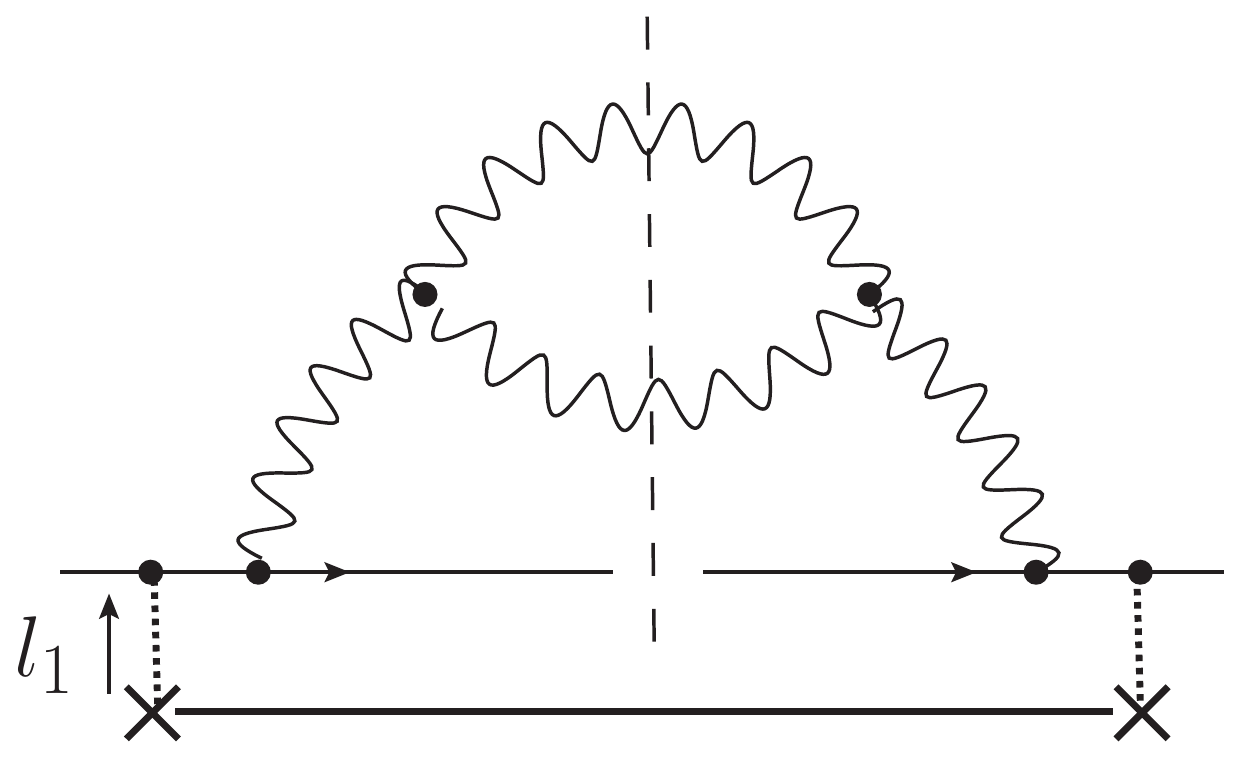}&
\includegraphics[height=2.5cm]{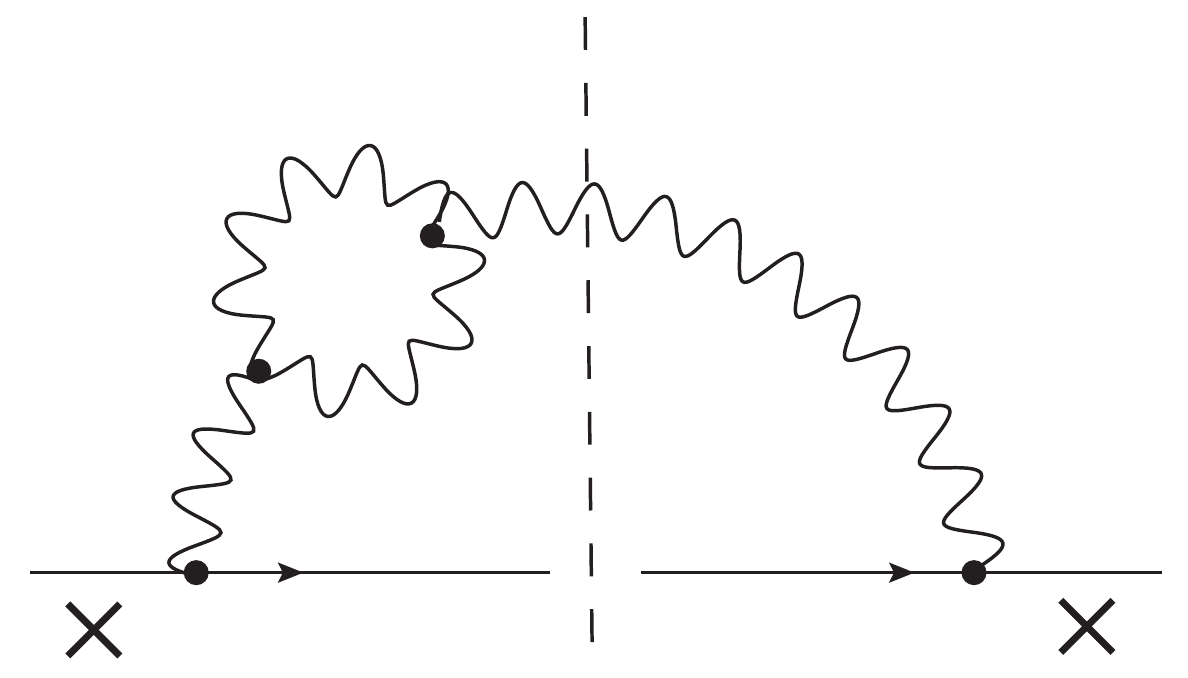}&
\includegraphics[height=2.5cm]{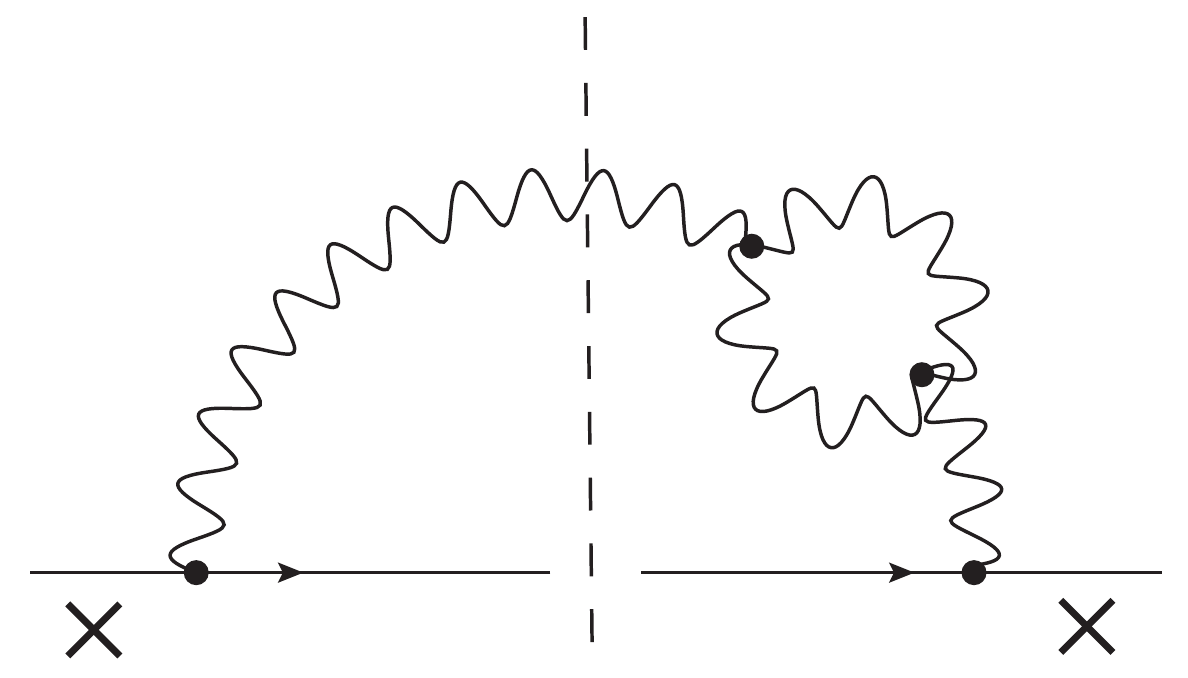}\\
$\graph{FF}{FF}$ & $\graph{FFF}{F}$ & $\graph{F}{FFF}$
\end{tabular}
\end{center}
\caption{\label{q2ffff}
Set of the causal graphs in which all vertices are in the final state.
One should also consider the graphs where some gluons are exchanged
instantaneously: We do not show them since 
they look exactly like those of Fig.~\ref{q2iiiis}
and~\ref{q2iiiiss}.
}
\end{figure}

There are 3 causal graphs
in which the gluon is emitted off the
quark at late times
both in the amplitude and in the complex conjugate amplitude.
They are represented in Fig.~\ref{q2ffff}.
Since the kinematics is the same for all these graphs,
it is again enough to address the energy denominators.

They read
\be
\begin{split}
D_{\graph{FF}{FF}}&=\frac{1}{(E_1 +E_2 )^2
\left(E_1 +E_2 -E  \right)^2}\\
D_{\graph{FFF}{F}}&=-\frac{1}{2i\varepsilon}
\frac{1}{{E  }^2
\left(E_1 +E_2 -E  \right)}
+\frac{1}{{E  }^3
\left(E_1 +E_2 -E  \right)}\\
&\  \ \ \ -\frac12
\frac{1}{{E  }^2
\left(E_1 +E_2 -E  \right)^2}\\
D_{\graph{F}{FFF}}&=\bar{D}_{\graph{FFF}{F}}.
\end{split}
\ee
After some rearrangements, their sum reduces to
\be
D_{\graph{FF}{FF}}+D_{\graph{FFF}{F}}+D_{\graph{F}{FFF}}
=\frac{1}{{E }^2
(E_1+E_2)^2}
+\frac{2}{{E }^3(E_1+E_2)},
\label{eq:resq2ffff}
\ee
namely it is identical to Eq.~(\ref{eq:resq2iiii}).
In order to take into account the instantaneous-exchange graphs,
we need to multiply the denominators by the respective 
factors
\be
F_{\graph{FF}{FF}}^2=\left(\frac{E_1+E_2}
{{E }}\right)^2,
\
F_{\graph{FFF}{F}}=1+\frac{2i\varepsilon}{E }
\ \ \text{and}\ \ F_{\graph{F}{FFF}}=\bar F_{\graph{FFF}{F}}.
\ee
Then we find that the sum of all instantaneous-exchange graphs
and of the causal graphs with the $(--)$ polarization for the gluons
that couple to the quark
has the factor
\be
F_{\graph{FF}{FF}}^2 D_{\graph{FF}{FF}}+F_{\graph{FFF}{F}}D_{\graph{FFF}{F}}
+F_{\graph{F}{FFF}}D_{\graph{F}{FFF}}=0,
\ee
and hence vanishes.


\paragraph*{One $qg$ vertex in the initial state, 
one in the final state.}

\begin{figure}
\begin{center}
\begin{tabular}{ccc}
\includegraphics[width=4cm]{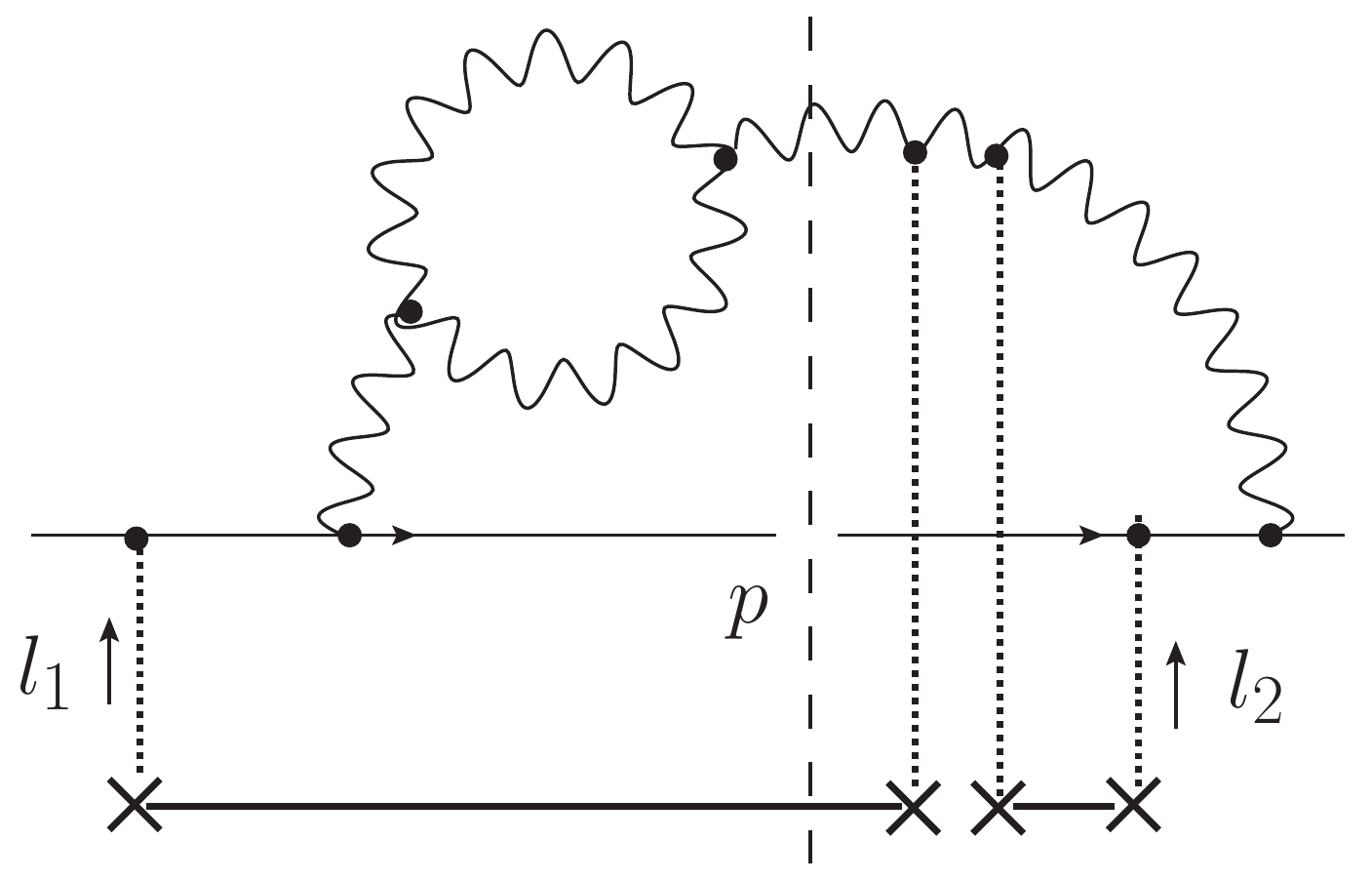}&
\includegraphics[width=4cm]{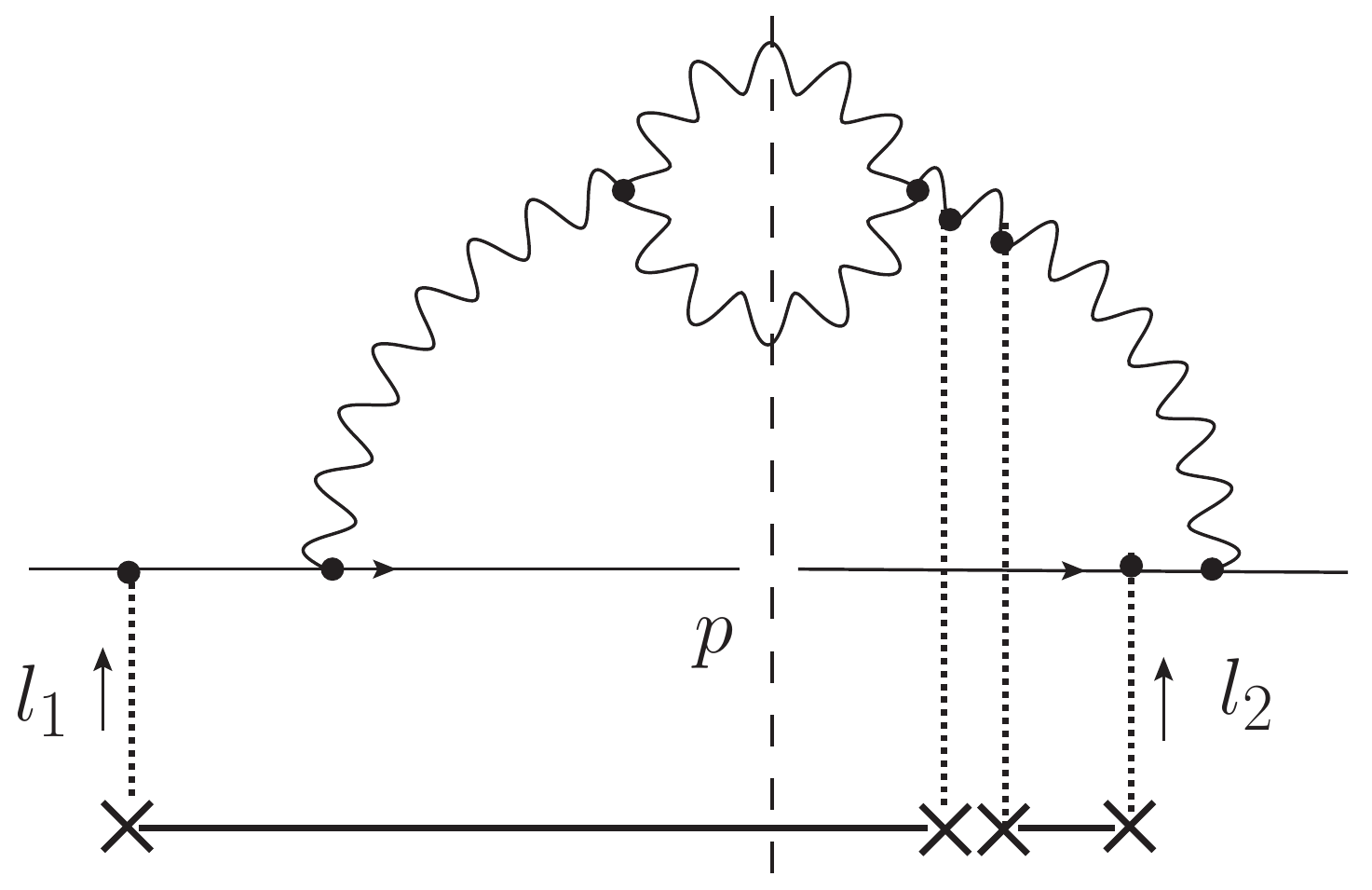}&
\includegraphics[width=4cm]{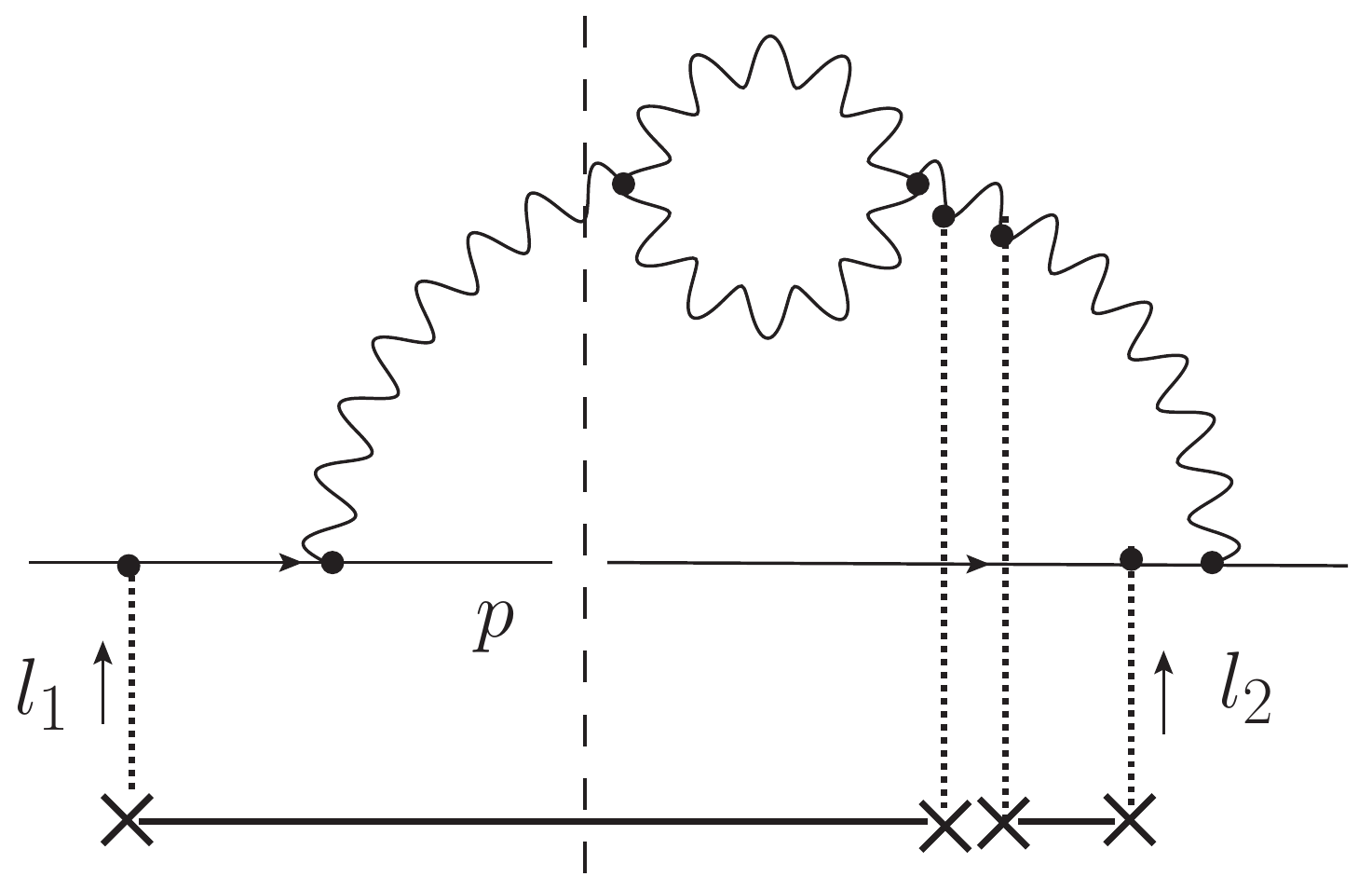}\\
$\graph{FFF}{I}$&$\graph{FF}{IF}$&$\graph{F}{IFF}$
\end{tabular}
\end{center}
\caption{\label{q2fff(i)}
Causal graphs in which one $qg$ vertex is in the initial
state, and one in the final state.
This list should be completed with the instantaneous-exchange graphs
in correspondence with the two leftmost graphs,
and with the complex-conjugate ones.
}
\end{figure}

The causal graphs are represented in Fig.~\ref{q2fff(i)}.
The energy denominators read
\begin{multline}
D_{\graph{FFF}{I}}=-\frac{1}{2i\varepsilon}
\frac{1}{E   \bar E (E  -E_1 -E_2 )}
+\frac12\frac{1}{E   \bar E ^2(E  -E_1 -E_2 )}\\
+\frac32\frac{1}{{E  }^2 \bar E (E  -E_1 -E_2 )}
+\frac12\frac{1}{E   \bar E (E  -E_1 -E_2 )^2}
\end{multline}
for the leftmost graph,
\be
D_{\graph{FF}{IF}}=-\frac{1}{\bar E (E_1 +E_2 )
(E  -E_1 -E_2 )^2}
\ee
for the graph in which there are 2 gluons in the final state
(graph in the middle in Fig.~\ref{q2fff(i)}), and
\begin{multline}
D_{\graph{F}{IFF}}=\frac{1}{2i\varepsilon}
\frac{1}{E   \bar E (E  -E_1 -E_2 )}
-\frac12\frac{1}{E  \bar E ^2(E  -E_1 -E_2 )}\\
-\frac12\frac{1}{{E  }^2 \bar E 
(E  -E_1 -E_2 )}
+\frac12\frac{1}{E   \bar E (E  -E_1 -E_2 )^2}
\end{multline}
for the graph in which the loop is 
on the right of the cut
(rightmost graph in Fig.~\ref{q2fff(i)}).
The sum of these denominators reads
\be
D_{\graph{FFF}{I}}+D_{\graph{FF}{IF}}+D_{\graph{F}{IFF}}=
-\frac{1}{{E }^2 \bar E (E_1+E_2)^2}.
\label{eq:resq2fffi}
\ee
The instantaneous-exchange graphs are taken into account
with the help of the factors
\be
F_{\graph{FFF}{I}}=1+\frac{2i\varepsilon}{E },\
F_{\graph{FF}{IF}}=\frac{E_1+E_2}{E }.
\ee
Again, the contribution of these graphs summed with the
causal graphs which have the $(--)$ polarization vanishes, since
\be
F_{\graph{FFF}{I}}D_{\graph{FFF}{I}}+F_{\graph{FF}{IF}}D_{\graph{FF}{IF}}
+D_{\graph{F}{IFF}}
=0.
\ee

\begin{figure}
\begin{center}
\begin{tabular}{c}
\includegraphics[width=6cm]{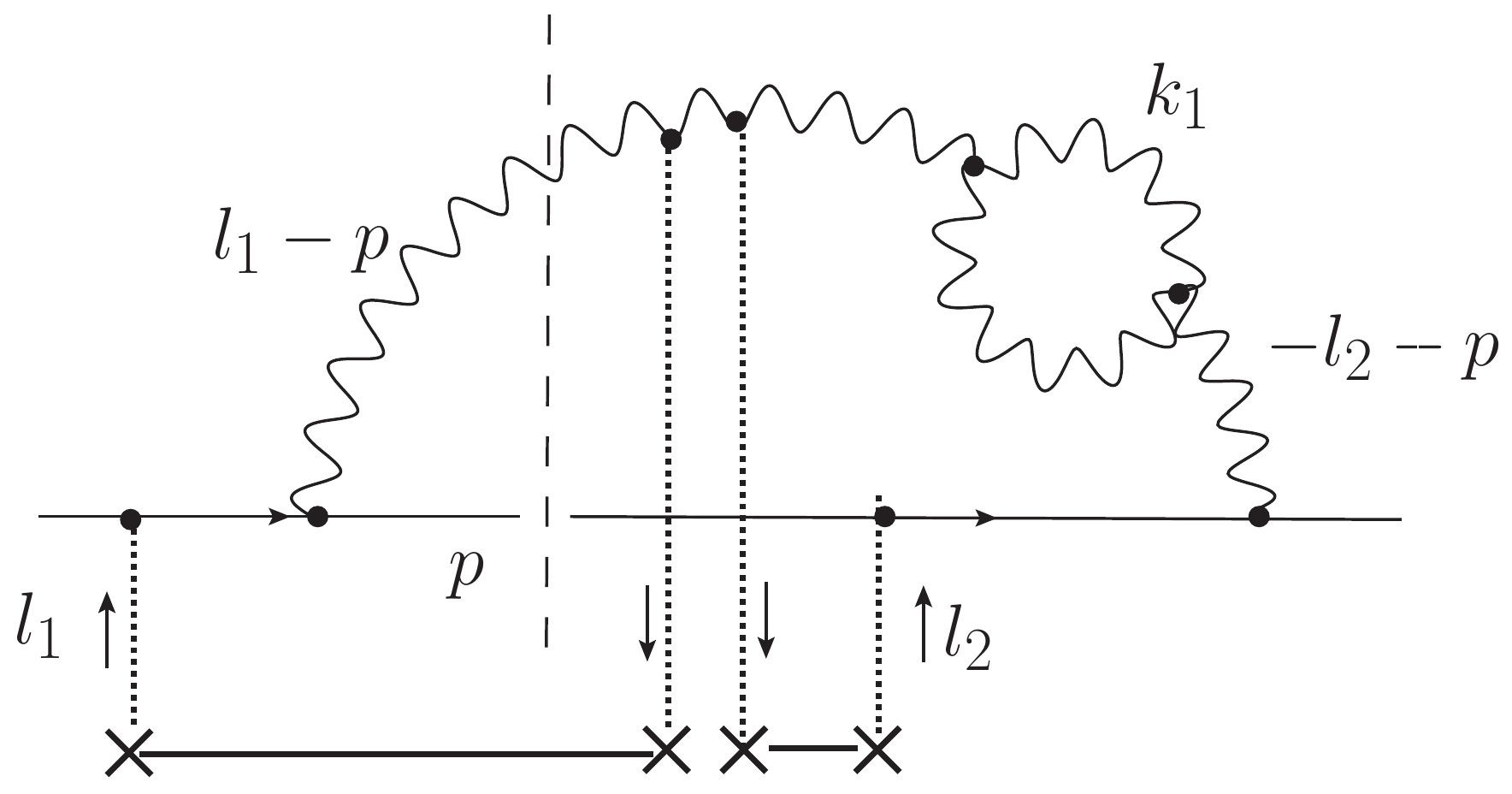}\\
$\graph{F}{III}$
\end{tabular}
\end{center}
\caption{\label{q2f(iii)}
Causal graph in which one $qg$ vertex is in the initial
state, one in the final state, and the gluon loop is
fully in the initial state.
}
\end{figure}
The graph in which the gluon loop is in the initial state
(Fig.~\ref{q2f(iii)})
has the following denominators:
\be
D_{\graph{F}{III}}=
-\frac{1}{E\bar E ^2(E_1+\bar E_2)}.
\label{eq:resq2fiii}
\ee
The sum of 
the instantaneous-exchange graph
with the $(--)$ polarization in the causal graph
once again vanishes.

\begin{figure}
\begin{center}
\begin{tabular}{cc}
\includegraphics[width=6cm]{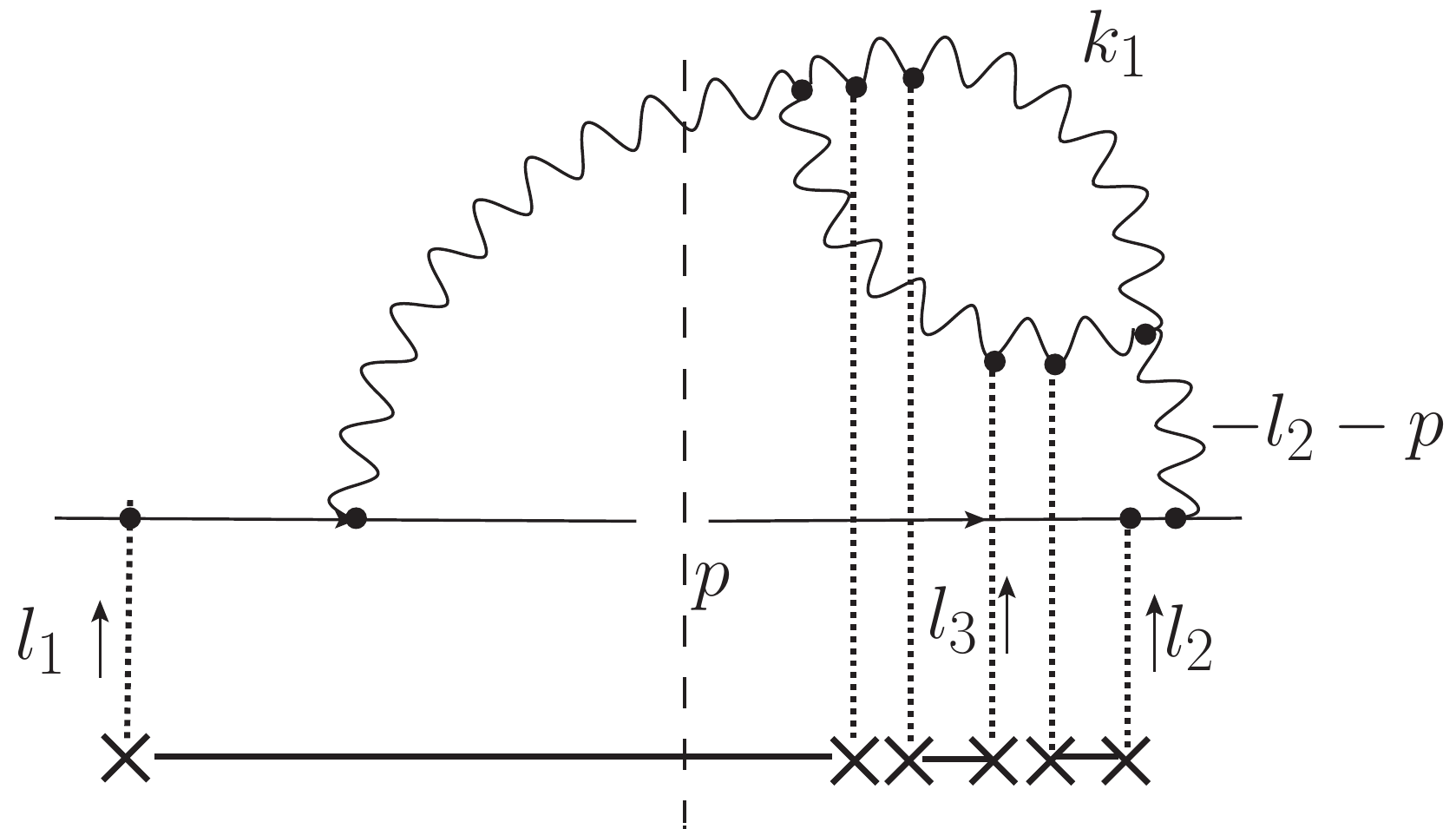}&
\includegraphics[width=6cm]{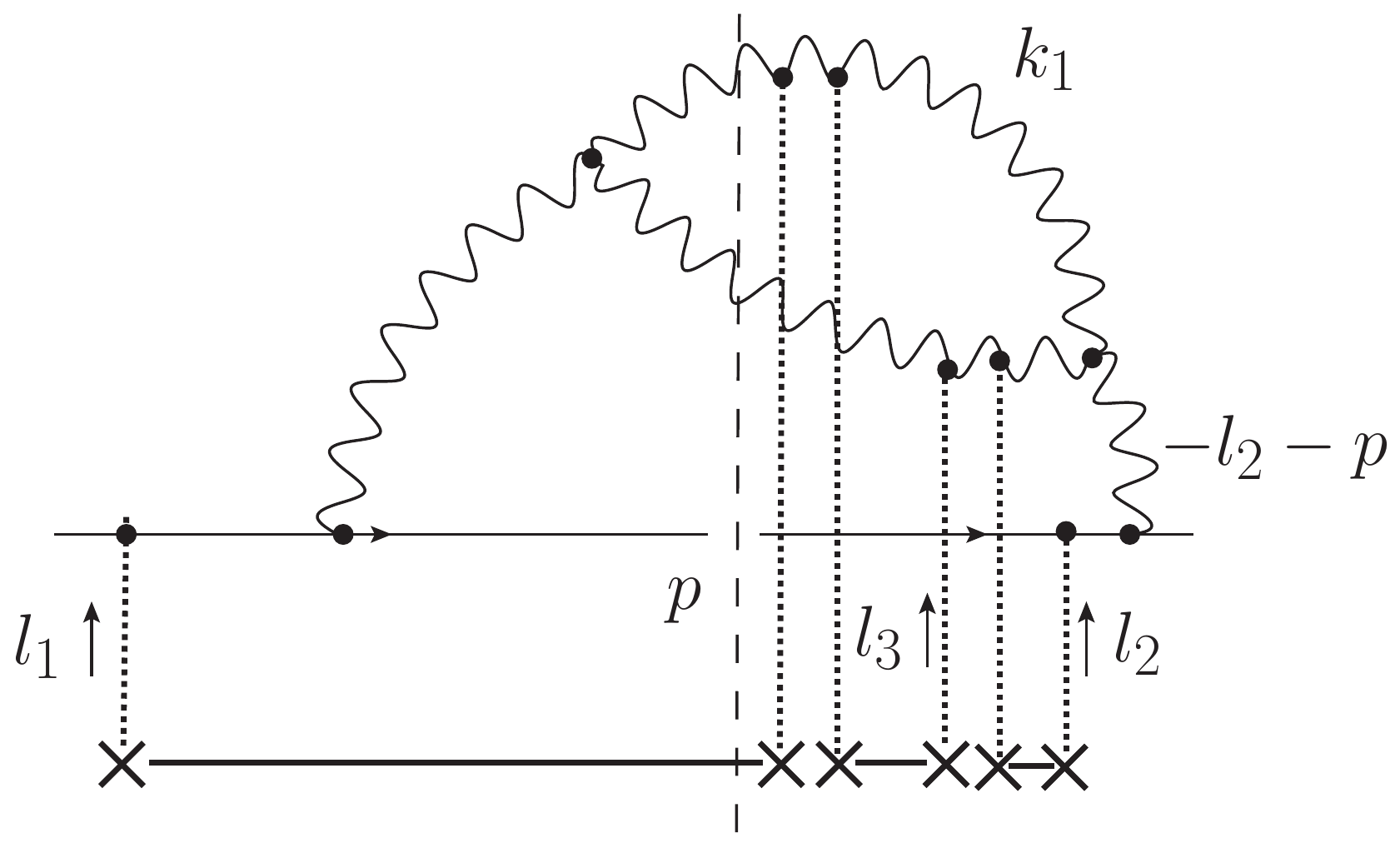}\\
$\graph{F}{IIF}$&$\graph{FF}{II}$
\end{tabular}
\end{center}
\caption{\label{q2f(iif)}
Causal graphs in which the gluon loop interacts.
There are in addition 2 instantaneous-exchange graphs (not shown).
}
\end{figure}

Finally, we address the graphs in Fig.~\ref{q2f(iif)}.
We find the following energy denominators:
\be
\begin{split}
D_{\graph{F}{IIF}}&=\frac{1}{E 
\bar E (E_1+\bar E_2)(E -E_1^{\prime\prime}-E_2^{\prime\prime})},\\
D_{\graph{FF}{II}}&=-\frac{1}{\bar E (E_1+\bar E_2)
(E_1^{\prime\prime}+E_2^{\prime\prime})
(E -E_1^{\prime\prime}-E_2^{\prime\prime})},
\end{split}
\ee
where
\be
E_{1}^{\prime\prime}=
\frac{(\vec k_1+\vec l_1+\vec l_3)^2}{2k_{1+}},\
E_{2}^{\prime\prime}=
\frac{(\vec p+\vec k_1+\vec l_3)^2}{2k_{2+}}.
\label{eq:defEffii}
\ee
Their sum reads
\be
D_{\graph{F}{IIF}}+D_{\graph{FF}{II}}=
-\frac{1}{E \bar E (E_1+\bar E_2)
(E_1^{\prime\prime}+E_2^{\prime\prime})}.
\label{eq:resq2ffii}
\ee
Again, the sum of the
$(--)$ components and of the instantaneous-exchange graphs
gives a null contribution,
due to the finiteness of the denominators and to the
fact that the $F$-factors
are of order $\varepsilon$.


\subsubsection{\label{sec:dipoles}Dipole graphs}

\begin{figure}
\begin{center}
\begin{tabular}{cc}
\includegraphics[width=4.5cm]{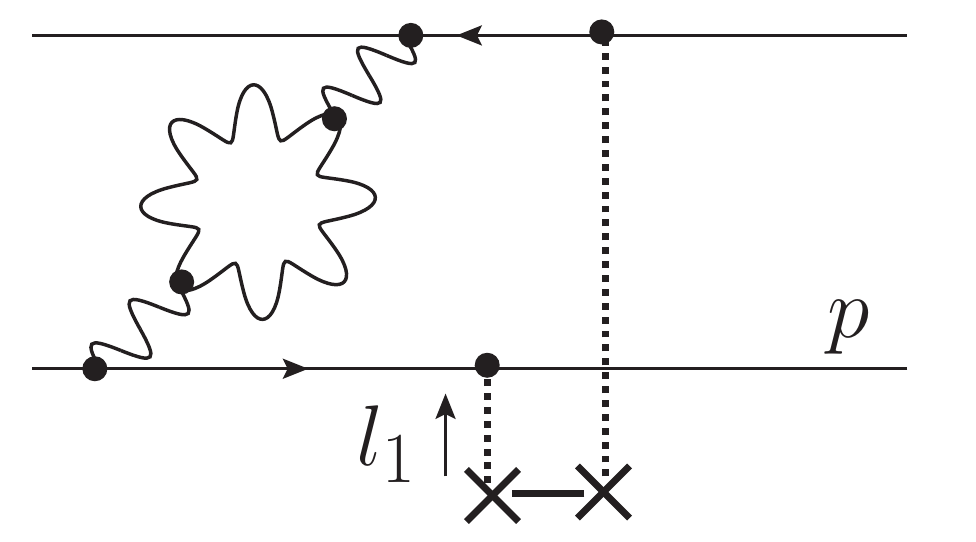}&
\includegraphics[width=4.5cm]{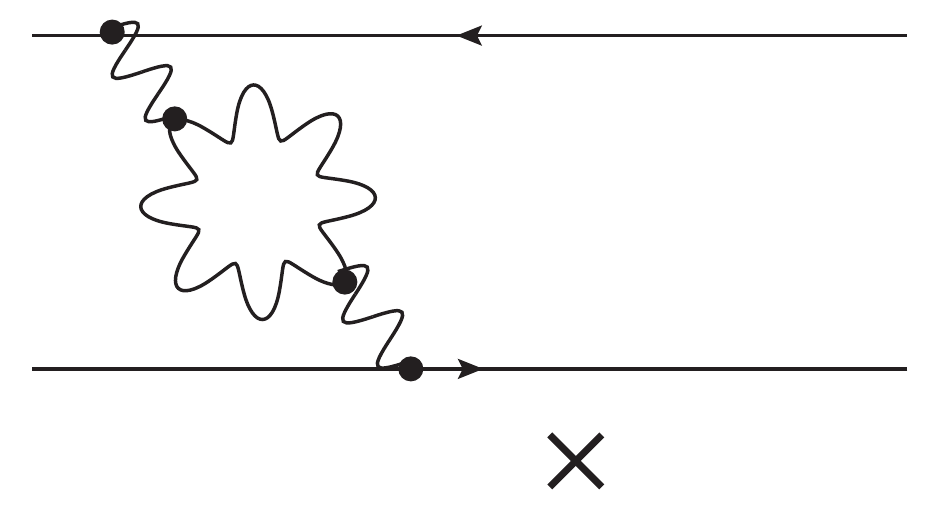}\\
$\alpha_1$ & $\alpha_2$\\
\end{tabular}
\end{center}
\caption{\label{q2da}
Virtual graphs (One dipole exactly scatters).
There would be 6 additional graphs in which
one or both gluons which couple to the quark/antiquark
are exchanged instantaneously. 
$\alpha_1$ and $\alpha_2$ only differ by the ordening of the
$qg$ and $\bar q g$ vertices.
We do not display the complex-conjugate graphs.
}
\end{figure}

\begin{figure}
\begin{center}
\begin{tabular}{cc}
\includegraphics[width=4.5cm]{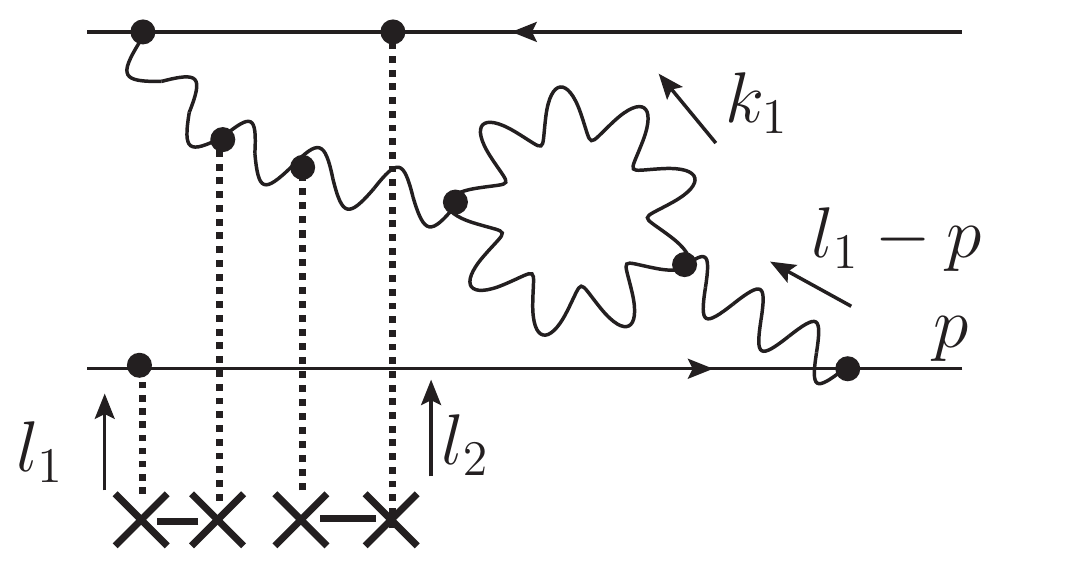}&
\includegraphics[width=4.5cm]{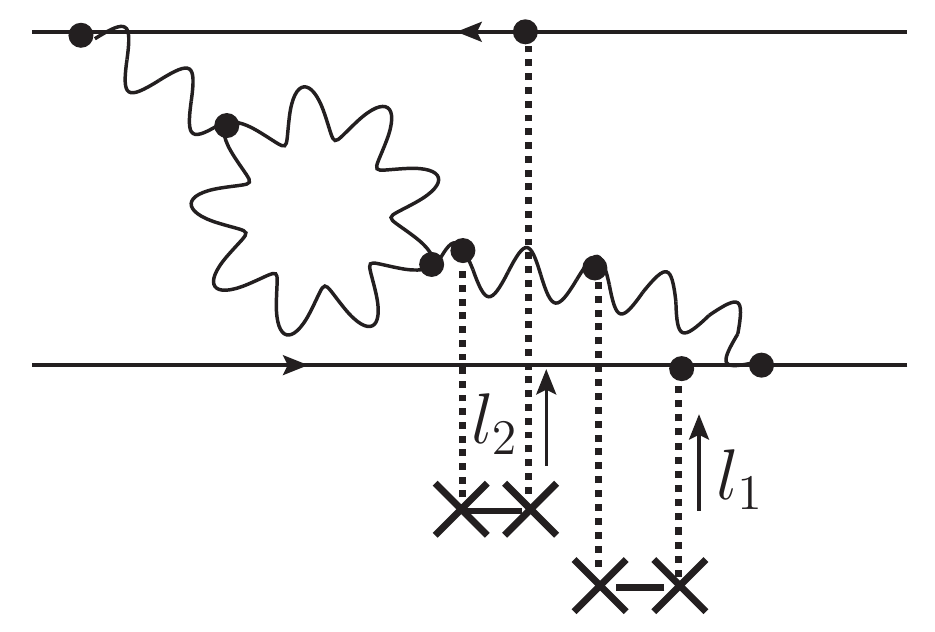}\\
$\beta_1$ & $\beta_2$
\end{tabular}
\end{center}
\caption{\label{q2db}Graphs where
there are 2 dipoles at the time of the interaction.
There would be 2 additional instantaneous-exchange graph,
and also the graphs in which the gluon couples to the quark in
the amplitude and to the antiquark in the complex-conjugate amplitude
(namely the complex-conjugate graphs).
The arrows on the gluons indicate the direction of the flow
of the transverse momentum.
}
\end{figure}

\begin{figure}
\begin{center}
\begin{tabular}{c}
\includegraphics[width=6cm]{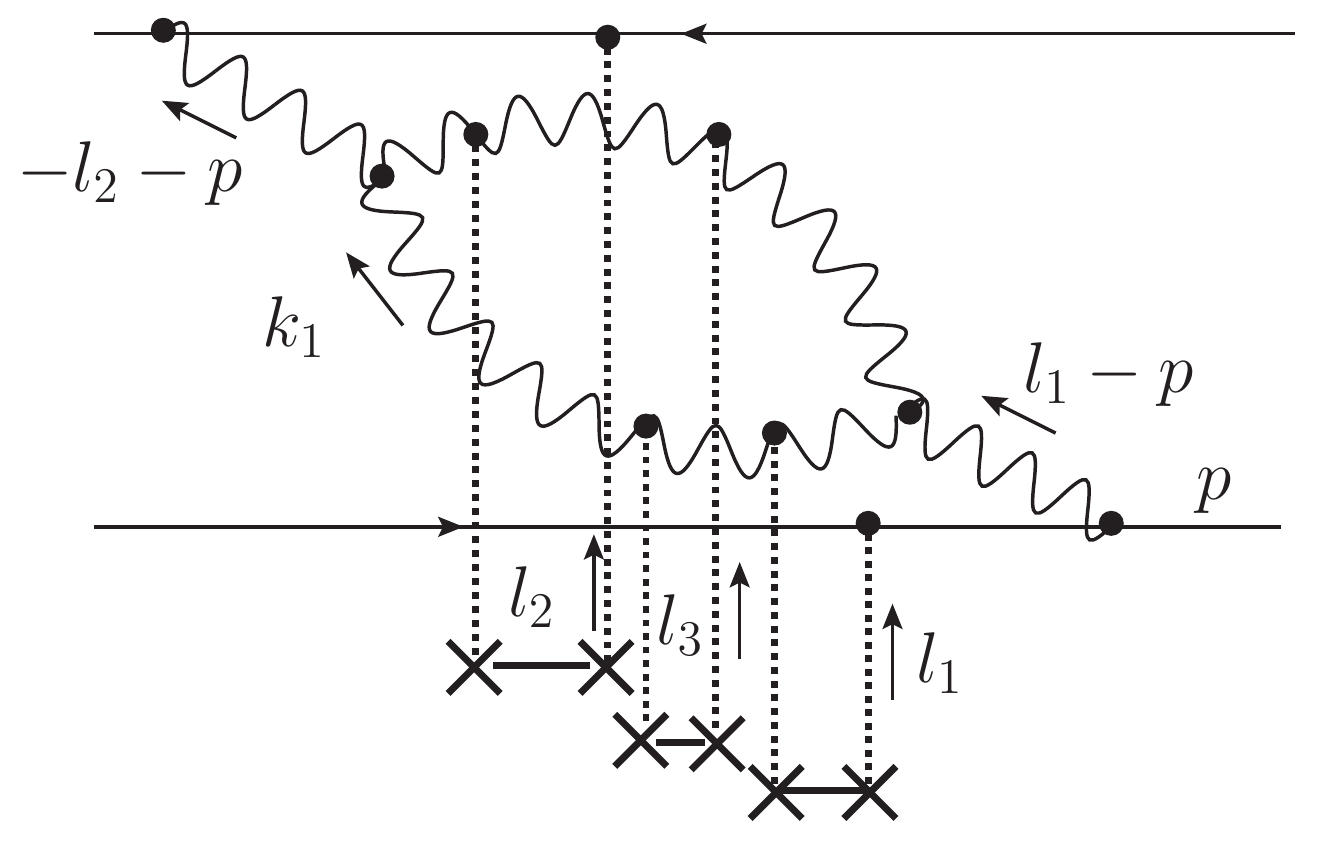}\\
$\gamma$
\end{tabular}
\end{center}
\caption{\label{q2dc}
Graph where there are 3 dipoles at the time of the interaction.
In the same manner as for the diagrams of Fig.~\ref{q2da}, there
would be 3 extra graphs with contact interactions instead of 
the external causal gluon propagators.
There are also the graphs in which the couplings of the
gluons to the quark and antiquark are interchanged
between the amplitude and the complex-conjugate amplitude.
}
\end{figure}

The full set of nontrivial causal dipole graphs is shown
in Fig.~\ref{q2da},\ref{q2db},\ref{q2dc}, up to graphs deduced
from the latter by obvious 
symmetries.
(We do not discuss the graphs in which the gluons both couple
to the same quark or antiquark, because the correspondence
with $p_\perp$-broadening graphs 
in which the evolution happens entirely
either in the amplitude or in the complex-conjugate amplitude
is then immediate).
We may label the flow of the momenta through the graphs
in such a way that it is the same as in the $p_\perp$-broadening
case.
This justifies a posteriori that only the energy denominators need
to be compared.

The expression for the energy denominators are
very easy to obtain for these graphs.
The graphs $\alpha$ in Fig.~\ref{q2da}
are virtual graphs, which contribute equally.
The energy denominators read
\be
D_{\alpha_1}+D_{\alpha_2}=-\frac{1}{2i\varepsilon}
\frac{1}{E ^2(E_1+E_2)}-\frac{2}{E ^3(E_1+E_2)}
-\frac{1}{E ^2(E_1+E_2)^2},
\ee
where $E $, $E_1$ $E_{2}$ are defined in the previous section.
To compute the sum of the $(--)$ polarization 
and of the instantaneous-exchange
graphs (not drawn), it is enough to multiply 
the previous denominators by the factor
\be
F=
\left(
1-\frac{E -i\varepsilon}{E }
\right)
\left(
1-\frac{E -3i\varepsilon}{E }
\right)\sim \varepsilon^2,
\ee
and thus the latter sum vanishes:
The $(--)$ polarization of the gluons does effectively
not contribute.

Similarly, the energy denominators for the graphs of Fig.~\ref{q2db}
read
\be
D_{\beta_1}=
\frac{1}{{E }^2 \bar E (E_1+E_2)^2},\
D_{\beta_2}=
\frac{1}{E {\bar E }^2 (E_1+\bar E_2)^2}.
\ee

Last, the graph in Fig.~\ref{q2dc} gives
\be
D_{\gamma}=
\frac{1}{E \bar E (E_1+\bar E_2)
(E_1^{\prime\prime}+E_2^{\prime\prime})}.
\ee
The $F$ factors for the two last classes of graphs
are proportional to $\varepsilon$: Hence
the $(--)$ polarization components compensate with the
instantaneous-exchange graphs also in the two previous cases.


\subsubsection{Proof of the correspondence}

The identification between the relevant graphs
is now quite straightforward.

The nonzero graphs 
in the case of $p_\perp$-broadening
are those of Fig.~\ref{q2iiii},~\ref{q2ffff},~\ref{q2fff(i)},
~\ref{q2f(iii)},~\ref{q2f(iif)} with the $(-\perp)$
polarization for the gluons which hook to the quark.
The associated denominators are given in
Eq.~(\ref{eq:resq2iiii}),(\ref{eq:resq2ffff}),%
(\ref{eq:resq2fffi}),(\ref{eq:resq2fiii}),%
(\ref{eq:resq2ffii}) respectively.

We naturally identify the dipole graphs $\alpha_1,\alpha_2$
of Fig.~\ref{q2da} with the $p_\perp$-broadening graphs
where all vertices are in the initial state of Fig.~\ref{q2iiii},
since they can be topologically related by bending over
the quark line to form the antiquark of the dipole.
It is easy to check the formal identity of the expressions for these
graphs:
The polarization and vertex factors are the same
in the $p_\perp$-broadening case and in the dipole case,
except for a minus sign difference which comes
from the $qg$ coupling becoming a $\bar q g$ coupling.
We notice that we may
also choose the interaction with the nucleus to be the same in
both cases, and label the momenta in the same way as for
$p_\perp$-broadening.
We see that the following identity between the energy denominators holds:
\be
D_{\graph{II}{II}}+D_{\graph{III}{I}}+D_{\graph{I}{IIII}}=
-\text{Re}\left(
D_{\alpha_1}+D_{\alpha_2}
\right).
\label{eq:q2equiv1}
\ee
The minus sign in the r.h.s.
is explained by the coupling to the antiquark.
The only remaining mismatch is an imaginary term
which cancels when the complex conjugate graphs are taken into account.

Bending over the quark on the right of the cut in 
the graphs of Fig.~\ref{q2ffff}, we see that we get the symmetric
of the dipole graphs of Fig.~\ref{q2da}, namely the complex conjugate graphs
(not represented).
The formal identity between the energy denominators reads
\be
D_{\graph{FF}{FF}}+D_{\graph{FFF}{F}}+D_{\graph{F}{FFF}}=
-\text{Re}\left(
\bar D_{\alpha_1}+\bar D_{\alpha_2}
\right).
\label{eq:q2equiv2}
\ee

In the same manner, we see that the graphs of Fig.~\ref{q2fff(i)}
are topologically related to $\beta_1$ in Fig.~\ref{q2db},
the one in Fig.~\ref{q2f(iii)} is related to $\beta_2$,
and the ones in Fig.~\ref{q2f(iif)} look equivalent to $\gamma$
in Fig.~\ref{q2dc}.
As a matter of fact,
the following identities are verified:
\be
\begin{split}
D_{\graph{FFF}{I}}+D_{\graph{FF}{IF}}+D_{\graph{F}{IFF}}&=-D_{\beta_1},\\
D_{\graph{F}{III}}&=-D_{\beta_2},\\
D_{\graph{F}{IIF}}+D_{\graph{FF}{II}}&=-D_\gamma.
\end{split}
\label{eq:q2equiv3}
\ee
Equations~(\ref{eq:q2equiv1}),(\ref{eq:q2equiv2}),(\ref{eq:q2equiv3})
prove the equivalence of $p_\perp$ broadening and dipoles
for the considered class of diagrams.

\begin{figure}
\begin{center}
\begin{tabular}{c}
\includegraphics[width=4cm]{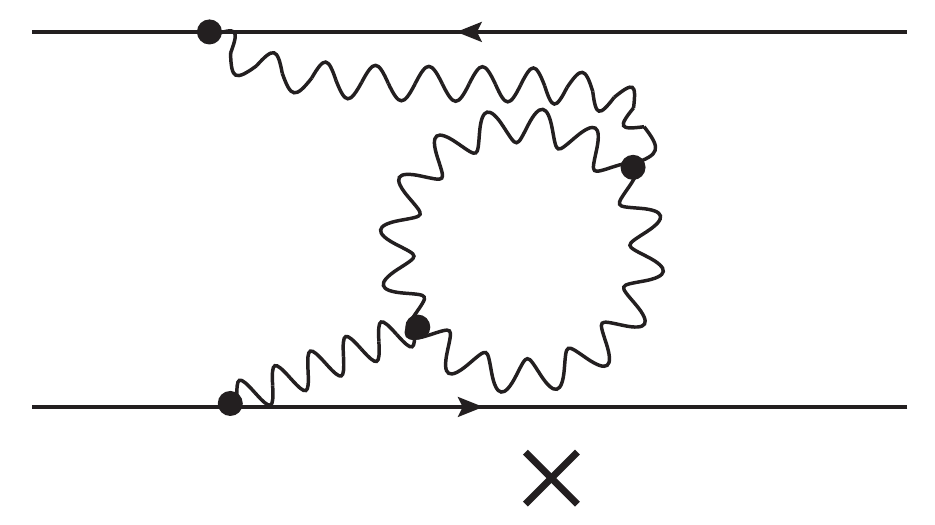}
\end{tabular}
\end{center}
\caption{\label{q2forbidden}
Dipole graph which would correspond
to the (vanishing) $p_\perp$-broadening graph
of Fig.~\ref{q2iifi}, but which is not allowed by momentum conservation.
}
\end{figure}

Note that the $p_\perp$-broadening graphs which cancel 
among themselves would
not have any dipole counterparts
allowed by the kinematics.
Indeed, for example the graphs of Fig.~\ref{q2iifi}
would correspond to a dipole graph such as the one shown in
Fig.~\ref{q2forbidden}, which is forbidden by momentum conservation.


\subsection{\label{sec:sec3qgv}Three quark-gluon vertices}

\begin{figure}
\begin{center}
\includegraphics[width=4cm]{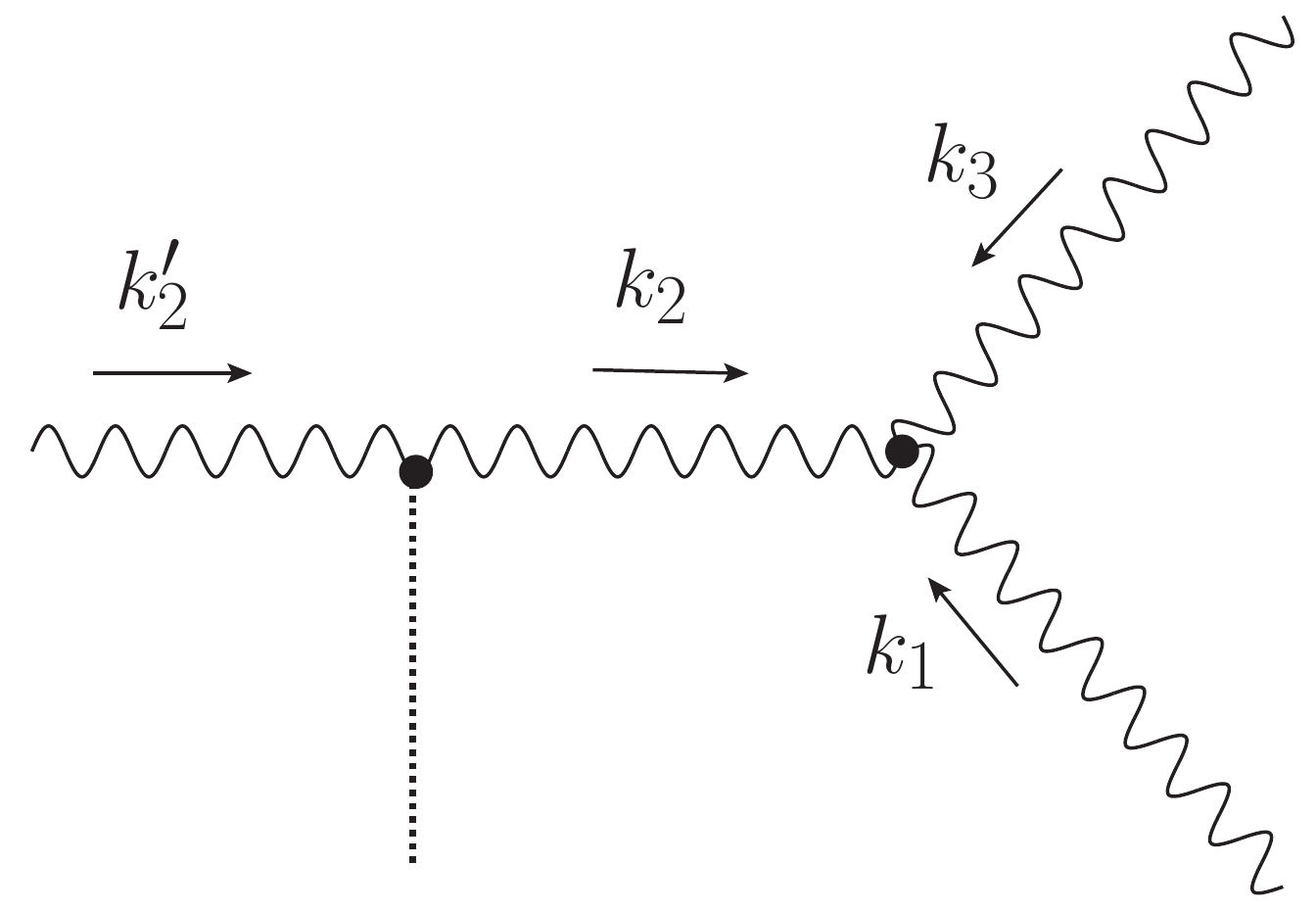}
\end{center}
\caption{\label{fig:3g}
Typical momentum configuration around the 3-gluon vertex.
The arrows show the direction of the transverse components
of the momenta. The ``$+$'' components are always directed
to the right.
The gluons which carry the momenta
$k_1$ and $k_3$ may also be exchanged instantaneously
in some graphs,
in which case the factors $E_1$ and $E_3$ 
which come with the polarization tensors of these gluons
(see Tab.~\ref{tab:3g})
are replaced
by $E_1-\Sigma_1$ and $E_3-\Sigma_3$, where $\Sigma_1$ and
$\Sigma_3$ are the energy denominators of the corresponding
causal graphs.
}
\end{figure}

We turn to the case in which there are 3 quark-gluon vertices.
The NLO graphs all have one 3-gluon vertex (like the one
in Fig.~\ref{fig:3g}), see e.g. Fig.~\ref{q3}
for a sample of
the $p_\perp$-broadening graphs and Fig.~\ref{q3d}
for the would-be equivalent dipole graphs.

Let us analyze the Lorentz structure of 
the upper part of 
these graphs.
The gluons entering the vertex may have ``$-$'' or ``$\perp$''
polarizations.
It turns out that there are only four possible configurations: 
Either all gluons
are transverse, or one of them (and only one) has a ``$-$'' polarization.
All other possibilities lead to vanishinig contributions.
The gluon which is polarized longitudinally may be exchanged
instantaneously
if the kinematical configuration of the diagram allows it;
Such configurations are
taken into account by a modification of the corresponding
polarization tensor, see Eq.~(\ref{eq:BLtrick}).

It is useful to write down the factors specific to the different
polarization configurations for the typical graph shown in 
Fig.~\ref{fig:3g}.
All other configurations of the momenta may be deduced 
from the one illustrated in Fig.~\ref{fig:3g}
by appropriate changes in the signs.
Note that we will keep the transverse momenta always
incoming to the 3-gluon vertex in order to simplify the
comparison of the graphs, at the expense of having sometimes
the ``$+$'' and ``$\perp$'' components 
of the momenta
flowing in opposite
directions
(in which case $d_{-\perp}(k)=-\vec k/k_+$).

We contract the 3-gluon vertex with the
gluon polarization tensors assuming that the 3 external legs of
the gluons represented by wavy lines couple
eikonally, thus always have the ``$-$'' polarization.
The gluons which are exchanged with the target 
(represented here by the dotted vertical line in Fig.~\ref{fig:3g})
also couple
eikonally to the leftmost gluon.
We temporarily
leave out the color factors, the coupling constants,
and the momentum factors~(\ref{eq:vertexfactor}) that come with each leg of
vertices since the latter may be put back in the
end and do not depend on the polarization.
There are four possible polarization configurations that
we need to distinguish. Either one gluon (and only one)
has the ``$-$'' polarization at the 3-gluon vertex, or all
have the ``$\perp$'' polarization.
To each of these configurations corresponds a specific factor $V$.

We present the factors $V$ for the four possible polarization
configurations in Tab.~\ref{tab:3g}.
The index of $V$ corresponds to the label of the 
gluon which enters the 3-gluon
vertex with a minus polarization, or is ``$\perp$''
when all gluons are transversely polarized.

\begin{table}
\begin{center}
\begin{tabular}{c|rl}
& \multicolumn{2}{c}{Vertex and polarization factors}\\
\hline
$V_1$    & \multirow{4}{*}{$\frac{2}{k_{1+}k_{2+}k_{3+}}\times$}
  &$-E_1(k_2+k_3)_+(\vec k_{2}^\prime\cdot \vec k_{3})$\\
$V_2$    &
  &$\frac{(k_1-k_3)_+}{2k_{2+}}(\vec k_{2}\cdot \vec k_{2}^\prime)
(\vec k_{1}\cdot \vec k_{3})$\\
$V_3$    &
  &$E_3(k_1+k_2)_+(\vec k_{1}\cdot \vec k_{2}^\prime)$\\
$V_\perp$ &
  &$\frac12\left[
    (\vec k_{2}\cdot \vec k_{3})(\vec k_{1}\cdot \vec k_2^\prime)
  -(\vec k_{1}\cdot \vec k_{2})
  (\vec k_{3}\cdot \vec k_2^\prime)
    \right]$
\end{tabular}
\end{center}
\caption{\label{tab:3g}%
Factors $V$
associated to the 3-gluon configuration shown in Fig.~\ref{fig:3g}
for the different polarization configurations of the gluons at
the 3-gluon vertex (see the main text).
The transverse momenta are always incoming, while all ``$+$''
components of the momenta are flowing from the left to the right.
If $k_2=k^\prime_2$, then $V_\perp=0$
and $V_2=2E_2(k_1-k_3)_+\vec k_1\cdot\vec k_3/(k_{1+}k_{2+}k_{3+})$.
}
\end{table}

Coming back to the full diagrams,
we have to distinguish the different possible time orderings
of the vertices.
Some orderings lead to real dipole graphs, that is, where there
are two dipoles 
in the wave function
at the time of the interaction,
while other orderings lead to virtual dipole graphs, which
contribute to the
renormalization of
the wave function of the initial dipole.

\subsubsection{Initial state/initial and final state}

\begin{figure}
\begin{center}
\begin{tabular}{cc}
\includegraphics[width=5cm]{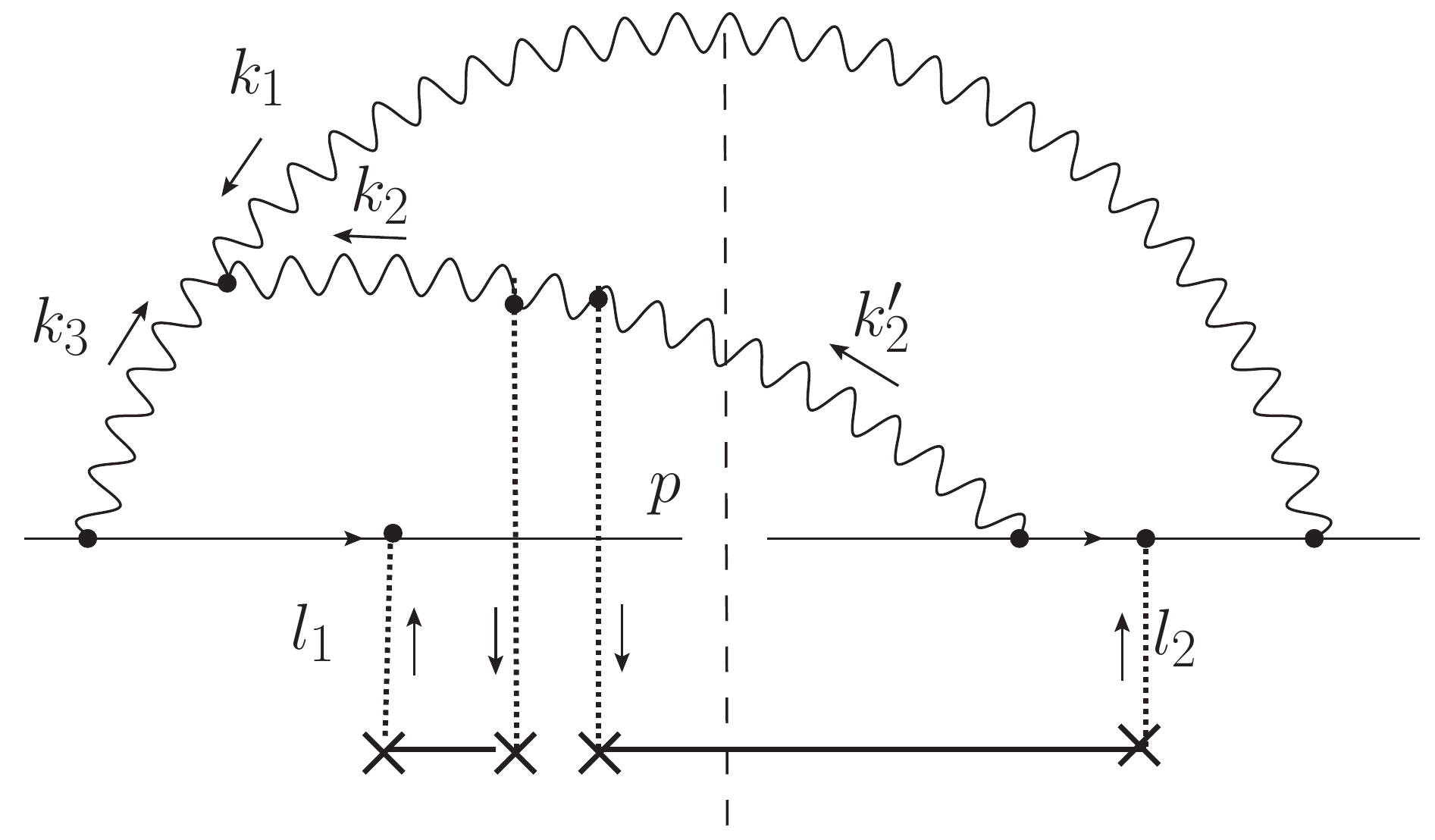}&
\includegraphics[width=5cm]{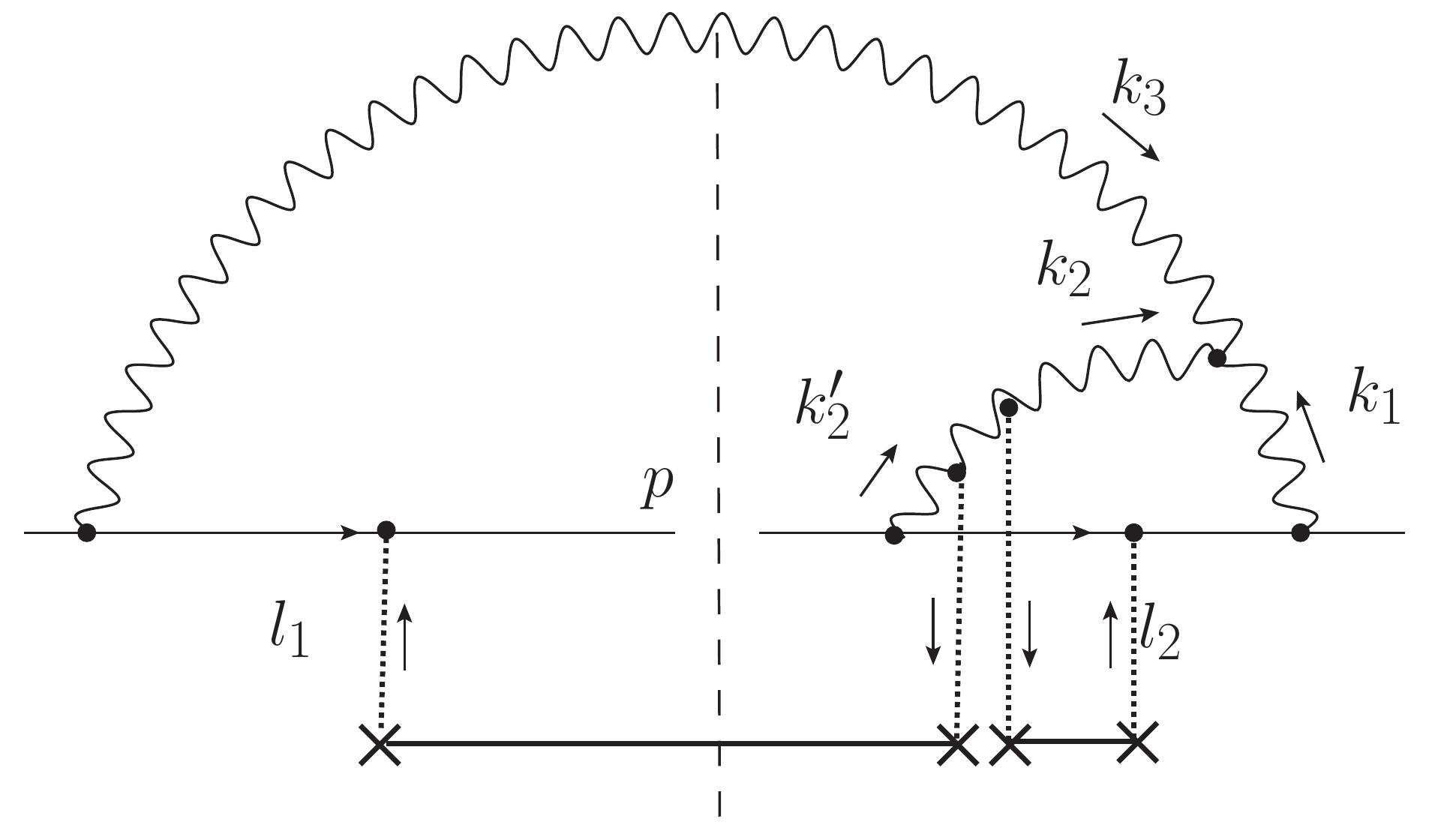}\\
$A$ & $B$
\end{tabular}
\end{center}
\caption{\label{q3}
$p_\perp$-broadening graphs in the case in which
two quark-gluon vertices are in the initial state
and one is in the final state.
The transverse momenta are directed according to the
arrows, while the ``$+$'' momenta are oriented to the
right.
We do not show the graph $A_3^\prime$ which is similar 
to $A$ but with gluon~3
exchanged instantaneously, and $B_1^\prime$, similar to $B$
but with gluon~1
exchanged instantaneously.
Momentum conservation implies that
$\vec k_2^\prime=\vec k_2+\vec l_1+\vec l_2$.
}
\end{figure}

\begin{figure}
\begin{center}
\begin{tabular}{ccc}
\includegraphics[width=4cm]{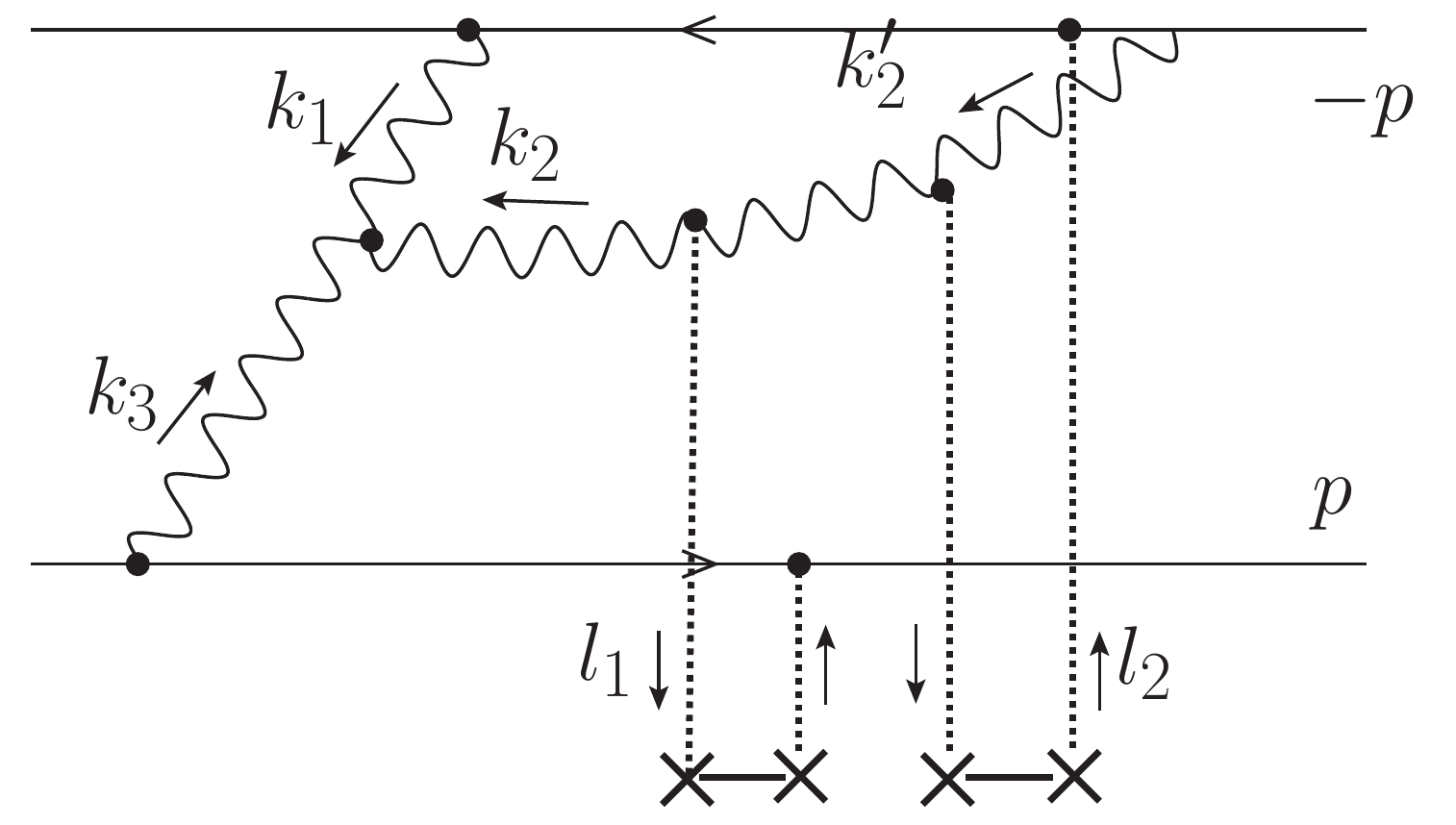}&
\includegraphics[width=4cm]{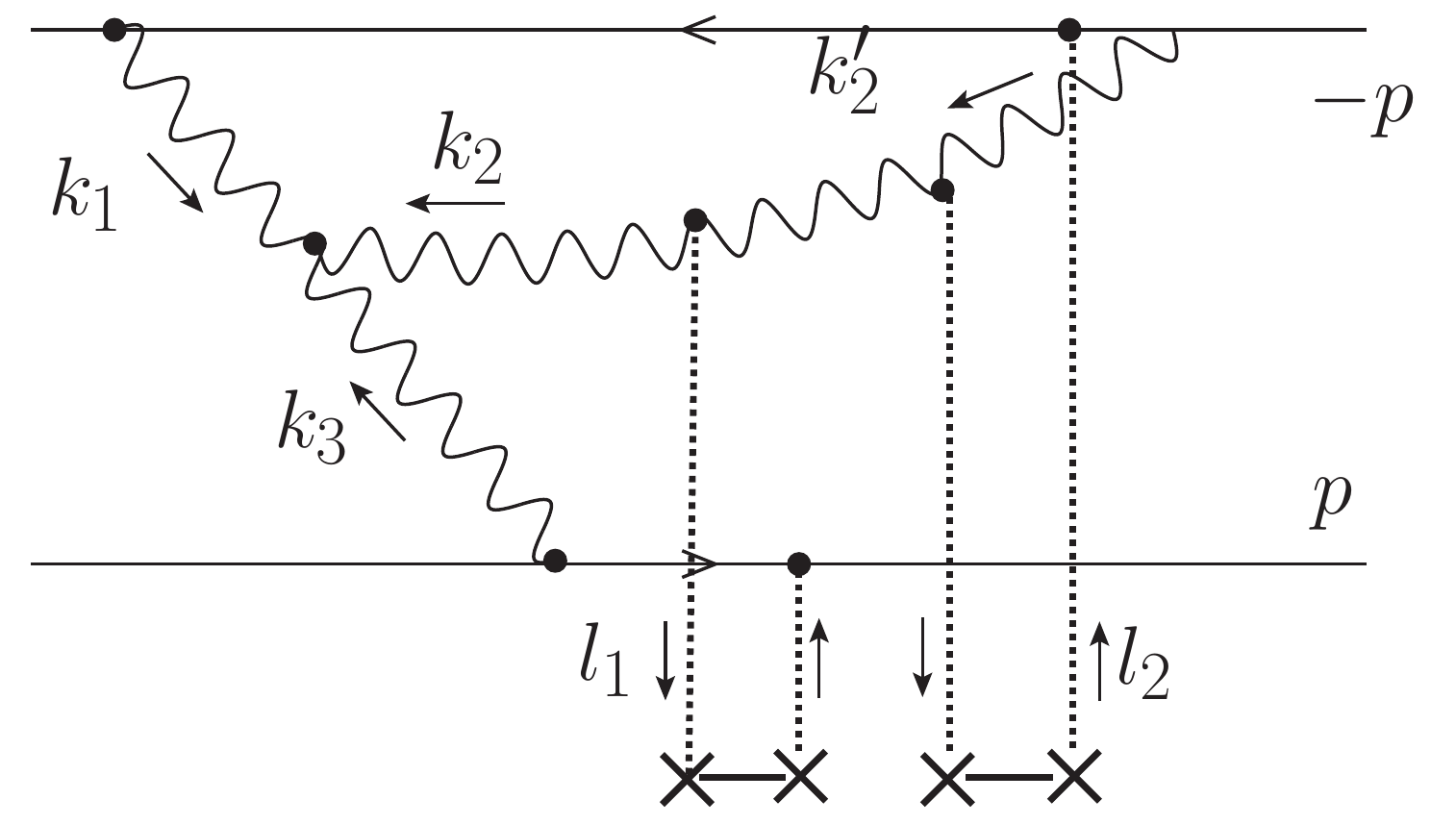}& 
\includegraphics[width=4cm]{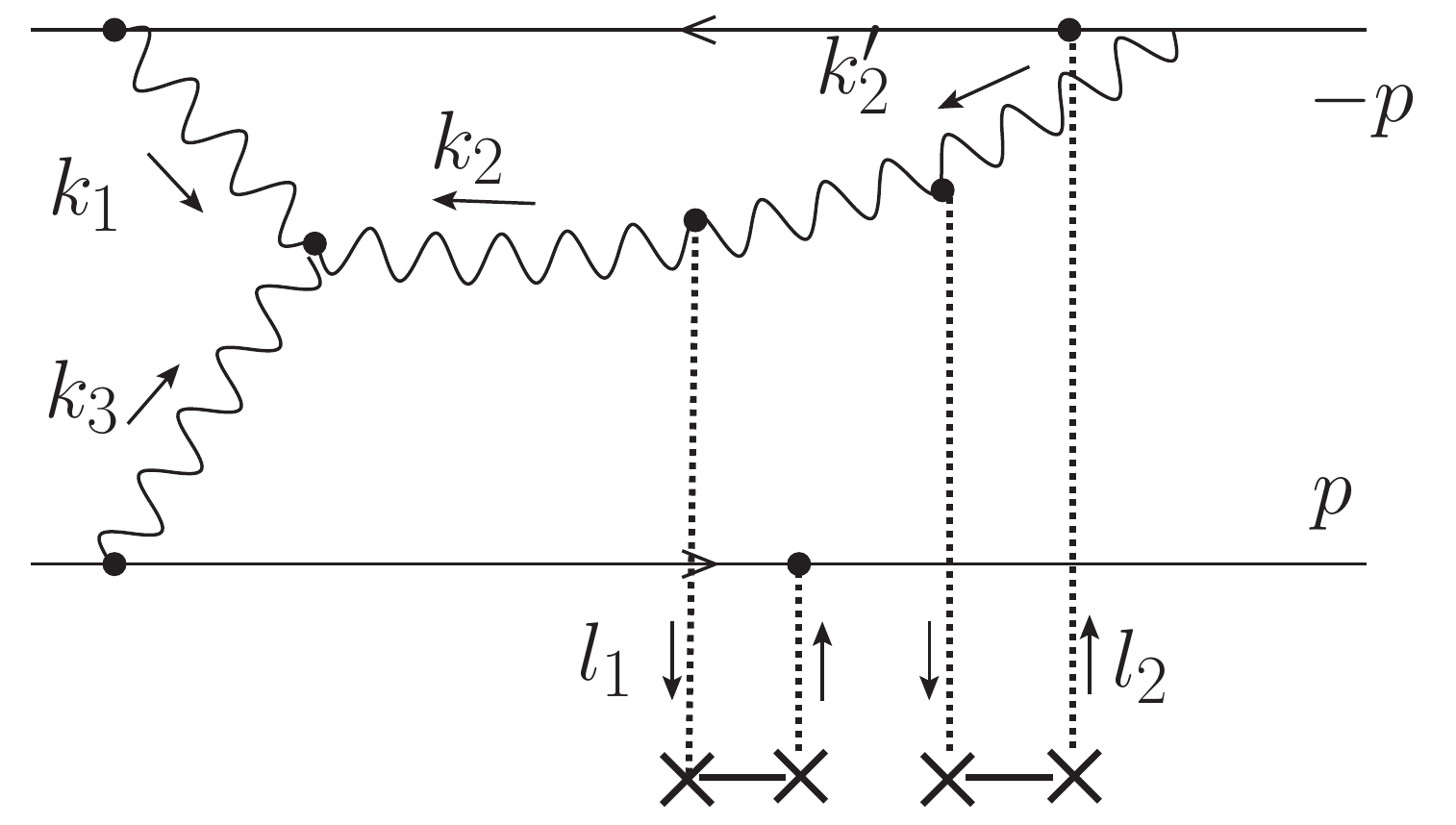}\\
$\alpha$ & $\beta$ & $\gamma$
\end{tabular}
\end{center}
\caption{\label{q3d}
Dipole graphs corresponding to the
$p_\perp$-broadening graphs in Fig.~\ref{q3}.
To this list one should add the graphs in which gluons~$1$ or~$3$
are exchanged instantaneously,
and distinguish two relative orderings of the
leftmost $qg$ and $\bar qg$ vertices
in $\gamma$, which we should draw as two different
graphs $\gamma^{(1)}$ and  $\gamma^{(2)}$.
}
\end{figure}

Let us start with graphs related to real dipole graphs, 
and first with
the simplest set of them, 
characterized by two of the $qg$ vertex and the 3-gluon vertex 
being
in the initial state, while the remaining $qg$ vertex
is in the final state (see Fig.~\ref{q3}).
On the dipole side, the corresponding process involves two dipoles
at the time of the interaction (Fig.~\ref{q3d}).

The most general case for the interaction with
the nucleus is captured by two scatterings such as in Fig.~\ref{q3}.
Higher-order scatterings add simple combinatorial factors.

The calculation of the different contributions 
using the previously edicted rules (Eq.~(\ref{eq:D})
for the energy denominators, Eqs.~(\ref{eq:eikq}) and~(\ref{eq:eikg})
for the eikonal vertices, 
Eq.~\ref{eq:vertexfactor} for additional factors associated to
all vertices, and
Tab.~\ref{tab:3g} for the 3-gluon vertex
dressed with the polarization tensors)
is straightforward.

We define the energy of the gluons as
\be
E_1=\frac{\vec k_1^2}{2k_{1+}},\
E_2=\frac{\vec k_2^2}{2k_{2+}},\
E^\prime_2=\frac{\vec k_2^{\prime 2}}{2k_{2+}}=
\frac{(\vec k_2+\vec l_1+\vec l_2)^2}{2k_{2+}},\
E_3=\frac{\vec k_3^2}{2k_{3+}}.
\ee
The calculations corresponding to the $A,\alpha$ 
and $B,\beta$
graphs 
(without including the lower part of the graphs
representing the coupling to the target)
are summarized in
Tab.~\ref{tab:Aa} and~\ref{tab:Bb}.
We distinguish the four different polarization configurations
(as in Tab.~\ref{tab:3g}) for each graph.
The two columns single out the energy denominators (which
are not polarization dependent) and the 3-gluon vertex and
polarization factors read off Tab.~\ref{tab:3g}
with the relevant kinematics.
In addition, there is an overall factor, which is identical for
all graphs: After averaging over the color and helicity
of the initial quarks, it reads
\be
\frac{8\pi^4\alpha_s^4 N_c^3 C_F}{(k_{1+}k_{2+}k_{3+})^2}.
\label{eq:overallfactor}
\ee
The $2/(k_{1+}k_{2+}k_{3+})$
factor in Tab.~\ref{tab:3g} has been incorporated 
in Eq.~(\ref{eq:overallfactor}).

In order to compute the contribution of 
these graphs to
$p_\perp$-broadening, we need to convolute the
content of Tab.~\ref{tab:Aa} and~\ref{tab:Bb} multiplied
by Eq.~(\ref{eq:overallfactor}) with some
weight which encodes the model for the interaction with
the target and which can be computed from the rules listed in
Sec.~\ref{sec:leadingorder}:
\begin{multline}
\left.\frac{dN}{d^2p}\right|_\text{
\begin{minipage}{3cm}
\tiny contribution of a\\
\tiny particular graph
\end{minipage}}=
-\int \frac{d^2\vec l_1}{\vec l_1^2}
\frac{d^2\vec l_2}{\vec l_2^2}
\left[
\frac{x g(x,\vec l_1^2)}{N_c^2-1}\rho L
\right]
\left[
\frac{x g(x,\vec l_2^2)}{N_c^2-1}\rho L
\right]\\
\times\int_0^{+\infty}dk_{1+}dk_{2+}dk_{3+}
\delta(\text{``$+$'' momentum})
\times\left[\text{overall factor}\right]\\
\times
\int \frac{d^2 \vec k_1}{(2\pi)^2}
\frac{d^2 \vec k_2}{(2\pi)^2}\frac{d^2 \vec k_3}{(2\pi)^2}
\delta^2(\vec k_1+\vec k_2+\vec k_3)
\delta^2(\text{global ``$\perp$'' momentum})\\
\times\left[
\text{right column of the corresponding table}
\right]
\label{eq:convolution}
\end{multline}
where, for example in the case of graph $A$,
the graph-dependent momentum conservation factors read
\be
\begin{split}
\delta(\text{``$+$'' momentum})&\equiv\delta(k_{1+}+k_{2+}-k_{3+})\\
\delta^2(\text{global ``$\perp$'' momentum})&
\equiv\delta^2(\vec k_{2}+\vec p)\\
\left[\text{overall factor}\right]
&=\text{Eq.~(\ref{eq:overallfactor})}
\end{split}
\ee
and the ``corresponding table'' is Tab.~\ref{tab:Aa}.

\begin{table}
\begin{center}
\begin{tabular}{c|c|c}
 & Energy denom.& 3-gluon vertex$\times$pola.\\
\hline
$A_1$ & 
\multirow{4}{*}{$-\frac{1}{E_1E^\prime_2 E_3(E_1+E_2)}$}
& $(\vec k_2^\prime\cdot\vec k_3)E_1(2k_3-k_1)_+$\\
$A_2$ & & $(\vec k_1\cdot\vec k_3)(\vec k_2\cdot\vec k_2^\prime)
\left[-(k_1+k_3)_+/(2(k_{3}-k_1)_+)\right]$\\
$A_3^*$ & & $0$\\
$A_\perp$ & & $\frac12\left[(\vec k_2\cdot \vec k_3)(\vec k_1\cdot \vec l )
-(\vec k_1\cdot \vec k_2)(\vec k_3\cdot \vec l)\right]$\\
\hline
$\alpha_1^*$ 
& \multirow{4}{*}{$\frac{1}{E_2E^\prime_2 E_3(E_1+E_2)}$}
& $(\vec k_2^\prime\cdot\vec k_3)(-E_2)(2k_3-k_1)_+$\\
$\alpha_2$ &
& $(\vec k_1\cdot\vec k_3)(\vec k_2\cdot\vec k_2^\prime)
\left[-(k_1+k_3)_+/(2(k_{3}-k_1)_+)\right]$\\
$\alpha_3^*$
& & $0$\\
$\alpha_\perp$ & & 
$\frac12\left[(\vec k_2\cdot \vec k_3)(\vec k_1\cdot \vec l)
-(\vec k_1\cdot \vec k_2)(\vec k_3\cdot \vec l)\right]$
\end{tabular}
\caption{\label{tab:Aa}%
Expressions for the graphs $A$ and $\alpha$
in Fig.~\ref{q3} and~\ref{q3d}.
$k^\prime_2=k_2+l$ (where $l=l_1+l_2$) and
according to momentum conservation,
$k_{2+}=k_{3+}-k_{1+}$.
The ``*'' superscript on the
graph labels means that an instantaneous exchange
graph
has been added.
The factor $2/(k_{1+}k_{2+}k_{3+})$ 
present in Tab.~\ref{tab:3g}
was left out and
integrated to the overall factor~(\ref{eq:overallfactor}).
The contribution of $A$ to the $p_\perp$-broadening
amplitude is obtained from the convolution given in
Eq.~(\ref{eq:convolution}).
}
\end{center}
\end{table}

\begin{table}
\begin{center}
\begin{tabular}{c|c|c}
& Energy denom.& 3-gluon vertex$\times$pola.\\
\hline
$B_1^*$ 
& \multirow{4}{*}{$\frac{1}{E_1E^\prime_2 E_3(E_2+E_3)}$}
& $0$\\
$B_2$  & & $(k_1\cdot k_3)(k_2\cdot k_2^\prime)
\left[-(k_1+k_3)_+/(2(k_{1}-k_3)_+)\right]$\\
$B_3$ & & 
$(k_1\cdot k_2^\prime)E_3(2k_1-k_3)_+$\\
$B_\perp$ & &
$-\frac12\left[(\vec k_2\cdot \vec k_3)(\vec k_1\cdot \vec l)
+(\vec k_1\cdot \vec k_2)(\vec k_3\cdot \vec l)\right]$
\\
\hline
$\beta_1^*$  
& \multirow{4}{*}{$\frac{1}{E_1E_2 E_2^\prime(E_2+E_3)}$}
& $0$\\
$\beta_2$ & & 
$(k_1\cdot k_3)(k_2\cdot k_2^\prime)
\left[(k_1+k_3)_+/(2(k_{1}-k_3)_+)\right]$\\
$\beta_3^*$
& & $(k_1\cdot k_2^\prime)E_2(2k_1-k_3)_+$\\
$\beta_\perp$ & & 
$\frac12\left[(\vec k_2\cdot \vec k_3)(\vec k_1\cdot \vec l)
-(\vec k_1\cdot \vec k_2)(\vec k_3\cdot \vec l)\right]$
\end{tabular}
\caption{\label{tab:Bb}%
The same as Tab.~\ref{tab:Aa} but for graphs $B$ and $\beta$ in 
Fig~\ref{q3} and~\ref{q3d}, for which momentum conservation
reads $k_{2+}=k_{1+}-k_{3+}$.
}
\end{center}
\end{table}
From the conservation of 3-momentum and from
the positivity of the ``+'' components, we 
note that the $A$ and $\alpha$ graphs have the same integration
range of the ``$+$'' components of the momenta, 
namely $0<k_{1+}<k_{3+}$, while in the case of $B$ and
$\beta$, $k_{1+}>k_{3+}$. As for the $\gamma$ graph,
$k_{1+}$ and $k_{3+}$ must be both integrated
from 0 to $+\infty$.
We may subtract the dipole graphs from the production graphs
whenever the kinematics are identical.
We find that 
only the ``2'' and ``$\perp$'' components are nonzero
in the differences $A-\alpha$ and $B-\beta$.
They are listed in Tab.~\ref{tab:diffg}
together with the nonzero components of $\gamma$.
We have specified the integration range and singled out
the factors which depend on the ``$\perp$'' and ``$+$'' components
of the momenta. Other factors are common to all graphs (see the 
caption of Tab.~\ref{tab:diffg}).
As for $\gamma$, we were careful to distinguish
two possible orderings of the $qg$ and $\bar qg$ vertices
$\gamma^{(1)}$ and $\gamma^{(2)}$ before adding
up their contributions, see the caption of Fig.~\ref{q3d}.

\begin{table}
\begin{center}
\begin{tabular}{c|c|c|c}
 & {Integration range} & 
\multicolumn{2}{c}{Momentum-dependent factors}\\
& & ``$\perp$'' & ``+''\\
\hline
$A_2-\alpha_2$ &\multirow{2}{*}{$0<k_{1+}<k_{3+}$}
&$\frac{(\vec k_1\cdot\vec k_3)(\vec k_2\cdot\vec k_2^\prime)}
{\vec k_1^2\vec k_2^2\vec k_2^{\prime 2}\vec k_3^2}$&
$\frac{k_{1+}+k_{3+}}{k_{1+} (k_{3+}-k_{1+}) k_{3+}}$
\\
$A_\perp-\alpha_\perp$ & &
 $
\frac{-(\vec k_2\cdot \vec k_3)(\vec k_1\cdot \vec l)
+(\vec k_1\cdot \vec k_2)(\vec k_3\cdot \vec l)
}{\vec k_1^2\vec k_2^2\vec k_2^{\prime 2}\vec k_3^2}$&
$\frac{1}{k_{1+} k_{3+}}$
\\
\hline
$B_2-\beta_2$ &\multirow{2}{*}{$k_{1+}>k_{3+}$}
& $\frac{(\vec k_1\cdot\vec k_3)(\vec k_2\cdot\vec k_2^\prime)}
{\vec k_1^2\vec k_2^2\vec k_2^{\prime 2}\vec k_3^2}$&
$\frac{k_{1+}+k_{3+}}{k_{1+} (k_{3+}-k_{1+}) k_{3+}}$
\\
$B_\perp-\beta_\perp$ & & 
$\frac{-(\vec k_2\cdot \vec k_3)(\vec k_1\cdot \vec l)
+(\vec k_1\cdot \vec k_2)(\vec k_3\cdot \vec l)
}{\vec k_1^2\vec k_2^2\vec k_2^{\prime 2}\vec k_3^2}$&
$\frac{1}{k_{1+} k_{3+}}$
\\
\hline
$\gamma_2$ &\multirow{2}{*}{$0<k_{1+}<+\infty$}
& $\frac{(\vec k_1\cdot \vec k_3)(\vec k_2\cdot \vec k_2^\prime)}
{\vec k_1^2 \vec k_2^2\vec k_2^{\prime 2}\vec k_3^2}$&
$\frac{k_{3+}-k_{1+}}{k_{1+} (k_{1+}+k_{3+}) k_{3+}}$
\\
$\gamma_\perp$ & &
 $\frac{-(\vec k_2\cdot \vec k_3)(\vec k_1\cdot \vec l)
+(\vec k_1\cdot \vec k_2)(\vec k_3\cdot \vec l)
}
{\vec k_1^2 \vec k_2^2 \vec k_2^{\prime 2}\vec k_3^2}$&
$\frac{1}{k_{1+} k_{3+}}$
\\
\end{tabular}
\caption{\label{tab:diffg}%
Differences between the contributions of graphs $A$,$\alpha$
and $B$, $\beta$ respectively, 
and nonzero contributions of graph $\gamma$.
We only keep the momentum-dependent factors, which we split
between a factor depending only on the transverse components,
and a factor depending only on the longitudinal ones.
(The overall constant, which gathers the coupling and
color factors, reads $64\pi^4\alpha_s^4 N_c^3 C_F$).
}
\end{center}
\end{table}

We are now in position to prove the equivalence between
the production and the dipole processes.
We need to compare the results
of the integration of the expressions for the
different graphs over the
$k_{1+}$ variable.
We see in Tab.~\ref{tab:diffg} that $A-\alpha$ and $B-\beta$
have exactly the same expressions: The difference is in the
integration ranges in $k_{1+}$, which turn out to be 
complementary on the positive real axis $[0,+\infty[$.

Let us show in detail that the analytical continuation
of $\gamma_2$ to the negative $k_{1+}$ region
has exactly
the same expression as $A_2-\alpha_2$ and $B_2-\beta_2$, once
the change of variable $k_{1+}\rightarrow -k_{1+}$ is
done in order to go back to positive $+$ factors.
\begin{figure}
\begin{center}
\includegraphics[width=9cm]{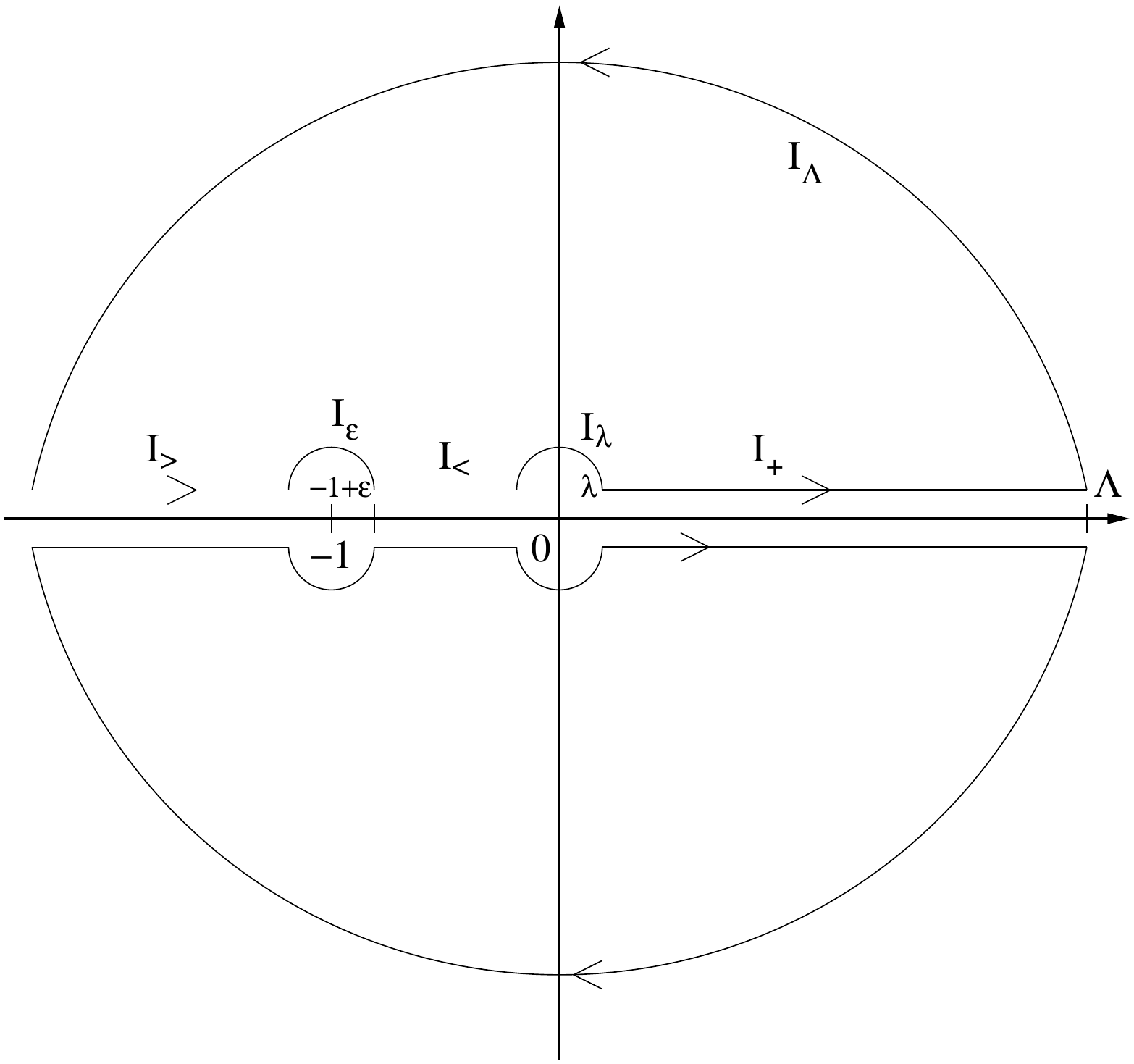}
\end{center}
\caption{\label{xplane}
Complex plane of the $x$-variable in the integrand
of Eq.~(\ref{eq:gI+}).
}
\end{figure}
To this aim, it is enough to focus on the ``+''
components
in Tab.~\ref{tab:diffg} (see the rightmost column)
since 
the labeling conventions have been chosen such that
the other factors remain identical for all graphs.
We define $x\equiv k_{1+}/k_{3+}$ in order to simplify the equations
and write the integral associated to $\gamma_2$
as
\be
\gamma_2\rightarrow I_+=\int_\lambda^\Lambda dx\frac{1-x}{x(1+x)}.
\label{eq:gI+}
\ee
We go to the complex plane of the $x$-variable
and sum the integrand over
the double contour shown in Fig.~\ref{xplane}.
The integrals over the semi-circles 
$I_\Lambda$, $I_\lambda$, $I_\varepsilon$ and their symmetrics
give a finite contribution,
but their sum vanishes.
From the Cauchy theorem, 
the remaining terms are then related by the equation
\be
2(I_++I_<+I_>)=0.
\label{eq:sumcontours}
\ee
The first term $I_+$ is proportional to $\gamma_2$.
The two other terms read
\be
\begin{split}
I_<&=\int_{-1+\varepsilon}^{-\lambda}dx\frac{1-x}{x(1+x)}=
-\int_{\lambda}^{1-\varepsilon}dx\frac{1+x}{x(1-x)}\rightarrow -(A_2-\alpha_2)\\
I_>&=\int_{-\Lambda}^{-1-\varepsilon}dx\frac{1-x}{x(1+x)}=
-\int_{1+\varepsilon}^\Lambda dx\frac{1+x}{x(1-x)}\rightarrow -(B_2-\beta_2)
\end{split}
\ee
After the limit $\varepsilon, \lambda\rightarrow 0$ and
$\Lambda\rightarrow +\infty$ is taken,
 Eq.~(\ref{eq:sumcontours})
shows the equivalence between the $p_\perp$-broadening diagrams and
the dipole diagrams.
The same argument applies to the ``$\perp$'' component
and leads to the same conclusion.

Note using the double contour in Fig.~\ref{xplane}
is equivalent to taking a principal-part prescription for
the pole at $x=-1$.
Originally, this pole was shifted off the real axis
by the regularization chosen for the energy denominators.
But since the imaginary terms anyway cancel
after one has included all relevant graphs, these two choices
are equivalent.

Of course, in the present case, the integrals $I_+$,
$I_<$ and $I_>$ are simple enough that
we can also afford to check directly the
identity
\be
\int_0^{k_{3+}-\varepsilon} dk_{1+} (A_{2,\perp}-\alpha_{2,\perp})
+\int_{k_{3+}+\varepsilon}^{+\infty} dk_{1+} (B_{2,\perp}-\beta_{2,\perp})
=\int_0^{+\infty} dk_{1+} \gamma_{2,\perp},
\label{eq:proofequivalence}
\ee
which holds
once appropriate cutoffs have been introduced
to regularize the infrared and ultraviolet divergences
at $k_{1+}\rightarrow 0$ and $k_{1+}\rightarrow +\infty$
respectively. But the analytical continuation 
argument presented above
appears
more general.

Having performed these detailed checks in this simple case,
we will be able to compare the graphs
in all other cases by 
going through the following steps:
\begin{enumerate}
\item We write the expression for all graphs for all possible
polarizations of the gluons, with conventions
for the momenta that are such that the same transverse structure
is kept for all graphs 
(This step is summarized in
Tab.~\ref{tab:Aa} and~\ref{tab:Bb} in the
case just discussed).
\item We choose 2 variables among the 3 ``+'' components
$k_{1+}$, $k_{2+}$, $k_{3+}$
and use momentum conservation 
in order to
express one of them ($k_{2+}$ here)
in terms of the other two.
We fix one of the remaining variables ($k_{3+}$ here).
We group all graphs according to the range of integration of the
leftover variable ($k_{1+}$).
In each such group, we add up the expressions for the
$p_\perp$-broadening graphs and subtract the ones for the
dipole graphs
(Tab.~\ref{tab:diffg}).
\item In the class of graphs for which the
range of integration is unrestricted
(graph $\gamma$ here, see Tab.~\ref{tab:diffg}), 
we invert the sign of the variable on which we integrate ($k_{1+}$ here),
and add an overall ``$-$'' sign to take into account the
change of sign in the integration element.
\item
The equivalence 
between $p_\perp$-broadening and dipole amplitudes
is true if the expressions 
obtained after this transformation
are the same
as the expressions found for all other classes of graphs
(Eq.~(\ref{eq:proofequivalence}) in our example).
\end{enumerate}
This procedure is a practical formulation of
the analytical continuation method
detailed above, which enables one to check
at first sight the equivalence.


\subsubsection{Initial state/final state}

We now turn to the case in which 
one quark-gluon vertex in the amplitude 
is in the initial state, while the two other quark-gluon
couplings on the right of the cut
are in the final state.

\begin{figure}
\begin{center}
\begin{tabular}{cc}
\includegraphics[width=5cm]{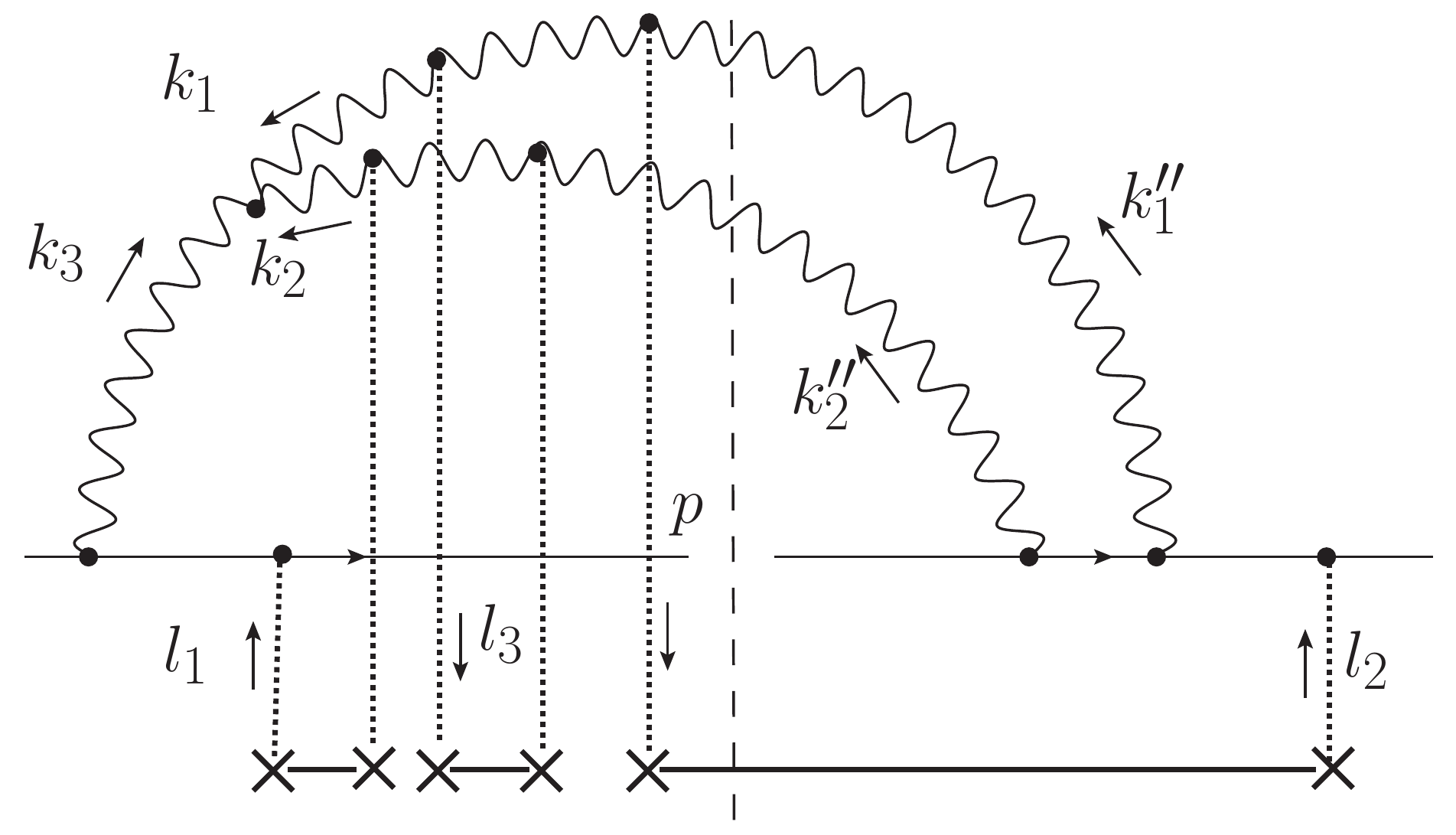}&
\includegraphics[width=5cm]{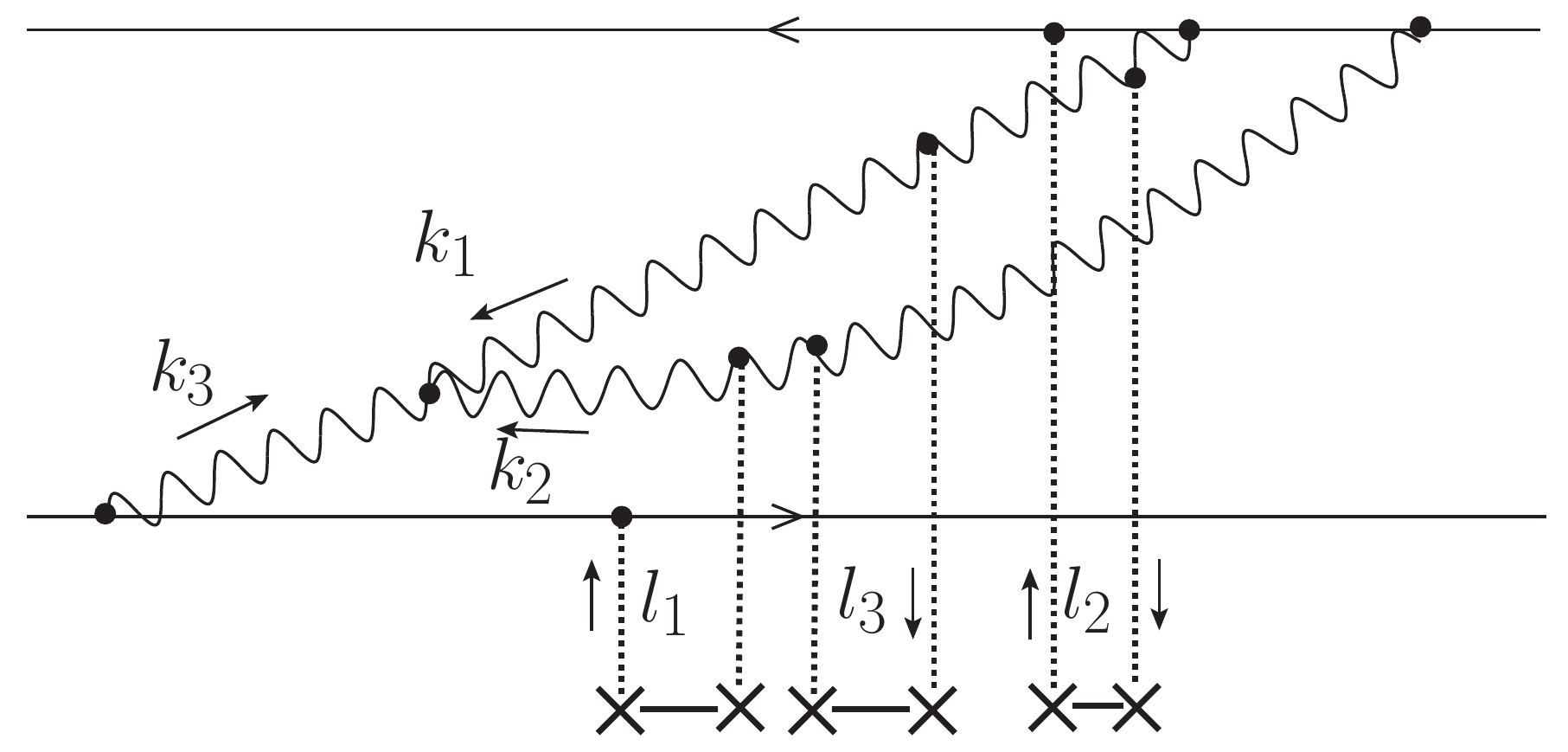}
\end{tabular}
\end{center}
\caption{\label{q3r3}
$p_\perp$-broadening 
graph for the configuration
in which one quark-gluon coupling and the 3-gluon vertex
are in the initial state, while the remaining two $qg$ vertices
are in the final state (left)
and the equivalent dipole graph (right).
}
\end{figure}

In the simplest case, the 3-gluon vertex is also
in the initial state.
This is shown in Fig.~\ref{q3r3}.
This case is actually straightforward.
On one hand, 
the energy denominators, for $p_\perp$-broadening as well as for
dipoles, read
\be
D=\frac{1}{
E^{\prime\prime}_2 E_3
(E_1+E_2)
(E^{\prime\prime}_1+E^{\prime\prime}_2)
}.
\ee
On the other hand, all other terms are identical for
both graphs.

\begin{figure}
\begin{center}
\begin{tabular}{ccc}
\includegraphics[width=4cm]{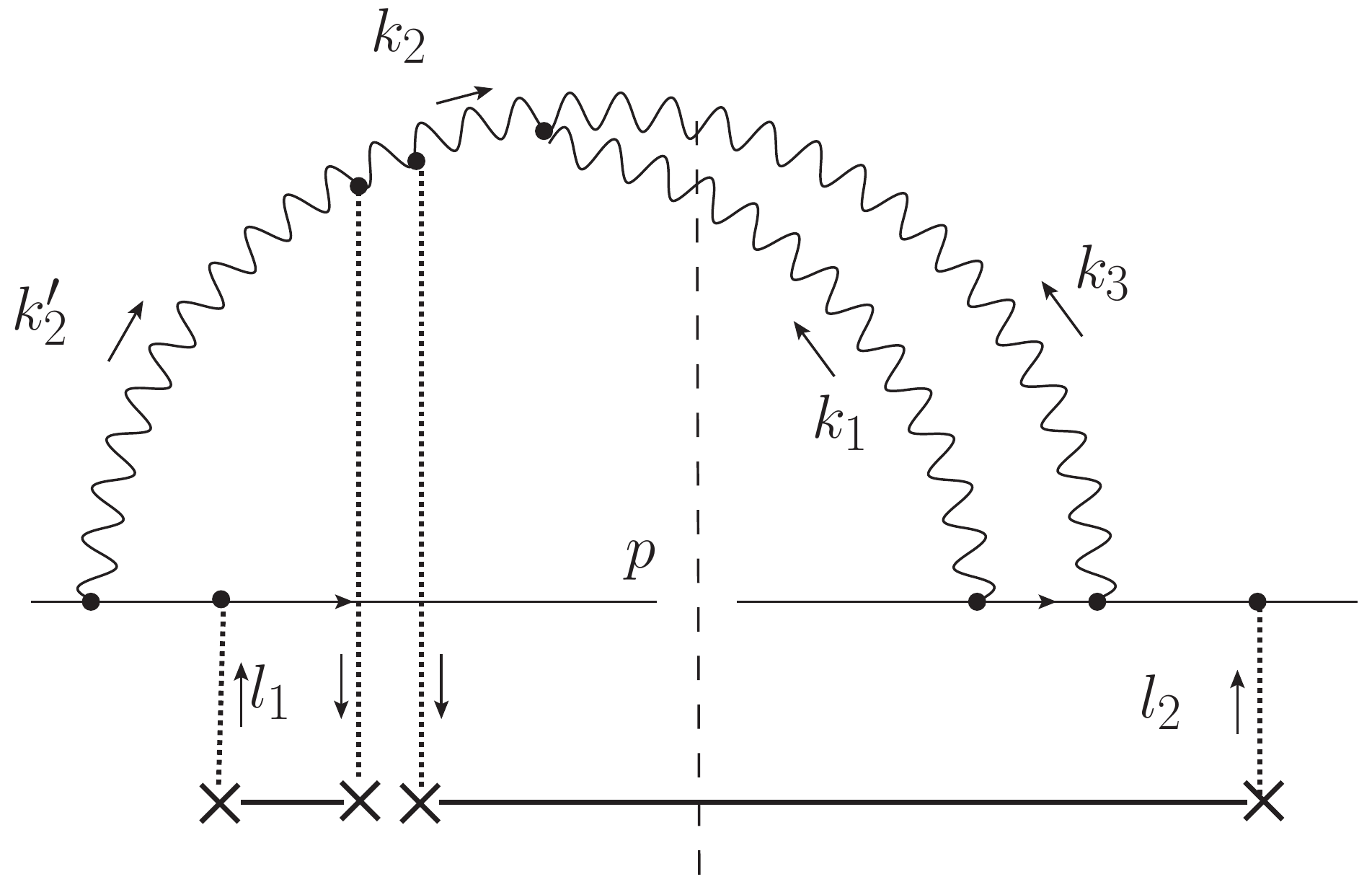}
&\includegraphics[width=4cm]{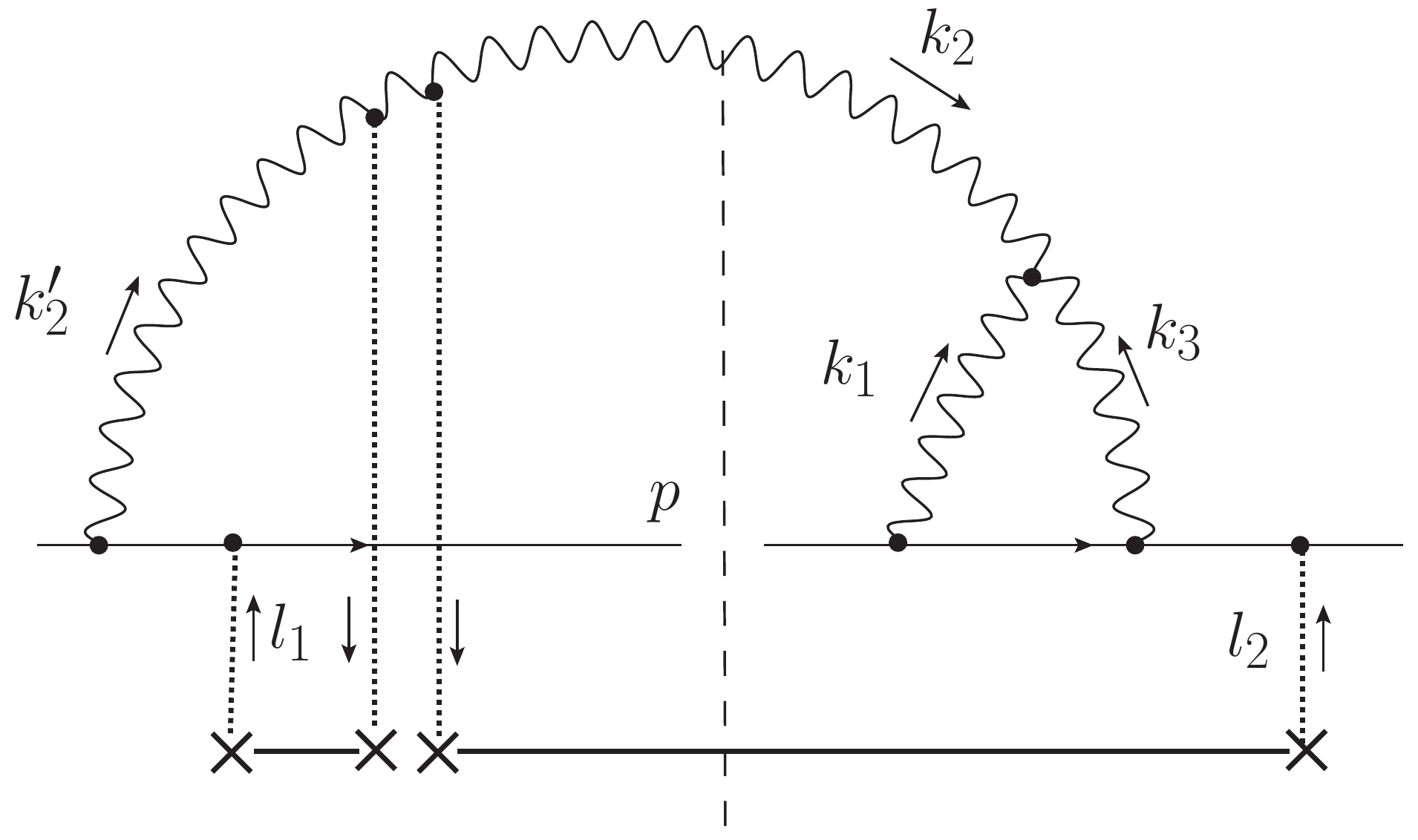}
&\includegraphics[width=4cm]{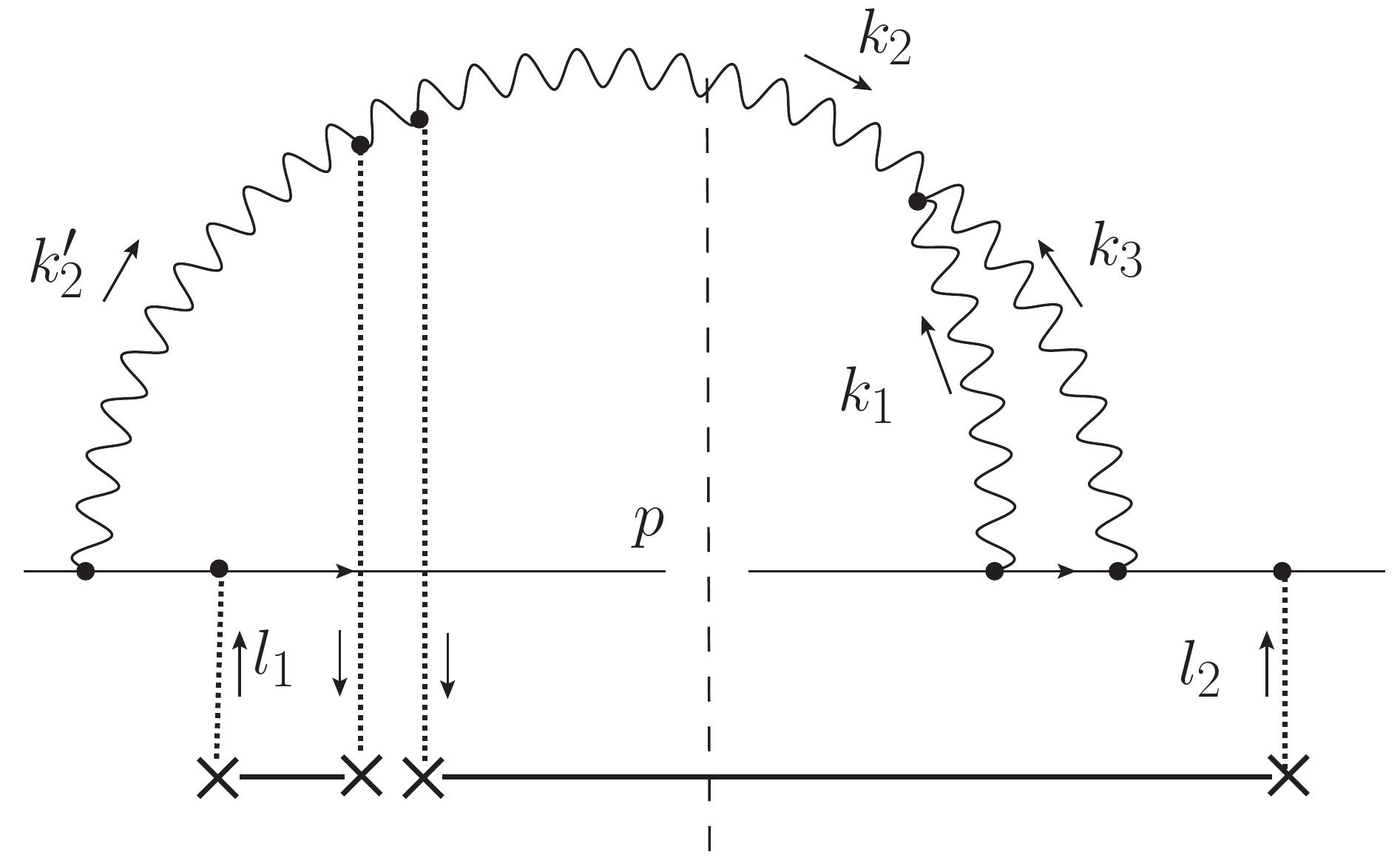}
\\
$A$ & $B$ & $C$
\end{tabular}
\end{center}
\caption{\label{q31}
$p_\perp$-broadening graphs for the configuration
in which all interactions but one quark-gluon coupling
occur in the final state (The instantaneous-exchange graphs are not shown).
}
\end{figure}

\begin{figure}
\begin{center}
\begin{tabular}{cc}
\includegraphics[width=5cm]{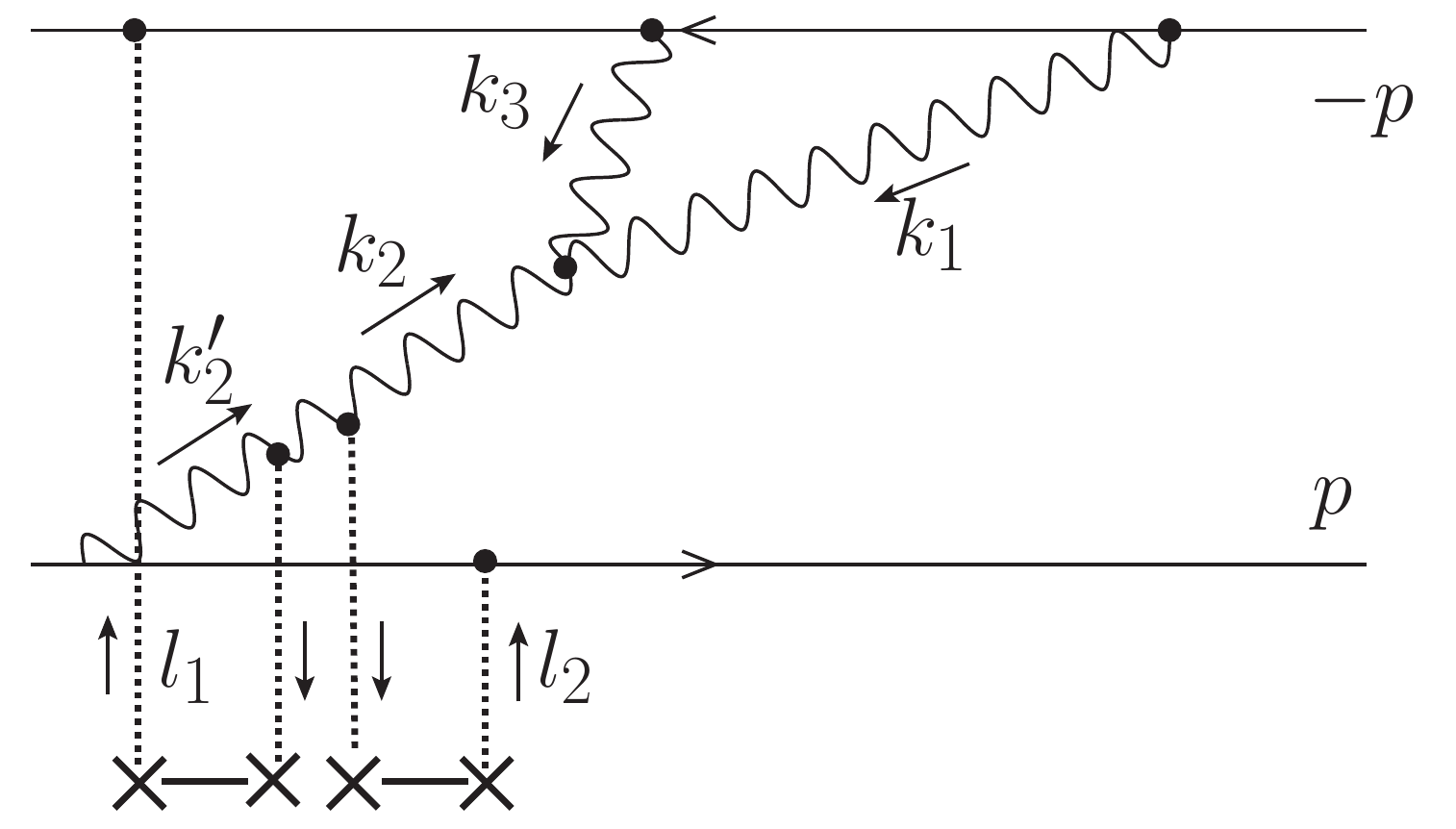}
&\includegraphics[width=5cm]{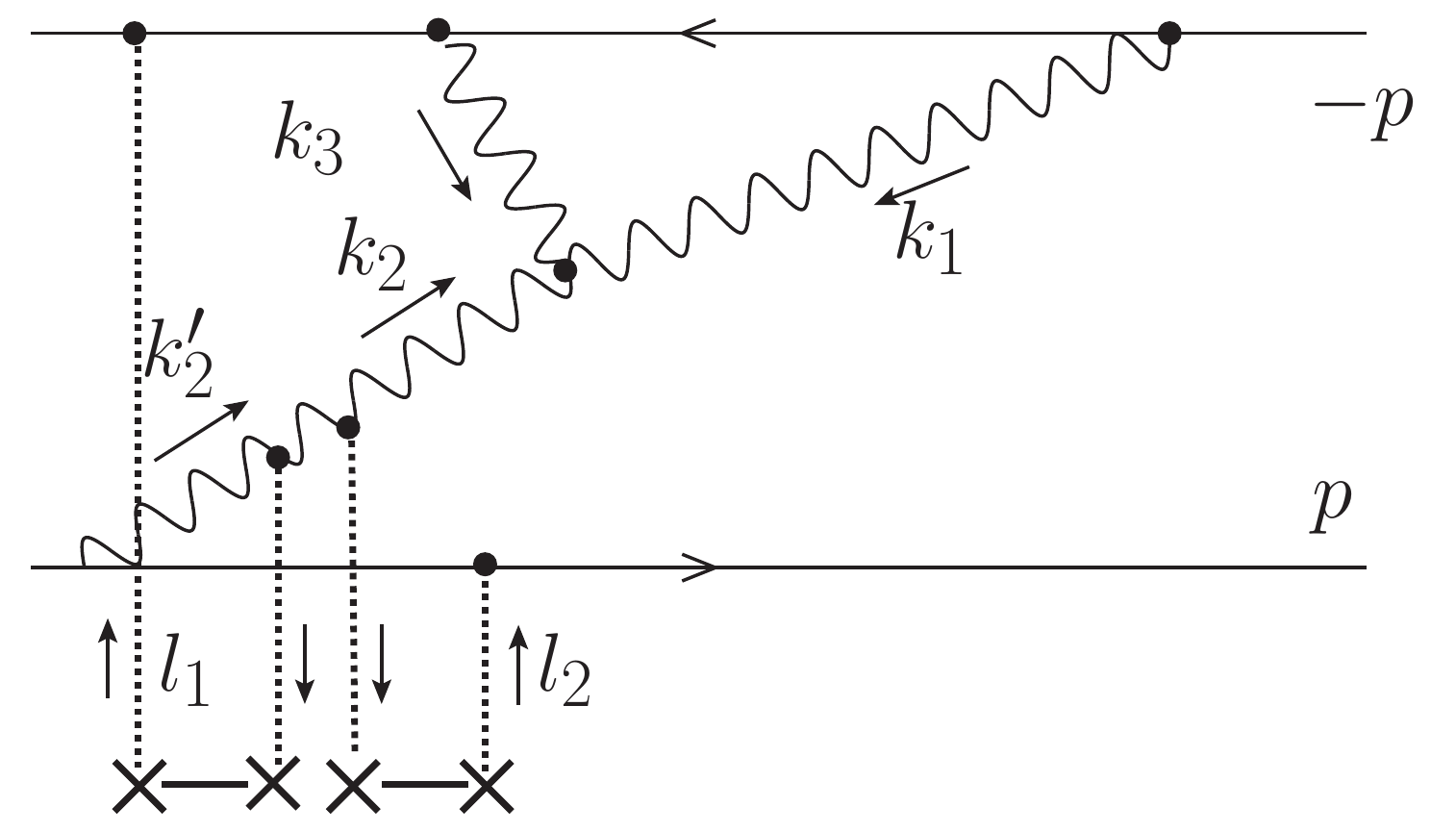}
\\
$\alpha$ & $\beta$
\end{tabular}
\end{center}
\caption{\label{q3d1}
Dipole graphs corresponding to the $p_\perp$-broadening graphs
of Fig.~\ref{q31} (The instantaneous-exchange graphs are not shown).
}
\end{figure}

The graphs in which the 3-gluon vertex
is in the final state are shown in Fig.~\ref{q31} 
(and Fig.~\ref{q3d1} for
the corresponding dipole graphs).
We may group the graphs $A$, $C$ and $\alpha$
in Fig.~\ref{q31} and~\ref{q3d1}
since they share the
same kinematics, while we treat separately the graphs $B$ and $\beta$.
We do not present the evaluation of all diagrams separately 
for this case (step~1 in the procedure outlined above)
since
their expression is very similar to the previous case.
We find that the expressions for $A+C-\alpha$ and for $B$
are identical once we express everything in terms of the variables
$k_{2+}$ and $k_{3+}$.

The overall factor is again given by Eq.~(\ref{eq:overallfactor}).
The results for the nonzero components of $A+C-\alpha$, $B$ and $\beta$
are shown in Tab.~\ref{tab:diffg1}.

\begin{table}
\begin{center}
\begin{tabular}{c|c|c|c}
 & Integration range & \multicolumn{2}{c}{Momentum-dependent factors}\\
 & & pure ``$\perp$'' & ``+'' and ``$\perp$''\\
\hline
$A_2+C_2-\alpha_2$, $B_2$ &\multirow{3}{*}{
\begin{minipage}[c]{3cm}{\centerline{
$k_{3+}<k_{2+}<+\infty$,}
\centerline{$0<k_{2+}<k_{3+}$}}
\end{minipage}}
&$\frac{(\vec k_1\cdot \vec k_3)(\vec k_2\cdot \vec k_2^\prime)}
{\vec k_1^2 \vec k_2^2 \vec k_2^{\prime 2}}$&
$\frac{1}{\vec k_2^2/k_{2+}-\vec k_3^2/k_{3+}}
\frac{k_{2+}-2k_{3+}}{(k_{2+}-k_{3+}) k_{2+}k_{3+}^2}$
\\
$A_3+C_3^*-\alpha_3^*$, $B_3$ &
&
$\frac{(\vec k_1\cdot \vec k_2^\prime)}
{\vec k_1^2 \vec k_2^{\prime 2}}$&
$\frac{1}{\vec k_2^2/k_{2+}-\vec k_3^2/k_{3+}}
\frac{2k_{2+}-k_{3+}}{(k_{2+}-k_{3+}) k_{2+}k_{3+}^2}$
\\
$A_\perp+C_\perp-\alpha_\perp$, $B_\perp$ & &
$\frac{(\vec k_2\cdot \vec k_3)(\vec k_1\cdot \vec l)
-(\vec k_1\cdot \vec k_2)(\vec k_3\cdot \vec l)
}
{k_1^2 k_2^2 k_2^{\prime 2}}$&
$\frac{1}{\vec k_2^2/k_{2+}-\vec k_3^2/k_{3+}}
\frac{1}{(k_{2}-k_3)_+ k_{3+}^2}$
\\
\hline
$\beta_2$ &\multirow{3}{*}{$0<k_{2+}<+\infty$}
& $\frac{(\vec k_1\cdot \vec k_3)(\vec k_2\cdot \vec k_2^\prime)}
{\vec k_1^2 \vec k_2^2 \vec k_2^{\prime 2} }$&
$-\frac{1}{\vec k_2^2/k_{2+}+\vec k_3^2/k_{3+}}
\frac{k_{2+}+2k_{3+}}{(k_{2+}+k_{3+}) k_{2+}k_{3+}^2}$
\\
$\beta_3^*$ &
&
$\frac{(\vec k_1\cdot \vec k_2^\prime)}
{\vec k_1^2 \vec k_2^{\prime 2}}$&
$-\frac{1}{\vec k_2^2/k_{2+}+\vec k_3^2/k_{3+}}
\frac{2k_{2+}+k_{3+}}{(k_{2+}+k_{3+}) k_{2+}k_{3+}^2}$
\\
$\beta_\perp$ & &
$\frac{(\vec k_2\cdot \vec k_3)(\vec k_1\cdot \vec l)
-(\vec k_1\cdot \vec k_2)(\vec k_3\cdot \vec l)
}
{k_1^2 k_2^2 k_2^{\prime 2}}$&
$-\frac{1}{\vec k_2^2/k_{2+}+\vec k_3^2/k_{3+}}
\frac{1}{(k_{2}+k_3)_+ k_{3+}^2}$
\\
\end{tabular}
\caption{\label{tab:diffg1}%
Sums and differences of the graphs 
of Fig.~\ref{q31} and~\ref{q3d1}
grouped according to kinematics.
The overall constant is $64\pi^4\alpha_s^4N_c^3C_F$.
}
\end{center}
\end{table}
It is clear that the analytical continuation of 
the components of the graph
$\beta$ in the variable $k_{2+}$ is equal to $A+C-\alpha$ for $k_{2+}>k_{3+}$
and to $B$ for $k_{2+}<k_{3+}$
as far as the real part is concerned.
The imaginary parts cancel with the complex conjugate graphs that also have
to be taken into account.


\subsubsection{Initial state/initial state}

\begin{figure}
\begin{center}
\begin{tabular}{cc}
\includegraphics[width=4cm]{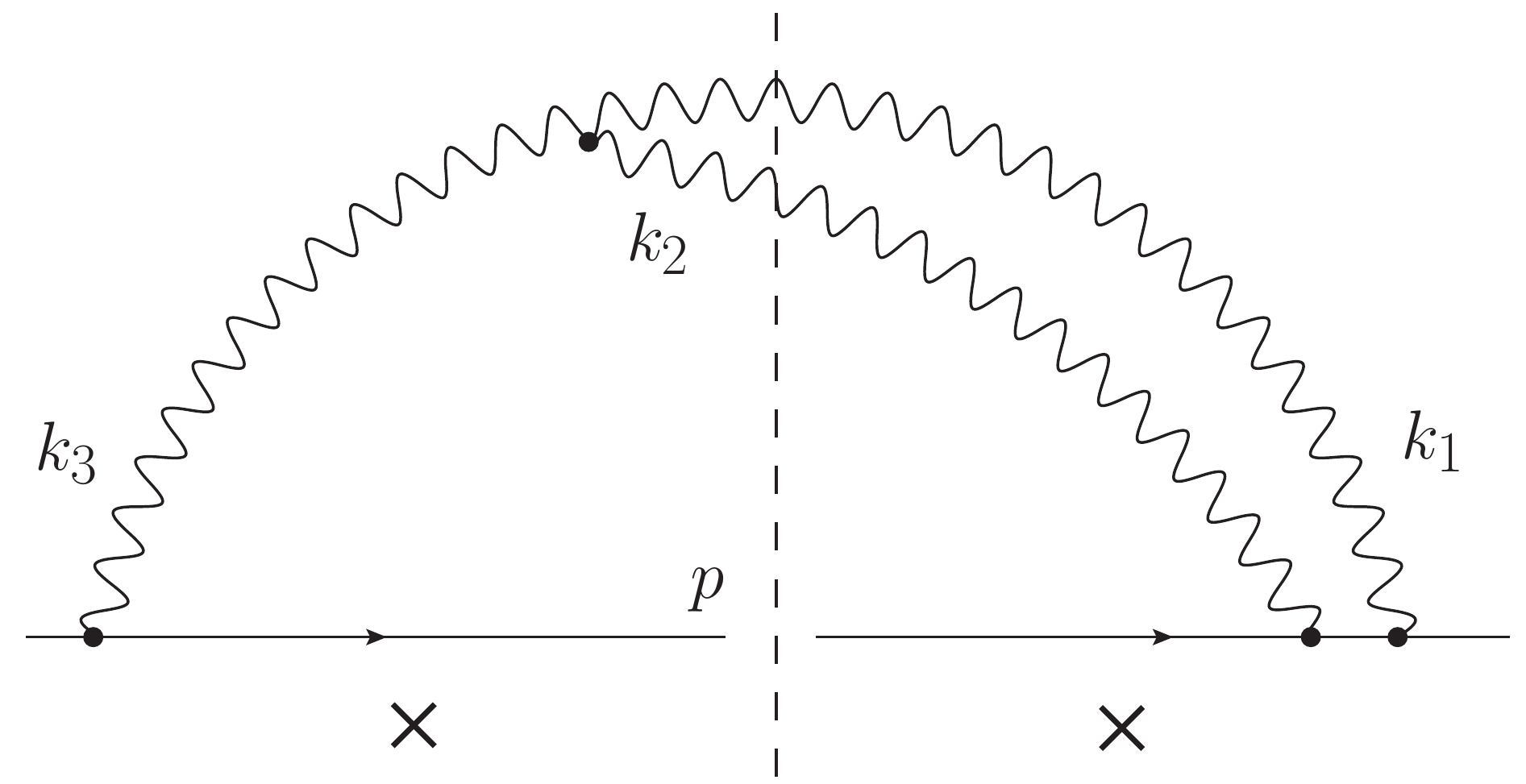}&
\includegraphics[width=4cm]{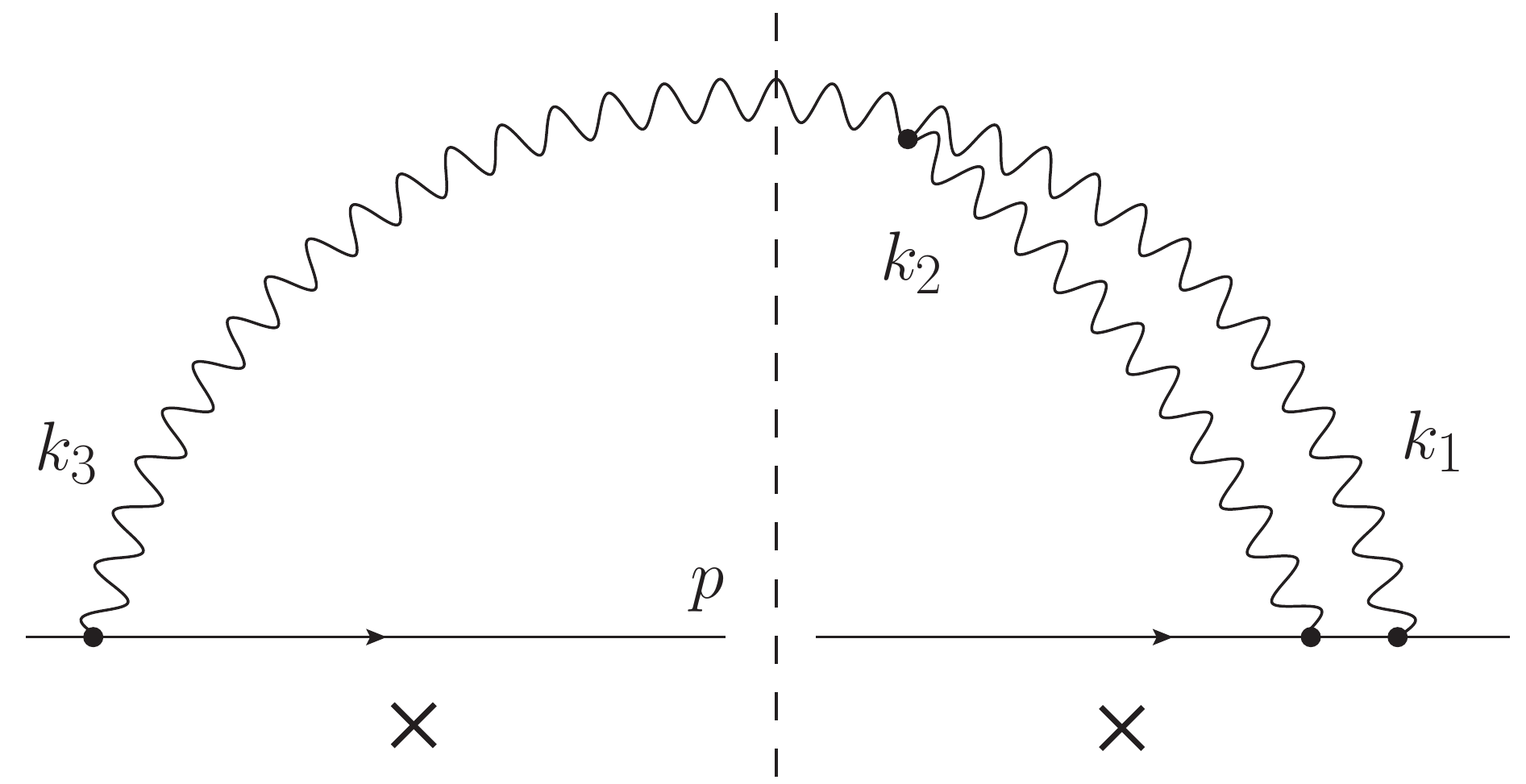}\\
\end{tabular}
\end{center}
\caption{\label{q3vnull}
Set of the $p_\perp$-broadening graphs for which all interactions occur
in the initial state but which cancel.
We have not drawn explicitly 
the gluons exchanged with the target.
}
\end{figure}

We turn to the graphs in which all vertices are in the
initial state.
First, there is the trivial case of the $p_\perp$-broadening
graphs in Fig.~\ref{q3vnull}
which cancel in exactly the same way e.g. as
the graphs of Fig.~\ref{q2iifi} cancelled.
There is no kinematically-allowed equivalent dipole graph.

\begin{figure}
\begin{center}
\begin{tabular}{ccc}
\includegraphics[width=4cm]{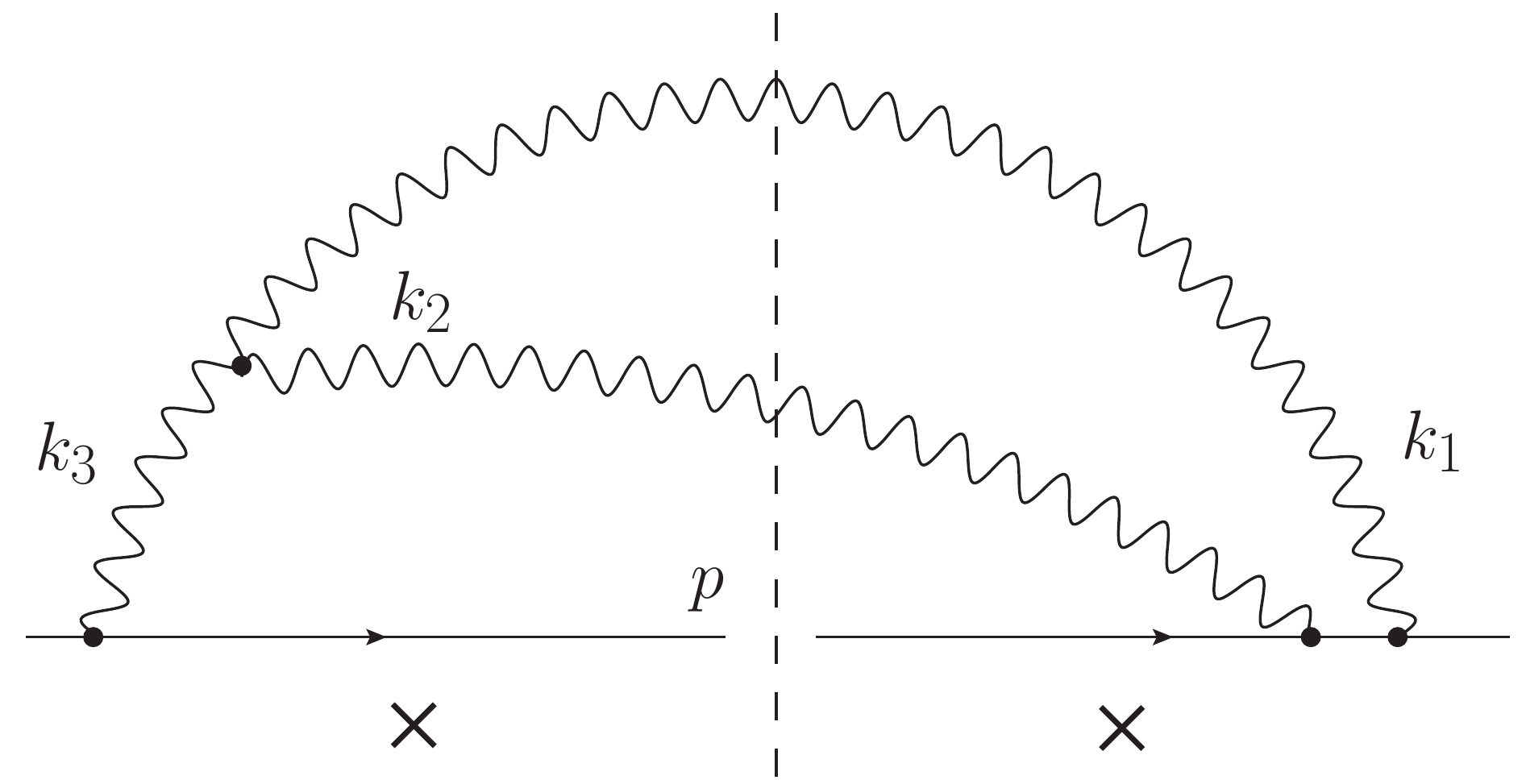}&
\includegraphics[width=4cm]{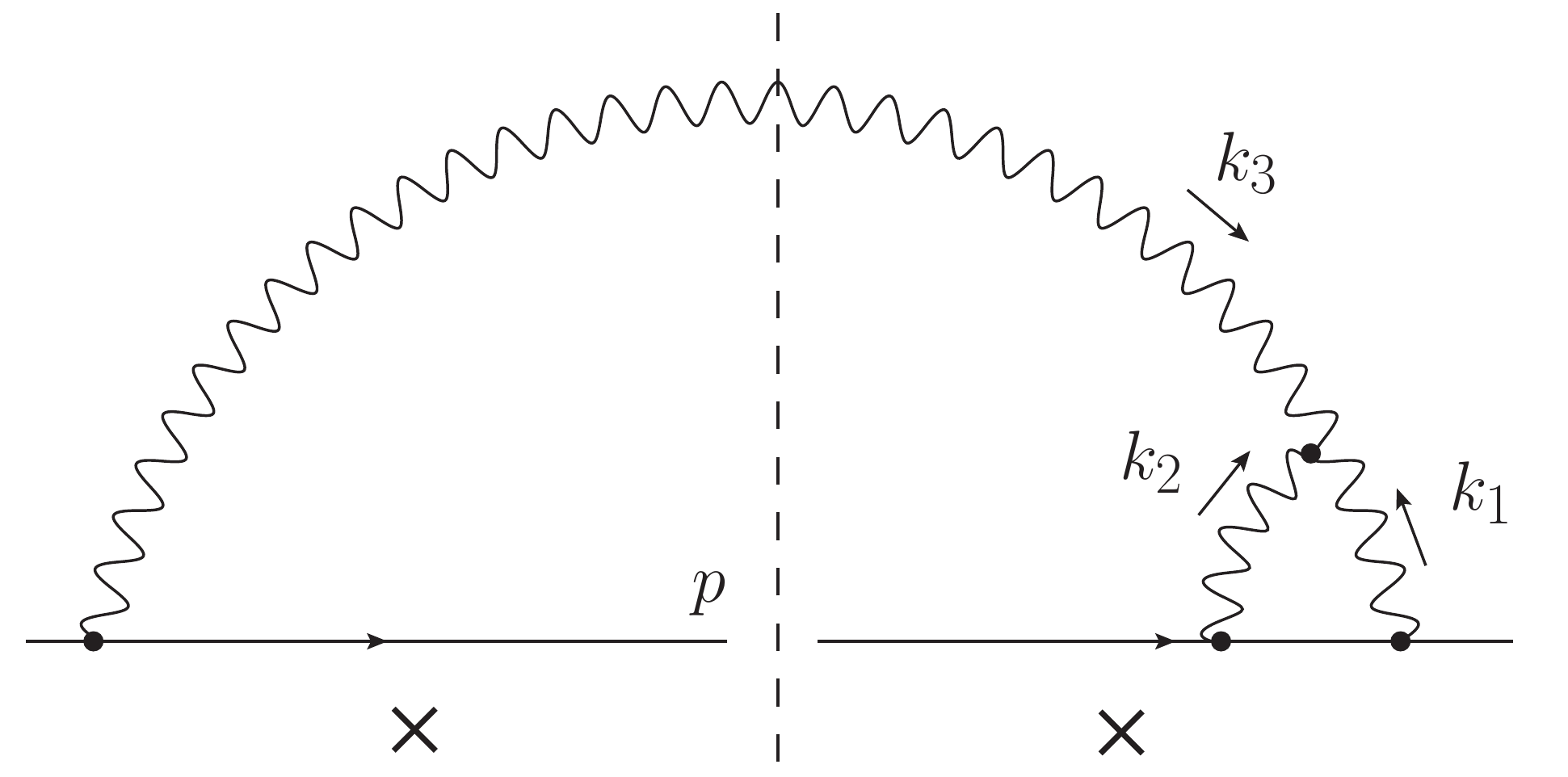}&
\includegraphics[width=4cm]{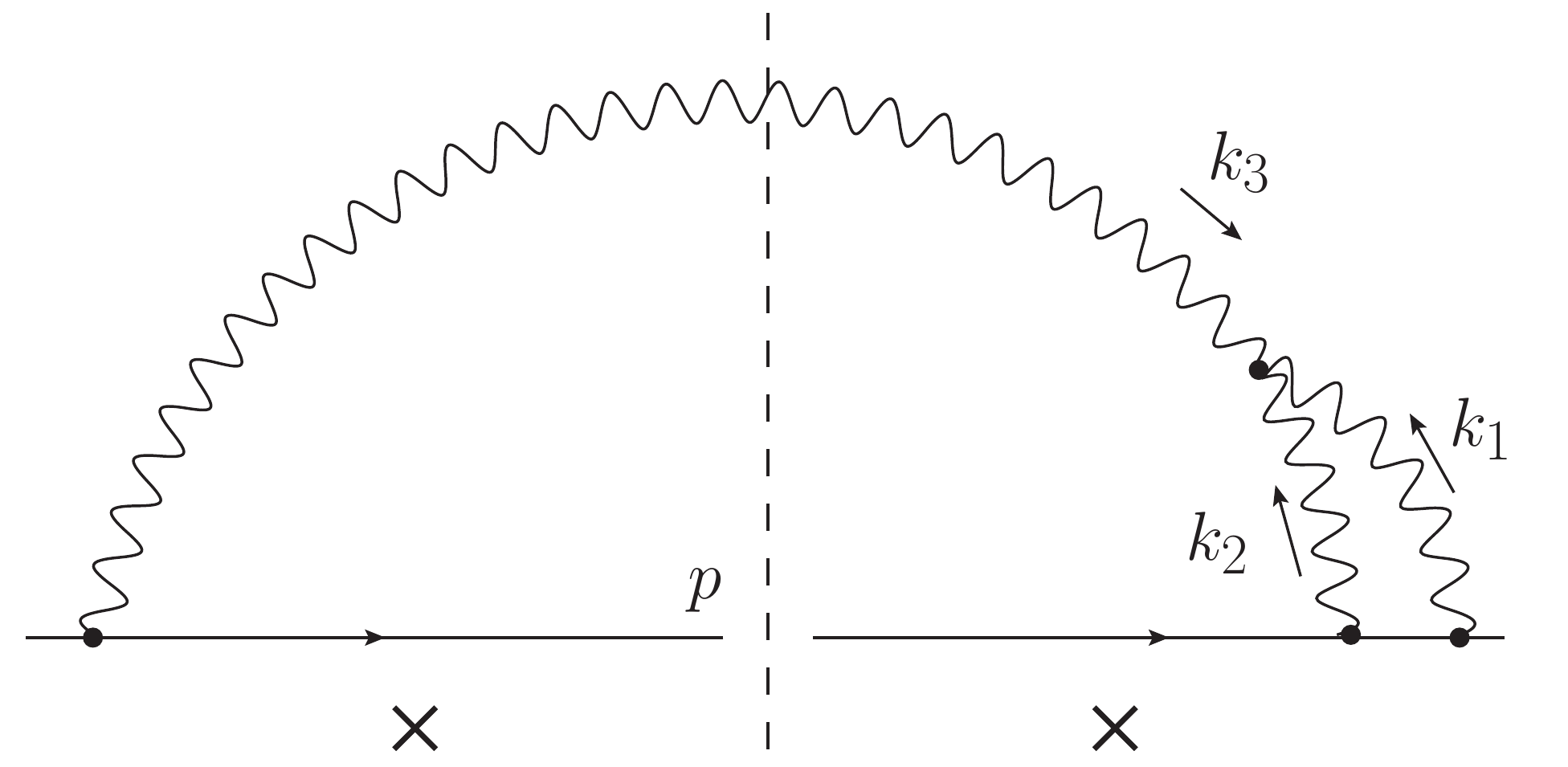}\\
$A$ & $B$ & $C$
\end{tabular}
\end{center}
\caption{\label{q3v}
The same as Fig.~\ref{q3vnull}
but for the non-cancelling graphs.
The gluons exchanged with the target (not drawn)
all couple to the quark line, and thus are not relevant to
the discussion of the equivalence.
}
\end{figure}

\begin{figure}
\begin{center}
\begin{tabular}{cccc}
 & & & 
\includegraphics[width=3cm]{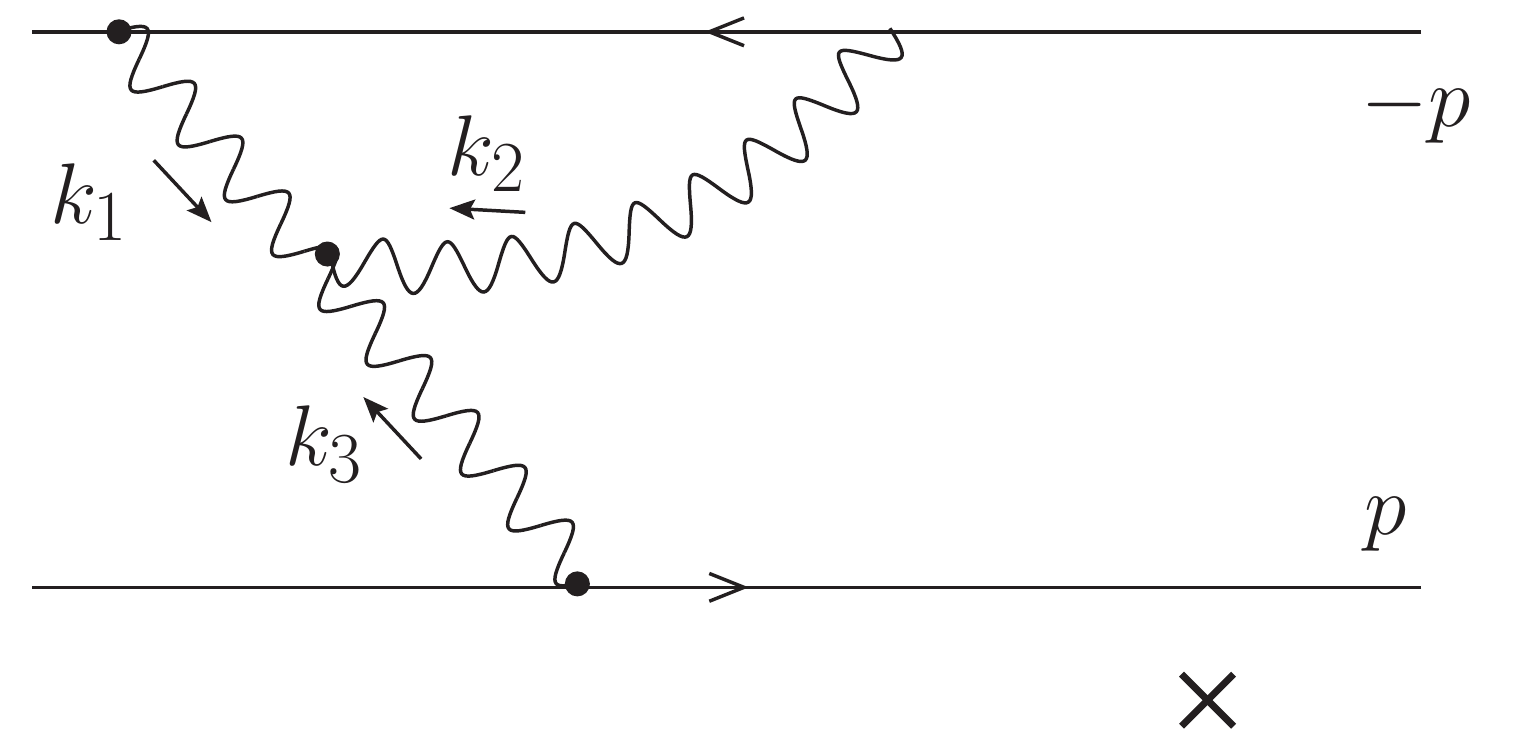}\\
\includegraphics[width=3cm]{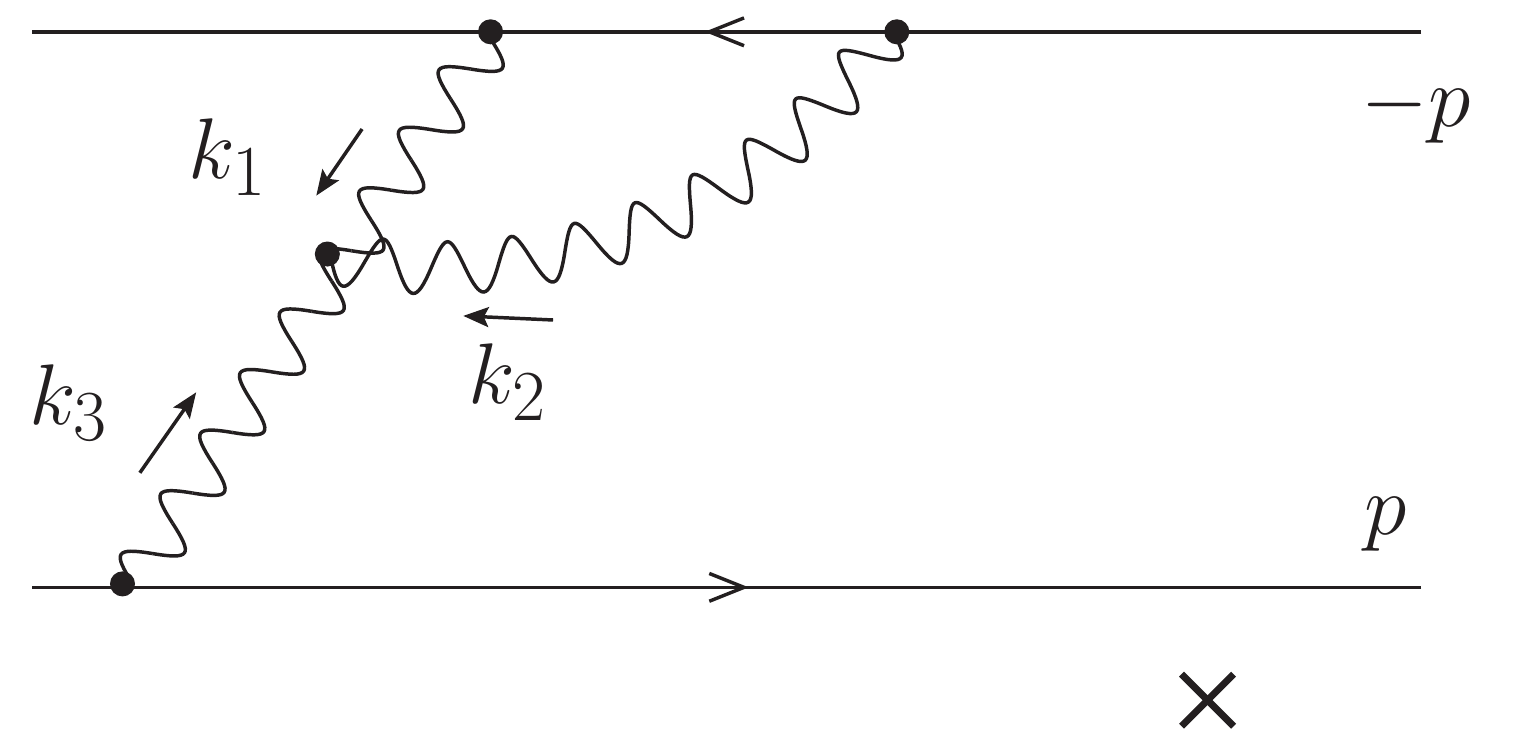}&
\includegraphics[width=3cm]{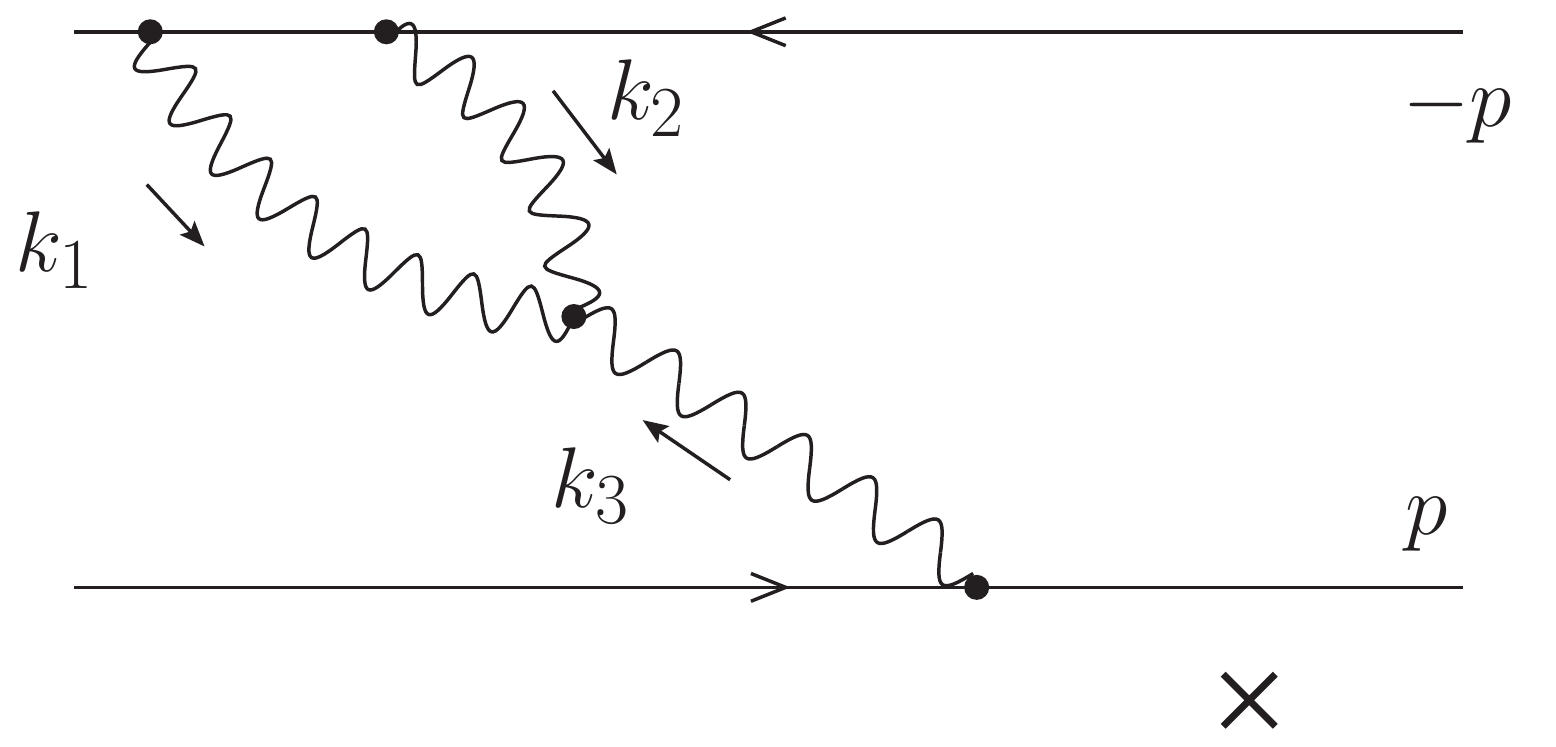}&
\includegraphics[width=3cm]{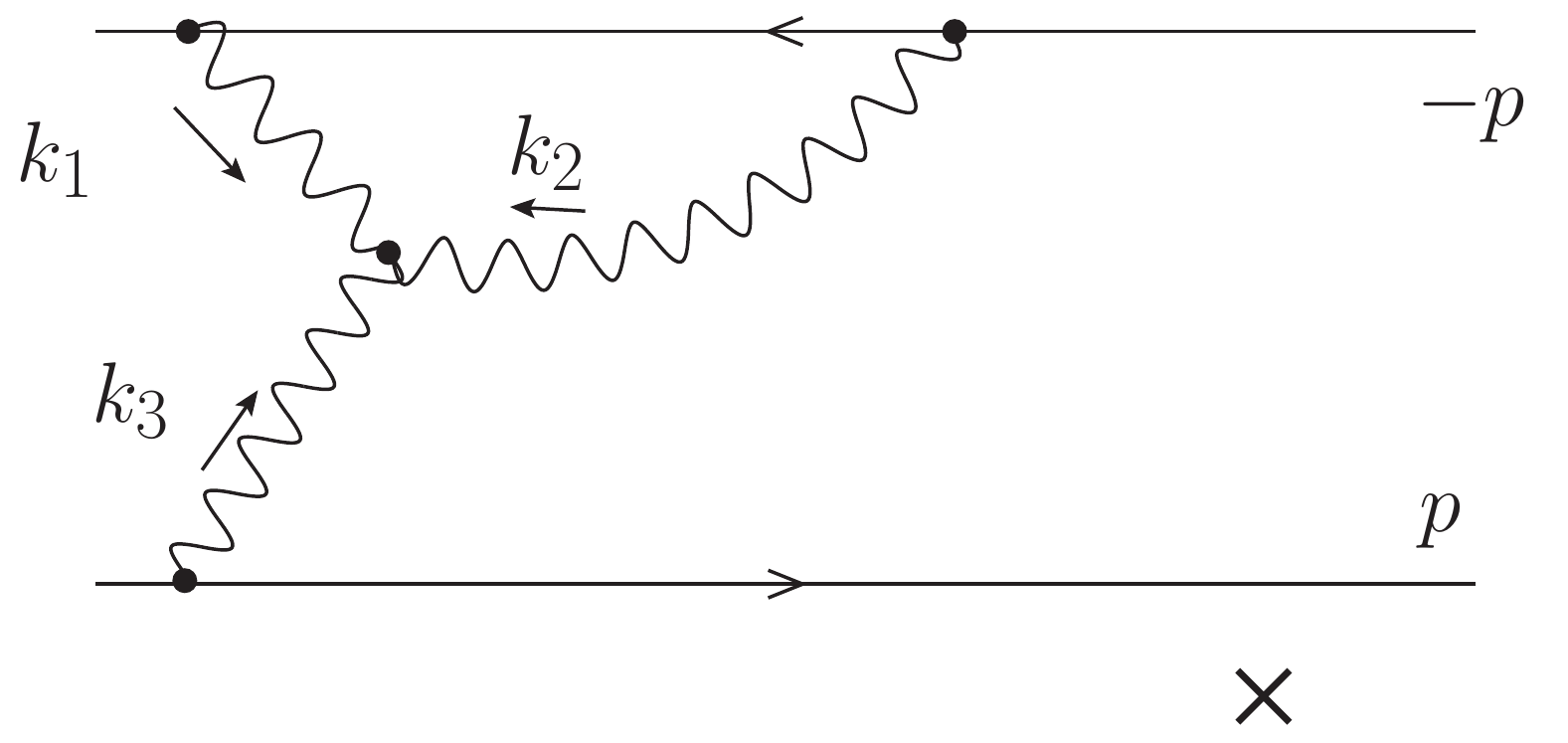}&
\includegraphics[width=3cm]{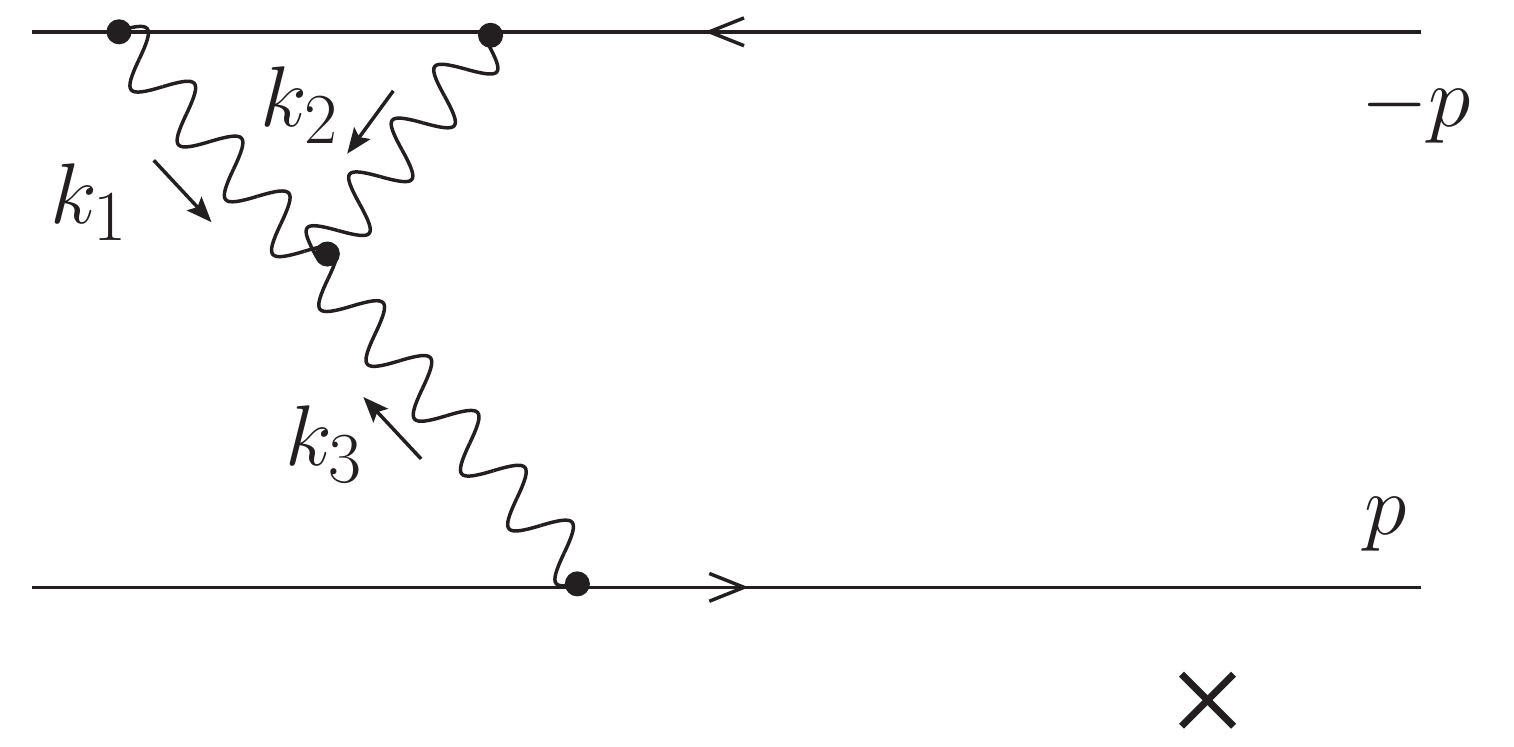}\\
$\alpha$ & $\beta$ & $\gamma$ & 
$\delta^{(1)}$ (up) and $\delta^{(2)}$ (down)
\end{tabular}
\end{center}
\caption{\label{q3dv}
Set of the dipole graphs
corresponding to the $p_\perp$-broadening graphs of Fig.~\ref{q3v}.
$\delta^{(1)}$ and $\delta^{(2)}$ differ by the ordering of the
absorption times of gluons $2$ and $3$.
$\gamma$ actually represents 2 graphs, which differ by the
order of the leftmost $\bar qg$ and $qg$ vertices.
}
\end{figure}

We turn to the more interesting graphs in Fig.~\ref{q3v}.
All interactions between the nucleus 
and one of the gluons cancel, hence the exchanged
gluons
necessarily hook to the quark line.
The corresponding dipole graphs are virtual (Fig.~\ref{q3dv}).

The evaluation of the $p_\perp$-broadening graphs
of Fig.~\ref{q3v} is straightforward.
The results are presented in Tab.~\ref{tab:ABCv}.

\begin{table}
\begin{center}
\begin{tabular}{c|c|c|c}
& Integration range & Energy denom.& 3-gluon$\times$pola.\\
\hline
$A_1$ & 
\multirow{3}{*}{$0<k_{1+}<k_{3+}$} 
& \multirow{3}{*}{$\frac{1}{E_1 E_3(E_1+E_2)^2}$}
& $(\vec k_2\cdot\vec k_3)E_1(2k_3-k_1)_+$\\
$A_2$ & & & $(\vec k_1\cdot\vec k_3)
(-E_2)(k_1+k_3)_+$\\
$A_3^*$ & & & $0$\\
\hline
$B_1^*$ &  \multirow{3}{*}{$k_{1+}>k_{3+}$} 
& \multirow{3}{*}{$\frac{1}{E_1 E_3^2(E_2+E_3)}$}
& 0\\
$B_2^*$ & & &$(\vec k_1\cdot\vec k_3)
E_3(k_1+k_3)_+$\\
$B_3$ & & &$(\vec k_1\cdot\vec k_2)
E_3(2k_1-k_3)_+$\\
\hline
$C_1$ & \multirow{3}{*}{$0<k_{1+}<k_{3+}$} 
& \multirow{3}{*}{$\frac{1}{E_1 E_3^2(E_1+E_2)}$}
&  $(\vec k_2\cdot\vec k_3)E_1(2k_3-k_1)_+$\\
$C_2^*$ & & &$(\vec k_1\cdot\vec k_3)
E_1(k_1+k_3)_+$\\
$C_3$ & & &$(\vec k_1\cdot\vec k_2)
E_3(k_3-2k_1)_+$\\
\end{tabular}
\caption{\label{tab:ABCv}%
Expressions of the graphs of Fig.~\ref{q3v}.
The momentum conservation reads
$k_{2+}=k_{3+}-k_{1+}$ for graphs $A$ and $C$, and
$k_{2+}=k_{1+}-k_{3+}$ for graph $B$.
These expressions must be multiplied by
the overall factor~(\ref{eq:overallfactorv}).
}
\end{center}
\end{table}

The dipole graphs in Fig.~\ref{q3dv}
bring in a new difficulty with respect e.g. to the real
graphs in Fig.~\ref{q3r3},\ref{q31},\ref{q3d1}:
The energy denominators diverge, 
thus it is crucial to keep the regulator $\varepsilon$ throughout.
The divergencies are not removed by instantaneous exchanges
as was the case for other graphs, but they
give imaginary contributions which are cancelled when
one adds the complex conjugate graphs.
The results are shown in Tab.~\ref{tab:abcdv}.

\begin{table}
\begin{center}
\begin{tabular}{c|c|c|c}
& Integration range & Energy denominators& 3-gluon vertex$\times$pola.\\
\hline
$\alpha_1^*$ & 
\multirow{6}{*}{$0<k_{1+}<k_{3+}$} 
& \multirow{3}{*}{$-\frac{1}{4i\varepsilon}
\frac{1}{(E_2-3i\varepsilon)(E_3-i\varepsilon)(E_1+E_2-2i\varepsilon)}$}
& $(\vec k_2\cdot\vec k_3)(2i\varepsilon-E_2)(2k_3-k_1)_+$\\
$\alpha_2$ & & & $(\vec k_1\cdot\vec k_3)
(-E_2)(k_1+k_3)_+$\\
$\alpha_3^*$ & & & $(\vec k_1\cdot\vec k_2)
(i\varepsilon)(k_3-2k_1)_+$\\
\cline{1-1}\cline{3-4}
$\beta_1$ & 
& \multirow{3}{*}{$-\frac{1}{4i\varepsilon}
\frac{1}{(E_1-i\varepsilon)(E_3-3i\varepsilon)(E_1+E_2-2i\varepsilon)}$}
& $(\vec k_2\cdot\vec k_3)(-E_1)(2k_3-k_1)_+$\\
$\beta_2^*$ & & &$(\vec k_1\cdot\vec k_3)
(2i\varepsilon-E_1)(k_1+k_3)_+$\\
$\beta_3^*$ & & &$(\vec k_1\cdot\vec k_2)
(3i\varepsilon)(2k_1-k_3)_+$\\
\hline
$\gamma_1^*$ & \multirow{3}{*}{$0<k_{1+}<+\infty$} 
& \multirow{3}{*}{$-\frac{1}{4i\varepsilon}
\frac{1}{(E_1-i\varepsilon)(E_3-i\varepsilon)(E_2-2i\varepsilon)}$}
&  $(\vec k_2\cdot\vec k_3)(i\varepsilon)(2k_3+k_1)_+$\\
$\gamma_2^*$ & & &$(\vec k_1\cdot\vec k_3)
(3i\varepsilon)(k_3-k_1)_+$\\
$\gamma_3^*$ & & &$(\vec k_1\cdot\vec k_2)
(-i\varepsilon)(2k_1+k_3)_+$\\
\hline
$\delta_1^{(1)*}$ & \multirow{6}{*}{$k_{3+}<k_{1+}<+\infty$} 
& \multirow{3}{*}{$-\frac{1}{4i\varepsilon}
\frac{1}{(E_1-i\varepsilon)(E_2+E_3-2i\varepsilon)(E_2-3i\varepsilon)}$
}
&  $(\vec k_2\cdot\vec k_3)(i\varepsilon)(2k_3-k_1)_+$\\
$\delta_2^{(1)}$ & & & $(\vec k_1\cdot\vec k_3)
E_2(k_1+k_3)_+$\\
$\delta_3^{(1)*}$ & & &$(\vec k_1\cdot\vec k_2)
(E_2-2i\varepsilon)(2k_1-k_3)_+$\\
\cline{1-1}\cline{3-4}
$\delta_1^{(2)*}$ & 
& \multirow{3}{*}{$-\frac{1}{4i\varepsilon}
\frac{1}{(E_1-i\varepsilon)(E_2+E_3-2i\varepsilon)(E_3-3i\varepsilon)}$
}
&  $(\vec k_2\cdot\vec k_3)(i\varepsilon)(2k_3-k_1)_+$\\
$\delta_2^{(2)*}$ & & & $(\vec k_1\cdot\vec k_3)
(2i\varepsilon-E_3)(k_1+k_3)_+$\\
$\delta_3^{(2)}$ & & &$(\vec k_1\cdot\vec k_2)
(-E_3)(2k_1-k_3)_+$\\
\end{tabular}
\caption{\label{tab:abcdv}%
Expressions of the graphs of Fig.~\ref{q3dv}.
The momentum conservation reads
$k_{2+}=k_{3+}-k_{1+}$ for graphs $\alpha$ and $\beta$,
$k_{2+}=k_{1+}+k_{3+}$ for graph $\gamma$,
and $k_{2+}=k_{1+}-k_{3+}$ for graph $\delta$.
These expressions must be multiplied by
the overall factor~(\ref{eq:overallfactorv}).
The contribution to $p_\perp$-broadening is obtained by
convoluting with the weight given in Eq.~(\ref{eq:convolution}).
}
\end{center}
\end{table}

The overall factor, including the averaging over the color
and helicity of the quark, reads
\be
-\frac{32\pi^4\alpha_s^4 N_c C_F^3}{(k_{1+}k_{2+}k_{3+})^2},
\label{eq:overallfactorv}
\ee
which is identical to~(\ref{eq:overallfactor}) in the large-$N_c$
limit ($C_F\simeq N_c/2$) except for the sign.

Again, we may put together the graphs with the
same kinematics.
This is shown in Tab.~\ref{tab:diffgv}.
We have left out some imaginary terms, namely
\be
\begin{split}
\text{Im}(A_1+C_1-\alpha_1^*-\beta_1)
&\propto
\frac{1}{\varepsilon \vec k_3^2(\vec k_1^2(k_{3+}-k_{1+})
+\vec k_2^2 k_{1+})}
\frac{(2k_3-k_1)_+}{k_{1+}(k_3-k_1)_+ k_{3+}}\\
\text{Im}(A_2+C_2^*-\alpha_2-\beta_2^*)
&\propto
\frac{1}{\varepsilon \vec k_3^2(\vec k_1^2(k_{3+}-k_{1+})
+\vec k_2^2 k_{1+})}
\frac{(k_1+k_3)_+}{k_{1+}(k_3-k_1)_+ k_{3+}}
\end{split}
\ee
which are divergent. 
The instantaneous-exchange graphs are not enough
to remove all divergencies, as they do in other cases.
But these terms do not play any r\^ole
in our discussion since we must
eventually add all complex conjugate
graphs together.

\begin{table}
\begin{center}
\begin{tabular}{c|c|c|c}
 & Integration range & \multicolumn{2}{c}{Momentum-dependent factors}\\
 & & ``$\perp$'' & ``+''\\
\hline
$\text{Re}%
(A_1+C_1-\alpha_1^*-\beta_1)$, $B_1^*-\delta_1^*$ &\multirow{3}{*}{
\begin{minipage}[c]{3cm}{\centerline{
$0<k_{1+}<k_{3+}$,}
\centerline{$k_{3+}<k_{1+}<+\infty$}}
\end{minipage}}
&$\frac{\vec k_2\cdot \vec k_3}
{\vec k_1^2 \vec k_2^2 \vec k_3^{2} }$&
$\frac{(2k_{3}-k_{1})_+}{(k_{1}-k_{3})_+ k_{1+}k_{3+}}$
\\
$\text{Re}%
(A_2+C_2^*-\alpha_2-\beta_2^*)$, $B_2^*-\delta_2^*$ &
&
$\frac{\vec k_1\cdot \vec k_3}
{\vec k_1^2 \vec k_2^{2} \vec k_3^3}$&
$\frac{3(k_{1}+k_{3})_+}{(k_{1}-k_{3})_+ k_{1+}k_{3+}}$
\\
$A_3^*+C_3-\alpha_3^*-\beta_3$, $B_3-\delta_3^*$ &
&
$\frac{\vec k_1\cdot \vec k_2}
{\vec k_1^2 \vec k_2^{2}\vec k_3^2}$&
$\frac{(2k_{1}-k_{3})_+}{(k_{1}-k_{3})_+ k_{1+}k_{3+}}$
\\
\hline
$\gamma_1^*$ &\multirow{3}{*}{$0<k_{1+}<+\infty$}
& $\frac{\vec k_2\cdot \vec k_3}
{\vec k_1^2 \vec k_2^2 \vec k_3^{2}}$&
$-\frac{(k_{1}+2k_{3})_+}{(k_{1}+k_{3})_+ k_{1+}k_{3+}}$
\\
$\gamma_2^*$ &
& $\frac{\vec k_1\cdot \vec k_3}
{\vec k_1^2 \vec k_2^2 \vec k_3^{2} }$&
$\phantom{-}\frac{3(k_{1}-k_{3})_+}{(k_{1}+k_{3})_+ k_{1+}k_{3+}}$
\\
$\gamma_3^*$ &
&
$\frac{\vec k_1\cdot \vec k_2}
{\vec k_1^2 \vec k_2^{2}\vec k_3^2}$&
$\phantom{-}\frac{(2k_{1}+k_{3})_+}{(k_{3}+k_{1})_+ k_{1+}k_{3+}}$
\end{tabular}
\caption{\label{tab:diffgv}%
Sums and differences of the graphs 
of Fig.~\ref{q3v} and~\ref{q3dv}
grouped according to kinematics.
The overall constant to be added reads 
$-64\pi^4\alpha_s^4 N_c C_F^3$.
}
\end{center}
\end{table}


\subsubsection{Final state/final state}

\begin{figure}
\begin{center}
\begin{tabular}{ccc}
\includegraphics[width=4.5cm]{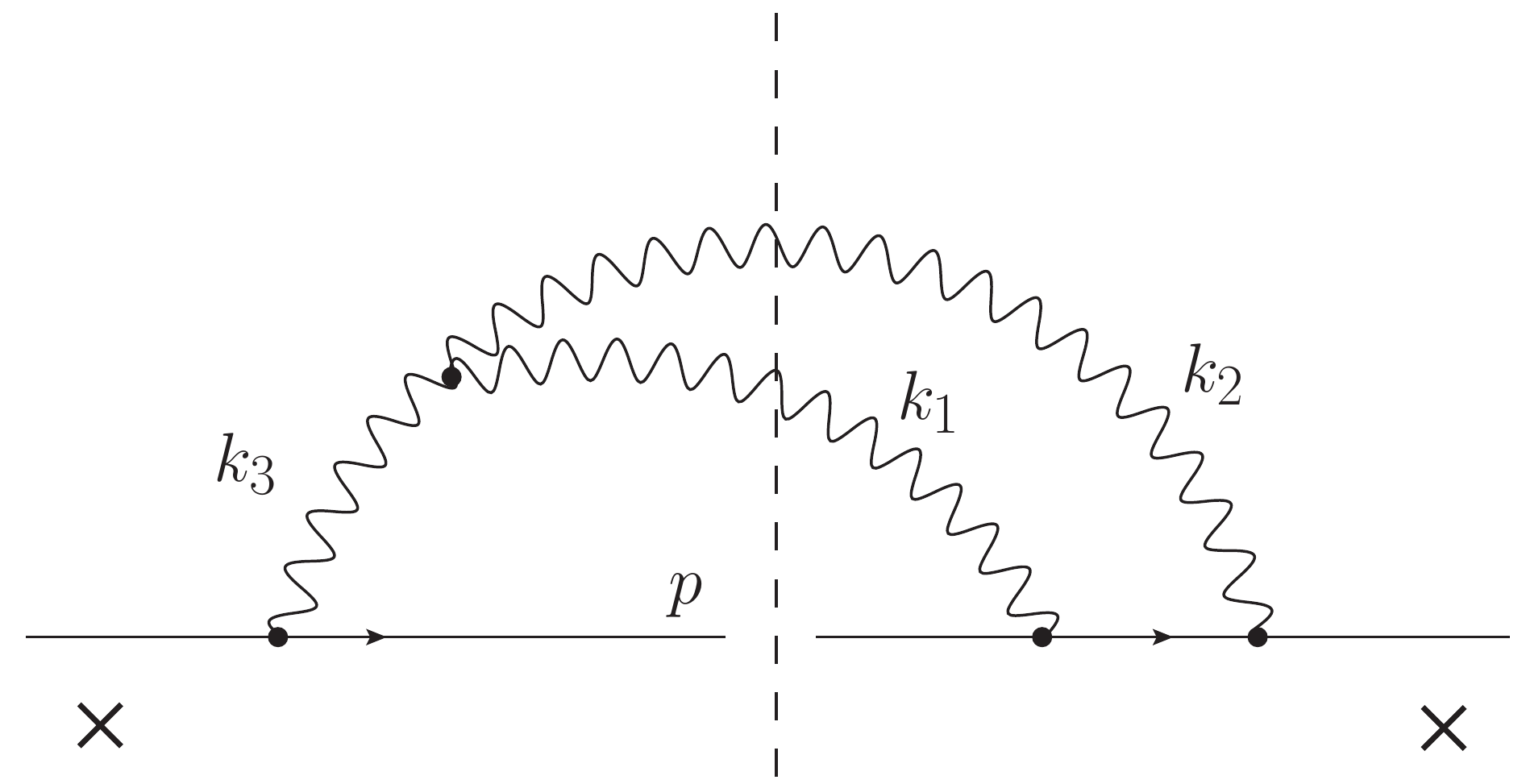}&
\includegraphics[width=4.5cm]{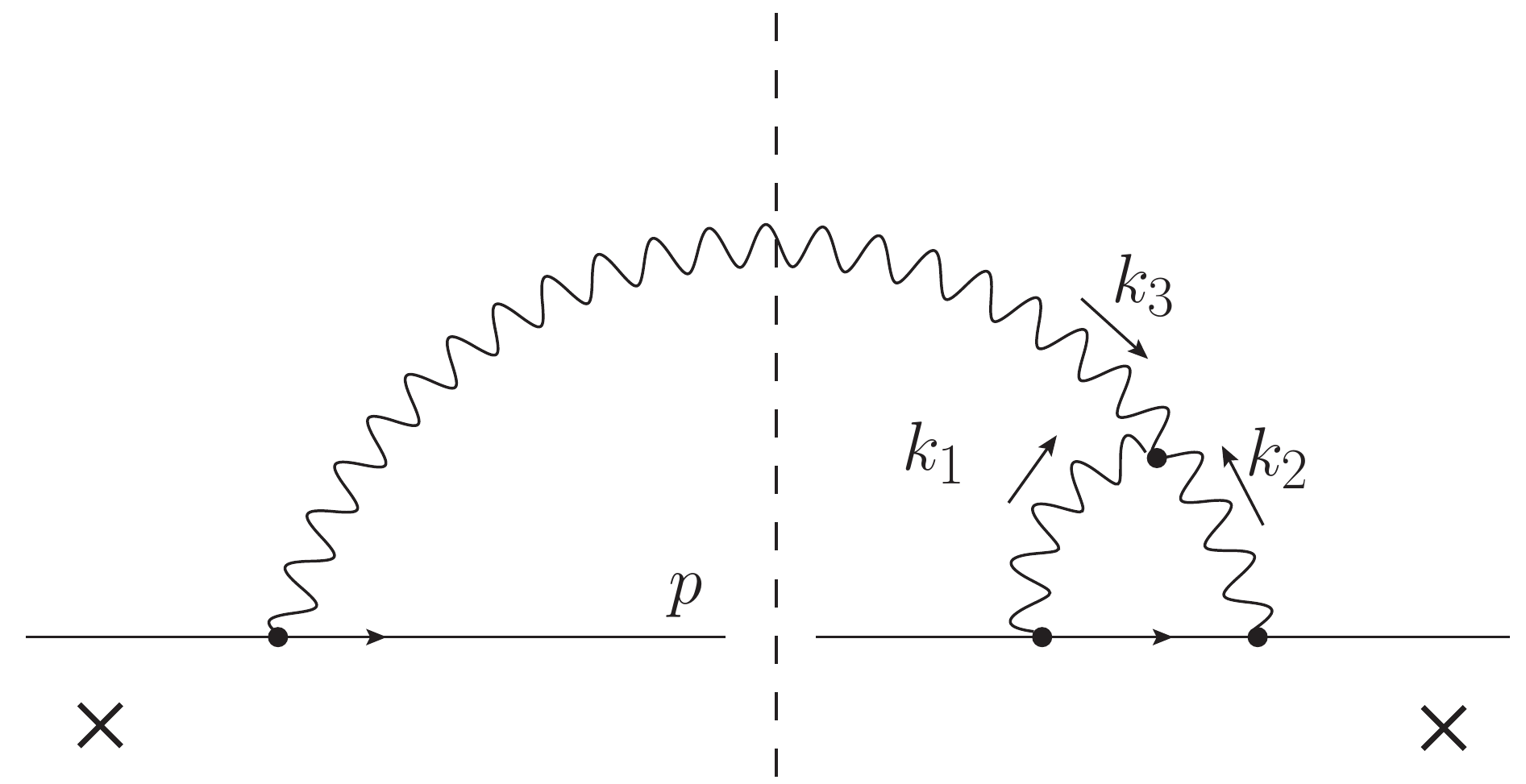}&
\includegraphics[width=4.5cm]{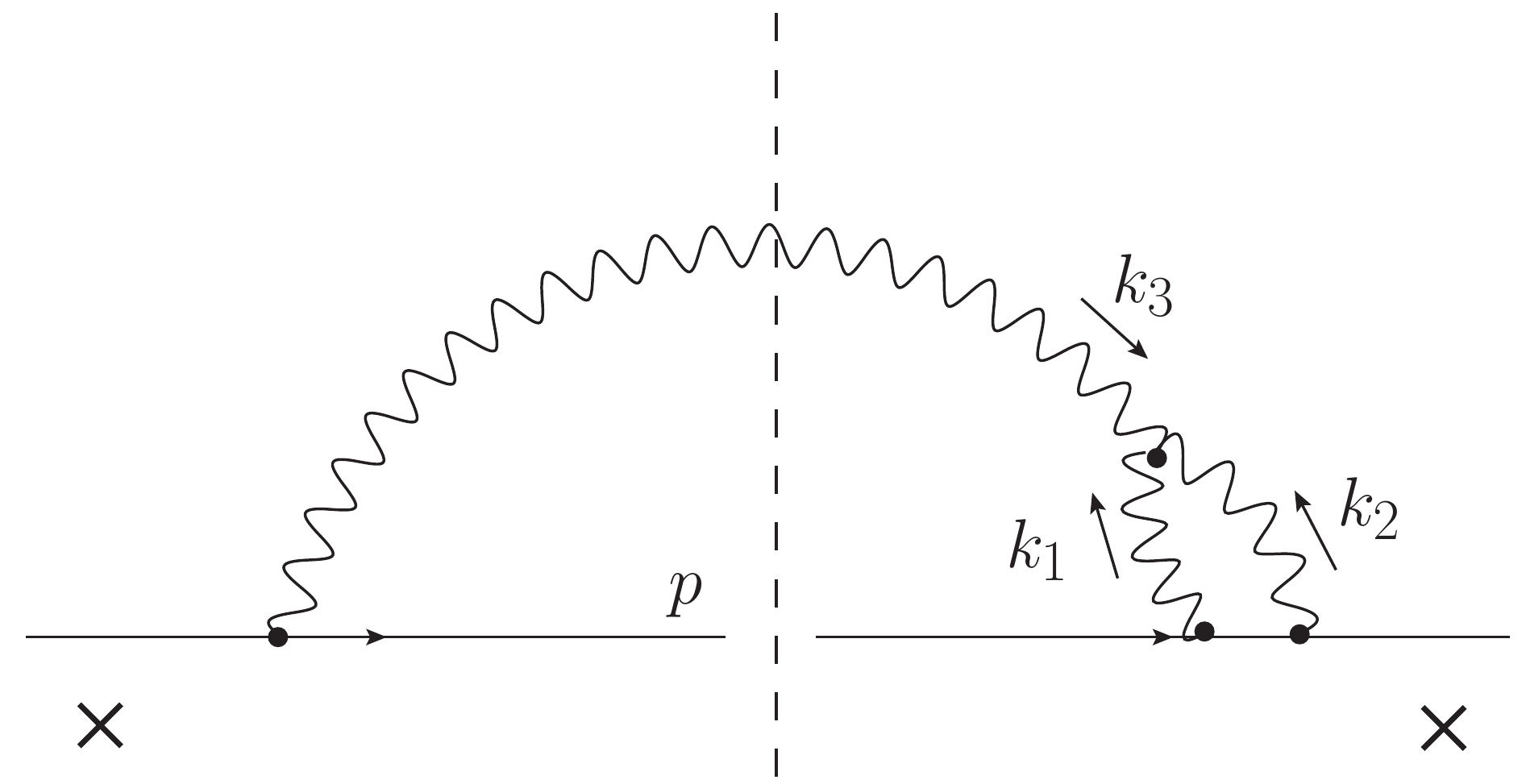}\\
$A$ & $B$ & $C$
\end{tabular}
\end{center}
\caption{\label{q3v2}
$p_\perp$-broadening graphs in which all radiation
is in the final state.
}
\end{figure}

We now move on to the case in which all interactions are in the
final state (Fig.~\ref{q3v2}). To make the comparison to dipoles
easier, we exchange the labeling of gluons~1 and~2
in such a way that gluon~2 is always the one whose coupling to
the quark is closer to the interaction with the target
than gluon~1.
The weights of the graphs $A,B,C$ are shown in Tab.~\ref{tab:ABCv2}.

\begin{table}
\begin{center}
\begin{tabular}{c|c|c|c}
& Integration range & Energy denominators& 
3-gluon vertex$\times$pola.\\
\hline
$A_1$ & 
\multirow{3}{*}{$0<k_{2+}<k_{3+}$} 
& \multirow{3}{*}{$\frac{1}{E_1(E_1+E_2)^2(E_1+E_2-E_3)}$}
& $(\vec k_2\cdot\vec k_3)(-E_1)(k_2+k_3)_+$\\
$A_2$ & & & $(\vec k_1\cdot\vec k_3)
E_2(2k_3-k_2)_+$\\
$A_3^*$ & & & $(\vec k_1\cdot\vec k_2)(E_1+E_2)(k_3-2k_2)_+$\\
\hline
$B_1^*$ &  \multirow{3}{*}{$k_{3+}<k_{2+}<+\infty$} 
& \multirow{3}{*}{$\frac{1}{E_1 E_3^2(E_2-E_3)}$}
& 0\\
$B_2^*$ & & &$(\vec k_1\cdot\vec k_3)
E_3(k_2-2k_3)_+$\\
$B_3$ & & &$(\vec k_1\cdot\vec k_2)
E_3(2k_2-k_3)_+$\\
\hline
$C_1^*$ & \multirow{3}{*}{$0<k_{2+}<k_{3+}$} 
& \multirow{3}{*}{$\frac{1}{E_3^2(E_2-E_3)(E_1+E_2-E_3)}$}
&  $(\vec k_2\cdot\vec k_3)(E_2-E_3)(k_2+k_3)_+$\\
$C_2$ & & &$(\vec k_1\cdot\vec k_3)
E_2(2k_3-k_2)_+$\\
$C_3$ & & &$(\vec k_1\cdot\vec k_2)
E_3(k_3-2k_2)_+$\\
\end{tabular}
\caption{\label{tab:ABCv2}%
Expressions for the graphs of Fig.~\ref{q3v2}
where all interactions are in the final state.
The overall factor is given by Eq.~(\ref{eq:overallfactorv}).
}
\end{center}
\end{table}

The corresponding dipole graphs are just the complex conjugates
of the graphs in Fig.~\ref{q3dv}, i.e. the mirror-symmetric graphs.
Thus they do not need to be computed again and we may directly
group the graphs according to the kinematics.
This is shown in Tab.~\ref{tab:diffgv2}.
Note that this time, $B$ has the same kinematics as~$\gamma$.

\begin{table}
\begin{center}
\begin{tabular}{c|c|c}
 & Integration range & \multicolumn{1}{c}{Momentum-dependent factors}\\
 & & ``+'', ``$\perp$''\\
\hline
$A_1+C_1^*-\alpha_1^*-\beta_1$, $B_1^*-\gamma_1^*$ &\multirow{3}{*}{
\begin{minipage}[c]{3cm}{\centerline{
$0<k_{2+}<k_{3+}$,}
\centerline{$k_{3+}<k_{2+}<+\infty$}}
\end{minipage}}
&
$-\frac{(k_{2}+k_{3})_+}{(k_{3}-k_{2})_+ k_{2+}k_{3+}}$
\\
$A_2+C_2-\alpha_2-\beta_2^*$, $B_2^*-\gamma_2^*$ &
&
$\left[1+\frac{4}{(\vec k_2^2 k_{3+})/(\vec k_3^2 k_{2+})-1}\right]
\frac{(2k_{3}-k_{2})_+}{(k_{3}-k_{2})_+ k_{2+}k_{3+}}$
\\
$A_3^*+C_3-\alpha_3^*-\beta_3^*$, $B_3-\gamma_3^*$ &
&
$\left[-1+\frac{4}{1-(\vec k_3^2 k_{2+})/(\vec k_2^2 k_{3+})}\right]
\frac{(k_{3}-2k_{2})_+}{(k_{3}-k_{2})_+ k_{2+}k_{3+}}$
\\
\hline
$\delta_1^*$ &\multirow{3}{*}{$0<k_{2+}<+\infty$}
&
$\frac{(k_{2}-k_{3})_+}{(k_{3}+k_{2})_+ k_{2+}k_{3+}}$
\\
$\delta_2^*$ &
&
$\left[1-\frac{4}{(\vec k_2^2 k_{3+})/(\vec k_3^2 k_{2+})+1}\right]
\frac{(2k_{3}+k_{2})_+}{(k_{3}+k_{2})_+ k_{2+}k_{3+}}$
\\
$\delta_3^*$ &
&
$\left[-1+\frac{4}{1+(\vec k_3^2 k_{2+})/(\vec k_2^2 k_{3+})}\right]
\frac{(k_{3}+2k_{2})_+}{(k_{3}+k_{2})_+ k_{2+}k_{3+}}$
\end{tabular}
\caption{\label{tab:diffgv2}%
Sums and differences of the graphs 
of Fig.~\ref{q3v2} and~\ref{q3dv}
grouped according to kinematics.
The transverse momentum-dependent factors are not shown since
they are the same as in Tab.~\ref{tab:diffgv} (third column).
Only the real part of the graphs are shown.
The global constant reads again
$-64\pi^4\alpha_s^4 N_c C_F^3$.
}
\end{center}
\end{table}

We also see that although the calculation is similar to
the case in which all interaction vertices are at early
times, the resuls shown in Tab.~\ref{tab:diffgv2}
are quite more complicated than the ones shown
in Tab.~\ref{tab:diffgv}.

But again, it is clear that the analytical continuation
of $\delta$ in the $k_{2+}$ variable is equal to 
$A+C-\alpha-\beta$ and $B-\gamma$ in their respective
integration domains, which proves the equivalence
between $p_\perp$-broadening and dipole amplitudes
also in this configuration.


\subsection{Four quark-gluon vertices}

We address the case in which there are four quark-gluon vertices
in the evolution of the wave function, 
linked by two bare gluons.

Writing the expressions for these graphs is quite straightforward
since we treat all vertices eikonally: The two gluons both carry
the $(--)$ polarization, and so the overall factors
need not be discussed since they are identical 
for all graphs of similar topologies
(except
for the minus signs already discussed
in several occasions which stem from the couplings of the
gluons to the antiquark in the dipole case).
Only the energy denominators are relevant for the comparison.

We name the $p_\perp$-broadening graphs according to the
chronology of the four $qg$ vertices, distinguishing the
graphs in which the loops are nested by a ``hat''.

A relevant way to classify the graphs is to distinguish
between interference graphs between
the initial and final states, and graphs in which the gluons
are emitted and reabsorbed either in the
initial and final state.


\subsubsection{Interference graphs between initial and final state}

\paragraph*{Two $qg$ vertices in the initial state, two in the final state.}

\begin{figure}
\begin{center}
\begin{tabular}{c}
\includegraphics[width=7cm]{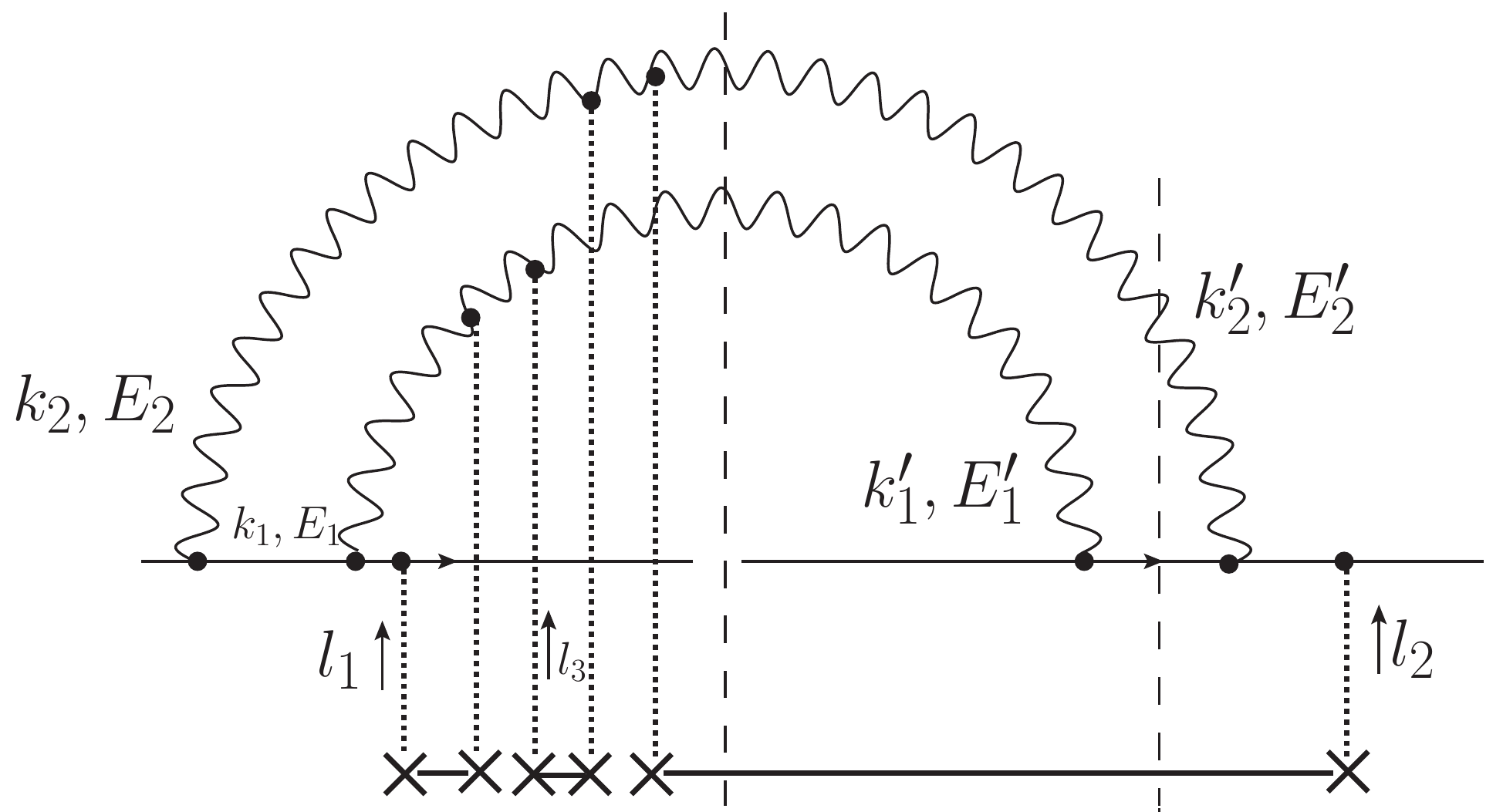}\\
$\graphr{II}{FF}$ (left),$\graphr{IIF}{F}$ (right)
\end{tabular}
\end{center}
\caption{\label{q4ii(ff)}
Four quark-gluon vertex graph.
This is an interference graph between the initial and final state,
with 2 couplings in each.
There are two possible nontrivial cuts.
}
\end{figure}
\begin{figure}
\begin{center}
\begin{tabular}{cc}
\includegraphics[width=6cm]{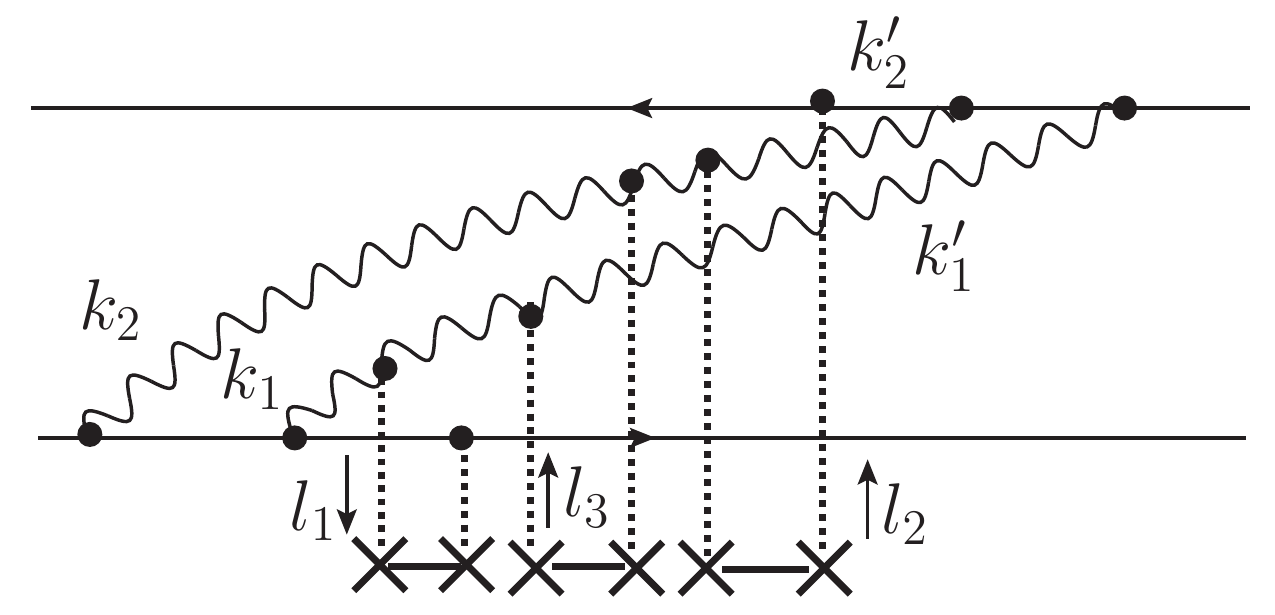}&
\includegraphics[width=4.5cm]{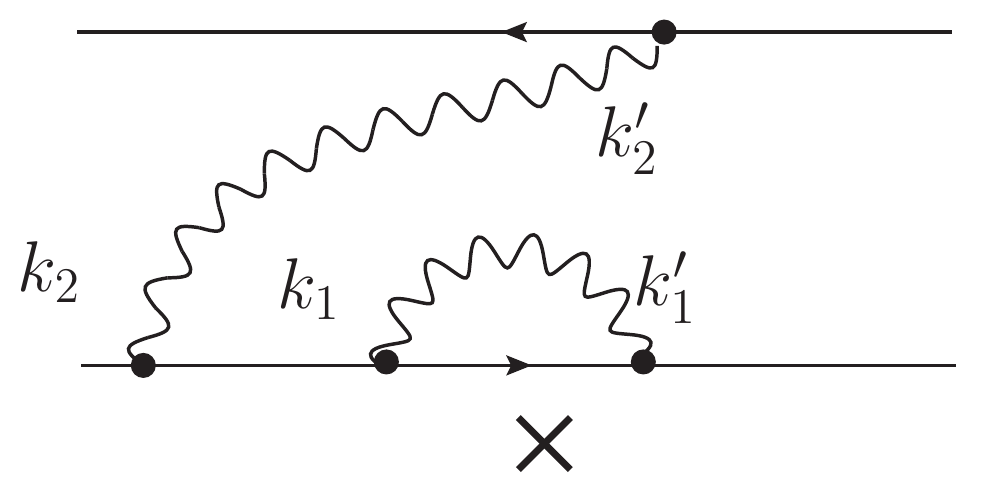}\\
$\gamma_1$&$\gamma_2$
\end{tabular}
\end{center}
\caption{\label{q4d3r}
Dipole graphs corresponding 
respectively
to the left and right cuts of
the $p_\perp$-broadening graph in Fig.~\ref{q4ii(ff)}.
As for $\gamma_2$, there are two possible orderings
between the $\bar q g$ and $qg$ vertices in the complex-conjugate
dipole amplitude.
}
\end{figure}

We first address the graph in Fig.~\ref{q4ii(ff)}
which has two $qg$ vertices in the initial state,
and two at late times.
There may be either two gluons (leftmost cut) or 
one gluon (rightmost cut)
in the final state.
These graphs are topologically related to the dipole graphs 
$\gamma_1$ and $\gamma_2$ respectively, shown in Fig.~\ref{q4d3r},
which both have 3 dipoles at the time of the interaction.
We are going to show that the expressions for these
graphs are indeed identical.

The energy denominators read
\be
\begin{split}
D_\graphr{II}{FF}&=
\frac{1}{E^\prime_1 E_2(E_1+E_2)(E_1^\prime+E_2^\prime)}
=D_{\gamma_1}
\\
D_\graphr{IIF}{F}&=
-\frac{1}{E^\prime_1E^\prime_2 E_2(E_1+E_2)}
=-\sum D_{\gamma_2}
\end{split}
\label{eq:idiiffgamma}
\ee
The $\sum$ sign recalls that $\gamma_2$ in Fig.~\ref{q4d3r}
actually represents two lightcone perturbation theory graphs,
differing by the ordering of some vertices.
The definition of the energies may be inferred by applying
momentum conservation to the graphs in Fig.~\ref{q4ii(ff)} 
and~\ref{q4d3r}. The only important point is that
the routing of the momenta be the same in the
$p_\perp$-broadening and dipole cases.
The ``$-$'' sign in the second line 
of Eq.~(\ref{eq:idiiffgamma})
corresponds to
the change of one gluon coupling from a quark to an antiquark
(while for the first graph, two such $qg$ couplings become
$\bar q g$ couplings, in such a way that the associated minus
signs cancel each other).

Note that the first equality in  Eq.~(\ref{eq:idiiffgamma}) 
is straightforward: It is a simple graph-to-graph
correspondence. We shall not address exhaustively
all such trivial cases in what follows.

\begin{figure}
\begin{center}
\begin{tabular}{c}
\includegraphics[width=10cm]{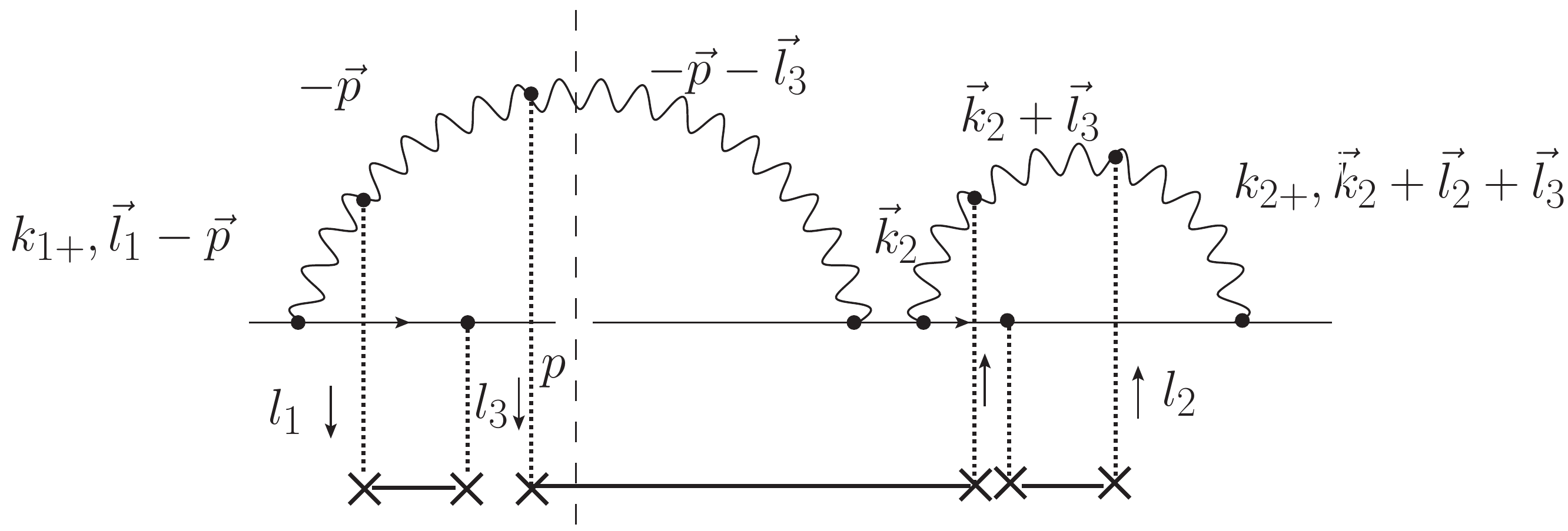}\\
$\graph{I}{IFF}$
\end{tabular}
\end{center}
\caption{\label{q4i(ffi)}
A $p_\perp$-broadening diagram whose identification with the corresponding
dipole graph requires analytical continuation.
A similar diagram (not drawn) would have the cut 
passing on the rightmost gluon, but
the calculation would be the same.
}
\end{figure}

\begin{figure}
\begin{center}
\includegraphics[width=8cm]{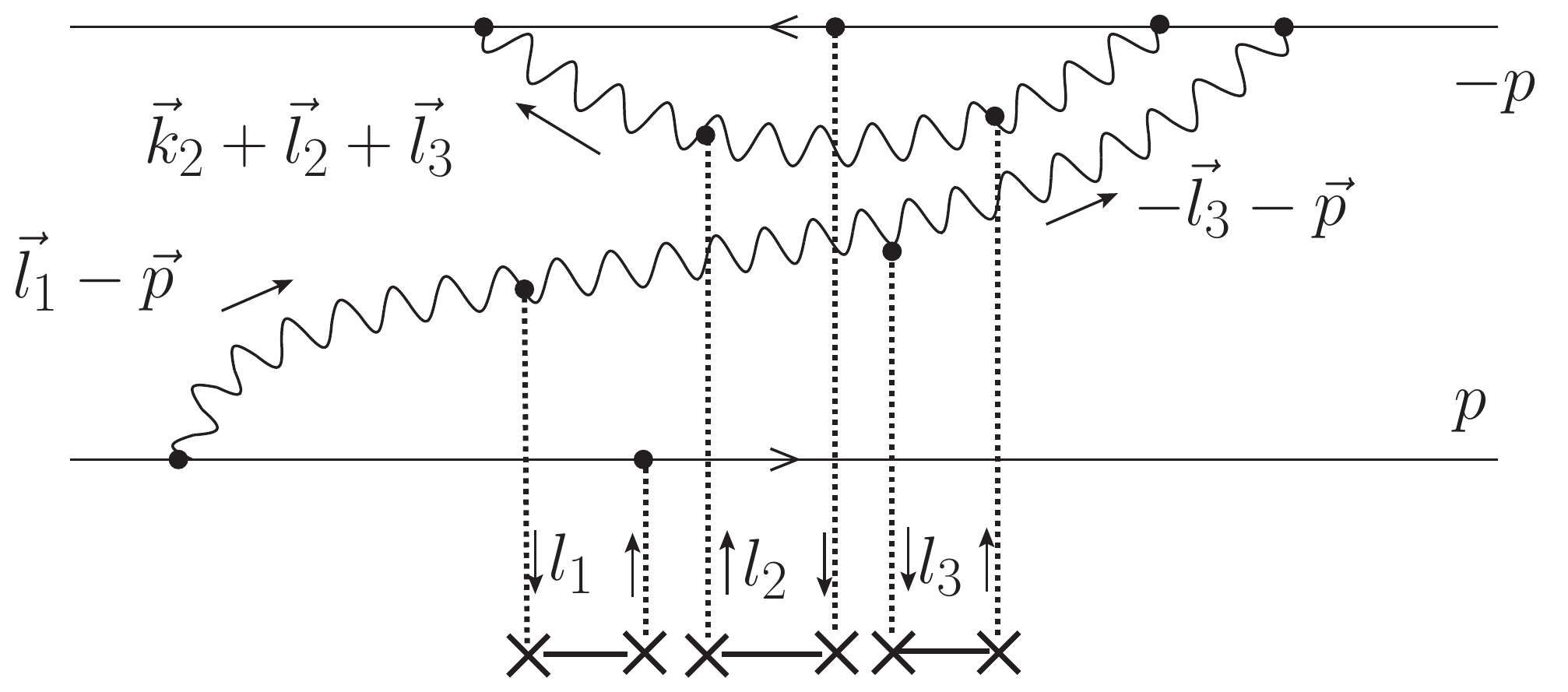}
\end{center}
\caption{\label{q4di(ffi)}
Dipole graph that corresponds to the
$p_\perp$-broadening graph of Fig.~\ref{q4i(ffi)}.
The + components $k_{1+}$ and $k_{2+}$ are always directed to
the right. The direction of the 
transverse components is indicated by the
arrows above the gluon lines.
There are actually two graphs, which are distinguished by
the respective ordering of the $qg$ and $\bar q g$
vertices in the amplitude (leftmost vertices).
Note that this graph is related to $\gamma_2$
in Fig.~\ref{q4d3r}
by a simple symmetry.
}
\end{figure}

We now consider the graph shown in Fig.~\ref{q4i(ffi)}
(as well as its dipole partner in Fig.~\ref{q4di(ffi)}).
We need to write the full expressions
of the contribution of the graphs 
to the amplitudes since as we will discover,
analytical continuation is needed.
This is also the opportunity to show how the complete
calculation goes in the four $qg$-vertex case.

The energy denominators 
read
\be
D_{\graph{I}{IFF}}=\frac{1}{(E_1-i\varepsilon)(E^\prime_1-i\varepsilon)
(E^\prime_1-E_2-2i\varepsilon)(E^\prime_2+i\varepsilon)}
\label{eq:q4denom}
\ee
where
\be
E_1=\frac{(\vec l_1-\vec p)^2}{2k_{1+}}\ ,\
E^\prime_1=\frac{(\vec p+\vec l_3)^2}{2k_{1+}}\ ,\
E_2=\frac{\vec k_2^2}{2k_{2+}}\ ,\
E^\prime_2=\frac{(\vec k_2+\vec l_2+\vec l_3)^2}{2k_{2+}}
\ee
(see Fig.~\ref{q4i(ffi)}).
The polarization factors for the gluons read
\begin{multline}
\left[d_{-\perp}(l_1-p)d_{\perp\perp}(-p)d_{\perp -}(-p-l_3)\right]
\left[d_{-\perp}(k_2)d_{\perp\perp}(k_2+l_3)d_{\perp -}(k_2+l_2+l_3)\right]\\
=
\left[\frac{(\vec p-\vec l_1)\cdot(\vec p+\vec l_3)}{k_{1+}^2}\right]
\left[\frac{\vec k_2\cdot(\vec k_2+\vec l_2+\vec l_3)}{k_{2+}^2}\right].
\label{eq:q4pola}
\end{multline}
The vertices are all eikonal.
After having performed the sum over the colors of the
gluons and averaged over the color of the quark, 
the associated factor reads
\be
-\frac{16\pi^5\alpha_s^5}{k_{1+}k_{2+}}
N_c^4 C_F.
\label{eq:q4vertices}
\ee
The factors (\ref{eq:q4denom}),(\ref{eq:q4pola})
and~(\ref{eq:q4vertices}) must be convoluted with 
the gluon densities as
\be
-\frac13\int 
\frac{dk_{1+}}{(2\pi)^3}
\frac{dk_{2+}d^2\vec k_2}{(2\pi)^3}
\frac{d^2\vec l_1}{\vec l_1^2}
\frac{d^2\vec l_2}{\vec l_2^2}
\frac{d^2\vec l_3}{\vec l_3^2}
\frac{xg(x,\vec l_1^2)}{N_c^2-1}
\frac{xg(x,\vec l_2^2)}{N_c^2-1}
\frac{xg(x,\vec l_3^2)}{N_c^2-1}
(\rho L)^3
\ee
(Momentum conservation was used to get rid of 
the integration over $\vec k_{1}$).
The $\frac13$ factor comes from the ordering
of the times at which the gluons interact with the nucleus.
We may perform the integration over the + component of the
loop momentum $k_2$ explicitly.
The result reads
\begin{multline}
\left.\frac{dN}{d^2\vec p}
\right|_{\mbox{\footnotesize Fig. \ref{q4i(ffi)}}}
=\frac13
\frac{\alpha_s^2 N_c^3}{(N_c^2-1)^2}
\int
\frac{d^2\vec l_1}{\vec l_1^2}
\frac{d^2\vec l_2}{\vec l_2^2}
\frac{d^2\vec l_3}{\vec l_3^2}\\
\times\left[{\alpha_s xg(x,\vec l_1^2)}\right]
\left[{\alpha_s xg(x,\vec l_2^2)}\right]
\left[{\alpha_s xg(x,\vec l_3^2)}\right]
(\rho L)^3\times
(\vec p-\vec l_1)\cdot(\vec p+\vec l_3)\\
\times\int
\frac{dk_{1+}}{k_{1+}^3}
\frac{1}{(E_1-i\varepsilon)(E^\prime_1-i\varepsilon)}
\left[
\ln\frac{\lambda}{\zeta k_{1+}}
+i\pi
\right]
\int \frac{d^2\vec k_2}{2\pi}
\frac{\vec k_2\cdot(\vec k_2+\vec l_2+\vec l_3)}
{\vec k_2^2(\vec k_2+\vec l_2+\vec l_3)^2}
\label{eq:dndpq4}
\end{multline}
The remaining integral over $k_{1+}$
may also be performed.
Instead, we stress that
the content of the square brackets in the last line
is the result of the integration
\be
I_+=\int_\lambda^\Lambda dk_{2+}
\frac{\zeta k_{1+}}{k_{2+}(k_{2+}-\zeta k_{1+}-i\eta)}
\label{eq:I+}
\ee
where $\zeta=\vec k_2^2/(\vec p+\vec l_3)^2$, 
and $\eta=4\varepsilon \zeta k_{1+}^2/\vec k_2^2$
is a small positive regulator.
The IR and UV cutoffs
are eventually
sent to the limits $\lambda\rightarrow 0$ 
and $\Lambda\rightarrow +\infty$.

We now address the corresponding dipole diagram
shown in Fig.~\ref{q4di(ffi)}.
We are going to show that its evaluation leads to
the same expression as in Eq.~(\ref{eq:dndpq4}).
We observe that we may label the transverse momenta
of the gluons in such a way that
only one factor looks different with respect
to the previous diagram, namely $I_+$.
Indeed, the overall
difference 
between the two graphs
is in the expression for the energy denominators, which
now read (compare to Eq.~(\ref{eq:q4denom}))
\be
\frac{1}{(E_1-i\varepsilon)(E^\prime_1+i\varepsilon)
(E^\prime_1+E_2+2i\varepsilon)(E^\prime_2-i\varepsilon)}.
\ee
Then $I_+$ is replaced by
\be
I^\prime_+=
-\int_\lambda^\Lambda 
dk_{2+}\frac{\zeta k_{1+}}{k_{2+}(k_{2+}+\zeta k_{1+}+i\eta)},
\ee
whose explicit expression reads
\be
I^\prime_+=\ln\frac{\lambda}{\zeta k_{1+}}=I_+-i\pi.
\label{eq:Iprime+2}
\ee
Hence the dipole amplitude differs from the $p_\perp$-broadening
amplitude by an imaginary term, which cancels when
one adds the complex conjugate graphs.

We shall now present an alternative way to view the connection
between the $p_\perp$-broadening
diagram and the dipole diagram essentially similar
to the analytical continuation presented before.
We remind that this
method has the advantage 
that it does not require the explicit evaluation
of the integrals, hence it is generalizable
to cases in which the latter
is not possible.
\begin{figure}
\begin{center}
\includegraphics[width=9cm]{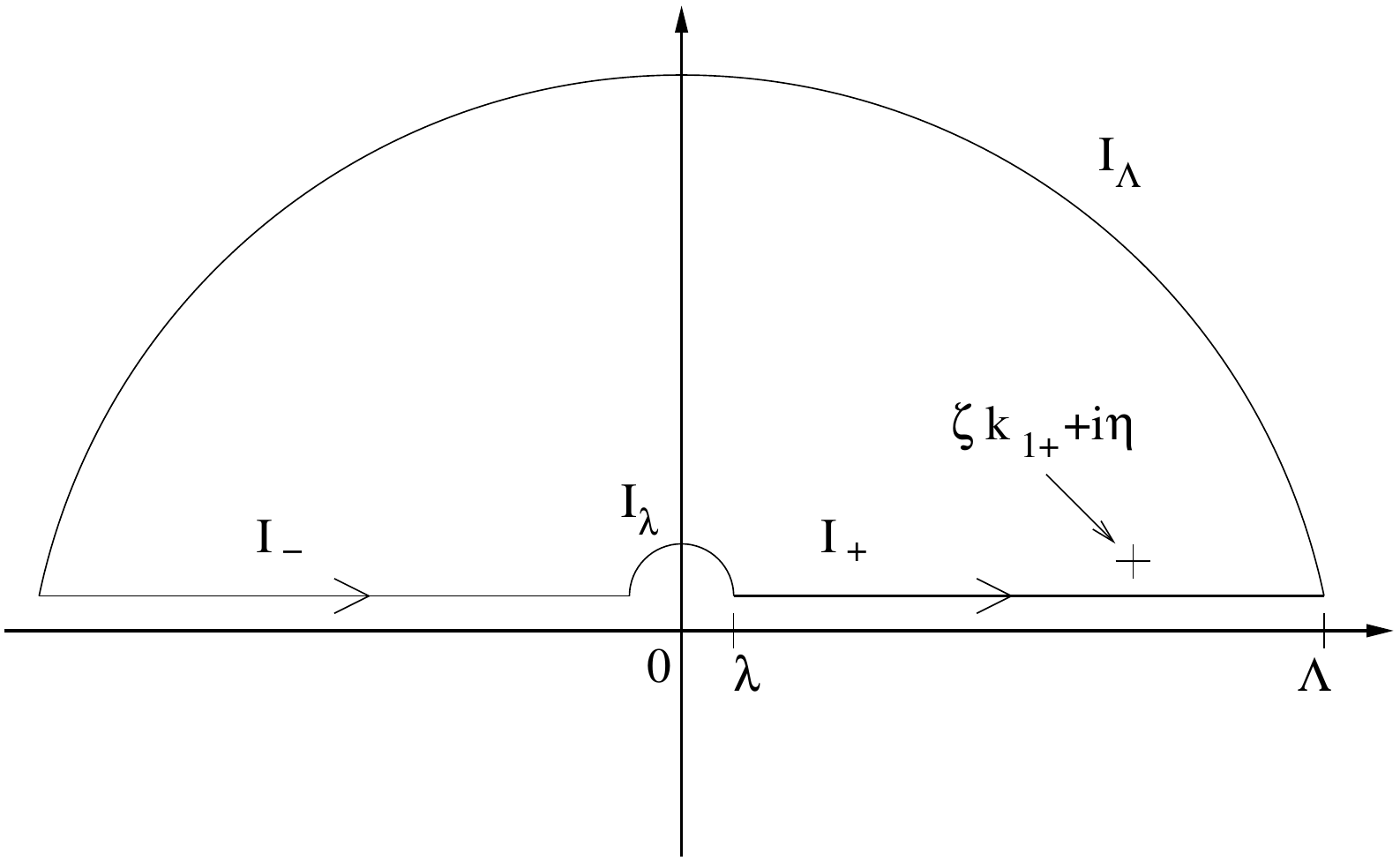}
\end{center}
\caption{\label{k2+plane}
Integration contour in the $k_{2+}$ plane.
The closed contour is divided in 4 sections:
$I_\lambda$ is the small half circle of radius $\lambda$ oriented
clockwise, $I_+$ the part $[\lambda,\Lambda]$ of the real axis,
$I_\Lambda$ the large half circle of radius $\Lambda$ 
oriented anticlockwise
and $I_-=[-\Lambda,-\lambda]$.
The position of the pole $\zeta k_{1+}+i\eta$ is
also shown.
}
\end{figure}
It is based on the observation that the only difference between
the two formulations is a change of the
direction of the $k_{2+}$ momentum.
We are going to interpret this change as an analytical
continuation of the integral $I_+$.
We start with the expression of $I_+$ in Eq.~(\ref{eq:I+}).
The $k_{2+}$ plane is represented in the plot of Fig.~\ref{k2+plane}.
The integral in  Eq.~(\ref{eq:I+}) is represented by the branch 
along the positive axis of the contour 
in the figure.
The Cauchy theorem on the full contour reads
\be
I_++I_\Lambda+I_-+I_\lambda={2i\pi}
\label{eq:sumcontour}
\ee
where the right-hand side is the contribution of the pole at
$k_{2+}=\zeta k_{1+}+i\eta$.
$I_\Lambda$ is zero when $\Lambda$ goes to infinity since the integral
is convergent in the ultraviolet,
and it is easy to see that
$I_\lambda=i\pi$ when $\lambda$ goes to zero.
Finally,
\be
I_-=\int_{-\Lambda}^{-\lambda}\frac{dk_{2+}}{k_{2+}}
\frac{\zeta k_{1+}}{k_{2+}-\zeta k_{1+}-i\eta}
=\int_{\lambda}^{\Lambda}\frac{dk_{2+}}{k_{2+}}
\frac{\zeta k_{1+}}{k_{2+}+\zeta k_{1+}+i\eta}\\
=-I^\prime_+
\label{eq:I-Iprime+}
\ee
Thus, from Eq.~(\ref{eq:sumcontour}) and~(\ref{eq:I-Iprime+}), we get
\be
I_+=i\pi+I^\prime_+
\ee
which is of course the same equation as~(\ref{eq:Iprime+2})
and thus leads to the same conclusion as before.

Note that here, we chose to stick to the adiabatic
regularization while we used the principal part prescription
in Sec.~\ref{sec:sec3qgv}.
The difference is in imaginary terms which anyway
cancel when all complex conjugate graphs are 
taken into account.


\paragraph*{Three gluons in the initial state, one in the final state.}

\begin{figure}
\begin{center}
\begin{tabular}{c}
\includegraphics[width=7cm]{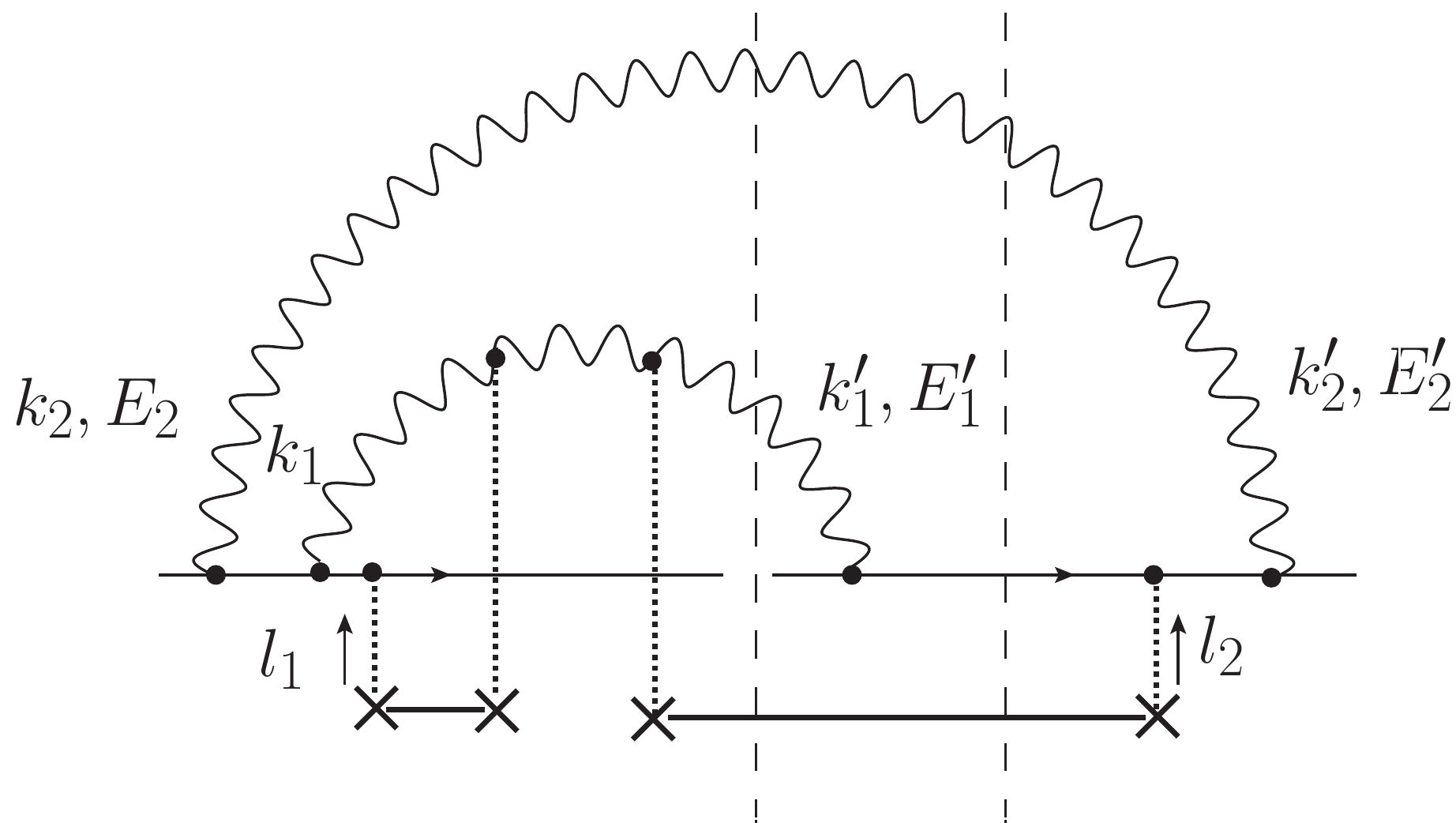}\\
$\graphr{II}{IF}$ (left cut),
$\graphr{IIF}{I}$ (right cut)
\end{tabular}
\end{center}
\caption{\label{q4iiif1}
$p_\perp$-broadening
graphs with three $qg$ vertices at early times and one at late times,
and two gluons (left cut) or one (right cut) in the final state.
}
\end{figure}
\begin{figure}
\begin{center}
\begin{tabular}{cc}
\includegraphics[width=4cm]{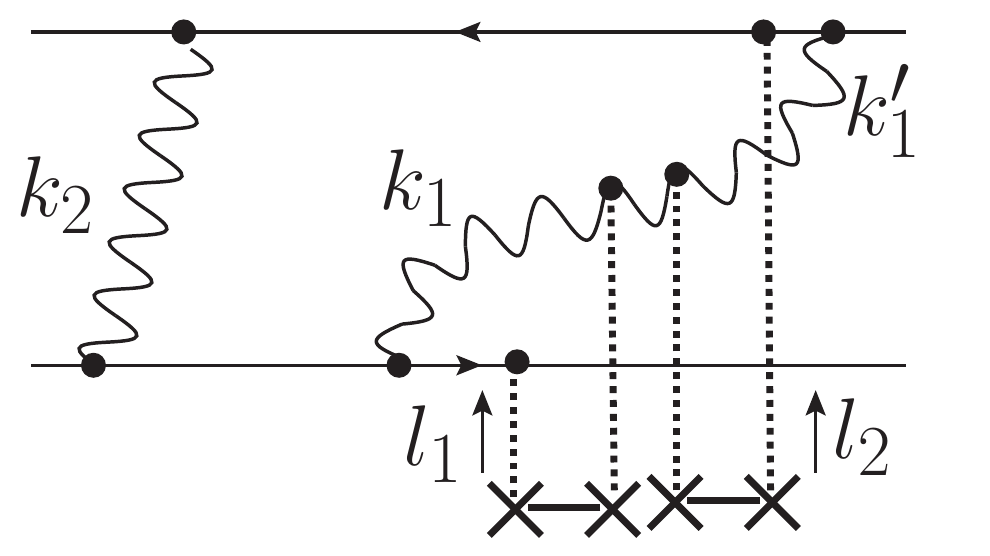}&
\includegraphics[width=4cm]{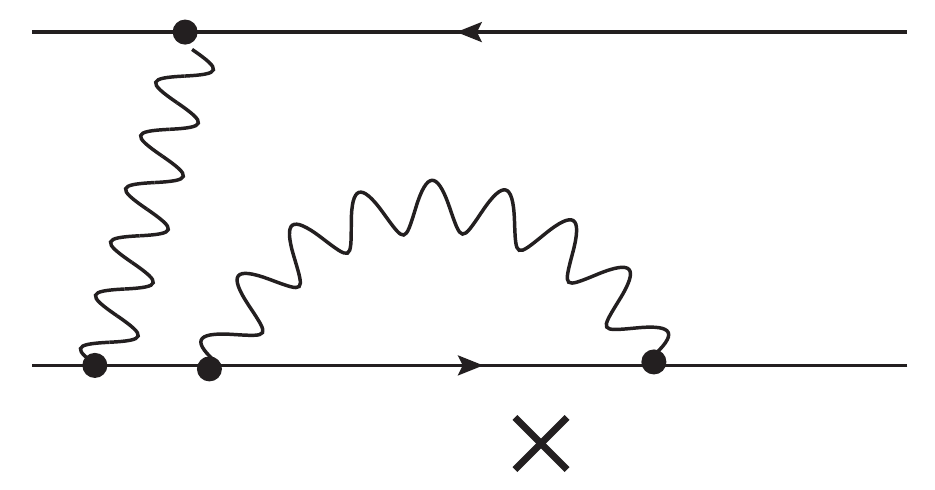}\\
$\beta_1$&$\beta_2$
\end{tabular}
\end{center}
\caption{\label{q4d1r1}
Dipole graphs that correspond to the two cuts of the $p_\perp$-broadening
graph of Fig.~\ref{q4iiif1}.
For each of these graphs, the graphs corresponding to different
orderings of the $\bar q g$ vertices with respect to the $qg$ vertex in
the amplitude are understood, and this also includes the instantaneous
exchange of gluon 2.
}
\end{figure}

Let us address the case in which there is only one $qg$
vertex after the time of the interaction with the nucleus
in the $p_\perp$-broadening case (which corresponds
to 2 real dipoles).
The relevant graph (with its two possible cuts) 
is represented in Fig.~\ref{q4iiif1}, and the 
topologically equivalent
dipole graphs
are shown in Fig.~\ref{q4d1r1}.
We find that the energy denominators for these graphs read
\be
D_\graphr{II}{IF}=-D_\graphr{IIF}{I}=-\frac{1}{E_1^\prime E_2^2(E_1+E_2)}
=\sum D_{\beta_1}=\sum D_{\beta_2}.
\label{eq:relb1b2}
\ee
Again, the $\sum$ signs recall that we have included all relevant
graphs, see the caption of Fig.~\ref{q4d1r1}.
The minus sign in front of $D_\graphr{IIF}{I}$
eventually gets cancelled 
when one goes to dipoles by
the minus sign of the $\bar qg$ vertex in $\beta_2$.
Thus the correspondence is verified also for these graphs, namely
\be
\graphr{II}{IF}\longleftrightarrow \sum \beta_1\ ,
\
\graphr{IIF}{I}\longleftrightarrow \sum \beta_2\ .
\ee
This is almost a graph-to-graph identity,
up to the relative position of some vertices
in the dipole case.

\begin{figure}
\begin{center}
\begin{tabular}{cc}
\includegraphics[width=5cm]{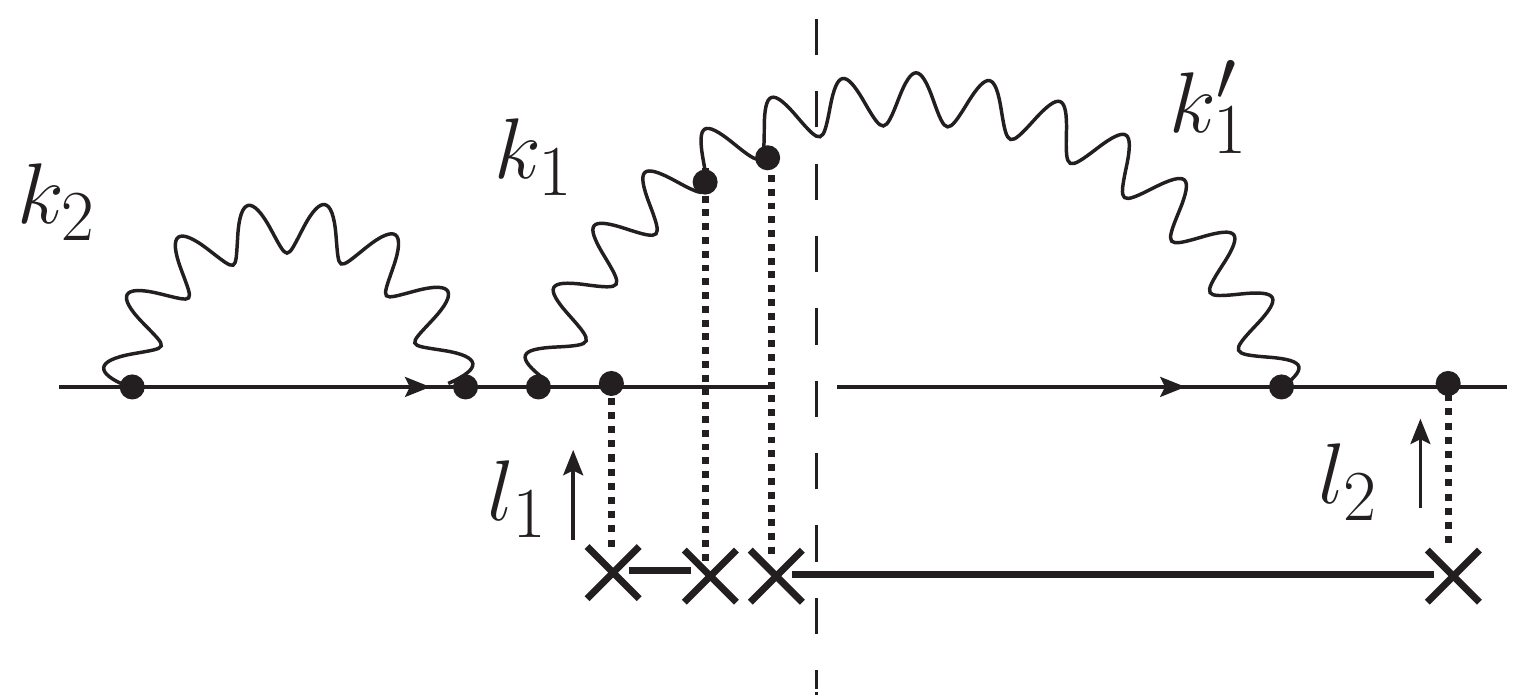}&
\includegraphics[width=5cm]{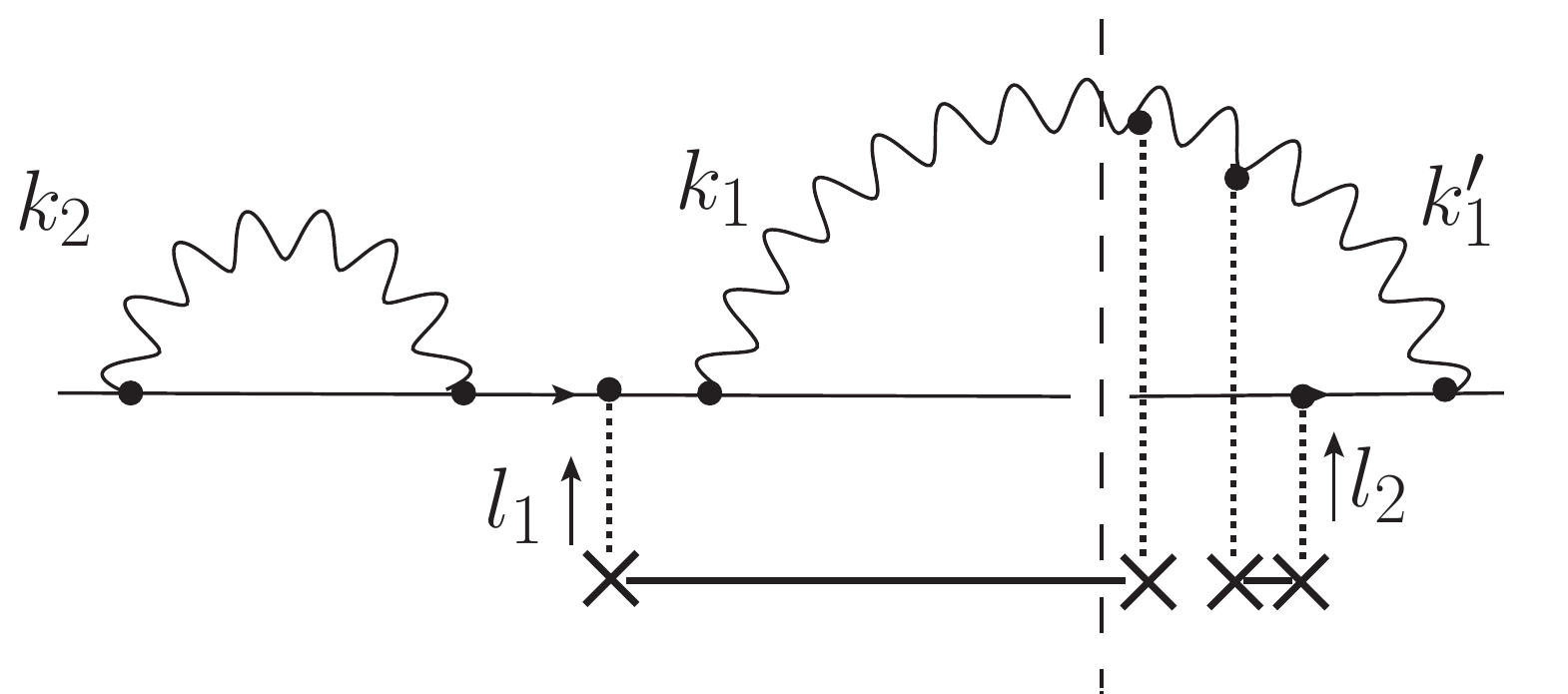}\\
$\graph{III}{F}$ & $\graph{IIF}{I}$
\end{tabular}
\end{center}
\caption{\label{q4iiif2}
Graphs for which 3 quark-gluon vertices are in the initial
state, one in the final state, and the gluon loops are disjoint.
}
\end{figure}

We move on to the case in which the loops are disconnected.
The relevant graphs are shown in Fig.~\ref{q4iiif2} for
$p_\perp$ broadening, and in Fig.~\ref{q4d1r1} for their dipole
equivalent $\beta_3$ and $\beta_4$.
Interestingly enough, there is no graph-by-graph correspondence
although the topologies of $\graph{III}{F}$ and
$\graph{IIF}{I}$ seem identical to the topologies of $\beta_3$ and
$\beta_4$ respectively.

\begin{figure}
\begin{center}
\begin{tabular}{cc}
\includegraphics[width=4cm]{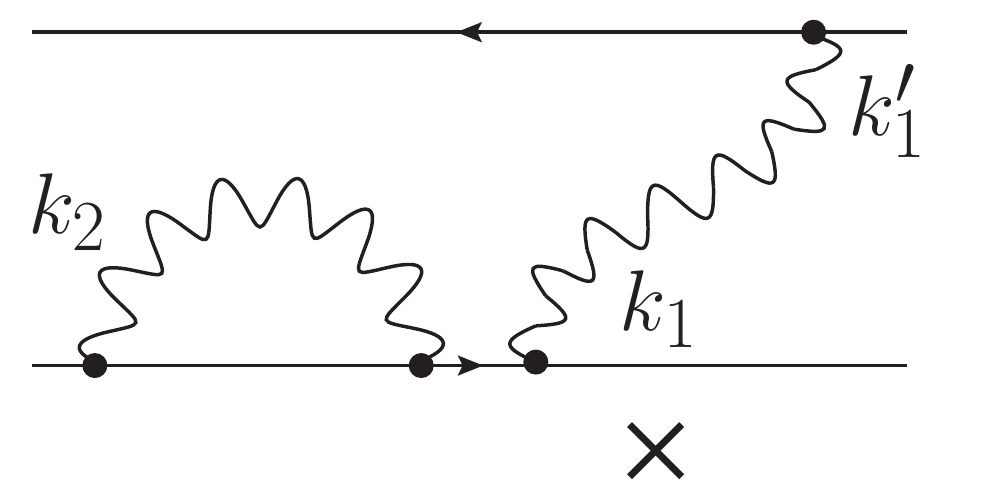}&
\includegraphics[width=4cm]{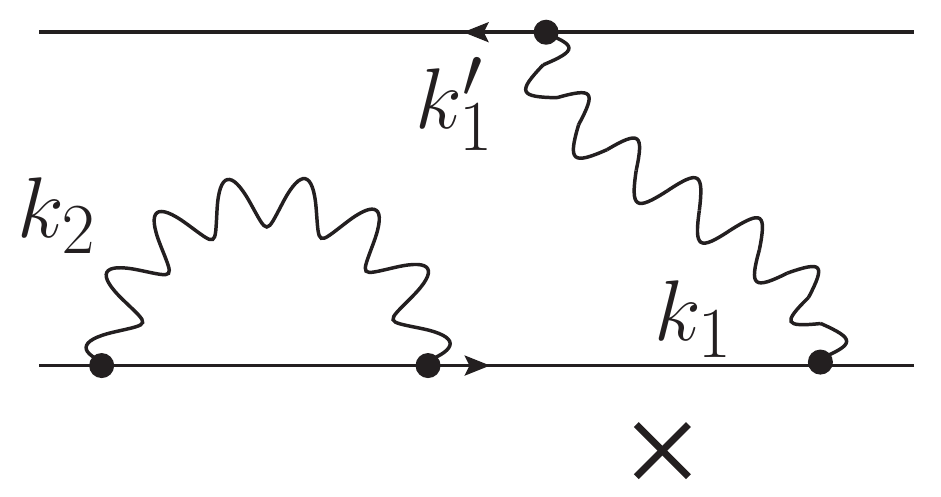}\\
$\beta_3$&$\beta_4$
\end{tabular}
\end{center}
\caption{\label{q4d1r2}
Dipole graphs that correspond to the $p_\perp$-broadening graphs
of Fig.~\ref{q4iiif2}.
The $\bar q g$ vertex in graph $\beta_4$
may arrive at any time in $]-\infty,0[$; Only
the graph in which it is posterior to the gluon
loop is shown.
}
\end{figure}
Indeed, the denominators read
\be
\begin{split}
D_\graph{III}{F}&=\frac{1}{2i\varepsilon (E_1-3i\varepsilon)
(E^\prime_1-i\varepsilon)(E_2-i\varepsilon)},\\
\text{while}\ \ \ \ 
D_{\beta_3}&=-\frac{1}{2i\varepsilon (E_1-3i\varepsilon)
(E^\prime_1+i\varepsilon)(E_2-i\varepsilon)}.
\end{split}
\label{eq:reliiifb3}
\ee
The difference in the
global sign is explained by the $qg\rightarrow \bar qg$ change
when one goes to dipoles,
but there is an extra sign difference in the 
energy denominator
involving $E^\prime_1$ which hampers the identification of $\graph{III}{F}$
and $\beta_3$.
On the other hand, the remaining diagrams have the following
denominators:
\be
\begin{split}
D_\graph{IIF}{I}&=\frac{1}{2i\varepsilon (E_1+i\varepsilon)
(E^\prime_1+i\varepsilon)(E_2-i\varepsilon)},\\
\sum D_{\beta_4}&=-\frac{1}{2i\varepsilon (E_1+i\varepsilon)
(E^\prime_1-i\varepsilon)(E_2-i\varepsilon)},
\end{split}
\label{eq:reliifib4}
\ee
and thus we see that the sum of all graphs satisfy the
identity
\be
D_\graph{III}{F}+D_\graph{IIF}{I}
=\frac{1}{i\varepsilon E_1 E^\prime_1 E_2}
+\frac{1}{E_1^2 E^\prime_1 E_2}
+\frac{1}{E_1 E^\prime_1 E_2^2}
=-(D_{\beta_3}+\sum D_{\beta_4}).
\ee
Hence the correspondence holds between
the sum of the $p_\perp$-broadening
graphs $\graph{III}{F}+\graph{IIF}{I}$
(left-hand side)
and  the sum of the dipole graphs
$\beta_3+\sum {\beta_4}$ (right-hand side).


\paragraph*{One $qg$ vertex in the initial state,
three in the final state.}

\begin{figure}
\begin{center}
\begin{tabular}{cc|cc}
\includegraphics[width=3cm]{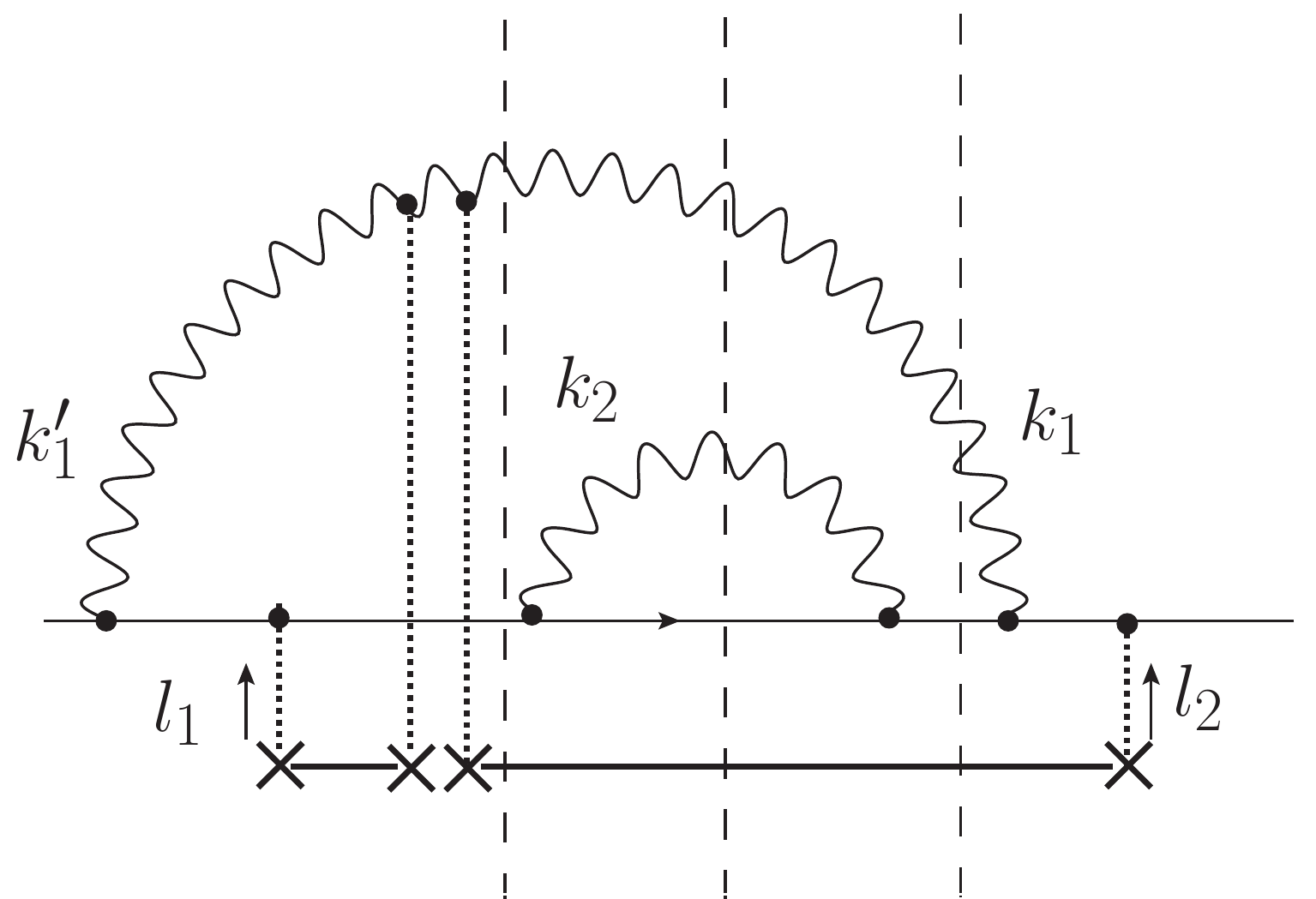}&
\includegraphics[width=3cm]{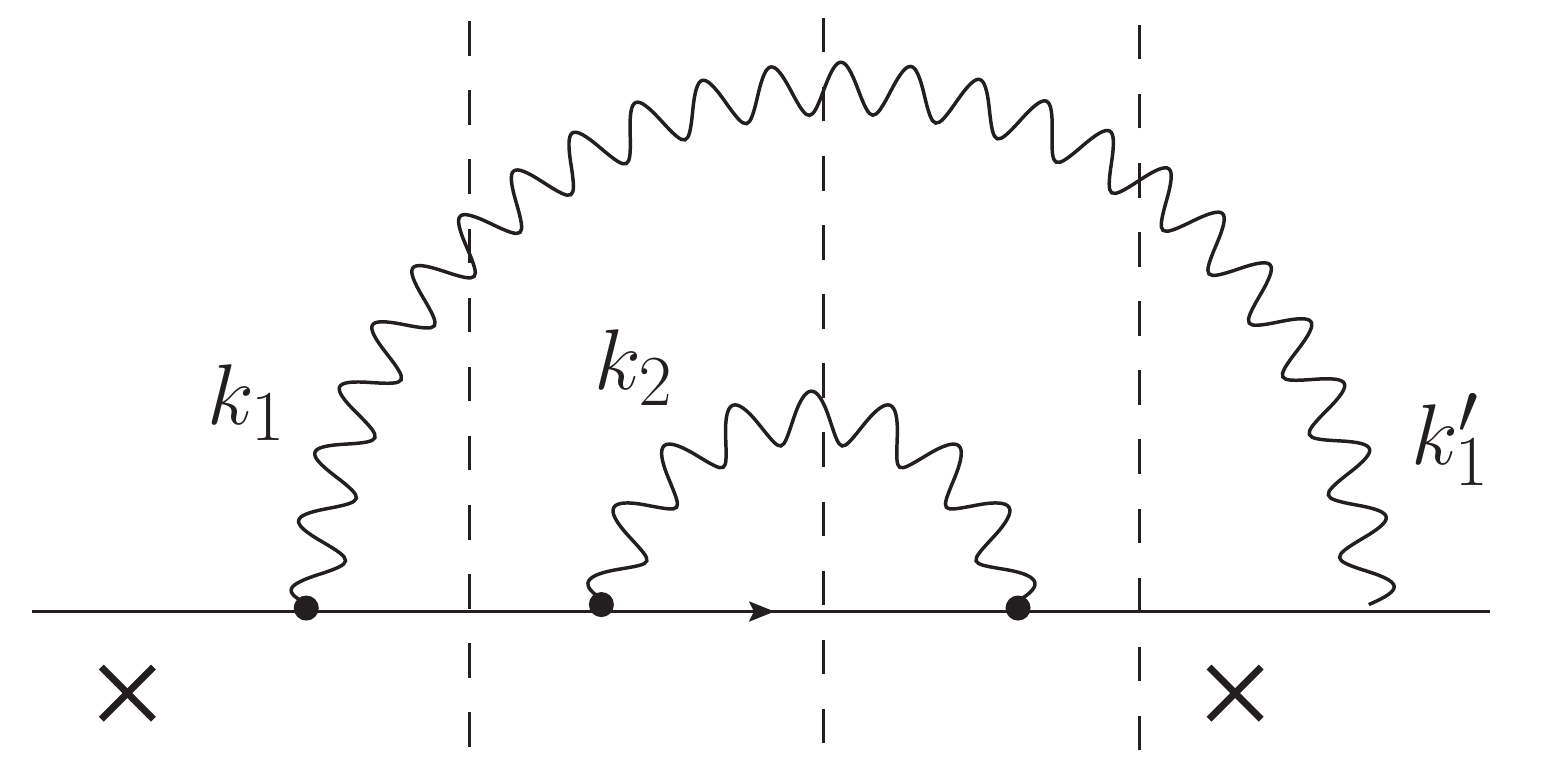}&
\includegraphics[width=3cm]{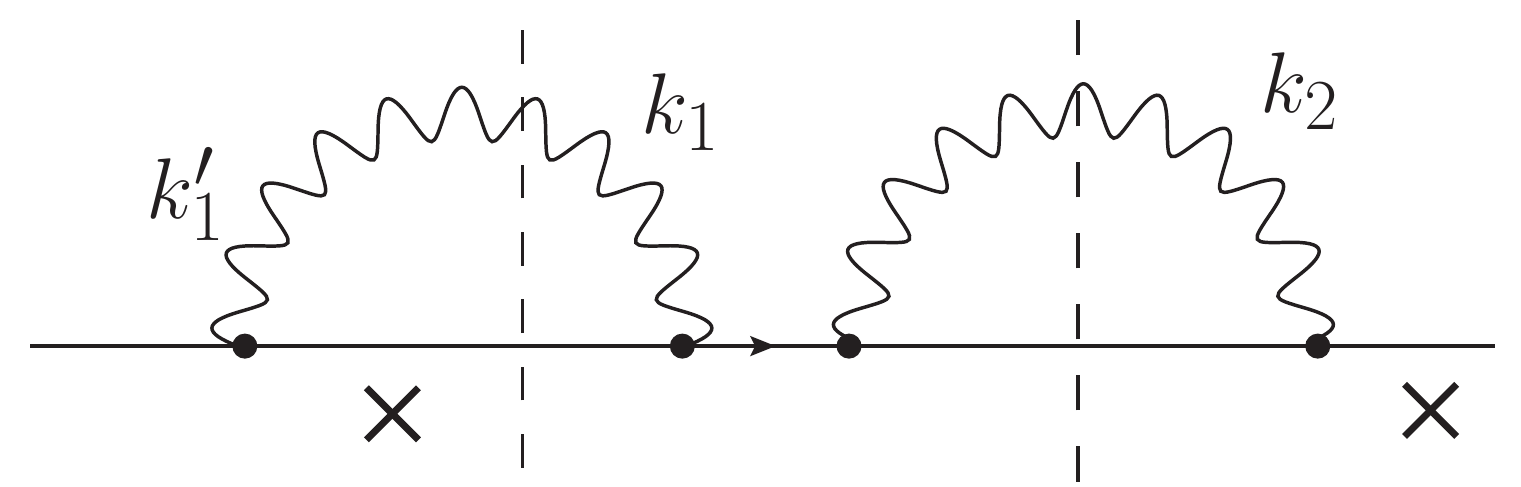}&
\includegraphics[width=3cm]{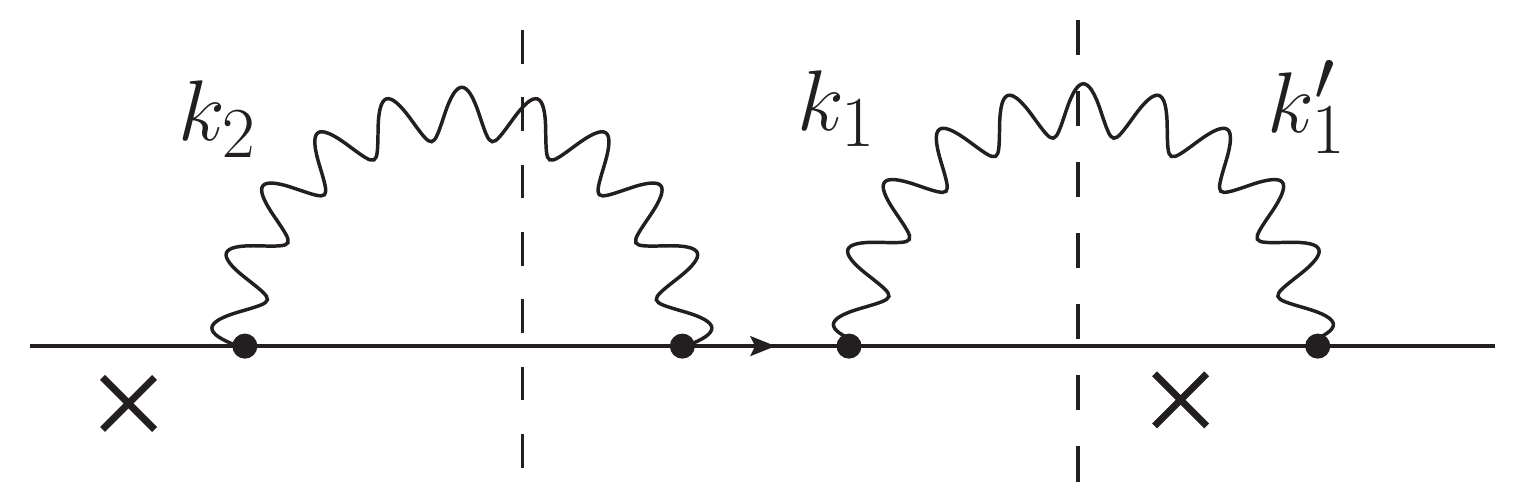}
\\
$\graphr{I}{FFF}$,$\graphr{IF}{FF}$,$\graphr{IFF}{F}$&
$\graphr{F}{IFF}$,$\graphr{FF}{IF}$,$\graphr{FFF}{I}$&
$\graph{I}{FFF}$,$\graph{IFF}{F}$&
$\graph{F}{IFF}$,$\graph{FFF}{I}$
\end{tabular}
\end{center}
\caption{\label{q4ifff}
$p_\perp$-broadening graphs for the case in which
there are 3 $qg$ vertices in the final state.
}
\end{figure}

\begin{figure}
\begin{center}
\begin{tabular}{ccc}
\includegraphics[width=3.5cm]{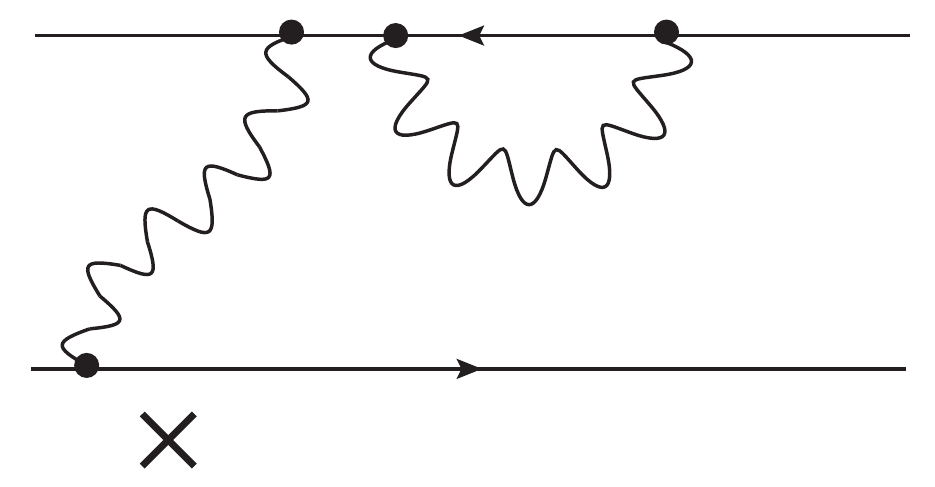}&
\includegraphics[width=3.5cm]{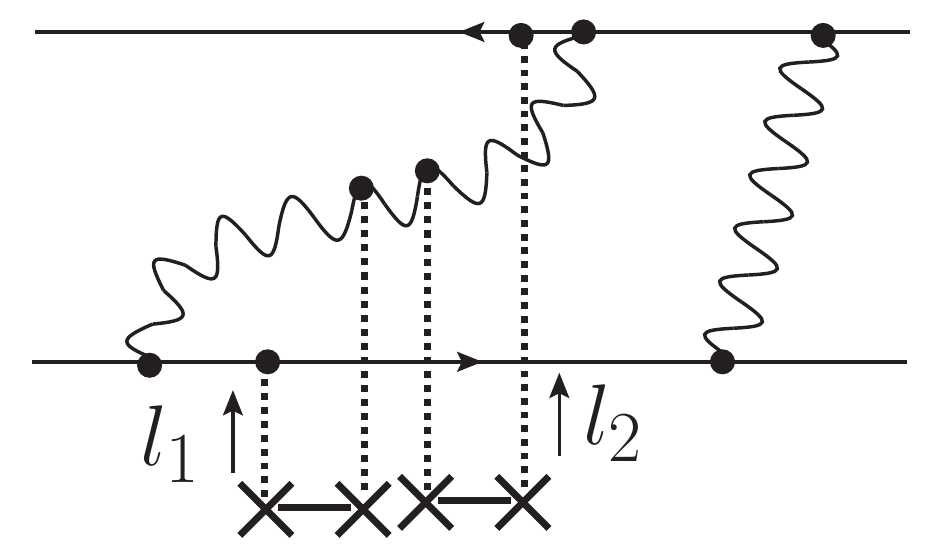}&
\includegraphics[width=3.5cm]{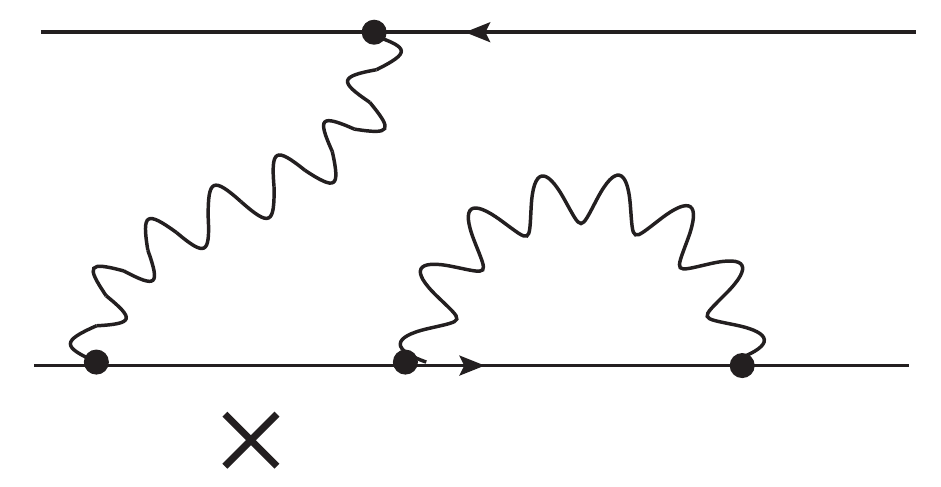}
\end{tabular}
\end{center}
\caption{\label{q4d2r1}
Dipole graphs which are topologically equivalent
to $\graphr{I}{FFF}$, $\graphr{IF}{FF}$
and $\graphr{IFF}{F}$ respectively in Fig.~\ref{q4ifff}
(Similar graphs would correspond to $\graphr{F}{IFF}$,
$\graphr{FF}{IF}$ and $\graphr{FFF}{I}$).
}
\end{figure}

\begin{figure}
\begin{center}
\begin{tabular}{cc}
\includegraphics[width=3.5cm]{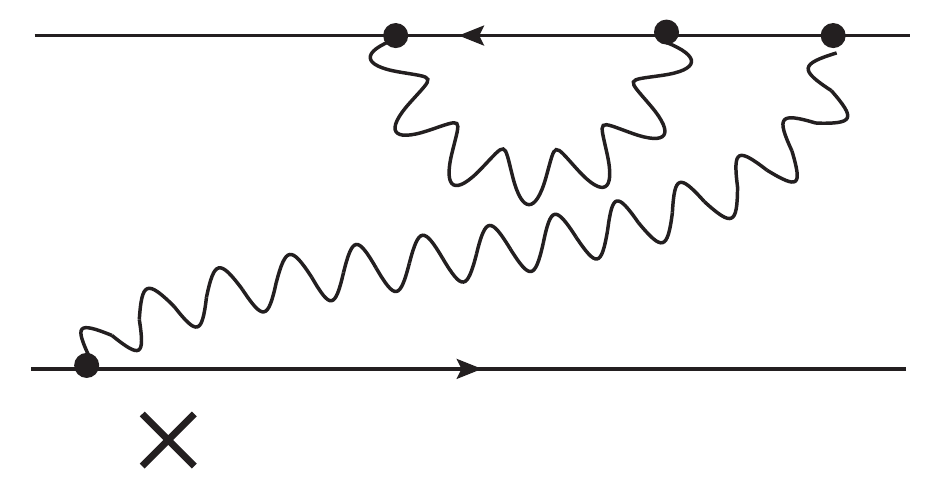}&
\includegraphics[width=3.5cm]{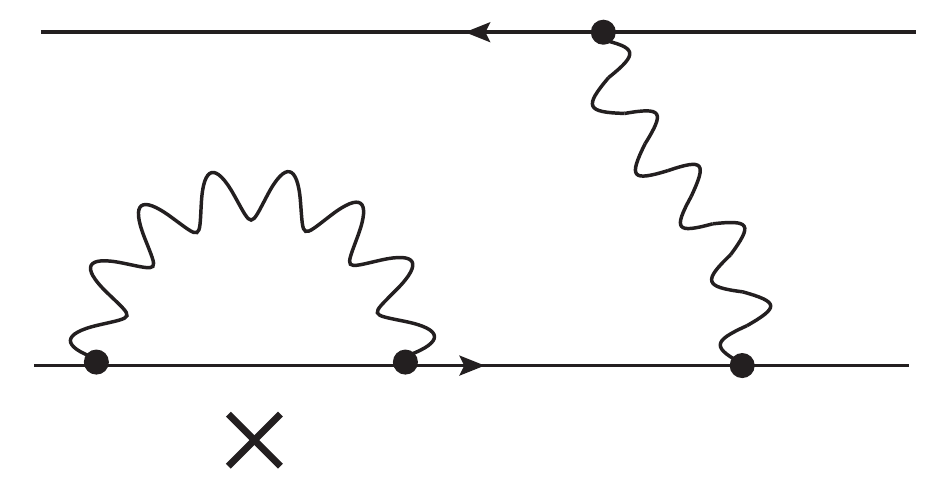}
\end{tabular}
\end{center}
\caption{\label{q4d2r3}
Dipole graphs topologically equivalent to
$\graph{I}{FFF}$ and $\graph{IFF}{F}$
in Fig.~\ref{q4ifff}.
}
\end{figure}

As for the case in which there are 3 $qg$ vertices
in the final state (Fig.~\ref{q4ifff}), one finds 
similar relations with dipole graphs of the type of
those represented in Fig.~\ref{q4d2r1} and Fig.~\ref{q4d2r3}.
We may
for example check that
\be
D_\graphr{IF}{FF}=\bar D_\graphr{II}{IF},\
\ee
which proves the relation between
$\graphr{IF}{FF}$ and $\bar \beta_1$.
As for the graph $\graph{IFF}{F}$,
\be
D_\graph{IFF}{F}=-\frac{1}{E_1^\prime E_2^2(E_2-E_1)}
\ee
which is similar to $D_\graphr{IIF}{I}$
up to signs, most notably in front of $E_1$ in the
denominator.
Analytical continuation
in the $k_{2+}$ variable enables one to identify
$\graph{IFF}{F}$ to $\graphr{IIF}{I}$,
\be
D_\graph{IFF}{F}
\underset{\text{\tiny continuation in $k_{2+}$}}{\longleftrightarrow}
\bar D_\graphr{IIF}{I}\ ,
\ee
and thus also to $\bar\beta_2$ using Eq.~(\ref{eq:relb1b2}).
In the same way,
\be
D_\graphr{IFF}{F}
\underset{\text{\tiny continuation in $k_{2+}$}}{\longleftrightarrow}
\bar D_\graph{IIF}{I}\ ,\ \
D_\graphr{FFF}{I}\longleftrightarrow\bar D_\graph{III}{F}.
\ee
The relation with 
$\bar \beta_3$, $\bar \beta_4$ 
then follows from Eq.~(\ref{eq:reliiifb3}),(\ref{eq:reliifib4}).


\subsubsection{No interference between initial and final states}

\begin{figure}
\begin{center}
\begin{tabular}{c|ccc}
\includegraphics[width=3cm]{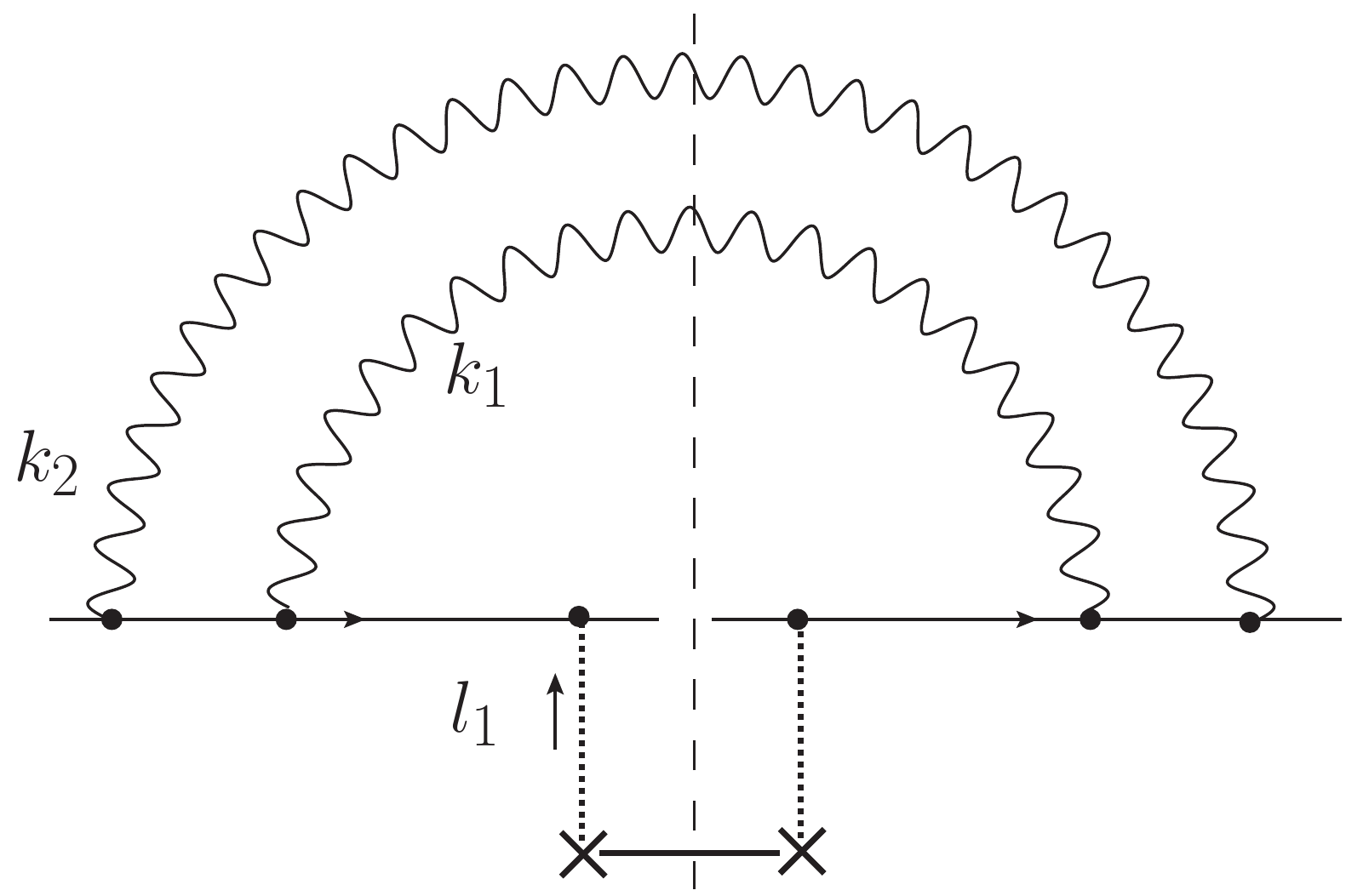}&
\includegraphics[width=3cm]{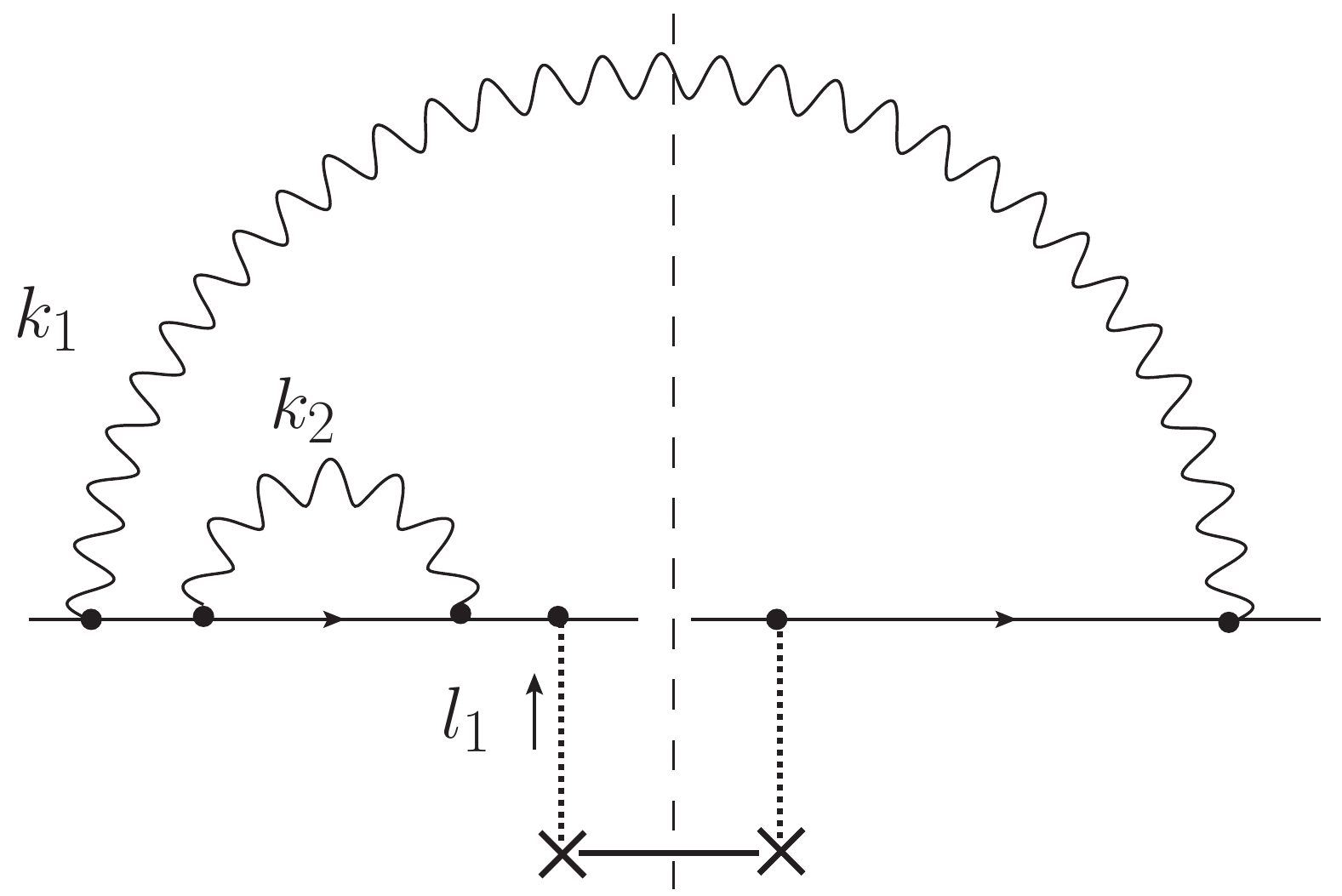}&
\includegraphics[width=3cm]{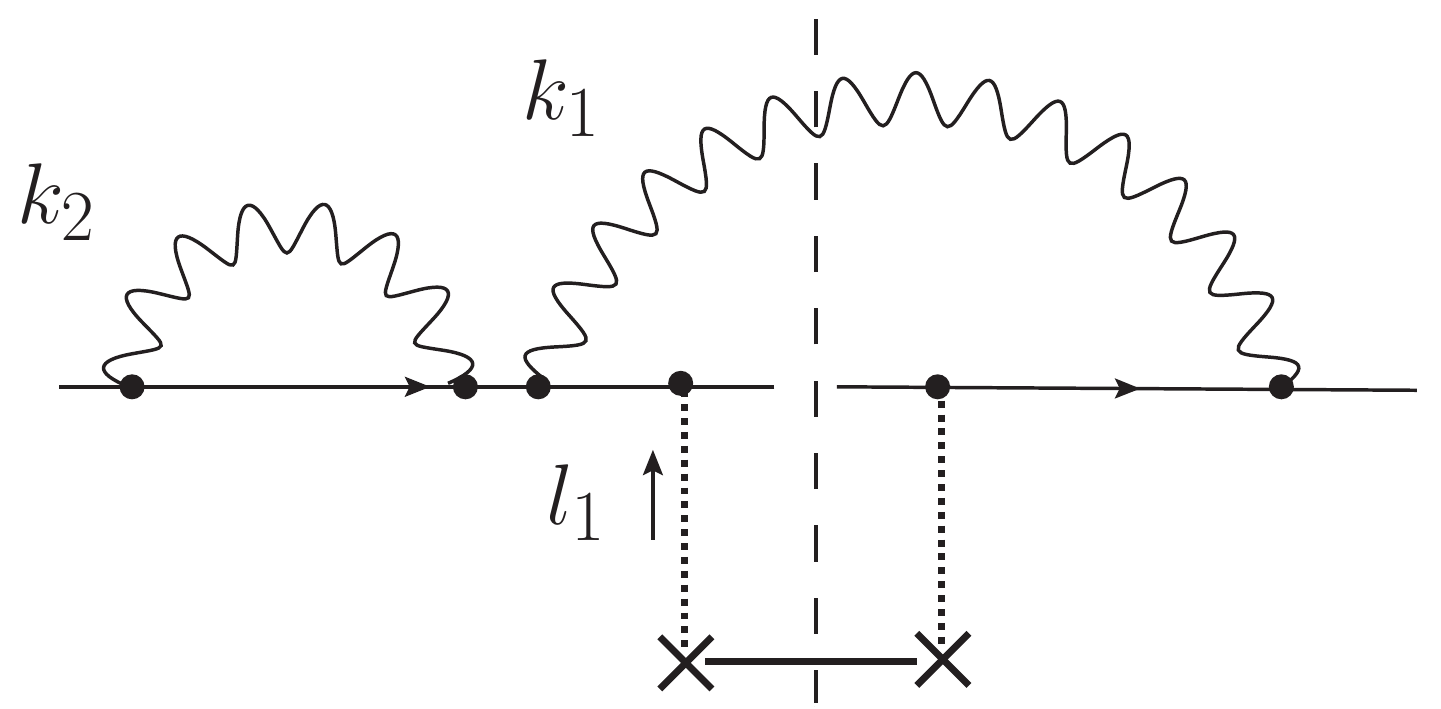}&
\includegraphics[width=3cm]{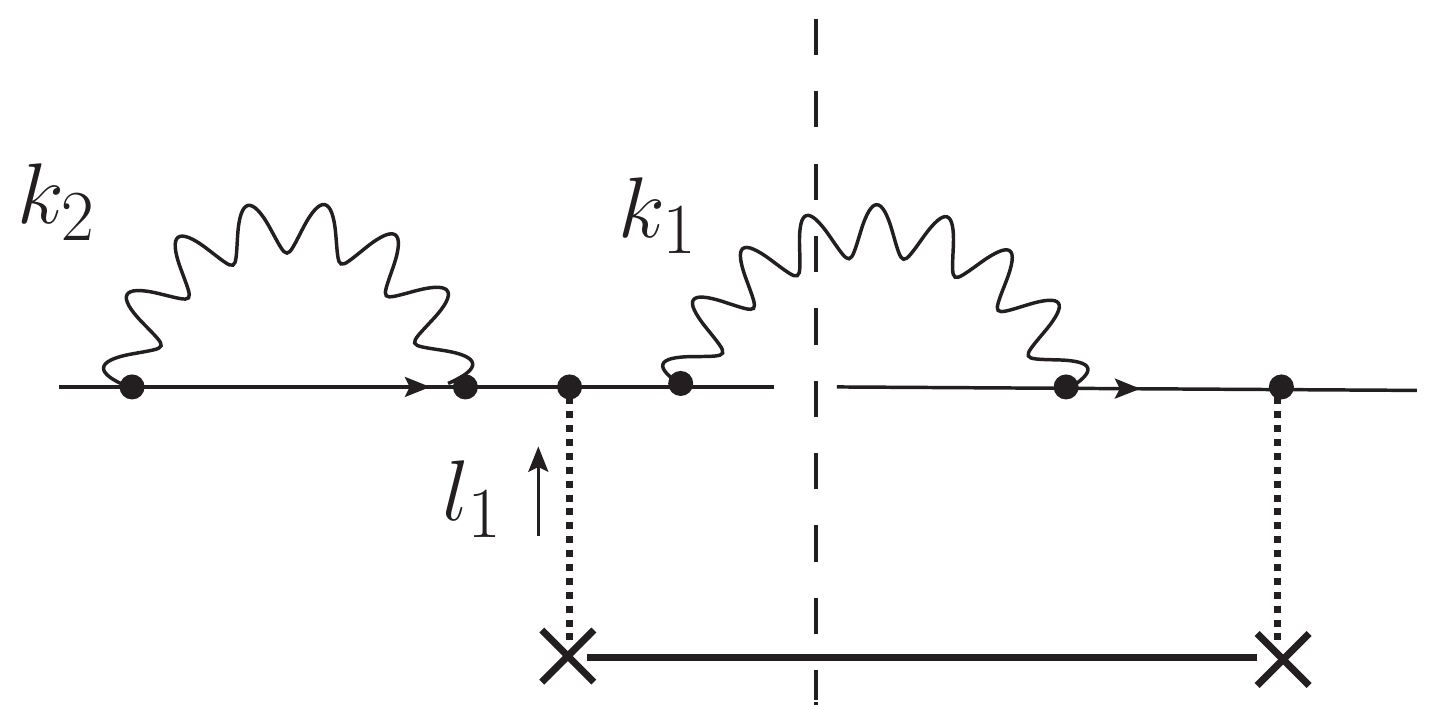}\\
$\graphr{II}{II}$ & $\graphr{III}{I}$ & $\graph{III}{I}$
& $\graph{IIF}{F}$
\end{tabular}
\end{center}
\caption{\label{q4iiii}
Graphs in which the gluons never connect $qg$ vertices in the initial
state to vertices in the final state.
It turns out that the 3 rightmost graphs have to be considered
together for the correspondence with dipoles to work.
}
\end{figure}

\begin{figure}
\begin{center}
\begin{tabular}{c|ccc}
\includegraphics[width=3.cm]{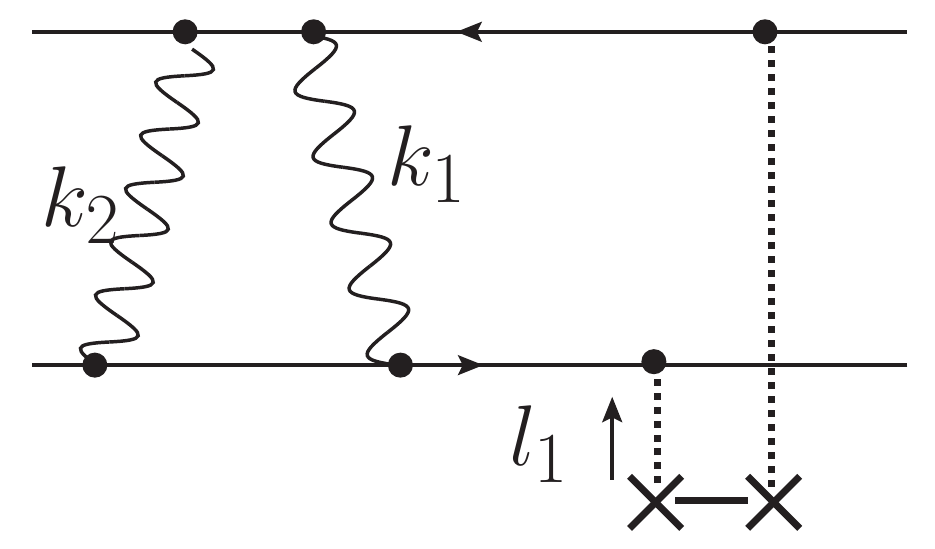}&
\includegraphics[width=3.cm]{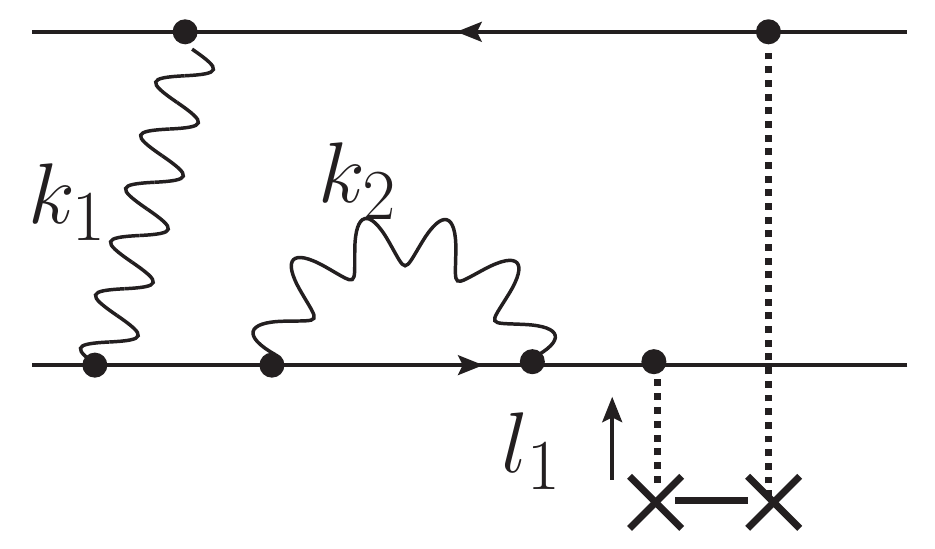}&
\includegraphics[width=3.cm]{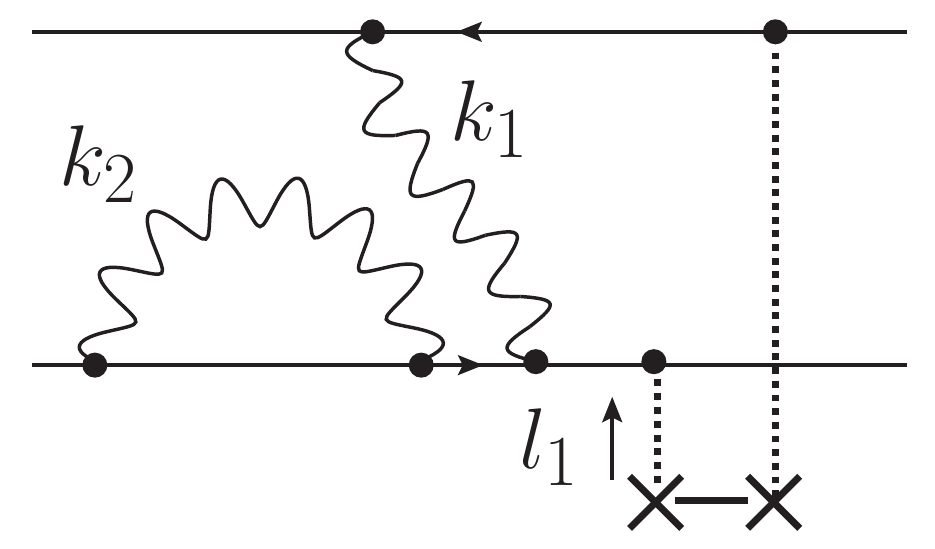}&
\includegraphics[width=3.4cm]{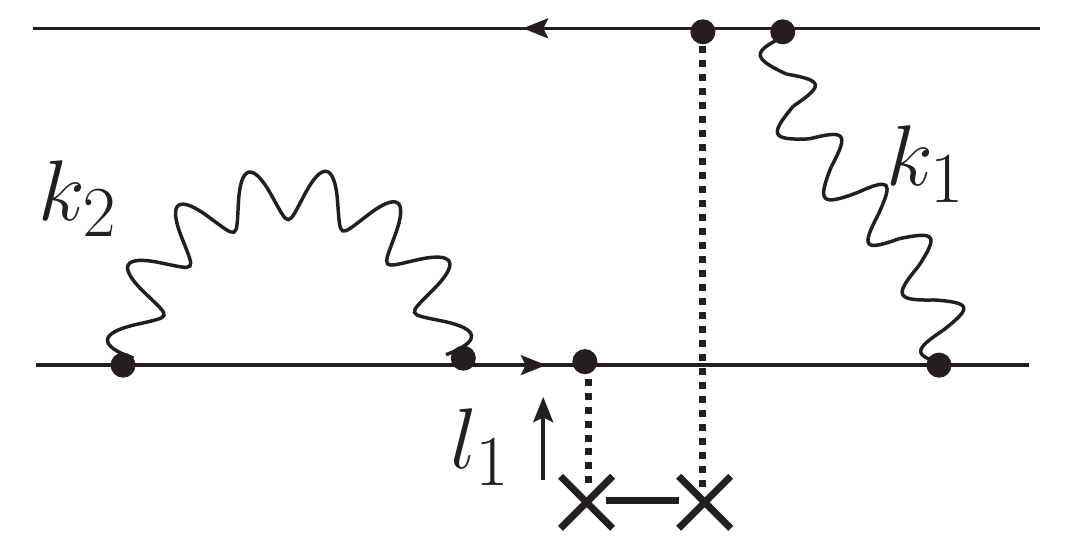}\\
$\alpha_1$&$\alpha_2$&$\alpha_3$&$\alpha_4$
\end{tabular}
\end{center}
\caption{\label{q4dv}
Dipoles graphs that correspond to the $p_\perp$-broadening graphs
of Fig.~\ref{q4iiii}.
As usual, this set has to be supplemented by graphs exhausting
all possible
orderings of the vertices, including instantaneous exchanges.
The leftmost graph corresponds to the leftmost graph in Fig.~\ref{q4iiii}
while the set of the 3 rightmost graphs 
corresponds to the set of the
3 rightmost graphs in Fig.~\ref{q4iiii}.
}
\end{figure}

Finally, we consider the case in which there is no gluon linking 
an initial-state vertex to a final-state vertex.
The simplest of these graphs is the leftmost graph in 
Fig.~\ref{q4iiii}, for which a graph-to-graph
identification with the leftmost graph in Fig.~\ref{q4dv}
holds (provided one includes all possible time
orderings of the $qg/\bar q g$ vertices, which implies to take
into account also instantaneous exchanges in Fig.~\ref{q4dv}).
The energy denominators read
\be
D_\graphr{II}{II}=\frac{1}{E_2^2(E_1+E_2)^2}=\sum D_{\alpha_1}.
\ee

We now turn to a more tricky case, namely the 3 rightmost graphs
in Fig.~\ref{q4iiii}.
These 3 graphs correspond to the 3 rightmost graphs in Fig.~\ref{q4dv}.
Indeed, the sum of the energy denominators reads
\be
\begin{split}
\text{Re}\left(D_\graphr{III}{I}+D_\graph{III}{I}+D_\graph{IIF}{F}\right)
&=-\frac{1}{E_1^2 E_2^2}-\frac{1}{E_1^2 E_2(E_1+E_2)}\\
&=-\text{Re}\sum \left(D_{\alpha_2}+D_{\alpha_3}+D_{\alpha_4}
\right).
\end{split}
\ee
It may look surprising that
there is no graph-by-graph correspondence in this case,
and especially that one needs to consider together purely
initial-state graphs such as $\graphr{III}{I}$
and a graph where one gluon is in the initial state and
one in the final state ($\graphr{IIF}{F}$).
However, we checked that
not including the latter would lead to a mismatch
between $p_\perp$-broadening and dipoles by a term $1/(2 E_1^3 E_2)$.

\begin{figure}
\begin{center}
\begin{tabular}{ccc}
\includegraphics[width=4cm]{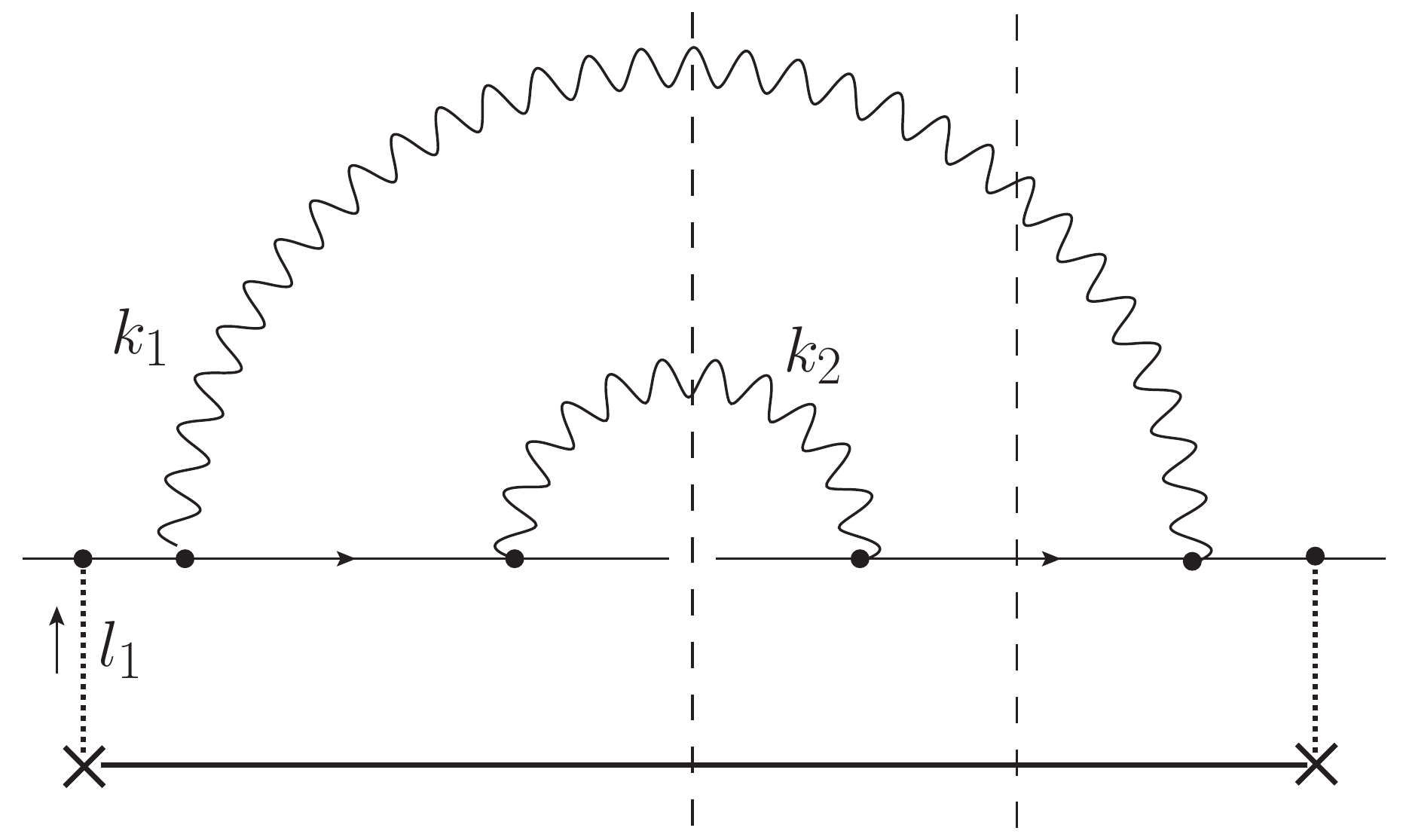}&
\includegraphics[width=4cm]{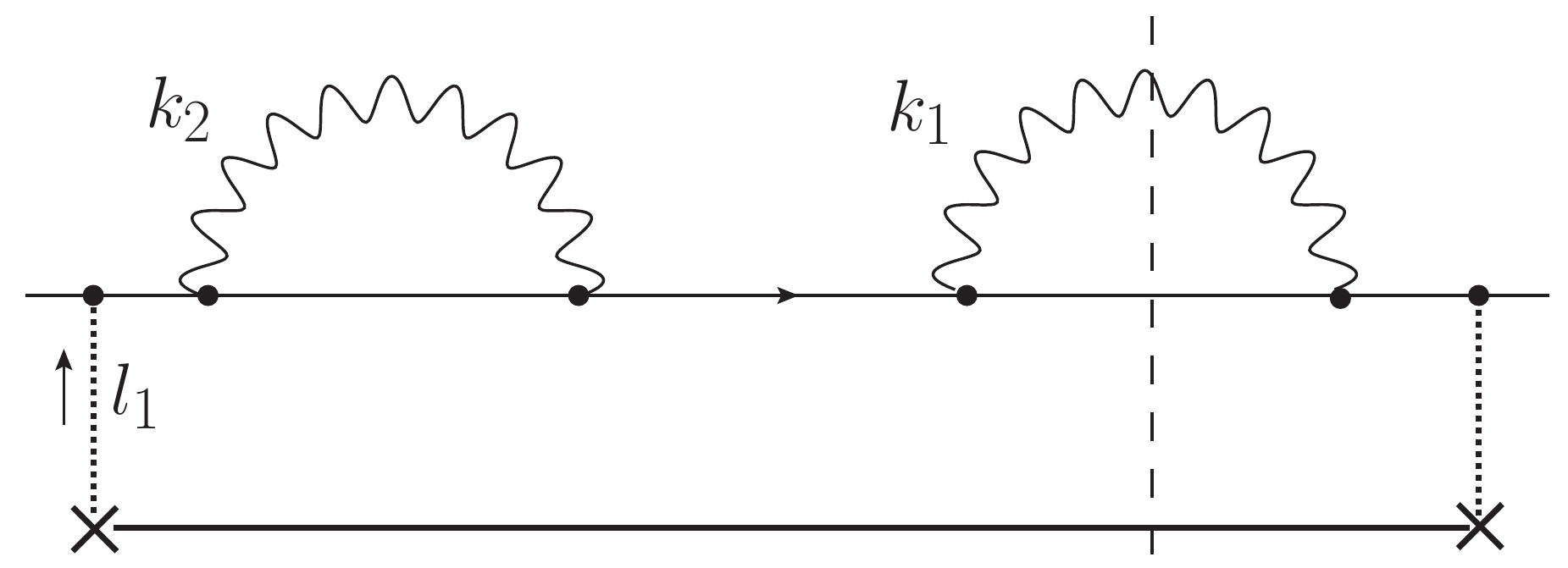}&
\includegraphics[width=4cm]{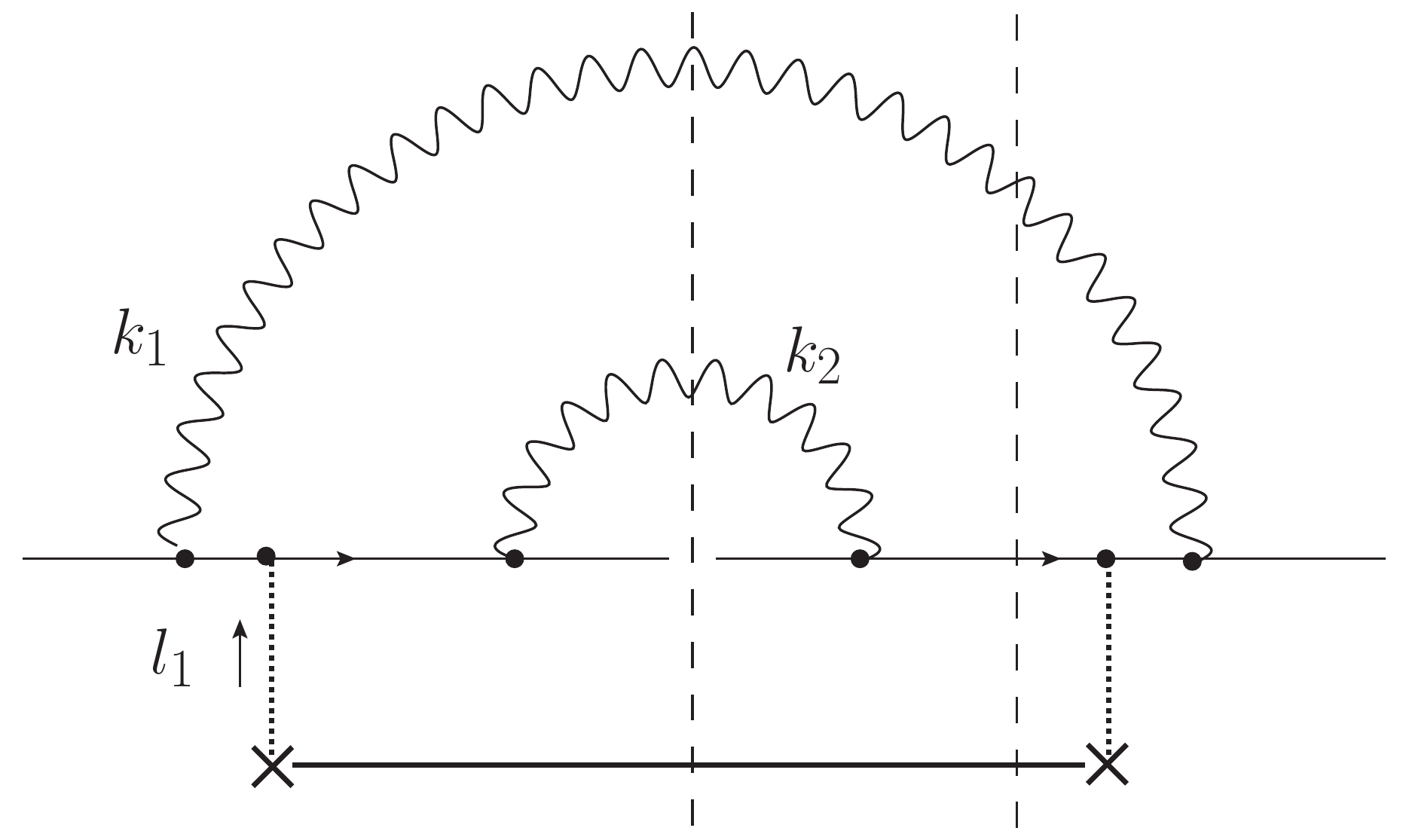}\\
$\graphr{FF}{FF}$,$\graphr{FFF}{F}$&
$\graph{FFF}{F}$&
$\graphr{IF}{IF}$,$\graphr{IFF}{I}$
\end{tabular}
\end{center}
\caption{\label{q4ffff}
Another set of graphs
similar to  Fig.~\ref{q4iiii}
in correspondence with the dipole graphs in Fig.~\ref{q4dv}
or with similar graphs.
One would also have the graphs in which the cut is on the left,
symmetric to the rightmost cut.
}
\end{figure}

We turn to the last class of graphs shown in Fig.~\ref{q4ffff}.
This case is solved by noting that these graphs may be mapped
to the ones shown in Fig.~\ref{q4iiii}, up to an analytical
continuation in the $k_{2+}$ variable.
We check the identities
\be
D_\graphr{FF}{FF}=D_\graphr{II}{II}\ ,\ \
D_\graphr{FFF}{F}
+D_\graph{FFF}{F}
+D_\graphr{IFF}{I}
\underset{\text{\tiny 
continuation in $k_{2+}$}}{\longleftrightarrow}
D_\graph{III}{I}
+D_\graphr{III}{I}
+D_\graph{IIF}{F}
\ee
and hence the corresponding identities with the
(sets of) dipole graphs. Other graphs 
which we do not review here
may be deduced from
these ones by simple symmetries.

We have left out of the discussion a few graphs
which are in exact one-to-one correspondence between
the two processes.
Up to these cases not discussed explicitly but trivial, 
the proof that the evolution with the energy
of $p_\perp$-broadening cross sections and of
dipole forward amplitudes is identical at
next-to-leading order
is now complete.


\section{\label{sec5}Extension to two-jet versus quadrupole amplitudes}

The specific process on which we have focussed so far,
namely $p_\perp$-broadening, is the simplest in the class of
production processes. However, it is not the easiest
to measure experimentally. Therefore, we wish to extend
our analysis to other production processes.

It has been shown 
that quadrupole structures appear 
in the computation of
observables which involve two particles (dijets) in the final state
\cite{JalilianMarian:2004da,Dominguez:2010xd,Dominguez:2011wm}.
The energy evolution of quadrupoles can be related to the 
BFKL evolution with saturation, see Ref.~\cite{Dominguez:2011gc}
(approximation schemes were worked out
in Ref.~\cite{Iancu:2011ns,Iancu:2011nj}).
The question is whether the dijet/quadrupole correspondence
is true when quantum evolution is included.

The extension of our discussion of the evolution
of $p_\perp$-broadening/dipole amplitudes
to the evolution of dijet/quadrupole amplitudes
is straightforward once
one notices that the fermion vertices and energies
entering the energy denominators
never play any r\^ole. 
Indeed, the former are represented by a mere
overall factor, which is identical for all graphs of a given class, 
and the latter do not even appear in the calculation,
thanks to the soft approximation.

\begin{figure}
\begin{center}
\begin{tabular}{cc}
\includegraphics[width=5cm]{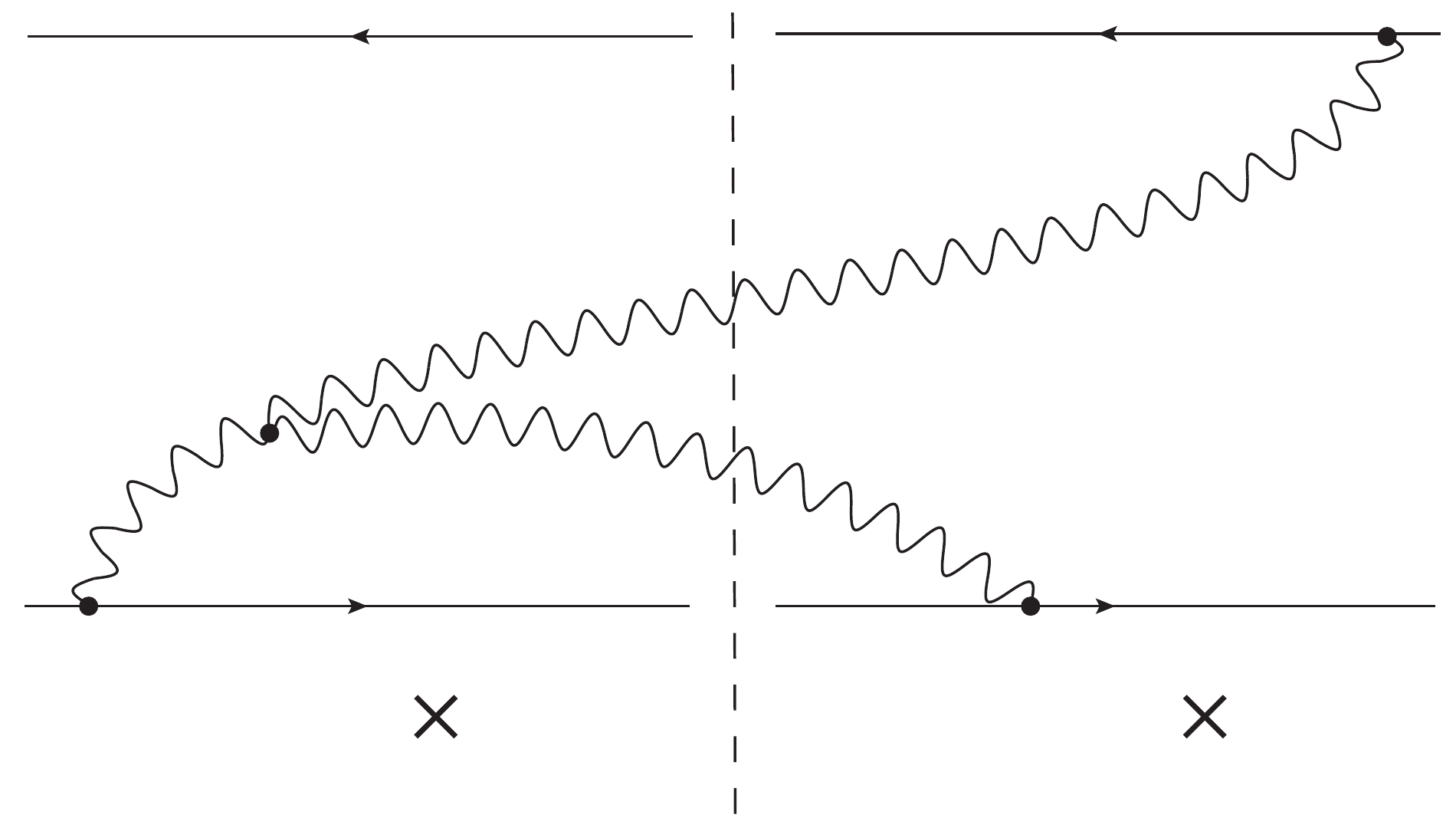}&
\includegraphics[width=5cm]{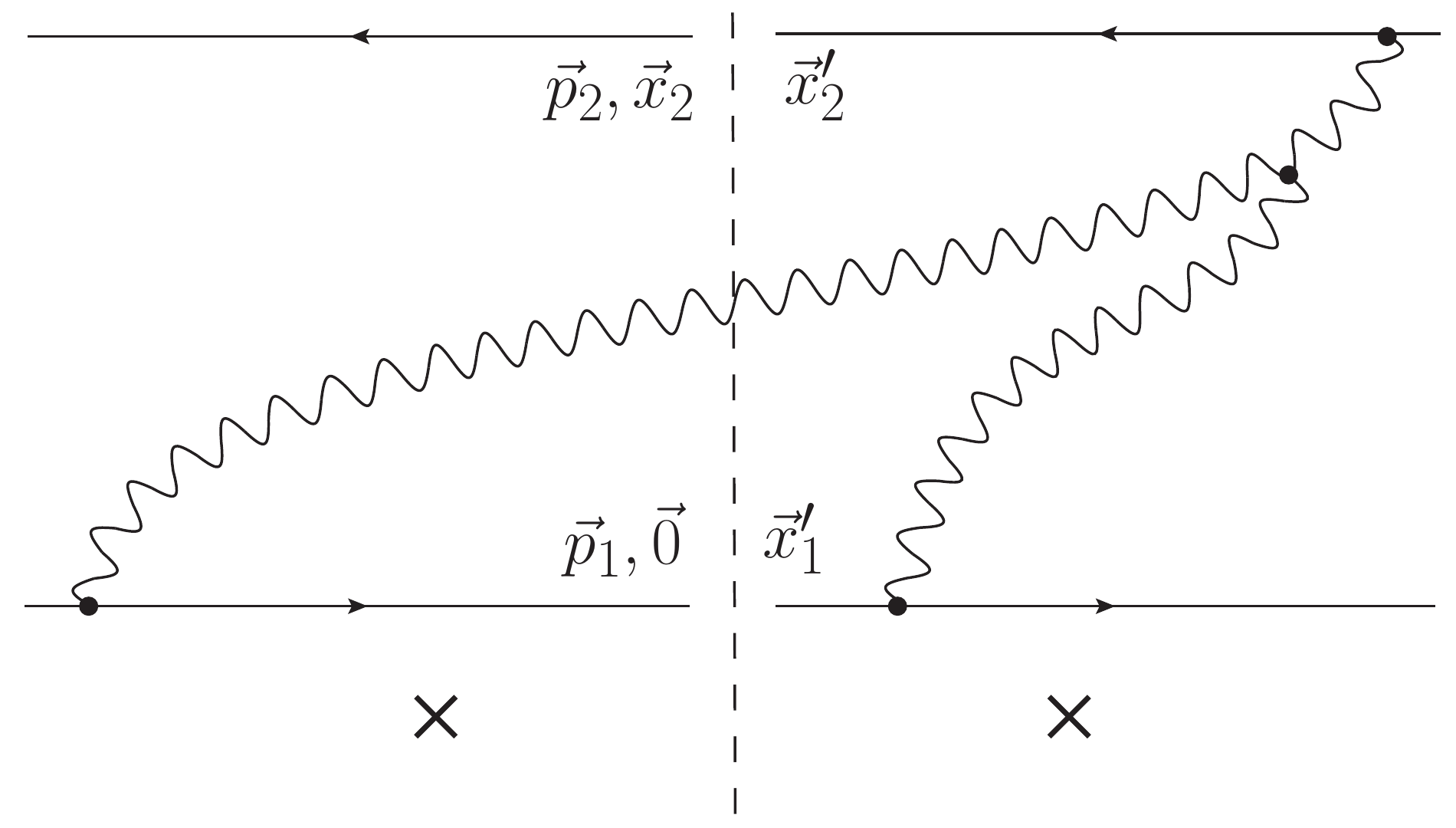}\\
$A$ & $B$
\end{tabular}
\end{center}
\caption{\label{dijet}
Two graphs contributing to the evolution of the
dijet cross section.
The incoming object is a $q\bar q$ dipole.
}
\end{figure}

\begin{figure}
\begin{center}
\begin{tabular}{ccc}
\includegraphics[height=2.5cm]{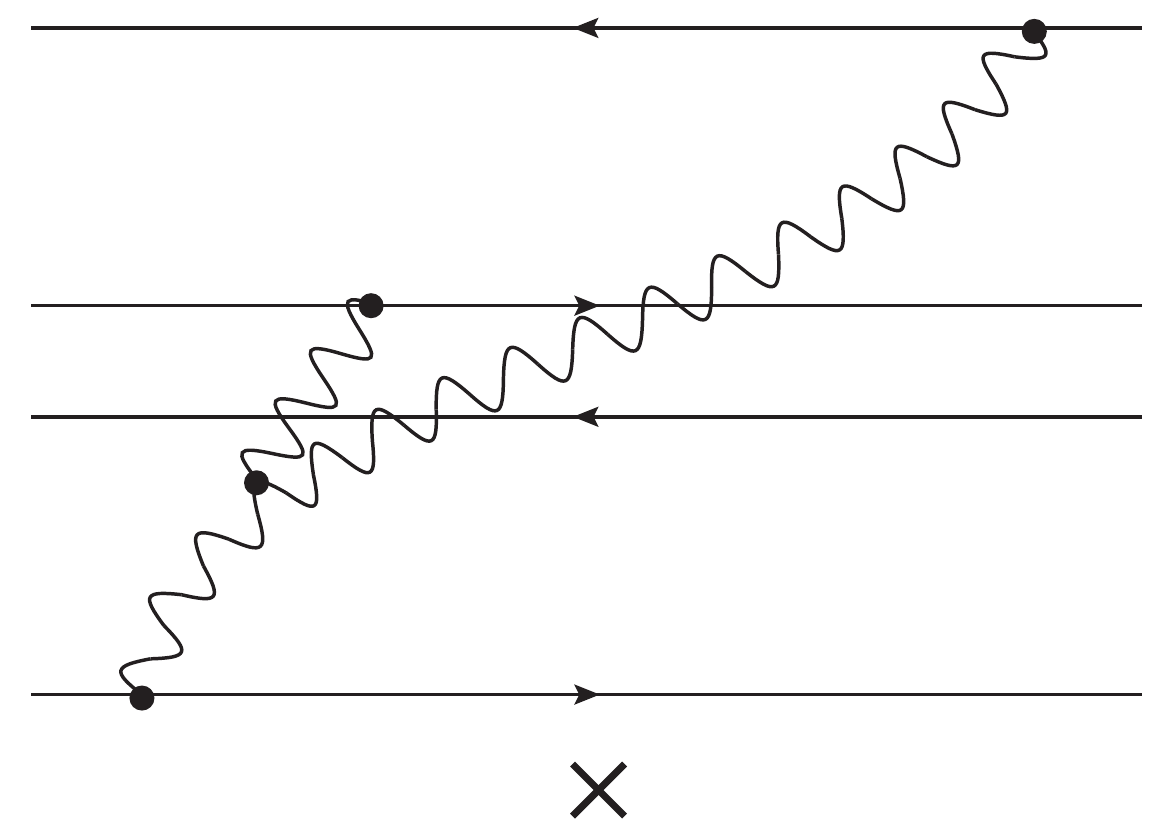}&
\includegraphics[height=2.5cm]{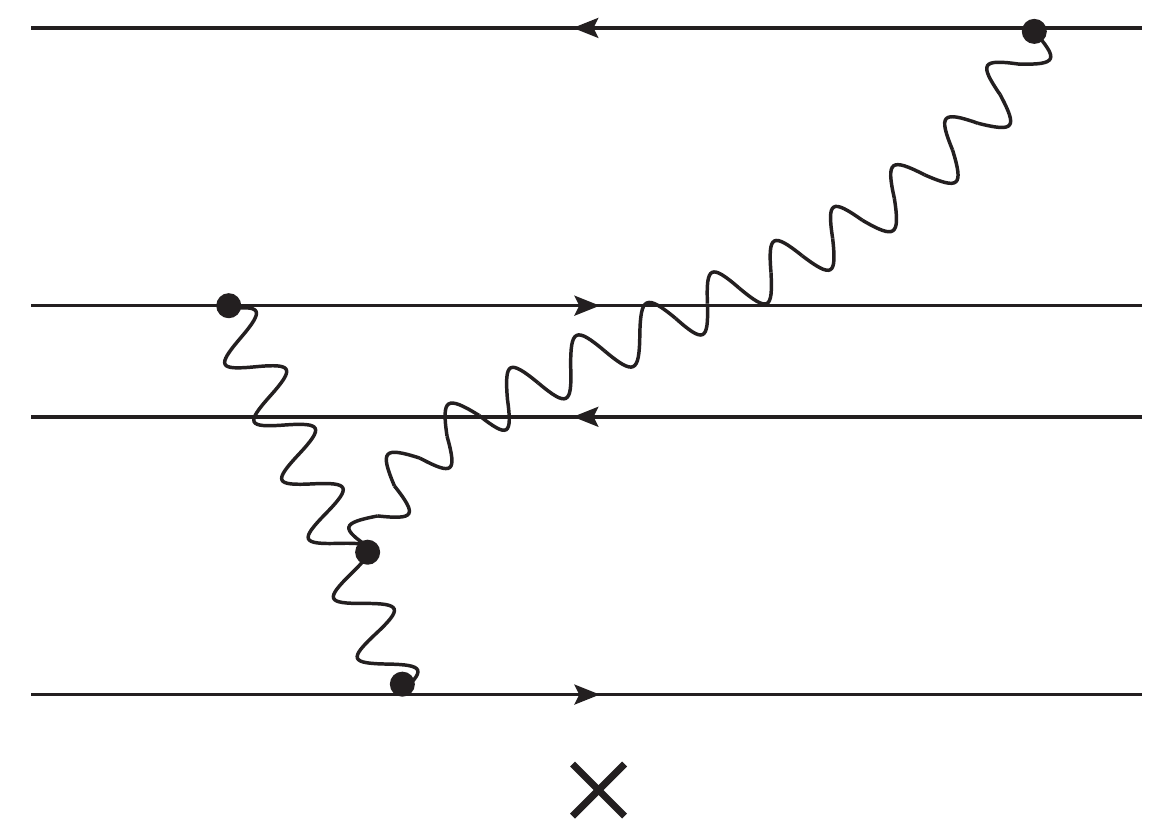}&
\includegraphics[height=2.5cm]{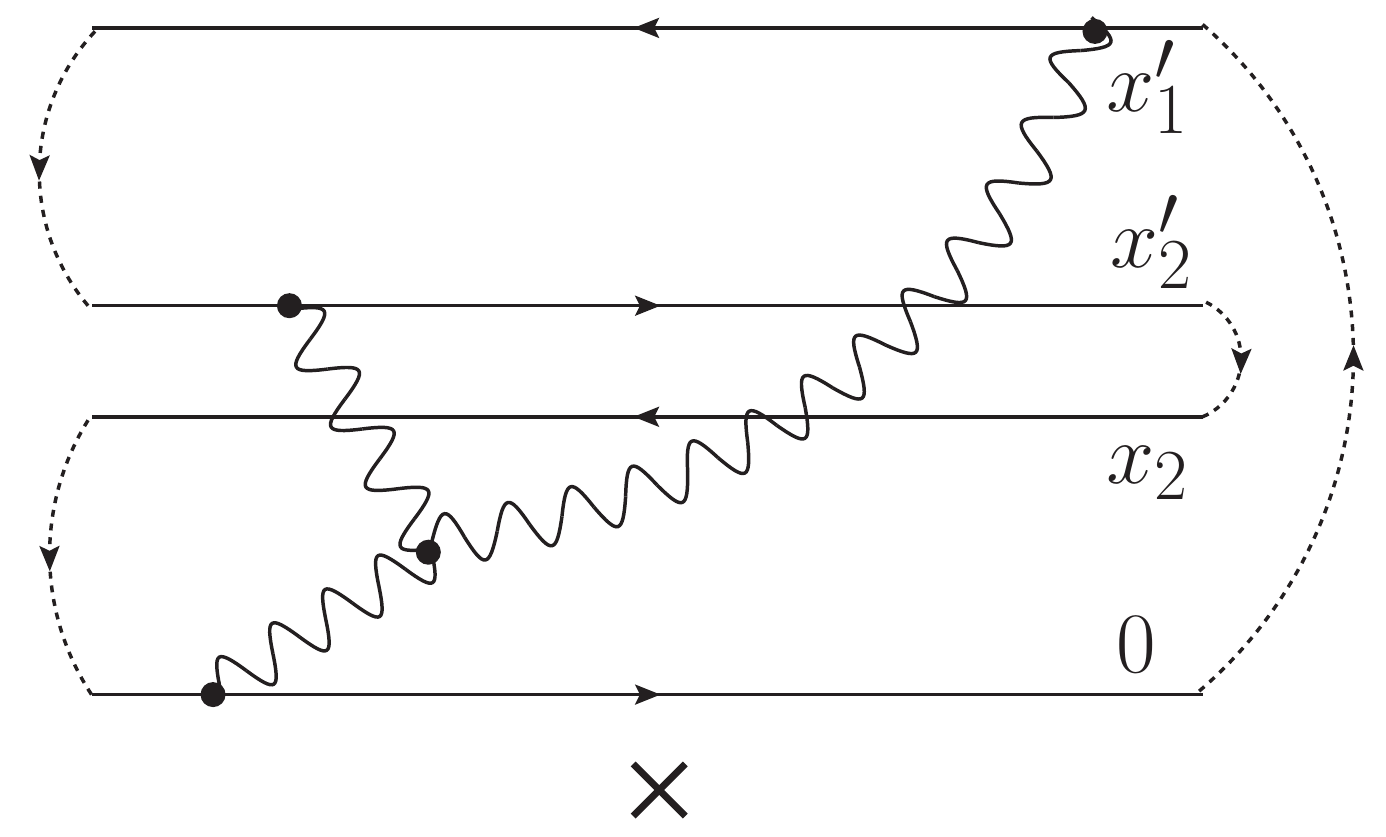}\\
$\alpha$&$\beta$&$\gamma$
\end{tabular}
\end{center}
\caption{\label{quadrupole}
Quadrupole graphs corresponding to the dijet graphs of
Fig.~\ref{dijet}. On the rightmost graph,
we have shown how the 
color lines close to form the
quadrupole.
}
\end{figure}

We illustrate the correspondence between two particular sets
of graphs in Fig.~\ref{dijet} and~\ref{quadrupole},
which are similar to Fig.~\ref{q3} and~\ref{q3d} respectively.
The relevant part of the dijet diagrams may be modeled by
an incoming quark-antiquark dipole which subsequently
evolves by gluon emission (Fig.~\ref{dijet}).

Let us denote by $\psi$ the wave function of the dipole,
and $S_4$ the $S$-matrix element for the scattering of the
quadrupole.
Equation~(\ref{eq:correspondence}) is to be replaced by the
following formula (see Fig.~\ref{dijet},\ref{quadrupole}
for the notations):
\begin{multline}
\frac{dN}{d^2 p_1 d^2 p_2}
=\int \frac{d^2x^\prime_1 d^2x_2 d^2x^\prime_2}{(2\pi)^2}
\psi(x_2)\bar\psi(x^\prime_2-x_1^\prime)
e^{i\vec p_1\cdot \vec x^\prime_1-i\vec p_2\cdot (\vec x_2-\vec x^\prime_2)}\\
\times
S_4(0,x_2,x^\prime_2,x^\prime_1),
\label{eq:dijetquad}
\end{multline}
which now holds at next-to-leading order in the logarithms
of the energy.




\section{\label{sec6}Conclusion}

We have shown that the identity~(\ref{eq:correspondence})
between the amplitude for $p_\perp$-broadening and the
dipole cross section holds at next-to-leading order level.
The relation may be extended to dijet cross sections
versus quadrupole amplitudes (Eq.~(\ref{eq:dijetquad})), 
and probably to other production
processes.
In other words, we have proved that the next-to-leading order
BFKL equation describes the energy evolution of these
production processes.

Our demonstration was based on a systematic (and quite awkward)
inspection of
all lightcone perturbation theory 
graphs which contribute to these respective processes.
We have however avoided the full computation of the graphs since our
goal was to prove the correspondence in the most general possible
way.

The way the matching occurs turns out to be very subtle.
In order to see it,
we grouped the graphs according to their topologies (number
of vertices of a given type, chronology of the different
interactions) on the $p_\perp$-broadening side and on the dipole
side. The time integrations for the dipole
graphs are more constrained than for the $p_\perp$-broadening graphs
(there are 4 independent integration regions in the latter case,
and only 2 in the former case; see the discussion of the
Keldysh-Schwinger formalism in Sec.~\ref{sec2}). This first apparent mismatch
is generally solved by subtle cancellations between graphs,
and by considering appropriate ensembles of graphs
(including, in particular, contact interactions
where gluons are exchanged instantaneously,
which turn out to play a crucial r\^ole). In some cases however,
irreducible differences in the 
flow of longitudinal momenta result in differences in
the energy denominators that can be resolved only by analytical 
continuation in the relevant longitudinal momentum variable.
We found a number of pecularities: 
For example, keeping the consistency of the regularization
(namely the adiabatic $\varepsilon$ parameter,
see Eq.~(\ref{eq:D})) throughout the calculations
turned out to be
surprisingly important.

Despite our efforts, we have not found
a simpler argument to explain the correspondence more systematically.
Such an argument would be crucial for example 
to be able to understand
whether the correspondence is still true at next-to-next-to-leading
order and beyond.

A full next-to-leading order calculation
of production processes
requires
also to take into account the next-to-leading order
corrections to the parton distribution functions and to the fragmentation
functions.
In particular, one has to release the eikonal approximation at the
quark-gluon vertices, which we have not done here. 
Progress in this direction has recently been reported, see
Ref.~\cite{Chirilli:2011km,Chirilli:2012jd}:
In these respects, our work may be seen as complementary
to the latter studies.


\section*{Acknowledgements}

This work was partly funded by 
the US Department of Energy,
by the ``consortium physique des deux infinis'' (P2I, France),
and by the ``LABEX physique des deux infinis et des origines'' 
(P2IO, France).
We thank Prof. Yuri Kovchegov for helpful discussions
and useful comments on the manuscript,
as well as Dr. Fabio Dominguez, 
and Dr. Bowen Xiao.
The Feynman diagrams were drawn using Jaxodraw 
\cite{Binosi:2003yf,Binosi:2008ig}.


\end{document}